\documentclass[oneside]{usiinfthesis}

\usepackage{lipsum}
\usepackage[draft]{fixme}
\usepackage{tikz}
\usetikzlibrary{automata}
\usepackage{xcolor}

\newcommand{\circled}[1]{\tikz[baseline=(myanchor.base)] \node[circle,fill=.,inner sep=1pt] (myanchor) {\color{-.}\bfseries\footnotesize #1};}

\newcommand{\squared}[1]{%
	\tikz[baseline=(myanchor.base)] 
	\node[rectangle,fill={rgb,255:red,255;green,175;blue,175},inner sep=2pt,minimum size=10pt,align=center] (myanchor) {\color{black}\bfseries\footnotesize #1};}

\usepackage{todonotes}
\setuptodonotes{inline}

\usepackage{multirow}
\usepackage{multicol}
\usepackage{array,colortbl}
\usepackage{diagbox}
\usepackage{booktabs}
\usepackage{siunitx}
\usepackage{stfloats}

\usepackage{threeparttable, tablefootnote}
\usepackage{makecell}
\usepackage{amsmath}
\usepackage{bm}
\usepackage{arydshln}
\usepackage{pifont}
\usepackage{marvosym}
\usepackage{soul}
\usepackage{cleveref}
\usepackage{enumitem}
\usepackage[normalem]{ulem} 

\RequirePackage{scrlfile}

\usepackage{algorithmicx}
\usepackage{algpseudocode}
\usepackage{listings}

\usepackage[section]{placeins} 

\usepackage{float}  
\usepackage{subcaption}
\newcommand\mycaption[2]{\caption[#1]{#1 \newline\small{#2}}}
\newcommand\mycaptiontitle[1]{\caption[#1]{#1}}

\newcommand\mycaptionoflisting[1]{\captionof{lstlisting}[#1]{#1}}

\lstdefinelanguage{algebra}
{morekeywords={import,sort,constructors,observers,transformers,axioms,if,
		else,end},
	sensitive=false,
	morecomment=[l]{//s},
}

\usepackage[skipabove=4mm, skipbelow=4mm]{mdframed}
\usepackage{thmtools}
\definecolor{border}{RGB}{82,86,89}
\definecolor{background}{RGB}{250,250,250}
\mdfdefinestyle{custom-box}{
	roundcorner=50pt,
	linewidth=1pt,
	linecolor=border,
	backgroundcolor=background,
	leftmargin=0pt,
	rightmargin=0pt,
	innerleftmargin=1em,
	innerrightmargin=1.5em,
	innertopmargin=1em,
	innerbottommargin=1em,
	font=\small,
}

\mdfdefinestyle{ndef-frame}{
	linewidth=1pt, %
	linecolor= gray, %
	backgroundcolor= gray!10,
	topline=false, %
	bottomline=false, %
	rightline=false,%
	leftmargin=0pt, %
	innerleftmargin=1em, %
	innerrightmargin=1.5em, 
	rightmargin=0pt, %
	innertopmargin=.8em,%
	innerbottommargin=.8em, %
	font=\small,
} 

\usepackage[most]{tcolorbox}
\newtcolorbox{custombox}{
	colback=background,
	colframe=border,
	arc=5pt, 
	boxrule=1pt,
	left=1em,
	right=1.5em,
	top=1em,
	bottom=1em,
	fontupper=\small,
}

\newtcolorbox{liberatoria}{
	breakable,
	colframe=black,
	colback=white,
	boxrule=0.5pt,
	leftrule=.5pt,
	toprule=.5pt,
	rightrule=.5pt,
	bottomrule=.5pt,
	boxsep=5pt,
	left=10pt,
	right=10pt,
	top=10pt,
	bottom=10pt,
	fontupper=\small,
	sharp corners,
	width=\textwidth,
	before={\hypersetup{colorlinks=true, urlcolor=blue, linkcolor=black}} 
}

\usepackage{xurl}
\PassOptionsToPackage{hyphens}{url}

\usepackage{pdfpages}

\usepackage{highlight}

\usepackage{silence}
\usepackage{mathtools} 
\usepackage{tikz}
\usetikzlibrary{matrix}

\usetikzlibrary{arrows,plotmarks,decorations.markings,trees,shapes}
\tikzstyle{n}=[ellipse,draw=black!100,fill=black!10,line width=.7pt,minimum width=1cm,align=center,minimum height=.5cm]
\tikzstyle{nodo}=[ellipse,draw=black!100,fill=black!30,line width=.7pt,minimum width=1.2cm,text width=3cm,align=center,minimum height=.5cm]
\tikzstyle{Qnodo}=[ellipse,draw=black!100,fill=black!10,line width=.7pt,minimum width=1.2cm,text width=3cm,align=center,minimum height=.5cm]
\tikzstyle{arco}=[draw=black!80,line width=1pt, postaction={decorate}, decoration={markings,mark=at position 1.0 with {\arrow[ draw=black!80,line width=.7pt]{>}}}]
\tikzstyle{decision} = [rectangle, draw, fill=black!100,text=white, text width=4.5em, text badly centered, node distance=3cm, minimum height=3em]
\tikzstyle{block} = [rectangle, draw, fill=blue!20, text width=5em, text centered, rounded corners, minimum height=3em]
\tikzstyle{line} = [draw, -latex']
\tikzstyle{cloud} = [draw, ellipse,fill=red!20, node distance=3cm, minimum height=2em]

\usepackage{tocbibind}
\usepackage{acro}
\usepackage{longtable,supertabular,tabularray}
\DeclareAcronym{AT}{short={AT}, long={Algorithmic Thinking}}
\DeclareAcronym{CT}{short={CT}, long={Computational Thinking}}
\DeclareAcronym{CS}{short={CS}, long={Computer Science}}
\DeclareAcronym{CT-cube}{short={CT-cube}, long={Computational Thinking cube}}
\DeclareAcronym{CAT}{short={CAT}, long={Cross Array Task}}
\DeclareAcronym{CTP}{short=CTP, long=Computational Thinking Problem}
\DeclareAcronym{BCTt}{short=BCTt, long=Beginner Computational Thinking test}
\DeclareAcronym{CTt}{short=CTt, long=Computational Thinking test}
\DeclareAcronym{cCTt}{short=cCTt, long=competent Computational Thinking test}
\DeclareAcronym{BPAt}{short=BPAt, long=Basic Programming Abilities test}
\DeclareAcronym{R2T2}{short=R2T2, long=Remote Rescue with Thymio II}
\DeclareAcronym{STEM}{short=STEM, long={Science, Technology, Engineering and Math}}
\DeclareAcronym{RL}{short=RL, long={Reinforcement Learning}}
\DeclareAcronym{ITAS}{short=ITAS, long={Intelligent Tutoring and Assessment System}}
\DeclareAcronym{IAS}{short=IAS, long={Intelligent Assessment System}}
\DeclareAcronym{PPO}{short=PPO, long={Proximal Policy Optimization}}
\DeclareAcronym{UX}{short=UX, long={User Experience}}
\DeclareAcronym{HCI}{short=HCI, long={Human-Computer Interaction}}
\DeclareAcronym{ANOVA}{short=ANOVA, long={Analysis of Variance}}
\DeclareAcronym{EMMs}{short=EMMs, long={Estimated Marginal Means}}
\DeclareAcronym{TE}{short=T\&E, long={Trial and Error}}
\DeclareAcronym{OLS}{short=OLS, long={Ordinary Least Squares}}
\DeclareAcronym{REML}{short=REML, long={Restricted Maximum Likelihood}}
\DeclareAcronym{LRT}{short=LRT, long={Likelihood Ratio Test}}
\DeclareAcronym{HG}{short=HG, long={HarmoS grade}}
\DeclareAcronym{BH}{short=BH, long={Benjamini-Hochberg}}
\DeclareAcronym{HSD}{short=HSD, long={Honestly Significant Difference}}
\DeclareAcronym{IRT}{short=IRT, long={Item Response Theory}}
\DeclareAcronym{BN}{short=BN, long={Bayesian Network}}
\DeclareAcronym{BKT}{short=BKT, long={Bayesian Knowledge Tracing}}
\DeclareAcronym{CPT}{short=CPT, long={Conditional Probability Table}}
\DeclareAcronym{HMM}{short=HMM, long={Hidden Markov Model}}
\DeclareAcronym{VE}{short=VE, long={Variable Elimination}}
\DeclareAcronym{SCM}{short=SCM, long={Structural Causal Model}}

\acsetup{list/template = tabularray}
\definecolor{codegray}{rgb}{0.5,0.5,0.5}
\definecolor{codepurple}{rgb}{0.53,0.20,0.87}

\definecolor{codeyellow}{rgb}{1,0.92,0.23}
\definecolor{codegreen}{rgb}{0.29,0.68,0.31}
\definecolor{codeblue}{rgb}{0.13,0.59,0.95}
\definecolor{codered}{rgb}{0.95,0.26.6,0.21}

\lstdefinelanguage{pseudocode}{
	backgroundcolor=\color{gray!10},
	commentstyle=\footnotesize\itshape\sffamily\color{codegray},
	numbers=left, 
	numberstyle=\scriptsize, 
	morekeywords={for, each, from, to, do, end, CSD, COD},
	keywordstyle=\bfseries, 
	stepnumber=1, 
	numbersep=8pt, 
	xleftmargin=2em,
	frame=lines, 
	breaklines=true, 
	showstringspaces=false, 
	morecomment=[l]{\#}, 
	columns=spaceflexible,
	literate=
	*
	{CSD}{{{\color{codepurple}CSD}}}{3}
	{COD}{{{\color{codepurple}COD}}}{3},
}

\lstdefinelanguage{cat}{
	backgroundcolor=\color{gray!10},
	basicstyle=\footnotesize\ttfamily,
	commentstyle=\footnotesize,
	numbers=left,
	numberstyle=\scriptsize,
	stepnumber=1,
	numbersep=8pt,
	xleftmargin=2em,
	showstringspaces=false,
	morekeywords={GO, PAINT, COPY, MIRROR, FILL_EMPTY},
	keywordstyle=\bfseries, 
	breaklines=true,
	frame=lines,
	morecomment=[l]{\#},
	columns=spaceflexible,
	literate=
	*
	{PAINT}{{{\color{codepurple}PAINT}}}{5}
	{GO}{{{\color{codepurple}GO}}}{2}
	{FILL_EMPTY}{{{\color{codepurple}FILL\_EMPTY}}}{10}
	{MIRROR}{{{\color{codepurple}MIRROR}}}{6}
	{COPY}{{{\color{codepurple}COPY}}}{4},
}

\title{Towards an intelligent assessment system for evaluating the development of algorithmic thinking skills} 

\subtitle{An exploratory study in Swiss compulsory schools} 
\author{Giorgia Adorni} 
\advisor{Prof. Luca Maria Gambardella} 
\coadvisor{Prof. Alberto Piatti} 
\Day{7} 
\Month{February} 
\Year{2025} 
\place{Lugano} 
\programDirector{Prof. Walter Binder \& Prof. Stefan Wolf} 

\committee{%
	\committeeMember{Prof. Cesare Alippi}{Università della Svizzera Italiana, Switzerland}
	\committeeMember{Prof. Monica Landoni}{Università della Svizzera Italiana, Switzerland}
	\committeeMember{Prof. Engin Bumbacher}{Haute école pédagogique Vaud, Switzerland}
	\committeeMember{Ph.D. Marc Lafuente Martínez}{ }
} 

\openepigraph{Preoccupied with a single leaf... \linebreak you won't see the tree. \linebreak Preoccupied with a single tree... \linebreak  you'll miss the entire forest. \linebreak Don't be preoccupied with a single spot. \linebreak See everything in it's entirety... effortlessly. \linebreak That is what it means to truly ``see''. }{Takuan Soho}

\makeindex 

\graphicspath{{images/}}

\makeatletter
\@openrightfalse  
\makeatother
\begin{document}
	\sloppy
	\maketitle 
	
	\frontmatter 
	
	\begin{abstract}
		The rapid digitalisation of contemporary society has profoundly impacted various facets of our lives, including healthcare, communication, business, and education.
		The ability to engage with new technologies and solve problems has become crucial, making computational thinking (CT) skills, such as pattern recognition, decomposition, and algorithm design, essential competencies.
		In response, Switzerland has undertaken considerable research and initiatives aimed at integrating CT into the educational system, preparing students for the digital age.
		
		This research aims to contribute to these efforts by developing a comprehensive framework for large-scale assessment of CT skills throughout the Swiss compulsory education system, with a particular focus on algorithmic thinking (AT), which pertains to the ability to design algorithms.
		To achieve this, we first developed a competence model that captures the situated and developmental nature of CT, enabling the design of activities tailored to varying cognitive abilities, learner age, and the contexts in which they occur. 
		This framework not only clarifies how the characteristics and components of these activities influence the development of CT competencies but also provides guidance for effectively assessing them. 
		
		A key contribution of this research is the development of an assessment activity to measure AT skills on a large scale.
		The activity is designed in two variants: one uses non-digital artefacts (unplugged format) and provides a manual expert assessment, while the other relies on digital artefacts (virtual format), automating the assessment process.
		To provide a more comprehensive evaluation of students' competencies, we developed an Intelligent Assessment System based on Bayesian Networks with noisy gates, which offers real-time probabilistic assessment for each skill rather than a single overall score.
		
		The results of this study indicate that the proposed instrument can measure AT competencies across different age groups and educational contexts in Switzerland, demonstrating its applicability for large-scale use. 
		The findings suggest that AT competencies exhibit a progressive development, with no overall gender differences, though variations are observed at the school level. 
		Several factors, including the type of artefact-based environment and the situated context, influenced AT performance significantly. 
		These results underscore the importance of creating assessment tools that are both accessible and adaptable to various contexts. 
		Additionally, they highlight the need for careful and nuanced interpretation of the data, considering the diverse factors that may impact student performance and the validity of the assessments across different settings.
		In conclusion, this instrument holds significant potential for integration into real classroom settings, providing a scalable solution for assessing AT skills across a wide range of educational environments.

	\end{abstract}
	
	
	\begin{acknowledgements}
		
		I want to express my profound thanks to my dissertation supervisors, \textbf{Luca Maria Gambardella} and \textbf{Alberto Piatti}, for their guidance, mentorship, and constructive feedback throughout the entire research process, which has helped me grow both academically and professionally. 
		
		My sincere gratitude goes to the members of my committee.
		\textbf{Engin Bumbacher}, who provided insightful academic advice and collaborated with me in discussing the definition of one of the theoretical frameworks I developed. 
		\textbf{Monica Landoni}, who offered support and guidance on some of my works and included me in the TecLadies initiative.
		\textbf{Marc Lafuente Martinez}, though I had limited direct contact with him, suggested relevant literature that contributed to my research. 
		\textbf{Cesare Alippi}, also one of the professors I assisted during my doctoral studies, provided valuable feedback from a unique engineering perspective, complementing the insights of others.

		%
		
		I would also like to thank my academic peers and colleagues who have supported me throughout my doctoral journey.
		The EPFL group, including \textbf{Francesco Mondada}, \textbf{Laila El-Hamamsy}, \textbf{Kunal Massé}, and \textbf{Jérôme Brender}, despite the geographical distance, has always been available for discussions, offering support, and advice that significantly contributed to the success of my work.
		The School of Education group at FHNW, including \textbf{Dorit Assaf} and \textbf{Elia Lutz}, for their contributions to the project and their assistance with data collection in the German-speaking cantons of Switzerland.
		%
		The DFA group, including \textbf{Lucio Negrini}, who introduced me to ROTECO and supported the project, \textbf{Moro Thierry} for his assistance with the platform installation, setup and testing, and \textbf{Masiar Babazadeh} who supported me during the Swiss TecLadies initiative.
		The IDSIA group, including \textbf{Francesca Mangili} and \textbf{Alessandro Antonucci} for their support in understanding the theories behind the probabilistic models, \textbf{Claudio Bonesana} and \textbf{David Huber} for their help with the implementation of these models, as well as their support with pre-existing libraries and technical resources, \textbf{Giovanni Profeta} for providing valuable UX design feedback for my platform, and \textbf{Sandra Mitrovic} for involving me in Swiss TecLadies and in other collaborations that unfortunately have not materialised yet.
		\textbf{Igor Artico}, initially supported me on a personal level and later offered valuable academic advice and assistance with one of my recent papers.
		Finally, the students I supervised for their Bachelor thesis project, including \textbf{Simone Piatti} and \textbf{Volodymyr Karpenko}, who contributed to the implementation of part of the platform are credited in the related paper, and \textbf{Samuel Corecco} with whom I co-authored another paper.

		I would also like to acknowledge the \textit{Swiss National Science Foundation} (SNSF), under the National Research Program 77 (NRP-77) Digital Transformation (No: 407740\_187246), for financially supporting my work, and express my gratitude to all the \textit{participants} in my research, whose time and valuable contributions were crucial to the success of this project.

		On a more personal note, I would like to thank my family. My father, \textbf{Stefano}, my strongest pillar and my grandparents, \textbf{Ariodante} and \textbf{Giulia}, always a source of inspiration. My brother, \textbf{Ario}, and my mother, \textbf{Stefania}, contributing in their own ways. My aunt, \textbf{Lorenza}, always there for me like a second mother, and the rest of my family for their presence in my life. A special mention to \textbf{Kira}, my dog, whose unconditional love and companionship have been a constant source of joy.
		
		A special thank you to my partner \textbf{Lorenzo}. Despite our relatively short time together, he has been by my side every step of the way, supporting me in countless ways. His kindness, sincerity, and sense of humour never fail to brighten my days. His love, patience, and unwavering belief in me have kept me going through the most challenging times.

		A mention goes to my office mates who have shared my everyday work life: Hubi, Filip, Ele, Tuls, Koppány and Ambro; the people from the 4th floor: Alle, Luca, Taimoor, Alex; and the incredible friends I’ve made at IDSIA: Step, Franca, Matte, Nicho, Elia, Luca, Simo, Milad, Angelo, Dario, and Davide.
		
		I would also like to thank those who, at different moments, have been part of this journey and enriched it: Ted, Lambe, Lollo, Nicola, Pietro, Marco, Claudio, Luca, Palla and all the members of the university choir.
		
		A big thank you to the friends who, even from a distance, have been there for me throughout: Teo, Nassim, Lore Benatti, Lore Bini, Fede and Chri.
		
		Finally, heartfelt thanks to the friends who have been with me from the very beginning: Beba and Manu.
		
		Thank you all for your friendship, support, and love.

	\end{acknowledgements}
	
	\tableofcontents 
	\let\cleardoublepage\clearpage
	\listoffigures 
	\let\cleardoublepage\clearpage
	\listoftables 
	
	\printacronyms[name=List of Acronyms,display=all]
	\addcontentsline{toc}{chapter}{\acrolistname}
	
	\mainmatter
	\addtocontents{toc}{\protect\setcounter{tocdepth}{2}}
	\part{Introduction and Related works}\label{part:intro_and_literature}
	\chapter{Introduction}\label{chap:introduction}

This chapter outlines the context and objectives of this doctoral research, presenting key challenges and ethical considerations related to the study.

\section{Research context}

This doctoral research is part of a project funded by the Swiss National Science Foundation (SNSF) under the National Research Programme ``Digital Transformation'' (NRP 77). This programme investigates the interrelationships and specific effects of digital change in Switzerland, with a focus on understanding its impact on various sectors, including education.
Specifically, the project is situated within the ``Education, Learning and Digital Change'' module, which examines how digitalisation influences educational content, skill acquisition, and lifelong learning processes while also identifying challenges and strategies to manage the transformation of the education system.



\subsection{Switzerland's education system} \label{sec:intro_swiss_schools_organisation}

\begin{table*}[ht]
	\renewcommand{\arraystretch}{1.5}
	\footnotesize\centering
	\mycaption{Swiss compulsory education system according to the HarmoS Agreement.}{
		For each linguistic region -- German (DE), French (FR) and Italian (IT) -- are shown the stages of education, represented by three key educational cycles (preschool, primary, and lower secondary), along with the corresponding HarmoS Grades (HGs) and ages.
		}\label{tab:harmos}
	\setlength{\tabcolsep}{.9mm}
	\begin{tabular}{l|>{\centering\arraybackslash}m{.9cm}|>{\centering\arraybackslash}m{.9cm}|>{\centering\arraybackslash}m{.9cm}|>{\centering\arraybackslash}m{.9cm}|>{\centering\arraybackslash}m{.9cm}|>{\centering\arraybackslash}m{.9cm}|>{\centering\arraybackslash}m{.9cm}|>{\centering\arraybackslash}m{.9cm}|>{\centering\arraybackslash}m{.9cm}|>{\centering\arraybackslash}m{.9cm}|>{\centering\arraybackslash}m{.9cm}|>{\centering\arraybackslash}m{.9cm}}
		\hline
		\textbf{HG} & \textbf{0} & {\textbf{1}}    & {\textbf{2}}    &{\textbf{3}}    &{\textbf{4}}    &{\textbf{5}}    &{\textbf{6}}    &{\textbf{7}}    &{\textbf{8}}    &{\textbf{9}}    &{\textbf{10}}    & {\textbf{11}}           \\ 
		{Age} & {3-4} & {{4-5}}    & {{5-6}}    &{{6-7}}    &{{7-8}}    &{{8-9}}    &{{9-10}}    &{{10-11}}    &{{11-12}}    &{{12-13}}    &{{13-14}}    & {{14-15}}  \\ 
		
		\hline
		{DE} & &\multicolumn{2}{c|}{Kindergarten}            
		& \multicolumn{6}{c|}{Primarschule} & \multicolumn{3}{c}{Sekundarstufe I} \\ 
		\cline{2-13} 
		{FR} && \multicolumn{4}{c|}{Cycle 1 (primaire)}                 & \multicolumn{4}{c|}{Cycle 2 (primaire)} & \multicolumn{3}{c}{Cycle 3 (secondaire 1)}  \\ 
		\cline{2-13} 
		IT & \multicolumn{3}{c|}{Scuola dell'infanzia}         & \multicolumn{5}{c|}{Scuola elementare} & \multicolumn{4}{c}{Scuola media} \\ \hline
	\end{tabular}
\end{table*}

Switzerland’s education system, characterised by its decentralisation and multilingual environment, which includes four official languages -- German, French, Italian, and Romansh --, ensures that education is tailored to the needs of each canton while maintaining a degree of coherence through the Intercantonal Agreement on Harmonisation of Compulsory Education, known as HarmoS Agreement \citep{harmos,unesco2012international}.

The structure of compulsory education typically spans 11 years, beginning at age 4 and ending at 15 (16 for students who repeat a year), and is divided into preschool, primary, and lower secondary levels, with subtle variations across linguistic regions (see \Cref{tab:harmos}).
The system offers a broad curriculum at all levels that covers essential subjects such as languages, mathematics, natural sciences, and physical education. 
While the overarching framework remains consistent, regional autonomy allows for variations in how specific subjects are integrated into local curricula.

Recent shifts, particularly in response to the COVID-19 pandemic, have accelerated the focus on digital competencies, now recognised as a crucial skill for students.
However, the inclusion and emphasis on these competencies vary across linguistic regions, each adopting unique approaches to their development and assessment.
In the German-speaking regions, the ``Lehrplan 21'' incorporates digital competencies alongside core subjects such as mathematics, science, and social studies \cite{lehrplan}. 
In the French-speaking regions, the ``Plan d’études romand (PER)'' emphasises the application of technological knowledge and Computer Science (CS) principles \cite{per}. 
Similarly, in the Italian-speaking canton of Ticino, the ``Piano di Studio della scuola dell'obbligo ticinese'' integrates digital skills to promote critical thinking, problem-solving, and creative engagement with technology \cite{cantone2022piano}.
 
In conclusion, Switzerland’s educational framework is adapting to both regional diversity and the growing significance of digital literacy, ensuring that students are equipped with the essential skills to thrive in the digital age. This ongoing evolution reflects the country’s commitment to fostering a well-rounded education system that prepares students for the challenges and opportunities of the future while respecting regional autonomy and cultural contexts.

\section{Research motivation}

With the growing impact of technology, it has become increasingly important for individuals to develop skills to effectively use and handle new technologies and engage with problem solving processes, known as Computational Thinking (CT) skills.
These competencies have been recognised as fundamental in curricula worldwide and are considered key components of students’ academic and professional success.
Despite the development of numerous educational approaches in recent years, a significant gap remains in terms of replicable, scalable, and easily applicable assessment tools and protocols to evaluate computational thinking skills on a large scale.
In Switzerland, efforts have been made to integrate CT into education, but the country’s linguistic and cultural diversity presents challenges for implementing uniform educational strategies, requiring adaptable, context-sensitive approaches.
This research aims to address these challenges by developing adaptable strategies, protocols, and instruments for assessing CT skills among compulsory school students in Switzerland. The goal is to create tools that are both easy for teachers to adopt and capable of supporting semi-automated, large-scale assessments.

\section{Research objectives}

\subsection{Project overview}

The SNSF project ``Assessing the Development of Computational Thinking Skills Through an Intelligent Tutoring and Assessment System'' contributes to a large-scale assessment of CT competencies among Swiss students, with the following specific goals:
\begin{enumerate}[noitemsep,nolistsep]
	\item Developing an age-based competence model for CT: The project aims to define a clear and adaptable model that categorises CT competencies according to age groups, ensuring its applicability across different educational levels and contexts.
	
	\item Creating standardised assessment instruments: A set of standardised problems will be identified and developed to assess CT skills in students of various ages, allowing for consistent and reliable measurement of CT capabilities.
	
	\item Developing a state-of-the-art Intelligent Tutoring and Assessment System (ITAS): The project will develop an advanced probabilistic ITAS that not only assists students in solving CT problems but also measures their performance and skill development in real-time. This system will enable semi-automatic, large-scale monitoring of CT skills in classrooms across different regions.
	
	\item Validating the framework and measuring effectiveness: The project will test the developed methodology in real classroom settings, particularly in the cantons of St. Gallen, Vaud, and Ticino, which have distinct educational approaches to CT. By collecting data on the effectiveness of the implemented strategies, the project aims to refine and validate the framework for improving CT skill development in Swiss students.
\end{enumerate}

The project is a collaborative effort involving the Dalle Molle Institute for Artificial Intelligence (IDSIA USI-SUPSI), where my research is based, the Federal Institute of Technology in Lausanne (EPFL), the University of Teacher Education of Southern Switzerland (SUPSI-DFA), the University of Applied Sciences Northwestern Switzerland (Fachhochschule Nordwestschweiz, FHNW), and other leading academic and research institutions.


\subsection{Doctoral research focus}

Within the broader framework of CT assessment, this doctoral research specifically focuses on Algorithmic Thinking (AT), a subcomponent of CT concerned with solving problems through step-by-step procedures known as algorithms.
AT was chosen as the focal construct because it underpins the development of structured reasoning, decomposition, and procedural thinking, skills essential for computational problem-solving. Research indicates that AT serves as a foundation for broader CT competencies, making its assessment particularly relevant for understanding how young learners develop computational skills. By concentrating on AT, this study aims to refine assessment methods that can capture its progression and impact in early education.
Unlike the overall project, which includes both tutoring and assessment components, this thesis concentrates solely on the assessment aspect. The key objectives of the research are:
\begin{enumerate}[noitemsep,nolistsep]
	\item Developing an age-based competence model for CT: Aligning with the broader project, this objective seeks to categorise CT competencies based on age, ensuring adaptability across educational settings.
	
	\item Developing a large-scale assessment instrument for AT: Designing an activity that effectively measures students’ AT skills, irrespective of age or educational background, independent of age or educational background, using the competence model established in Objective 1.
		
	\item Developing an Intelligent Assessment System (IAS): Implementing a probabilistic system to monitor students' AT skills in real-time during problem-solving activities, enabling large-scale, semi-automated assessment without integrating tutoring mechanisms.
	
	\item Examining AT competencies in Swiss educational settings: Testing assessment tools and the IAS in real-world classrooms to explore how AT competencies develop across different age groups in the Swiss educational landscape, and the contextual factors influencing their progression. 
\end{enumerate}

\section{Research questions}
This doctoral thesis is guided by the following research questions, which span both the fields of Education and Computer Science (CS):
\begin{itemize}[noitemsep,nolistsep]
	\item[RQ1.] How can a competence model for CT be defined to assess skills across different age groups and educational contexts? (Education)
	
	
	\item[RQ2.] How can an activity and related instruments be developed to assess AT competencies on a large scale across different age groups and educational contexts, and what characteristics should they have to ensure their effectiveness and validity? (Computer Science \& Education)
	
	\item[RQ3.] How can a probabilistic IAS be designed and integrated into the instrument for assessing AT skills across different age groups and educational contexts? (Computer Science)
	
	\item[RQ4.] What are the key AT competencies in the Swiss educational landscape, how do they develop across age groups, and what demographic or contextual factors are associated with variations in these competencies? (Education)
\end{itemize}
By addressing these research questions, this thesis contributes to both educational research and computational assessment, bridging theoretical and practical insights from both domains.

\section{Research challenges}

Following the definition of the research objectives and research questions, several challenges have emerged during this doctoral project, significantly shaping its scope and methodology.

\subsection{Diversity of the population}\label{sec:contextual_factors_intro}
First, the diversity of the Swiss educational landscape presented a complex challenge. The project aimed at assessing AT skills across all compulsory school levels in Switzerland required careful navigation of regional, linguistic, and pedagogical differences.
Such elements affect not only the delivery of educational content but also the design and implementation of assessment tools, which need to be adaptable to different contexts.
The heterogeneity of the student population, spanning different ages, educational settings, and activity domains, further complicates this challenge, necessitating a comprehensive approach to analysis.

To address this, we considered several factors that may influence the study’s outcomes, categorising them into two main groups: demographic factors, which reflect the individual and social characteristics of the participants, and contextual factors, which relate to the environment in which the study takes place and are influenced by external circumstances rather than the individual participants.

%

\paragraph{Demographic factors}
Demographic factors are essential for achieving all research objectives, including developing the competence model, assessment instrument, and exploring results.
In the context of this study, we considered three key demographic factors:
	\begin{itemize}[noitemsep,nolistsep]
	\item \textit{Canton:} Switzerland’s linguistic diversity influences educational practices and students' experiences. By categorising participants based on their {canton} of residence, we can explore how variations in educational systems, specifically across the German and Italian-speaking regions, impact AT development and assessment.

	\item \textit{Gender:} While gender differences do not directly influence the development of the competence model or assessment instrument, they are crucial for analysing findings. Considering gender helps identify possible disparities in AT performance and engagement, ensuring a comprehensive understanding of student experiences.
	
	\item \textit{Age category:} The broad age span of participants (3–16 years) necessitates an assessment tool that is both versatile and developmentally appropriate. To ensure suitability across cognitive stages, we categorised students into four age groups: 3–6, 7–9, 10–13, and 14–16 years old. 
	This categorisation was informed by both psychological and educational considerations. Developmentally, these groups align with cognitive stages outlined in Piaget's theory \citep{piaget1964development}, ensuring that assessments correspond to typical cognitive milestones. Additionally, they map onto the Swiss educational cycles, covering early childhood education, primary education, and secondary education. This dual alignment allows for a structured analysis of AT skill development across key educational transitions.
\end{itemize}

\paragraph{Contextual factors}
Contextual factors primarily contribute to the objectives related to testing the assessment instrument and exploring AT competencies.
These factors provide a broader and more precise context for data analysis by considering the educational settings in which students interact with AT content. The key contextual factors examined in this study include:
	\begin{itemize}[noitemsep,nolistsep]
	\item \textit{Educational context:} The study takes into account the different school types (preschool, primary, and lower secondary education) and HGs, which define the 11 specific levels within Switzerland’s federal education system. Additionally, the influence of individual schools and class sessions is considered, recognising that each school has distinct student compositions, resources, and teaching strategies that may affect AT learning and assessment.
	
	\item \textit{Activity domains:} The format in which educational activities are conducted significantly impacts student engagement and performance. We differentiate between unplugged and virtual formats, each offering distinct interaction methods that influence the development of the competence model and assessment instrument. These formats also shape students' experiences and learning outcomes, making them a crucial factor in the study’s broader analysis.
	
	\end{itemize}

%


Understanding the interplay of these demographic and contextual factors allows for a more nuanced analysis of AT competence development across Switzerland’s diverse educational landscape. The study’s methodological approach ensures that these factors are systematically accounted for, enhancing the validity and applicability of the findings.



\subsection{Sample size considerations}
Given the diversity of the population, estimating an appropriate sample size is essential to ensure meaningful and reliable results.
In this research, we referred to a simulation study by \citet{Pan2018} to estimate the number of participants needed to detect a mediation effect in our analysis. 
Based on effect size considerations, we estimated that a sample size of 200 to 350 participants would provide an 80\% probability of detecting meaningful effects.

Given the exploratory nature of this study, our primary goal was to assess the applicability of our assessment instrument across diverse age groups and educational contexts, rather than to test predefined hypotheses requiring formal power calculations. Consequently, we designed our study to include approximately 300 participants, accounting for an expected 20\% attrition rate due to factors like disengagement, logistical constraints, unsigned consent forms, or unforeseen circumstances, like absenteeism on the day of the activity. 

While power considerations may be relevant in the context of our statistical analysis, since we employed hierarchical linear modelling to account for the nested structure the data, the chosen modelling approach helped mitigating potential limitations related to sample size by appropriately handling variability across groups. Given the diversity of our sample and its alignment with previous studies of similar scope, we considered our sample size sufficient to detect meaningful trends and effects.

By addressing the challenges posed by population diversity, this study ensures that the assessment instrument is both robust and adaptable, capable of providing valuable insights into AT development across Switzerland’s educational system.

\subsection{Impact of the COVID-19 pandemic}

Another major challenge was the unprecedented outbreak of the COVID-19 pandemic, which significantly disrupted our plans for direct engagement with schools and participants. 
My doctoral research began in November 2020, and we faced difficulties accessing school classes to conduct the necessary experimentation during the initial stages.
During the first experimental study, conducted between March and April 2021, restrictions were still in place, severely impacting our ability to engage with participants in person. 
As a result, certain activities, such as co-designing activities in the early phase of the study, were not feasible. Additionally, during the data collection phase, we had to take extra precautions to ensure safety. These included measures such as wearing masks, maintaining physical distance, and sanitising materials regularly. While these adjustments allowed us to continue the research, they introduced constraints on the types of activities that could be conducted and required continuous attention to health protocols.

\subsection{Ethical considerations}\label{sec:ethics}
A final critical challenge in this study relates to the ethical responsibility of conducting research with young and vulnerable participants. 
To ensure participant protection, data confidentiality, and the integrity of the research process, rigorous protocols were developed in compliance with both national and international ethical standards \citep{petousi2020contextualising,aebi2021code}.
This study adhered to the EPFL Human Research Ethics Committee's (HREC) ethical guidelines and received approval (HREC No: 048-2023). 

\subsubsection{Informed consent and participant information procedures}

The informed consent and participant information procedures were designed to ensure clarity and transparency for all involved parties.
First, school directors and teachers were provided with an information sheet detailing the research project, the involved institutions and researcher, the experimental activities to be performed in the classrooms, as well as the nature of the data being collected, the data storage and access methods.
School directors and teachers were then asked to provide explicit authorisation for the research team to access the school and classrooms for the study.
Parents were also provided with the same information sheet, along with the contact details of the principal investigator, for any further inquiries.
They were asked to sign and return the consent form attached to the information letter.
The participant information sheet template and parental consent form are in \Cref{app:informative}.

In line with ethical standards, informed consent was obtained from all participants' parents or legal guardians. Only pupils with explicit consent were allowed to participate in the study. In addition, children were informed about the study before their participation and were given the option to decide whether to participate.

\subsubsection{Data collection and storage}
The data collected in this research include session details (e.g., canton, school, class), pupil information (e.g., gender, month of birth), and performance data. 
In compliance with Swiss and international guidelines, data were pseudonymised to protect participant identities by eliminating identifiable information, such as assigning unique indices to schools instead of using their actual names.

All data storage and communication channels are encrypted to maintain data integrity and confidentiality. To ensure data security during transmission, secure procedures were followed to transfer data to the local server.

Access to confidential data is restricted to authorised researchers affiliated with the project, and all data are stored on the partner research institutions' Swiss servers (Switch Drive), with access controlled through secure login credentials.
In line with long-term preservation, data are also stored indefinitely on Zenodo, a secure digital repository, ensuring open access and continued availability.



%
%

%

\section{Research contribution}
This dissertation contributes to the fields of Artificial Intelligence (AI), Human-Computer Interaction (HCI), and Education.
	\begin{itemize}[noitemsep,nolistsep]
		\item This work can advance the application of AI-based assessment in education by developing and evaluating a tool for measuring AT skills. The study demonstrates how AI can support automated evaluation, offering insights into student performance across various educational stages. It contributes to AI research by refining assessment methodologies, integrating AI techniques in educational evaluation, and exploring their implications for adaptive learning systems.
		
		\item The research contributes to HCI by employing a user-centred design approach to ensure that the AI-based assessment tool is accessible, usable, and pedagogically effective. Through iterative prototyping and participatory design, the study examines how interaction design principles influence digital assessment environments. The findings provide insights into usability, engagement, and the role of intelligent systems in facilitating AT development.
		
		\item At its core, this dissertation addresses a critical gap in educational assessment by proposing a structured framework for evaluating AT across different cognitive and educational stages. By aligning the assessment tool with developmental theories and educational curricula, the study contributes to understanding how AT skills evolve over time, how digital tools can support formative assessment, and how AI-driven approaches can complement traditional evaluation methods.
	\end{itemize}
By integrating AI, HCI, and educational research, this dissertation highlights the potential of AI-powered assessment tools to enhance learning processes while maintaining usability and accessibility through HCI principles. The study also critically reflects on the varying depth of exploration in each domain, emphasising their intersection and collective impact on the broader field of educational technology and digital assessment.

\section{Structure of the thesis}

The thesis is structured in five parts.
\Cref{part:intro_and_literature} includes the introduction and literature review, providing an overview of the study's context, objectives, and key concepts, followed by a discussion of previous research relevant to the topic.
\Cref{part:methodological_framework} outlines the methodological frameworks developed for the research, including the age-based competence model for CT.
\Cref{part:assessment_instruments} presents the instruments designed for assessing AT and the developed IAS.
\Cref{part:results} presents the results from the experimental studies, as well as from the application of the IAS to the collected data.
Finally, \Cref{part:dicussion} offers a discussion, addressing the research questions and concluding with an exposition of the study's limitations and suggestions for future work.

	\chapter{Related works}\label{sec:literature}

This chapter provides a comprehensive overview of CT in education, beginning with its significance and the challenges in defining CT skills. It then delves into AT as a core component of CT and explores the integration of CT and AT in educational contexts. The chapter continues by examining existing assessment tools and the challenges involved in assessing AT, before discussing IAS and the role of probabilistic graphical models in skill assessment.

\section[Computational Thinking (CT) in education]{Computational Thinking in education}

In an era defined by rapid technological advancement and increasing digitalisation, CT has gained significant attention across educational sectors as a foundational skill enabling students to engage with complex systems, address interdisciplinary challenges, and meet the demands of an evolving digital landscape \cite{weintrop2021assessing,wing2014computational,wing2006computational,kafai2020theory}. 
Over the past two decades, substantial research on various facets of CT has underscored its importance as a 21st-century competence, driving global efforts to integrate it into K-12 education \cite{weintrop2021assessing}.

The term CT was popularised by Jeannette Wing, who described it as \emph{``a fundamental skill for everyone, not just for computer scientists''} \cite{wing2006computational}. 
She proposed that CT should be integrated into every child’s analytical toolkit alongside reading, writing, and arithmetic. 
In particular, she defined CT as \emph{``the thought processes involved in formulating a problem and expressing its solution(s) in such a way that a computer – human or machine – can effectively carry out''} \cite{wing2014computational}.
Wing underscored that CT extends beyond computational tools, emphasising key processes such as abstraction, decomposition, and algorithmic design.

As highlighted by \citet{rapaport2015philosophy}, the term \emph{CT} comprises two essential components. 
The first, ``computational'', implies a focus on computation, which in turn involves the design and execution of algorithms. 
The second, ``thinking'', reflects the cognitive process required for problem-solving.
This perspective aligns with Wing’s view, suggesting that CT is inherently tied to algorithms rather than artificial agents (e.g., computers) \cite{wing2014computational}.
Consequently, CT encompasses both digital and unplugged tasks, where the solution involves an algorithm to accomplish a specific task.
%
CT is thus defined as a cognitive process involving interrelated skills such as \textit{decomposing} complex problems into manageable components, identifying \textit{patterns} and recurring structures, \textit{abstracting} relevant information from irrelevant details, and formulating \textit{algorithms} to devise effective solutions \cite{wing2006computational,Wing2008,wing2011research,DENNING2021,denning2019computational,Lodi2021,Tedre2016,papert1980,Papert2000,RomnGonzlez2024}.

\section[Algorithmic Thinking (AT) in education]{Algorithmic Thinking in education}

AT is a specific aspect of CT, focusing on the design of step-by-step procedures or algorithms to solve problems systematically and achieve specific outcomes  \cite{Syso2015,futschek2006algorithmic,Lodi2021,RomnGonzlez2024}.
While CT broadly encompasses skills like decomposition, pattern recognition, and abstraction, AT is more narrowly concerned with the development of algorithms, which are crucial for structured problem-solving in both human and computer contexts  \cite{seehorn2011csta,barr,poulakis2021computational,futschek2006algorithmic}. 

In educational settings, AT is recognised as a critical skill for fostering logical reasoning, creativity, and problem-solving \cite{yadav2014computational,yildiz2020ideal,amini2016bedside,jocz2023motivating}.
These skills extend beyond the realm of computer science (CS) to address broader interdisciplinary challenges, making AT an essential competency in the digital world across personal and professional domains \cite{wing2006computational,kules2016computational,Wing2017,weintrop2021assessing,Bers2022,bocconi2022reviewing,Webb2017}. 
Its integration into education has gained significant momentum, as it equips students with the ability to break down complex problems, devise sequential actions, structure systematic solutions, and comprehend foundational concepts like algorithms and data structures \cite{olkhova2022development,korkmaz2019adapting,oluk2016comparing,dagiene2016it,kong2019introduction}. 


The theoretical underpinnings of AT draw on early developmental theories from Piaget and Vygotsky, emphasising the role of active learning and social interactions in constructing knowledge during the early stages of childhood development  \cite{piaget1952origins,piaget1964development,piaget1983handbook,vygotsky1978mind}. 
Piaget's constructivist theory posits that children build knowledge through hands-on experiences and interactions with their environment, while Vygotsky's social constructivist theory adds that social interactions and cultural context significantly influence learning outcomes.
These perspectives align with the idea that engaging students in problem-solving and critical-thinking activities, such as those involved in AT, can significantly enhance cognitive development.

Modern research expands on these foundations, underscoring the importance of introducing AT concepts early in education to cultivate critical thinking, logical reasoning, and analytical skills \cite{kanaki2022age,Voronina2016,nikolopoulou2023stem,georgiou2021developing,Hsu2018,Jiang2022}. 
These skills are not only essential for STEM education but also have transferable applications across various disciplines \cite{weintrop2021assessing,Bers2022,bocconi2022reviewing,Webb2017}.
This growing emphasis reflects a broader understanding of education that values skills relevant to the demands of the 21st century \cite{ezeamuzie2021computational,scherer2019the,pilotti2022is,voogt2015computational}.


\section[Defining CT]{Defining Computational Thinking} 

Defining CT and AT is a significant challenge in both research and practice despite the growth in tools, activities, and curricula designed to teach it.
The absence of a universally accepted definition of CT and its relationship to AT represents an obstacle to its integration into educational standards and curricula \cite{weintrop2021assessing}.
Various definitions have emerged, each emphasising different aspects of CT, which has hindered the development of a coherent framework for developing and assessing CT and AT competencies \cite{lafuentemartinez2022assessing}. As a result, the field has struggled to move beyond an exploratory stage.

Unlike other introductory skills such as reading, writing, and arithmetic, research on CT and AT has not consistently considered its developmental nature. 
Specifically, existing literature rarely considers age or developmental progression when defining or discussing these competencies \cite{shute2017demystifying, tikva_mapping_2021}. 
Additionally, complex components such as abstraction, a key element often associated with CT \cite{wing2014computational, shute2017demystifying}, are considered beyond the cognitive capabilities of very young individuals. 
Moreover, many existing CT models primarily focus on internal cognitive processes, neglecting the situated nature of tasks that require CT, such as the social context and the artefactual environment in which these tasks occur.

The field has also struggled to establish a structured approach for assessing CT and AT due to the complexity of its components. Efforts to decompose CT into sub-dimensions, such as decomposition, generalisation, and pattern recognition \cite{brennan2012new, grover2017computational,RomnGonzlez2024}, have encountered difficulties because these dimensions are often interwoven, making it challenging to assess them independently \cite{lafuentemartinez2022assessing,RomnGonzlez2024}. 

As a result, there is a pressing need for a more structured and precise approach to defining CT and AT. Establishing a clear, comprehensive, and standardised definition would not only clarify the concept but also facilitate the design of effective interventions and assessment tools.
Additionally, this definition should explicitly consider the developmental nature of cognitive abilities, the age of learners, and the context in which CT and AT activities are situated. Such an approach would provide a more accurate foundation for understanding how these skills evolve throughout an individual's learning journey. This, in turn, would support the long-term integration into educational curricula, enabling its development and assessment in a consistent and meaningful way.

\subsection{A situated cognition perspective}\label{sec:situated_cognition_perspective}
The most widely recognised definitions of CT often emphasise the cognitive processes an individual should activate to solve a computational task \cite{shute2017demystifying}.
These definitions typically align with the classical view of cognition as an internal process occurring within a single individual, i.e., the ``thought processes'' in the definition of \citet{wing2014computational}.

An alternative perspective on CT emerges from theories of situated cognition, which view cognitive activities as inherently social and contextual \cite{roth2013situated,heersmink2013taxonomy}.
According to this view, cognitive activities are not isolated within the individual’s mind but are shaped by interactions with the environment and social context.
This perspective suggests that learning and knowledge construction occur through shared practices, where external cognitive artefacts play a crucial role in mediating thinking and problem-solving. 

In educational settings, as well as more broadly, CT and AT are often activated in environments that involve social interactions and rich artefactual contexts, aligning better with this situated approach to cognition. 
Theories of situated learning further support this notion, highlighting that learning is most effective when it occurs in authentic, meaningful contexts, where knowledge is co-constructed through engagement with the environment and community rather than through abstract, isolated instruction \cite{roth2013situated, heersmink2013taxonomy}.
In the case of AT, this perspective stresses that algorithmic problem-solving is often influenced by the surrounding context, tools, and social interactions that learners engage with.

\subsection{A developmental perspective} \label{sec:tall_three_worlds_mathematics}
Mathematical thinking is a multifaceted set of skills and attitudes widely recognised as a fundamental component of human thinking that evolves over time, beginning early in life and developing throughout education into adulthood \cite{tall2013humans}. 
A foundational example of this progression is the act of counting, which starts with basic enumeration and evolves into more complex concepts like the idea of ``number'' \cite{gelman1986childs, dehaene2011number}. 
Pupils continuously refine these skills through various experiences ranging from their first counting experiences with concrete objects in pre-primary school and even before \cite{benoit2004young, bruce2004one, sarnecka2008how}, through the development of counting strategies, symbolisation, automatisation, abstraction, and so on, up to axiomatisation of natural numbers and the development of very complex counting algorithms.

Tall's model of the three worlds of mathematics provides a framework to describe how mathematical thinking develops over time \cite{tall2006theory, tall2013humans, tall2020three}.
The model suggests that acquiring mathematical concepts and theories unfolds incrementally, starting with concrete, embodied experiences, moving through internalisation via symbolic concepts (procepts), and culminating in abstract axiomatisation. 
Tall argues that mathematical thinking progresses through three interconnected stages:
\emph{
	\begin{itemize}[noitemsep,nolistsep,leftmargin=.25in]
		\item an object-based conceptual-embodied world reflecting on the senses to observe, describe, define and deduce properties developing from thought experiment to Euclidean proof;
		\item an action-based proceptual-symbolic world that compresses action schemas into thinkable concepts operating dually as process and concept (procept);
		\item a property-based formal-axiomatic world focused on building axiomatic systems based on formal definitions and set-theoretic proof.'' 
	\end{itemize}
}
Each stage builds upon the experiences from the previous one, indicating a developmental progression rather than isolated stages. 
For example, in the development of counting and the concept of number, these stages can be mapped to various educational levels: in pre-primary school, reasoning is primarily rooted in the first world; in primary school, both the first and second worlds are present; in secondary school, reasoning shifts predominantly to the second world, with the introduction of the third world; and at the tertiary level, reasoning is primarily situated in the third world.

This developmental model of mathematical thinking closely parallels the progression of AT.
Just as mathematical concepts evolve from concrete experiences to abstract formalisation, AT follows a similar trajectory: 
	\begin{itemize}[noitemsep,nolistsep,leftmargin=.25in]
	\item Concrete execution: Young children engage in sequential actions, such as counting objects or following step-by-step instructions;
	\item Symbolic representation: As they develop, they learn to represent these sequences using symbols, diagrams, or structured notation;
	\item Abstract generalisation: Eventually, they internalise algorithmic structures, enabling them to design, analyse, and optimise problem-solving procedures independently of specific contexts. 
\end{itemize}
In Tall’s framework, the transition from the conceptual-embodied world to the proceptual-symbolic world mirrors the way learners move from performing concrete steps in an algorithm to recognising and manipulating these steps as symbolic entities. Similarly, the shift to the formal-axiomatic world aligns with the ability to construct, prove, and reason about algorithms in a rigorous and abstract manner.

This perspective has important implications for assessing AT: it suggests that an individual’s competence in AT should be evaluated in relation to their cognitive development, considering their ability to engage with different levels of abstraction.
Certain algorithmic constructs may be too complex for younger or less experienced learners, just as advanced mathematical reasoning requires foundational cognitive skills developed over time.
Thus, the developmental trajectory of mathematical thinking provides a valuable lens for understanding how AT emerges and evolves, informing both educational strategies and assessment frameworks for computational competencies.

%


\section[CT and AT integration in education]{Computational Thinking and Algorithmic Thinking integration in education}

\subsection{Global overview}
In the \textit{United States}, the emphasis on CT began in the early 2000s, particularly through initiatives like the Next Generation Science Standards (NGSS)\cite{schwarz2017helping,Bybee2014,wilkerson2017using,ngss} and the CS for ALL Students initiative \cite{csforall,obama}, which aimed to ensure CS education is accessible to all students from early education onward.
Other countries are also advancing in this area.
\textit{New Zealand}, \textit{Australia}, \textit{South Korea} and \textit{Japan} from 2015 have started integrating digital technologies, CT and AT across STEM subjects at all educational levels, focusing on themes like algorithms and problem-solving, making programming a compulsory subject \cite{newzeland,japan,australia,bocconi2016developing,park2016preparing}.
Similarly, \textit{Singapore}, under the 2014 Smart Nation initiative led by the Prime Minister to promote early programming exposure \cite{Hoe2016,smartnation}, launched a CT framework in 2016, introduced a computing subject focused on programming and algorithms in secondary schools by 2017, and mandatory CT and coding program for upper primary students by 2020 \cite{singapore,bocconi2016developing,bocconi2022reviewing}.
In \textit{Canada} (British Columbia), CT has been incorporated into middle school subjects, with plans for broader application at the secondary level \cite{BC, bocconi2016developing}.

\subsection{European context}
Several European countries have significantly advanced in integrating CT and AT into their compulsory education systems. 
While some have incorporated these skills across all compulsory educational levels, others have focused primarily on secondary education.
The degree of integration and the scope of the curricula reforms vary widely across the continent, with some countries adopting a holistic, cross-curricular approach, while others emphasise CS or technology education as separate subjects \cite{bocconi2016developing,bocconi2022reviewing}.

The pioneers in integrating CT and AT across both primary and secondary levels have significantly influenced the approaches of subsequent nations. 
Among them, \textit{England} was one of the earliest to make CT mandatory, incorporating it into its national curriculum in 2014 as a separate subject \cite{uk2013national}. 
\textit{France} followed closely, integrating CT within existing subjects such as mathematics and technology in 2015 \cite{france}.
\textit{Finland} incorporated CT and AT in 2016 as a cross-curricular theme, later extending its integration within subjects like mathematics, crafts, and environmental studies by 2022 \cite{finnish}.

In the years following these initial pioneering efforts, several other countries have embraced CT and AT, albeit at different rates and in various formats. 
Countries such as \textit{Malta}, \textit{Slovakia}, \textit{Poland}, \textit{Portugal}, \textit{Croatia}, \textit{Greece}, \textit{Austria}, and \textit{Hungary} have integrated CT/AT as a separate subject, primarily through informatics or CS courses, emphasising the importance of computational skills as a distinct area of study with dedicated instructional time \cite{poland,croatia,grece,hungary,malta,Slovakia2024,slovakia,AUSTRIA,austria2024}.
In contrast, \textit{Sweden}, \textit{Norway} and \textit{Lithuania} have opted to embed CT within existing subjects, such as mathematics, science, and the arts, promoting an interdisciplinary model that fosters CT across various academic domains \cite{norway,Lithuania,sweden}.
In \textit{Cyprus}, \textit{Luxembourg}, and \textit{Serbia}, CT is integrated into primary education primarily within other subjects, while in secondary education, it is structured as a separate subject, reflecting a flexible and context-specific approach to embedding CT across different educational levels \cite{cyprus,Luxembourg,serbia}.  

Despite notable advancements in various countries, several have achieved only partial integration of CT and AT. Specifically, \textit{Ireland}, \textit{Romania}, and \textit{Scotland} have incorporated these skills into secondary education, while formal integration at the primary level continues to be lacking \cite{ireland2,ireland,Romania,scotland}.

Additionally, several countries have made little to no progress in integrating CT and AT into their educational systems. In \textit{Denmark}, \textit{Slovenia}, \textit{Italy}, the \textit{Czech Republic}, the \textit{Netherlands}, and \textit{Spain}, the situation varies, with most of these countries at the drafting stage of curricula or strategic plans for future actions \cite{denmark,italy,Czech,slovenia,netherlands,bocconi2016developing,bocconi2022reviewing}.
For instance, Denmark has yet to integrate CT but has initiated a pilot program \cite{werner2012children}, while Italy recognises CT as a key topic but lacks formal integration in its national curriculum.

The situation in Belgium further illustrates this complexity, as integration depends on specific regions. In Flanders, CT has been integrated as part of a separate subject, while Wallonia plans to address it as a compulsory subject for primary and lower secondary schools \cite{belgium}.

\subsection{Swiss context}
The Swiss educational system has progressively integrated AT and CT into its curriculum, adapting to the specific needs of its diverse linguistic regions. These skills are embedded within various subjects, such as mathematics and CS, through activities like coding, algorithmic exercises, and robotics projects, ensuring that students acquire essential skills from an early age \cite{bocconi2016developing,bocconi2022reviewing}.

In the German-speaking region, the integration of CT began around 2014, with competencies such as coding and programming incorporated into the curricula of primary and lower secondary schools. At the upper secondary level, these skills are formalised within the national curriculum framework for non-vocational schools, ensuring a comprehensive acquisition of computational skills throughout the educational journey \cite{des2016lehrplan,lehrplan}.
In the French-speaking region, CT is taught through the Plan d'études romand (PER) under the subject MITIC (Média, Image, Technologie de l'Information et de la Communication), implemented since 2015 \cite{per}. Within the framework, students engage in activities that require them to analyse problems, devise solutions, and implement basic programs, reinforcing CT skills from early education onward. Additionally, the subject ``Media and Informatics'' introduces CT as a core component, fostering logical thinking and problem-solving abilities.
In the Italian-speaking region, AT and CT are integrated primarily through subjects like mathematics, with a strong emphasis on coding, problem-solving, and robotics. However, computer science is considered a transversal competency rather than a distinct subject. This approach encourages students to apply these skills across various disciplines, promoting interdisciplinary learning. While this may give the impression of a highly technological approach, the reality is that CT and AT are woven into the curriculum in a way that emphasises their applicability in different contexts rather than as standalone subjects \cite{cantone2015piano,cantone2022piano}.

\section[Assessing AT]{Assessing Algorithmic Thinking}

With growing recognition of AT as an essential component of compulsory education, there is an increasing need for reliable, scalable assessment instruments that can measure students' development across various age groups and educational settings on a large scale  \cite{lafuentemartinez2022assessing,grover2013computational,barr,voogt2012,giannina2020pisa}.

Research on AT assessment is limited and often contradictory \cite{tikva_mapping_2021, tang2020assessing, grover2014assessing}.
Many assessment tools exist, but each focuses on specific aspects, and none cover all educational and cognitive needs.
Current methods often neglect developmental stages, social contexts, and available resources \cite{brennan2012new,roman2017complementary,yadav2014computational,korkmaz2019adapting,korkmaz2017validity,lafuentemartinez2022assessing,roman2019combining,tsai2020the,roman2017cognitive,polat2021a,RomnGonzlez2024}.
The lack of standardised tools and diverse evaluation methods make it difficult to assess students' overall progress  \cite{ezeamuzie2021computational,scherer2019the,pilotti2022is,grover2017assessing,pisa,Fraillon2020}.

Empirical research has examined the effectiveness and challenges of various instruments used to assess AT. 
Traditional methods, such as written tests and multiple-choice questions, are widely used for their efficiency in covering broad topics, straightforward administration and grading; however, they may oversimplify the assessment by focusing on rote memorisation rather than deeper problem-solving skills \cite{garcia2022application,oyelere2022developing,simmering2019what,csernoch2015testing,campbell2012exploring}.
In contrast, open-ended tasks and problem-solving exercises offer richer evaluations by assessing reasoning and creativity, but they can be more challenging to grade \cite{csernoch2015testing}.

Unplugged methods, which involve hands-on, tangible activities, assess AT concepts effectively, particularly in environments without access to basic technology infrastructure \cite{brackmann2017development,del2020computational,bell_csu,kalelioglu2016framework,relkin2021learning, wohl2015teaching}.
These activities are especially useful for building a strong foundation in computational principles and for young children without prior programming experience \cite{unnikrishnan2016elephants,wallet2015information,brackmann2017development}, 
Empirical evidence from \citet{relkin2021learning} and \citet{brackmann2017development} supports the unplugged approach, showing improvements in students' CT skills after participating in unplugged computing instruction. Moreover, studies by \citet{tsarava2017training, tsarava2018training, delal2020developing, brackmann2016computational} have highlighted the positive effects of unplugged activities on motivation, engagement, and overall effectiveness, particularly in primary education.
Research by \citet{del2020computational, saxena2020designing} has shown that unplugged activities can significantly enhance CT skills. Further, \citet{delal2020developing} cited various studies confirming the development of CT skills through unplugged computing activities while also improving students' understanding of CS concepts and fostering greater interest in the subject \cite{bell2018cs, cortina2015reaching, lu2009thinking, rodriguez2016using,metin2020activity, hermans2017scratch}.
Specifically, \citet{relkin2021learning} found that unplugged programming activities help students achieve the highest levels of understanding in AT, logic predictions, and debugging concepts, further solidifying the effectiveness of unplugged methods in fostering foundational computational thinking.
However, unplugged activities are not ideal for large-scale assessments due to their hands-on, time-intensive nature, which can be challenging to scale for large groups of students and may require significant resources and coordination \cite{del2020computational,ElHamamsy2022}.

Digital methods, including programming assignments, coding challenges, and other computer-based activities, engage students in practical applications of algorithmic skills through interactive and individualised experiences.
While these methods are scalable and well-suited for large-scale assessments, they often require intensive grading efforts and may lack the physical engagement provided by unplugged activities \cite{sun2021comparing,roman2017cognitive,zapata2020computational,relkin2020techcheck}. 
Additionally, reliance on technology can pose barriers for students with limited access or those who prefer non-digital learning environments \cite{brackmann2017development,kalelioglu2016framework,bell2018cs,relkin2020techcheck,keith2019roles,mccormick2022computational}.


The current landscape of tools for assessing AT reveals a significant gap in the availability of instruments that can meet the diverse needs of learners and educational contexts.
One of the key shortcomings is the lack of tools that effectively integrate multimodal features. While unimodal tools, those offering a limited artefactual environment with a single method of interaction, may serve some students, they often fail to accommodate the diverse learning styles found in modern classrooms, limiting engagement and the accuracy of assessments. In contrast, multimodal tools, which provide various options for interaction, offer greater flexibility, allowing learners to choose the method that best suits their learning preferences, thus improving both engagement and assessment reliability.


Additionally, the target age range of a tool plays a pivotal role in its effectiveness. 
Tools designed for narrow age groups often fail to account for the diversity of developmental stages within a typical classroom. 
For example, the BPAt \cite{Mhling2015} and the cCTt \cite{ElHamamsy2022} are designed for specific ages, and while they provide valuable assessments of foundational skills, their applicability in projects with a broader student demographic is limited, as they do not cater to students outside of the targeted age groups. This restriction hampers their versatility in diverse classroom settings.
In contrast, instruments that cater to a broader age range ensure versatility and inclusivity, making them more adaptable to different classroom settings.
Tools like Scratch \cite{Maloney2004scratch,moreno2015dr,GroverPeaCooper2015} and Code.org \cite{Blockly} are widely used in educational settings and cater to a broad age range, allowing for engagement with students across different developmental stages. 
Code.org, for example, provides a wide variety of coding activities that can engage students from early primary school to high school. However, while these platforms are suitable for various age groups, the specific activities they offer are often tailored to specific age ranges, which limits their flexibility for cross-age assessments.
Similarly, Scratch is adaptable to many age groups but is typically used with younger learners for simpler projects, which limits its application for more advanced learners.
To address this limitation, instruments that cover a wide age range and are adaptable to different developmental stages are needed. 
These tools are better suited for longitudinal assessments, tracking students' progress over time and offering a comprehensive understanding of how their skills evolve across cognitive stages.

Finally, another significant gap in the current tools for assessing AT is their inability to address the full range of cognitive levels in skill development. According to the frameworks of \citet{CTF,bloom1956taxonomy}, cognitive skills progress through several stages, from foundational levels to more advanced stages.
The progression typically starts with basic levels, where students focus on recognising and understanding key concepts. At this stage, students may identify patterns, follow simple instructions, or understand basic algorithms.
For example, tools like the Basic programming abilities test (BPAt) \cite{Mhling2015}, the Computational Thinking test (CTt) \cite{roman2015computational}, the Beginners Computational Thinking test (BCTt) \cite{zapata2020computational,zapata2021} and the competent Computational Thinking test (cCTt) \cite{ElHamamsy2022} focus on foundational skills such as recognition and understanding, limiting their effectiveness for more advanced assessments.
As learners advance, they move to the next level, where they apply these basic concepts in more complex contexts. This stage involves using knowledge to solve problems, create algorithms, or engage in basic programming.
Tools like Scratch \cite{Maloney2004scratch,moreno2015dr,GroverPeaCooper2015} and Code.org \cite{Blockly} support students in applying their knowledge to create projects, such as games or animations, and apply algorithms in a more interactive environment.
The final stages of the progression are characterised by higher-order skills such as analysis, synthesis, and abstraction. At this stage, students have assimilated these competencies and should be able to evaluate complex algorithms, analyse problems deeply, and create more sophisticated solutions. 
Unfortunately, many existing tools fail to adequately address this advanced level of thinking, limiting their ability to assess long-term progression.

\section[Intelligent Assessment Systems (IASs)]{Intelligent Assessment Systems}

The challenges in assessing AT have been a long-standing issue in education.
Traditional manual assessment methods are time-consuming, inconsistent, and difficult to scale.
The time spent on assessments often outweighs the educational value, and inconsistent results lead to disparities in evaluating student performance \cite{qian2018using, romero2017computational}. 
Additionally, the subjectivity of these assessments makes them unreliable and non-standardised, raising concerns about their accuracy and fairness, especially when comparing large groups of students.
Furthermore, general educators often lack the technical expertise to assess the complex aspects of AT \cite{Ukkonen2024, mason2019preparing, yadav2014computational}. While some schools try to involve IT professionals in the assessment process, these experts are often unavailable in under-resourced areas.
This shortage further complicates the accurate assessment of AT. As a result, educators must rely on standardised tools, which, though consistent, often oversimplify the evaluation and fail to capture the full scope of AT skills \cite{DENNING2021}.

Given these challenges, IASs represent a promising solution that can improve and simplify AT evaluation in educational contexts  \cite{qian2018using,romero2017computational,stanja2022formative,guo2021evolution}. These systems would provide scalable, consistent, and objective assessments, addressing the limitations of traditional methods and standardised tools \cite{qian2018using,romero2017computational}. 
These technology-driven assessment tools, which are developed by experts and integrated into educational platforms, can be easily administered by teachers without specialised training.
They also allow students to work independently, providing real-time feedback and evaluations without the need for constant supervision by a teacher. 
This autonomy not only supports individualised learning but also ensures that students receive immediate insights into their progress. 
Furthermore, IAS can adapt to various cognitive levels and educational contexts, making them suitable for diverse learning environments and student needs. This flexibility ensures that students at different stages of learning can benefit from the system's adaptive features.


Despite the benefits of IAS, these systems are still evolving in their ability to assess complex AT skills comprehensively. 
While IASs excel in providing scalable, consistent feedback, they face limitations in monitoring long-term progress, particularly when evaluating higher-order cognitive skills and more nuanced problem-solving strategies \cite{stanja2022formative}. 
These systems are particularly effective for assessing basic and intermediate levels of AT, but they struggle to capture the development of more advanced skills over time.
Ongoing research is focusing on enhancing these systems to evaluate the full spectrum of AT competencies, with a particular emphasis on tracking how students' abilities evolve. By improving their ability to measure progress over extended periods, IAS can offer more meaningful insights into student development, thus providing a more complete picture of learning outcomes in AT.

\subsection{Learner modelling and competence profiling}


IASs collect data on a learner's performance while accomplishing a task and use this data to develop a competence profile based on a predefined model of the learner's knowledge and behaviour. 
As new knowledge is gathered through the learning activity, the competence profile is continuously updated. 
This allows for more personalised and adaptive assessments, as it reflects the learner's current state and progress.

A typical modelling approach is to define a \textit{learner model} that mathematically describes the learner’s competencies, represented by a hidden variable, referred to as \textit{skills}.
These skills influence observable actions or manifest variables, which are the learner’s behaviours while solving tasks. 
An \textit{assessment rubric} provides a structured way to evaluate a student’s performance. It typically includes a list of competence components to be assessed, qualitative descriptions of observable behaviours corresponding to various competence levels, and criteria for evaluating each component. 
This rubric explicitly defines the relationship between competencies and the learner’s observable actions, helping to formalise and codify the evaluation process. 
By specifying competence levels and corresponding behaviours, the rubric effectively guides assessments and enables accurate measurement of student skills.

\subsection{Probabilistic graphical models}


Several sources of uncertainty and variability can affect the relationship between the non-observable competencies and the corresponding observable actions. 
As a result, a deterministic approach cannot accurately model this relationship. Instead, probabilistic reasoning provides a more appropriate method for translating qualitative assessment rubrics into standardised, quantitative measures of student proficiency \cite{mayo2001bayesian}.
This approach accounts for the inherent uncertainties and variability in student performance, enabling more precise evaluations of learner skills and competencies.


In learner knowledge modelling, common probabilistic approaches include Bayesian Knowledge Tracing (BKT) \cite{corbett1994knowledge}, Item Response Theory (IRT) \cite{embretson2013item}, and Bayesian Networks (BNs) \cite{koller2009probabilistic}. 
These models help estimate a learner’s knowledge state based on observable data, offering a way to track learner progress and predict future performance. 
IRT has limitations when dealing with multiple skills, particularly in modelling skill correlations. In such cases, more expressive probabilistic formalisms are required.

\subsubsection{Bayesian-based learner modelling}
Among these alternatives, BNs represent a highly interpretable option and are frequently adopted in the implementation of IASs \cite{millan2000adaptive}.  
In their comprehensive review, \citet{desmarais2012review} recognised and presented BNs as the most general approach to modelling learner skills, highlighting their versatility and effectiveness in educational contexts.
\citet{mousavinasab2021intelligent} systematically reviewed 53 papers about IASs applications from 2007 to 2017, exploring the characteristics, applications, and evaluation methods, and found that a significant proportion of the reviewed papers employed BN techniques, highlighting their widespread adoption and success in modelling learner knowledge.
More recent works continue to support the use of BNs. For instance, \citet{hooshyar2018sits} developed a system designed to help students acquire problem-solving skills in computer programming using a BN model to track and assess their progress.
Additionally, \citet{xing2021automatic} introduced an automatic assessment method for evaluating students' engineering design performance, leveraging BNs for real-time evaluation.
Other studies, such as those by \citet{wu2020student} and \citet{rodriguez2021bayesian}, further advocate for the use of BNs in constructing IASs, demonstrating the ongoing interest in these techniques to enhance educational technologies and assessments.

BNs are directed graphical models that effectively represent complex relationships between multiple skills, enabling dynamic updates to the learner's knowledge state as new information becomes available.
One of the key advantages of BNs is their ability to model causal relationships, helping to understand how different learner behaviours influence learning outcomes.
The graphical nature of these models enhances their interpretability, making them accessible to domain experts who can use them to refine and elicit student models \cite{millan2000adaptive}. 
This interpretability is especially valuable in educational settings, as it allows educators and researchers to improve the design and functionality of IASs by understanding the underlying mechanics of the model.
%

	
Despite their advantages, BNs present several challenges in their design and implementation.
Not all BNs are easy to design, and creating an effective model requires a deep understanding of BN theory.
While BNs are often considered causal models, defining them is not always straightforward due to the complexity of the causal relationships involved and the presence of hidden causes, which complicate the causal dynamics.
Additionally, obtaining the network structure and parameters can be challenging. These parameters often need to be elicited through expert knowledge or inferred from large datasets, which may not always be available. Even when the structure of the learning model is well-defined, the task of eliciting and learning BN parameters can make the computation of inferences unmanageable.
The complexity of these models increases as the number of parameters grows, particularly with the addition of more arcs to the network. A large number of parameters not only complicates the inference process but can also slow down real-time computation, which is crucial for providing timely feedback in IASs.
This complexity might discourage practitioners from adopting them, as the effort involved in eliciting expert knowledge or gathering extensive datasets can be overwhelming. 
Additionally, managing the computational demands for real-time feedback can be a significant challenge, especially when dealing with many parameters.
For these reasons, managing the number of parameters is critical in ensuring efficient processing.


To address the challenges posed by the large number of parameters in BNs, some research has focused on reducing the model's complexity. 
One such approach is the use of \emph{noisy-OR gates} for a more compact parametrisation of the Conditional Probability Tables (CPTs) within the BN \cite{Antonucci2022Flairs}. 
This method reduces the exponential complexity of parameter elicitation, transforming it from a task that scales exponentially with the number of parent skills for each observable action into a more manageable linear process \cite{pearl1988probabilistic}.

To enable real-time assessment while still capturing the necessary complexity, it is essential to create learner models that are both accurate and computationally efficient. 
To address this challenge, general approaches have been developed for translating assessment rubrics into interpretable BN-based learner models \cite{softcom}. 
These models aim to be sufficiently simple to allow for fast computation and real-time feedback while maintaining enough complexity to accurately reflect the relationships between the learner’s skills and their observable actions.
This balance ensures that the model is computationally efficient and provides meaningful insights into the learning process.
Moreover, learner models based on assessment rubrics are more accessible to teachers, who are typically more familiar with them than probabilistic graphical models. Teachers can assess student competencies through realistic scenarios specifically designed for this purpose, allowing them to compare actual student performance with a model of competence outlined by an assessment rubric \cite{dawson2017assessment}.

BN models can be extended to incorporate hierarchies of competencies to account for the acquisition order of competencies and capture complex learner behaviours. 
While rubrics focus on specific behaviours in context, they indirectly model the interactions between skills by organising them into a hierarchy. 
This allows the model to assign higher probabilities to advanced competencies without assuming that mastering a higher-level skill automatically implies mastery of all lower-level skills. 
By integrating the constraints defined by the rubric through auxiliary nodes, the model can maintain its simplicity while better reflecting the complexity of learner progression. 
This approach eliminates the assumption of direct skill acquisition and improves the accuracy of assessments \cite{softcom}.

Moreover, considering that assessment rubrics typically focus only on the competence components being assessed, referred to as \textit{target skills}, this modelling approach may result in oversimplified learner models that fail to capture the full range of factors contributing to a learner's performance.
Failures are not always due to the absence of target skills but may stem from deficiencies in other skills required for the specific task, known as \textit{supplementary skills}.
To address this, BN models can be extended to incorporate supplementary skills alongside target skills. This requires expanding approaches like the noisy-OR model to represent both disjunctive (OR) and conjunctive (AND) relationships between skills \cite{Antonucci2022Flairs}. 
While the behaviours in the assessment rubric are often mutually exclusive (OR), the combination of supplementary and target skills should be modelled together (AND) to reflect the learner's ability to complete the task more accurately.

\section{Gaps in existing research and the contribution of this study}
This section addresses the gaps identified in the existing research on CT and AT in education, particularly in the context of assessing these competencies across different age groups and educational contexts. We highlight how this study aims to address these challenges and provide a more comprehensive framework for assessing and understanding the development of AT competencies.

\subsection[Defining a competence model for CT - RQ1]{Defining a competence model for Computational Thinking - RQ1}
As identified in the literature, one of the key challenges in assessing CT is the lack of a standardised and widely accepted definition. 
This lack of a common understanding makes it difficult to establish a universal competence model that can be applied across different contexts. 
Existing competence models often focus on isolated skills, ignoring how different CT skills interconnect and evolve over time. 
Furthermore, these models tend to be designed for specific age groups, making it challenging to understand how these skills develop from early childhood through to later stages of education. 
Additionally, many models overlook the contextual influences on the development of CT skills, particularly the role of social interactions and the use of tools.

To address these gaps, our approach begins with the formalisation of a comprehensive definition of CT that takes into account both its cognitive and contextual dimensions. 
This framework will provide a more holistic understanding of CT, considering not just the individual cognitive processes but also the social and contextual influences that shape its development. 
The details of this comprehensive model are presented in \Cref{chap:CT-cube}, which explores the CT-cube framework in depth, alongside \Cref{chap:CTP}, where we further elaborate on the application of this model in practice.

\subsection[Developing assessment instruments for AT - RQ2]{Developing assessment instruments for Algorithmic Thinking - RQ2}
To effectively assess AT competencies, it is crucial to have relevant problems, activities, and instruments that are capable of capturing the complexities of AT across different age groups and contexts. 
However, as highlighted in the literature review, existing assessment tools are often constrained by the same limitations as traditional competence models. They tend to focus on narrow skills, are age-specific, and overlook the situated nature of AT, particularly the role of social interactions and artefacts in learning.

In response to these challenges, we decided to develop a new assessment instrument aligned with our competence model for CT. The first step in this process involved the development of an unplugged activity called the Cross Array Task (CAT), which allowed us to gain insights into the complexities involved in evaluating AT across different age groups and contexts. 
However, several limitations emerged from this approach, including the variability in how instructions were interpreted and coded by different human administrators, the lack of immediate feedback for students, and the time-consuming nature of the assessment.

To address these issues, we transitioned to a digital version of the CAT, known as the virtual CAT. This new version automates the interpretation and codification of instructions, providing a more standardised approach to assessment.
Furthermore, the digital platform enables real-time feedback, enhancing the reflective process for students and allowing for large-scale implementation. 

While the virtual CAT improves upon the unplugged version in several key areas, it still presents some limitations, such as producing a single, task-specific score that does not capture the full scope of a student’s abilities. This highlights the ongoing need for a more comprehensive evaluation approach, which we aim to address through the integration of a probabilistic IAS in the next phase of our research.

Details on the development and evaluation of these assessment instruments are presented in \Cref{part:assessment_instruments}.

\subsection[Designing a probabilistic IAS for AT assessment - RQ3]{Designing a probabilistic Intelligent Assessment System for Algorithmic Thinking assessment - RQ3}
The limitations of the first version of the virtual CAT assessment instruments, such as the inability to capture a full range of skills across different tasks and contexts, necessitate the use of a more dynamic, data-driven approach. 
To address these shortcomings, we propose the integration of a probabilistic IAS, which evaluates student performance and tracks interactions across tasks over time. 
This method uses probabilistic models, such as BNs, to represent a learner’s competencies and assess their progress.
BNs offer several advantages, including the ability to model complex relationships between skills and competencies. However, as the number of competencies increases, the computational complexity of these models grows exponentially. To address this challenge, we adopt Noisy-Gate BNs, which reduce the number of parameters required, making the model more efficient and manageable. 
The integration of the probabilistic IAS provides a more detailed and holistic view of a student’s competence profile, offering real-time assessment and enabling long-term tracking of progress. This approach successfully addresses the completeness issue identified in the earlier assessments and provides a more comprehensive picture of student competencies. 
The implementation of this system is detailed in \Cref{chap:method_ias}, which covers the definition of the IAS model, and \Cref{sec:casestudy}, which discusses its adaptation to the CAT case.

\subsection[Understanding AT competencies in the Swiss educational context - RQ4]{Understanding Algorithmic Thinking competencies in the Swiss educational context - RQ4}
Despite the growing interest in AT in education, there is a lack of comprehensive studies on the specific competencies required for AT across different educational contexts, particularly in Switzerland. 
This gap is significant because understanding the key AT competencies in the Swiss educational landscape will provide valuable insights into how these competencies develop across age groups and what factors, such as demographic or contextual variables, are associated with variations in these competencies.

This study aims to address this gap by conducting a thorough analysis of AT competencies in Switzerland, which could potentially be generalised to other contexts. 
The findings from this analysis will not only contribute to the understanding of how AT competencies evolve across age groups but also help refine the assessment instruments we have developed. This is crucial for tailoring our tools to different educational contexts and ensuring their effectiveness across diverse learning environments. 
The results of this analysis are presented in \Cref{part:results}.

	
	\part{Methodological frameworks}\label{part:methodological_framework}
	\chapter[A framework for the design and the assessment of CT activities]{A framework for the design and the assessment of Computational Thinking activities}\label{chap:CT-cube}

\begin{custombox}
    {The content of this chapter has been adapted from the following article with permission of all co-authors and publishers:}
    \begin{itemize}[noitemsep,nolistsep,leftmargin=.25in]
        \item Piatti, A., \textbf{Adorni, G.}, El-Hamamsy, L., Negrini, L., Assaf, D., Gambardella, L., and Mondada, F. (2022). The CT-cube: A framework for the design and the assessment of computational thinking activities. \textit{Computers in Human Behavior Reports} \cite{piatti_2022}. 
    \end{itemize}
    As an author of this publication, my contribution involved:\\
    \textit{Conceptualisation, Methodology, Validation, Formal analysis, Investigation, Resources, Data curation, Writing -- original draft \& review \& editing, Visualisation.}
\end{custombox}


\section{Summary}
This chapter addresses RQ1 by proposing a competence model for CT that considers both developmental and situated aspects of CT.
Specifically, it explores how models from mathematical thinking can be used to model the evolution of CT, contributing to a theoretical foundation for assessing CT skills across different age groups and educational contexts. 
In particular, we extend a model introduced by David Tall for the development of mathematical thinking \citep{tall2006theory, tall2013humans, tall2020three} to conceptualise the evolution of CT skills. 
Furthermore, we adopt the situated cognition framework by \citet{roth2013situated}, emphasising the context-dependent and developmental nature of CT. 
This approach allows us to account for both cognitive progression and the role of real-world learning environments in shaping CT.

As a result, we present the Computational Thinking cube (CT-cube), a novel framework that supports the design, analysis and assessment of CT activities. It extends existing CT models to provide a life-long developmental perspective, from childhood to adulthood, while incorporating the situated nature of CT, recognising that skills emerge and evolve in different educational and practical settings. 

By introducing this framework, this chapter lays the theoretical foundation for defining a CT competence model (RQ1), which later informs the design of assessment tools and activities in subsequent chapters.



\section[Definition of CT]{Definition of Computational Thinking}
\label{sec:ct_definition}

In our view, consistently with the original definition of \citet{wing2014computational}, {CT} is not strictly related to the presence of artificial agents (e.g., computers) but to that of an \textit{algorithm}. 
We consider a {CT} problem to be an activity whose solution consists of an algorithm that an artificial or human agent should perform to realise a specific task, encompassing thus also unplugged tasks which require the use of algorithms to be realised. 


Our formulation of {CT} is also inspired by Roth's basic concepts of situated cognition \citep{roth2013situated}: cognitive activities are embodied, enacted and embedded in a situated cognitive system \citep{heersmink2013taxonomy}, consisting of a social context and/or an artefactual environment. 
In other words, cognitive activities do not occur in the head of a single individual but are shared social practices, mainly based on representations and manipulation of knowledge and information through external cognitive artefacts. 
We argue that this theoretical setting corresponds better to the concrete settings in which {CT} is activated (in education but also in general), which are usually characterised by the simultaneous presence of several persons (social setting) and a rich artefactual environment.
As such, we define CT as follows:
\begin{mdframed}[style=ndef-frame]
Computational Thinking (CT) is a \textbf{situated cognitive activity}, individual or collective, {consisting of three, eventually iterative, steps}: 
\begin{enumerate}[noitemsep,nolistsep,leftmargin=.25in]
    \item[1.1] \textbf{Problem setting}: recognising, understanding, contextualising, reformulating and/or modelling a problem within a specific context in such a way that its solution can be computed;
    \item[1.2] \textbf{Algorithm}: specifying a set of rules or instructions, or conceiving and representing a procedure that should be adopted or followed by an executor -- human, artificial and/or virtual agent -- in order realise a task that solves the problem;
    \item[1.3] \textbf{Assessment}: evaluating the quality and suitability of the obtained solution with respect to the original problem.
\end{enumerate}
\end{mdframed}

From this perspective, an individual's cognitive skills can be inferred by observing him solving a contextualised (computational) task together with other individuals in a given artefactual environment. In particular, the way an individual interacts with other individuals and his contribution to the collective reasoning, respectively their choice of (cognitive) artefacts and his ability to use them correctly and efficiently in the different steps, can be observed directly and used to assess his competence level with respect to the given task.
This definition allowed us to define a theoretical framework which explicitly considers the situated nature of CT.

The sound acquisition of complex computational concepts, such as parallelisation and iteration, requires a long-term learning path similar to that followed by individuals to internalise complex mathematical concepts, like counting. 
Extending Tall's model of mathematical thinking \cite{tall2020three}, presented in Section~\ref{sec:tall_three_worlds_mathematics} to include CT, we aim to identify the cognitive tools used across three realms in both CT and mathematical thinking. We hypothesise that CT occurs in three worlds of computation, characterised by different types of cognitive and representational artefacts:
\begin{mdframed}[style=ndef-frame]
\begin{enumerate}[noitemsep,nolistsep,leftmargin=.25in]
    \item[2.1] \textbf{Embodied}: based on embodiment and perception, in which {CT} is mainly focused on the solution of contextualised problems through ecological and iconic representational cognitive artefacts; 
    \item[2.2] \textbf{Symbolic}: based on the conception, description and application of procedures and rules for solving contextualised problems through symbolic (both formal and natural) cognitive artefacts;
    \item[2.3] \textbf{Formal}: based on the creation, generalisation and representation of algorithms through formal languages in order to define structures that can be applied for problem-solving in different, even yet unknown, contexts.
\end{enumerate}
\end{mdframed}


Regarding autonomy, individuals may exhibit varying degrees of involvement in relation to their social interactions and the context of situated cognition. These levels of autonomy are shaped by the collaborative and contextual nature of the activity, where individuals' roles and contributions depend on their interactions with others and the cognitive environment:
\begin{mdframed}[style=ndef-frame]
\begin{enumerate}[noitemsep,nolistsep,leftmargin=.25in]
    \item[3.1] \textbf{Inactive role}: the individual is not active because he is not able and/or motivated to realise the requested activity.
    \item[3.3] \textbf{Non-autonomous active role}: the individual is motivated for the activity and is able to realise the activity if helped, scaffolded or guided by other members of the situated cognitive system.
    \item[3.3] \textbf{Autonomous active role}: the individual is motivated for the activity and is able to 
    realise autonomously the activity, respectively to guide and/or inspire the other members of the 
	situated cognitive system in the realisation of their activities.
\end{enumerate}
\end{mdframed}


\section{The CT-cube}\label{sec:ct_cube}

\begin{figure}[htb]						
	\centering
	\includegraphics[width=.75\linewidth]{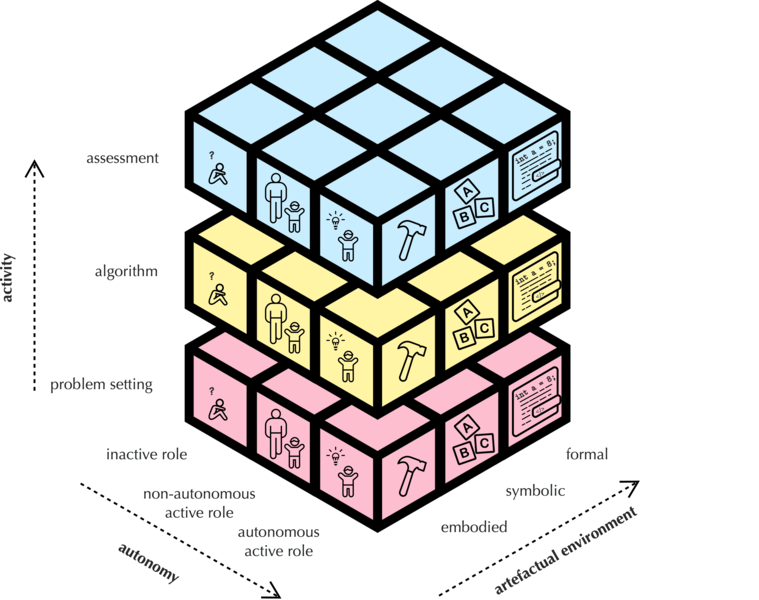}
	\mycaption{The CT-cube.}{This model considers the type of activity (problem setting, algorithm, assessment), the artefactual environment (embodied, symbolic, formal), and the autonomy (inactive role, non-autonomous active role, or autonomous active role).}\label{fig:ct-cube}
\end{figure}

We introduce the CT-cube, a three-dimensional framework that provides a comprehensive view of CT and accounts for all three aspects of the above definition.
It can be used alongside any CT model for the design, the analysis and the assessment of CT (see \Cref{fig:ct-cube}). 

At each moment of an activity, the situated cognitive system has three dimensions that apply to every individual:
\begin{mdframed}[style=ndef-frame]
\begin{itemize}[noitemsep,nolistsep,leftmargin=.25in]
    \item the type of cognitive \textbf{activity} that is being performed or that is required (problem setting, algorithm, assessment);
    \item the \textbf{artefactual environment} or the computational world in which the activities take place, characterised by the tools being used (embodied, symbolic, formal);
    \item the \textbf{autonomy} of the individual with respect to the other individuals in the situated cognitive system (inactive or passive role, active support role or supported individual activity, active leading role or autonomous individual activity).
\end{itemize}
\end{mdframed}
It is important to remark that the three dimensions are easily observable in practice, making this framework suitable for the assessment of {CT} skills.

The {CT-cube} combines two frames of {CT} that have been considered extensively in literature: cognitive {CT} and situated cognitive thinking \citep{kafai2020theory}, but while in situated {CT}, the focus is often concentrated on social and creative skills, the {CT-cube} adopts a situated cognition view, that consider the whole situated cognitive system, consisting of both the social context and the artefactual environment. 
In particular, given one or more components of {CT}, according to a given {CT} model, a task based on the selected components is designed and realised by explicitly structuring (i) the type of activity that is being performed, (ii) the artefactual environment and (iii) the social interactions and the level of autonomy, a priori and/or during the realisation of the activity. 


\subsection{Framework applications}

\paragraph{Design of CT activities}
The design of CT activities involves structuring the situated cognitive system in a way that promotes effective learning. In an educational context, this includes setting the social environment, organising the artefactual environment, and selecting appropriate problems for students to solve. 
By strategically guiding these elements, activities can be confined to a subset of the cells within the framework, ensuring a focused learning experience that aligns with desired outcomes. 
Moreover, the design process can influence the execution of the activities by actively shaping the situated cognitive system. This may involve opening or blocking access to specific cells within the framework, such as adding or removing cognitive artefacts from the environment, thus directing the participants' engagement with particular cognitive and representational tasks.

\paragraph{Assessment of CT activities}
The assessment of CT activities involves mapping the paths that individuals take within the situated cognitive system during their engagement with the task. This can be achieved through various methods, such as video analysis, to track which cells of the framework are visited over time. 
By observing the trajectory of these interactions, it is possible to assess individuals' competency levels across multiple dimensions, offering insights into both short- and long-term development of CT skills. %
This assessment process enables the comparison of individuals across different ages, educational levels, or developmental stages in relation to a specific task. 
Additionally, it allows for tracking individual progress over time, facilitating a comprehensive evaluation of the progression of CT competencies. 
The ability to observe these developmental patterns provides valuable feedback for understanding the trajectory of CT learning, from initial exposure to more advanced stages.




	\chapter[A framework for the analysis and design of CTPs]{A framework for the analysis and design of Computational Thinking Problems}\label{chap:CTP}
\begin{custombox}
    {The content of this chapter has been adapted from the following article with permission of all co-authors and publishers:}
    \begin{itemize}[noitemsep,nolistsep]
    \item \textbf{Adorni, G.}, Piatti, A., Bumbacher, E., Negrini, L., Mondada, F., Assaf, D., Mangili, F., and Gambardella, L. M. (2025). FADE-CTP: A Framework for the Analysis and Design of Educational Computational Thinking Problems. \textit{Technology,  Knowledge and Learning} \cite{adorni2023ctpframework}.
    \end{itemize}
    As an author of this publication, my contribution involved:\\
    \textit{Conceptualisation, Methodology, Validation, Formal analysis, Investigation, Resources, Data curation, Writing -- original draft \& review \& editing, Visualisation, Supervision.}
\end{custombox}


\section{Summary}

This chapter addresses RQ1 by proposing a CT framework for analysing and designing Computational Thinking Problems (CTPs), which are activities that require the application of CT to be solved. 
By defining CTPs and establishing a general set of core components and characteristics that define these problems, the chapter lays the groundwork for understanding the nature of CT competencies.
The characteristics of CTPs are connected to the CT-cube framework (see \Cref{chap:CT-cube}), particularly focusing on aspects of situated cognition, which highlight the importance of the context and environment in which CT skills are applied. 
 
The chapter discusses how the design and complexity of CTPs impact the skills that students can cultivate, offering insights into designing activities that are both scalable and effective for assessing AT competencies, providing a foundation for RQ2.

Next, a hierarchical structure of CT competencies, which organises them in a way that reflects their interrelated nature. This section supports RQ1 by refining the conceptual framework of CT competencies, which is critical for developing valid assessments.

Finally,the chapter concludes by providing a template for describing the profile of a CTP, which is based on the relationship between the characteristics of a CTP and CT competencies. 
This template can be adapted to create valid, context-sensitive assessments, thereby contributing to RQ2 and offering guidance for developing tools to assess AT competencies in educational contexts.

\section[Definition of CTP]{Definition of Computational Thinking Problem}\label{sec:ctp}

%

Our definition of CTP is grounded in the CT-cube theoretical framework (see \Cref{chap:CT-cube}), which integrates foundational concepts of CT from \citet{wing2006computational} with situated theories of learning by \citet{roth2013situated} and  \citet{heersmink2013taxonomy}, framing CT as a dynamic, adaptive process embedded in real-world contexts, rather than a fixed set of competencies.
Building on this integrated perspective, we define a CTP as a task that engages learners in applying CT skills to derive solutions within realistic environments, emphasising the influence of physical and social contexts on CT activities and the role of cognitive artefacts in supporting problem-solving.
%
%
The interplay of the type of activity, the artefactual environment, and the learner's autonomy, key dimensions of the CT-cube, collectively determines the characteristics of the CTP, shaping the competencies developed during engagement and influencing both the learning experience and the effectiveness of problem-solving approaches.

\subsection{{Components}}
\begin{figure}[!ht]
	\centering
	\includegraphics[width=.65\columnwidth]{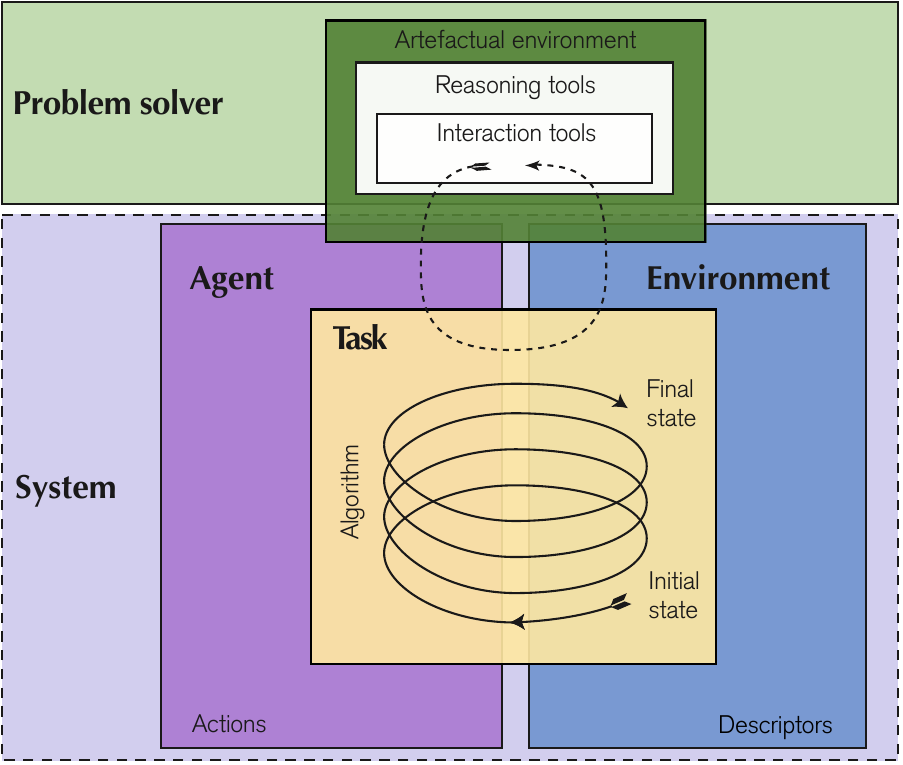}
	\mycaption{Components of a CTP.}
	{CTPs include (1) the problem solver (in green) characterised by the artefactual environment, i.e., the set of reasoning and interaction tools, (2) the system, which consists of an environment with its descriptors (in blue) and an agent with its actions (in violet), and (3) the task (in yellow) characterised by the set of initial states, algorithms and final states.}
	\label{fig:framework}
\end{figure}

We identified several components that constitute a CTP, illustrated in \Cref{fig:framework}.

The \textit{system} comprises the environment and the agent.
The \textit{environment} is a physical and/or a virtual external space, characterised by one or more variables, called ``descriptors'', which may change over time according to the dynamics of this space.
The \textit{agent} is a human, robotic or virtual being {that interacts with the environment by} performing ``actions'' to change the value of its descriptors and, therefore, alter the state of the environment.
An ``algorithm'' is a finite set of instructions an agent should follow to perform actions in the environment to solve the task.
Algorithms for different agents can take various forms: code for a virtual agent, behaviour for a robot, or a verbal or written set of instructions for a human.

The \textit{problem solver} is a human or group of people who can solve tasks that require the use of algorithms, such as designing, implementing, or communicating them to an agent to change the state of an environment. 
They have access to \textit{reasoning tools}, which are cognitive artefacts {that assist in thinking} about the task, such as whiteboards used to organise ideas and understand the logic of a problem or solution.
Some of these tools, known as \textit{interaction tools}, also allow the problem solver to interface with the system.
For example, a programming platform may serve as both a reasoning tool, enabling the problem solver to plan and design code, and an interaction tool, facilitating the execution of the algorithm and allowing the observation of its effect on the system.
Collectively, these tools form the \textit{artefactual environment}, which according to our definition, provided in \Cref{chap:CT-cube} and the model of the three worlds of mathematics by \citet{tall2006theory,tall2013humans,tall2020three}, can also be categorised in: ``embodied'', iconic representational or ecological tools, based on sensory perception and embodiment; ``symbolic'' tools, used to conceive and apply procedures and rules; and ``formal'' tools,  used to create, generalise and represent structures.

The \textit{task} is the activity that the problem solver performs to find one or more solutions to a CTP.
A solution is a combination of ``initial states'', ``algorithms'', and ``final states'' that meet the system's requirements for a particular environment, with its set of states, and a given agent, with its set of algorithms. 
The initial state is the starting configuration of the environment, while the final state is the state of the environment after the algorithm is performed.
For a solution to be valid, the algorithm must be executed on the initial state and then produce the final state.
{Each element that composes a task (initial state, algorithm, final states) can be ``given'' or is ``to be found''. 
	Based on the number and the epistemic nature of elements to be found, it is possible to divide tasks into six types.
	Those with a single objective are: 
	(1) \emph{find the initial state}: given the final state and the algorithm that produced it, the problem solver must infer the initial state on which the algorithm was applied;
	(2) \emph{find the algorithm}: given the initial and the final states, the problem solver must devise and describe an algorithm, or a part of it, that the agent can execute to transform the system from the initial to the final state;
	(3) \emph{find the final state}: given the initial state and an algorithm, the problem solver must derive the final state.
	Pairs of single-objective tasks form those with multiple objectives:
	(4) \emph{creation act}: a combination of find the algorithm and find the final state; 
	(5) \emph{application act}: a combination find the initial state and find the final state; 
	(6) \emph{project act}: a combination find the initial state and find the algorithm. 
}

\subsection{{Characteristics}}

\begin{figure}[!ht]
	\centering
	\includegraphics[width=\textwidth]{TKNL/framework-characteristics\_old.pdf}
	\mycaption{Template for defining components and characteristics of a CTP.}
	{The same colour scheme as in \Cref{fig:framework} is applied.}
	\label{fig:framework-components}
\end{figure}

{After defining the components of CTPs, we identify key characteristics that further clarify their nature. These attributes, along with their role, are illustrated in} \Cref{fig:framework-components}, {which serves as a template for CTP analysis.  
}

\begin{mdframed}[style=ndef-frame]
	\textbf{Problem domain} The category of an activity, determined by the nature of the agent and of the environment. 
\end{mdframed}
Three main categories of domains are commonly recognised in cognitive tasks, including: ``unplugged'' activities, which involve a human agent and a physical environment; ``robotic'' activities, in which the agent is a robot, and the environment is physical; and ``virtual'' activities, where both agent and environment are virtual, such as in a simulated scenario. 


\begin{mdframed}[style=ndef-frame]
	\textbf{Tool functionalities}
	The {artefactual environment's capabilities} enable the problem solver to {construct the algorithm}.
\end{mdframed}
{
	The functionalities we included in this categorisation are tailored for beginner-level CT education to introduce foundational algorithmic concepts, such as} ``variables'', ``operators'', ``sequences'', ``repetitions'', ``conditionals'', ``functions'', ``parallelism'' and ``events''. 
For example, a symbolic artefact, such as a block-based programming platform, may have many functionalities, such as sequences, repetitions, conditionals, etc. 
In contrast, the programming interface may have limited functionalities during a robotic activity, for example, it could only consent using operators (like moving forward) or events. 

\begin{mdframed}[style=ndef-frame]
	\textbf{System resettability}
	The property of a system to be restored to its initial state, either through the direct intervention of the problem solver on the system or indirectly via the reversibility of actions within the system.
\end{mdframed}
{Resettability can be ``direct'' when the problem solver can directly intervene on the system by manually returning the robot to its starting position and restoring the environment; or ``indirect'' when the problem solver can use a system-provided reset mechanism.
	If neither option is available, the system is ``non-resettable'', for example, when the problem solver can move the robot back to the starting position only through an algorithm, but any environmental alterations remain irreversible.}



\begin{mdframed}[style=ndef-frame]
	\textbf{System observability}
	The property of a system that allows the problem solver to observe the effects of the agent's actions in the environment and their impact on its state. 
\end{mdframed}
Systems can be classified as ``totally observable'' if every action and their effects are visible, {e.g., if the problem solver and the robot are in the same room, and all changes to the system state are visible in real-time;} ``partially observable'' when only the aggregate effects of a set of actions are visible, {e.g., if the problem solver can enter the room only at the end of the task and observe the final state of the system, without seeing the actions that led to it}; or ``not observable'', if none of the agent's actions or their results are visible, {e.g., if the problem solver cannot enter the room and must infer the system state from other information, such as sensor data.}
{It is worth noting that, in the unplugged domain, problem solver and agent can be the same entity. When they overlap, the system is totally observable.}


\begin{mdframed}[style=ndef-frame]
	\textbf{Task cardinality}
	The {relationship} between the number of given elements {and those} to be found to solve a task. 
\end{mdframed}
{CTPs can present three types of cardinality:} ``one-to-one'', ``many-to-one'' or ``many-to-many''. 
{In a one-to-one task, each provided element corresponds directly to one element to be found, e.g., if a single initial and a final states are given, a single algorithm has to be found. 
	In a many-to-one task, multiple given elements are intended to be resolved by a single solution element, e.g., if several initial states are provided, and the goal is to find a single algorithm that can transform each of these initial states into the same final state.
	In a many-to-many task, both the provided and target elements are multiple, requiring the problem solver to find various solutions. 
	For example, a task might provide multiple initial states and a single final state, and the solver would need to identify several algorithms, each capable of transforming one or more of the initial states into the specified final state.
	This type of task can be traced back to multiple many-to-one tasks.}

\begin{mdframed}[style=ndef-frame]
	\textbf{Task explicitness}
	{The level of detail in the presentation of the task's elements.} 
\end{mdframed}
{In a CTP, the given elements can be ``explicit'' if they are directly provided and immediately usable in the problem-solving process, or ``implicit'' if they are expressed with constraints that require further interpretation to be understood.
	For example, in a task where the problem solver must find the algorithm for a robot to turn on its lights after finding a ball, the ball's position can be given explicitly (e.g., coordinates) or implicitly (e.g., in the playground).}

\begin{mdframed}[style=ndef-frame]
	\textbf{Task constraints}
	The limitations or specific requirements that {the task elements to be found must meet to consider the solution valid.}
\end{mdframed}
{In a CTP, the elements to be found can be} ``unconstrained'' if they can be freely selected among all possible states and algorithms, with no limitations or specific requirements that need to be met to consider the solution valid; or ``constrained'' if they must belong to a restricted subset of states or algorithms.
{Referring to the same example presented to explain the task explicitness characteristic,} the algorithm to be found can be unconstrained if the robot can perform any action to find the ball ({e.g., moving randomly, using sensors, etc.}) or constrained if the programming platform limits the {robot's actions (e.g., restrict movement to specific directions, using only specific sensors).}

\begin{mdframed}[style=ndef-frame]
	\textbf{Algorithm representation}
	The mean by which an algorithm is given.
\end{mdframed}
An algorithm is considered ``manifest'' if directly expressed, while ``latent'' if not stated but should be inferred by the problem solver.
Manifest algorithms can be ``written'' if represented by an external {and persistently, like the code in a programming language}, or ``not written'' if communicated verbally or through other non-permanent means.

\section[Catalogue of CT competencies]{Catalogue of Computational Thinking competencies}\label{sec:skills}

\begin{figure*}[!h]
	\centering
	\includegraphics[width=.9\textwidth]{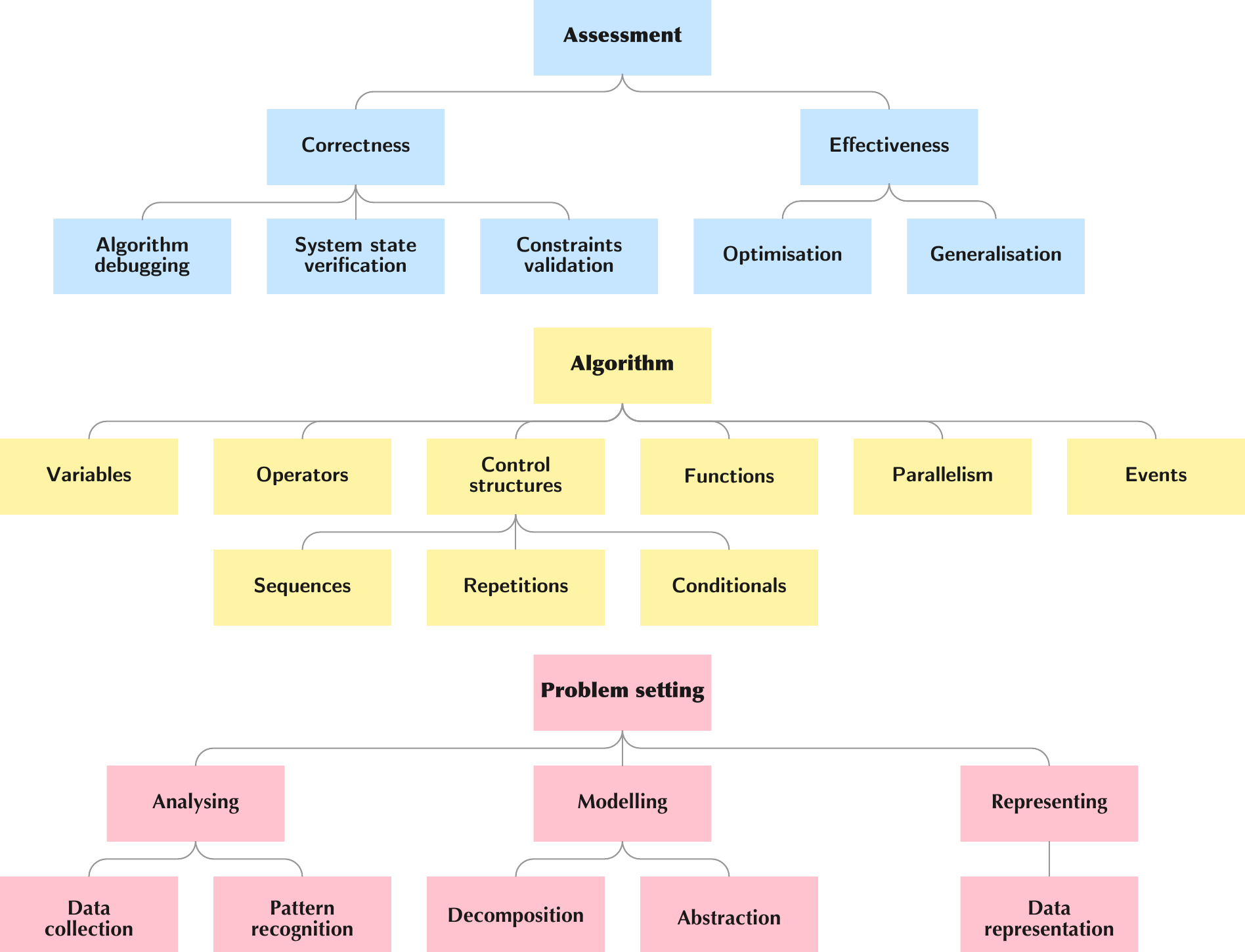}
	\mycaption{Taxonomy of CT competencies.}{
		The overall structure is based on CT-cube framework. The sub-skills are derived from validated CT models by \citet{brennan2012new,weintrop2016defining,shute2017demystifying}.
		The same colour scheme as in \Cref{fig:ct-cube} is applied.
	}
	\label{fig:taxonomy}
\end{figure*}
Alongside the definition of CTPs, their components, and their characteristics, we have developed a catalogue of CT competencies that are fundamental abilities students need to solve CTPs effectively. 
This integration is primarily theoretical, synthesising existing competencies frameworks to propose a structured perspective on the relationship between CT and CTPs.
To ensure a through approach, we drew from multiple state-of-the-art competency models and frameworks.
Our selection was informed by the literature reviews of \citet{tikva_mapping_2021} and \citet{bocconi2016developing,bocconi2022reviewing}, which provide a comprehensive overview of CT skills in compulsory education.

A key framework that guided the development of our catalogue is that of \citet{brennan2012new}, which categorises CT skills into {computational concepts}, {practices}, and {perspectives}. 
While this model is commonly referenced in literature, it primarily focuses on activities conducted in digital environments, such as programming and software development.
Although the CT aspects covered are essential, they do not encompass the broader range of CTPs we explore, including hands-on robotics and unplugged activities. To bridge this gap, we extended the framework to incorporate competencies applicable across these varied contexts, such as elements from the STEM taxonomy proposed by \citet{weintrop2016defining}, which includes data practices, modelling and simulation practices, computational problem-solving, and systems thinking practices.
We also referenced the work of \citet{shute2017demystifying}, which expands on the previous frameworks by offering a broader, more adaptable set of CT competencies with a focus on cognitive processes and applicability across diverse contexts.

To provide a comprehensive framework for CT competencies, we organised our catalogue into a hierarchy of skills and sub-skills (see \Cref{fig:taxonomy}).
This structure clarifies the relationships among competencies, making it easier to identify specific skills within broader categories and supporting a more precise and targeted approach for educators and researchers working with CT skill development and assessment.
The first layer of competencies (see \Cref{tab:h1} ) is based on the activity dimension of the CT-cube, while the lower layers (see \Cref{tab:h21,tab:h22,tab:h23}) are based on the frameworks of \citet{brennan2012new,weintrop2016defining,shute2017demystifying}.
While this framework provides a structured theoretical foundation, further empirical validation is necessary to confirm the relationships proposed.
\begin{table}[!ht]
	\renewcommand{\arraystretch}{1.2}
	\setlength{\tabcolsep}{2pt}
	\footnotesize
	\centering
	\begin{threeparttable}
		\mycaptiontitle{Core skills definition (level 1).}\label{tab:h1}
		\begin{tabular}{m{3.2cm}m{10.5cm}}
			\toprule
			{\textbf{Competence} (level 1)} & \textbf{Definition} \\ 
			\midrule
			{\textit{Problem setting}}\tnote{a} & 
			Recognise, understand, reformulate or model a {CTP} and its components so that its solution can be computed.
			\\ 
			{\textit{Algorithm}}\tnote{b} & 
			Conceive and represent a set of agent's actions that should be executed by a human, artificial or virtual agent to solve the task.
			\\ 
			{\textit{Assessment}}\tnote{c} & 
			Evaluate the quality and validity of the solution in relation to the original task.\\
			\bottomrule
		\end{tabular}
		\begin{tablenotes}
			\item[] These skills are based on the values of the activity dimension of our CT-cube framework.
			\item[a] See \Cref{tab:h21} for ``problem setting'' sub-competencies. 
			\item[b] See \Cref{tab:h22} for ``algorithm'' sub-competencies. 
			\item[c] See \Cref{tab:h23} for ``assessment'' sub-competencies. 
		\end{tablenotes}
	\end{threeparttable}
\end{table}

\begin{table}[!hb]
\renewcommand{\arraystretch}{1.2}
\setlength{\tabcolsep}{2pt}
\footnotesize
\centering
\begin{threeparttable}  
	\mycaptiontitle{Problem setting sub-skills definition (level 2 and 3).}\label{tab:h21}
\begin{tabular}{m{2cm}m{2.5cm}m{9cm}}
    \toprule
   \textbf{Competence} \newline (level 2)  & \textbf{Competence} \newline (level 3) & \textbf{Definition} \\ 
    \midrule
    {\textit{Analysing}} & & Collect, examine and interpret data about the system: environment descriptors and agent actions. \\ 
    \cdashline{2-3}[1pt/2pt]
    &{\textit{Data collection}} & Gather details about the system. \\  
    &{\textit{Pattern recognition}} & Identify similarities, trends, ideas and structures within the system.\\ 
    \hdashline[4pt/2pt]
    {\textit{Modelling}} & & Restructure, clean and update knowledge about the system.\\ 
    \cdashline{2-3}[1pt/2pt]
    &{\textit{Decomposition}} & Divide the original task into sub-tasks that are easier to be solved.\\ 
    &{\textit{Abstraction}} &Simplify the original task, focus on key concepts and omit unimportant ones.\\ 
    \hdashline[4pt/2pt]
    {\textit{Representing}} & & Illustrate or communicate information about system and task. \\ 
    \bottomrule
\end{tabular}
\begin{tablenotes}
	\item[] The skills listed are based on leading-edge competence models 
	( \citet{thalheim2000database,barr,wing2011research,brennan2012new,selby2013computational,selby2014can,csizmadia2015computational,angeli2016k,bocconi2016developing,weintrop2016defining,shute2017demystifying}).
	
\end{tablenotes}
\end{threeparttable}
\end{table}

\begin{table}[ht]
\renewcommand{\arraystretch}{1.2}
\setlength{\tabcolsep}{2pt}
\footnotesize
\centering
\begin{threeparttable}
\mycaptiontitle{Algorithm sub-skills definition (level 2 and 3).}
\label{tab:h22}
\begin{tabular}{m{2cm}m{2.5cm}m{9cm}}
	\toprule
	 \textbf{Competence} \newline (level 2)  & \textbf{Competence} \newline (level 3) & \textbf{Definition} \\ 
    \midrule
    {\textit{Variables}} &
    & Entity that stores values about the system or intermediate data.
    \\ 
    \hdashline[4pt/2pt]
    {\textit{Operators}} &
    & Mathematical operators (such as addition ($+$), subtraction ($-$) etc.), logical symbols (such as and (\&), or (|), and not (!)) or for comparison (such as equal to (==), greater than (>), and less than (<)), or even specific commands or actions (such as ``turn left'' or ``go straight'').
    \\ 
    \hdashline[4pt/2pt]
    {\textit{Control \newline structures}} &
    & Statements that define the agent actions flow's direction, such as sequential, repetitive, or conditional.
    \\ \cdashline{2-3}[1pt/2pt]
    &{\textit{Sequences}} 
    & Linear succession of agent actions. 
    \\ 
    &{\textit{Repetitions}} 
    & Iterative agent actions.
    \\ 
    &{\textit{Conditionals}} 
    & Agent actions dependent on conditions. 
    \\ \hdashline[4pt/2pt]
    {\textit{Functions}} &
    & Set of reusable agent actions producing a result for a specific sub-task.
    \\ \hdashline[4pt/2pt]
    {\textit{Parallelism}} &
    & Simultaneous agent actions.
    \\ \hdashline[4pt/2pt]
    {\textit{Events}} &
    & Variations in the environment descriptors that trigger the execution of agent actions.\\ 
    \bottomrule
\end{tabular}
\begin{tablenotes}
	\item[] The skills listed are based on leading-edge competence models
		(\citet{brennan2012new,bocconi2016developing,bocconi2022reviewing,shute2017demystifying,rodriguez2020computational,cui2021interplay}).
\end{tablenotes}
\end{threeparttable}
\end{table}

\begin{table}[hb]
\renewcommand{\arraystretch}{1.2}
\setlength{\tabcolsep}{2pt}
\footnotesize
\centering
\begin{threeparttable}
\mycaptiontitle{Assessment sub-skills definition (level 2 and 3).}\label{tab:h23}
\begin{tabular}{m{2cm}m{2.5cm}m{9cm}}
	\toprule
	 \textbf{Competence} \newline (level 2)  & \textbf{Competence} \newline (level 3) & \textbf{Definition} \\ 
    \midrule
    {\textit{Correctness}} && Assess whether the task solution is correct. \\ \cdashline{2-3}[1pt/2pt]
    &{\textit{Algorithm \newline debugging}} &  Evaluate whether the algorithm is correct, identifying errors and fixing bugs that prevent it from functioning correctly. \\ 
    &{\textit{System states \newline verification}} & Evaluate whether the system is in the expected state, detecting and solving potential issues.  \\
    &{\textit{Constraints \newline validation}} & Evaluate whether the solution satisfies the constraints established for the system and the algorithm, looking for and correcting eventual problems. \\\hdashline[4pt/2pt]
    {\textit{Effectiveness}} && Assess how effective is the task solution. \\ \cdashline{2-3}[1pt/2pt]
    &{\textit{Optimisations}} & Evaluate whether the solution meets the standards in a timely and resource-efficient manner, and eventually identify ways to optimise the performance.\\
    &{\textit{Generalisation}} & Formulate the task solution in such a way that it can be reused or applied to different situations.\\
    \bottomrule
\end{tabular}
\begin{tablenotes}
	\item[] The skills listed are based on leading-edge competence models
		(\citet{brennan2012new,bocconi2016developing,weintrop2016defining,shute2017demystifying}).
\end{tablenotes}
\end{threeparttable}
\end{table}

\FloatBarrier

\section[Profiling CTPs]{{Profiling Computational Thinking Problems}}\label{sec:link}
This section introduces our framework for profiling CTPs, building on our earlier discussions of CTP components and characteristics.
We defined specific relationships between CTP characteristics and CT competencies, outlining, for each skill, the set of characteristics essential for their development and those that inhibit it. All required characteristics must be present to develop a specific competence, and none may be inhibitory. 

While identifying the required and absent characteristics allows us to determine which competencies can technically be developed, this perspective is somewhat limited. 
Therefore, we also included characteristics that can enhance and support skill development beyond basic requirements, strengthening the overall framework for competency development.
For example, a manifest written algorithm can significantly facilitate the development of algorithmic skills at different levels of abstraction, such as repetitions, by helping learners understand how loops work, recognise them and practice their application, ultimately leading to assimilation \citep{bloom1956taxonomy,CTF}.

{The relationships between CTP characteristics and CT competencies are examined in detail in \Cref{app:tknl_appA_connections_supplementary},} where we outline (i) how various CTP characteristics influence the development of CT competencies and (ii) which CT skills are more frequently developed and/or employed when solving CTPs with specific traits. 
While this integration is theoretically informed by existing research, further empirical validation is necessary to refine these relationships. 

\begin{table*}[!ht]
	\centering
	\mycaption{Profiling template for CTPs.}{
		Rows represent the CT competencies in our catalogue and columns the CTPs characteristics, including tools' functionalities, system's property and task traits. 
		{The same colour scheme as in} \Cref{fig:ct-cube,fig:framework} {is applied}.
	}
	\label{tab:static-framework}
	\includegraphics[width=\textwidth]{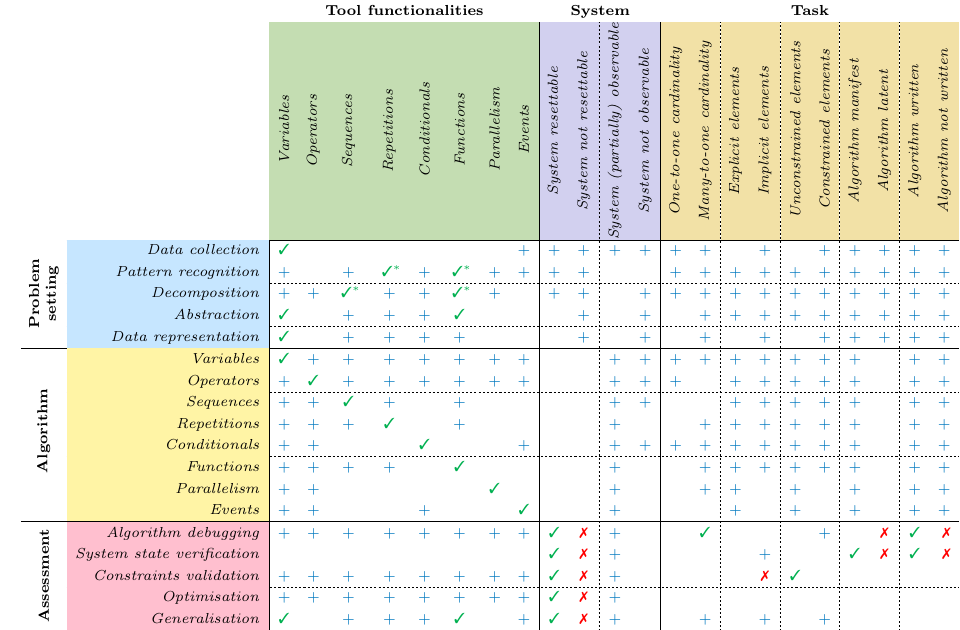}
	\\\vspace{1em}
	\footnotesize
	\begin{minipage}{\linewidth}
		{{\color[HTML]{00B050}\ding{51}} indicates that the characteristics is required for the development of the competence.}
		
		{{\color[HTML]{00B050}\ding{51}$^*$} indicates that at least one of several characteristics in a group is required for the development of the competence.
		}
		
		{{\color[HTML]{FF0000}\ding{55}} indicates that the characteristic prevents the development of the competence.}
		
		{{\color[HTML]{0078BF}$+$} indicates that the characteristic can support the development of the competence.}
		
		{Blank cells indicate that the characteristic is irrelevant to the development of the competence.}
	\end{minipage}
\end{table*}
\Cref{tab:static-framework} illustrates our framework and serves as a template for analysing and designing CTPs by creating a profile of each specific CTP.

\FloatBarrier
\subsection{Framework applications}

\paragraph{Analysis of CTPs}
The analysis of CTPs involves identifying the characteristics that define a task, enabling the determination of which competencies can be developed or assessed based on the presence of necessary traits and the absence of inhibitory ones. By evaluating the structural and contextual elements of a CTP, this methodology provides a detailed understanding of how different tasks align with specific CT competencies, aiding both in assessment and identifying areas for improvement.

\paragraph{Design of CTPs}
The design of CTPs involves selecting the specific competencies to be targeted for development and/or assessment. Once the desired competencies are identified, it is essential to determine the necessary characteristics to include in the task, ensuring that any elements that could inhibit skill development are excluded. Furthermore, supportive characteristics that may not be strictly required but could enhance the learning process can be selectively integrated, enriching the CTP and making it more effective in fostering the intended competencies. This approach ensures that the design of CTPs is purposeful, targeted, and conducive to achieving the desired outcomes.


	\chapter[Probabilistic modelling for IASs]{Probabilistic modelling for Intelligent Assessment Systems}\label{chap:method_ias}

\begin{custombox}
	{The content of this chapter has been adapted from the following articles with permission of all co-authors and publishers:}
	\begin{itemize}[noitemsep,nolistsep,leftmargin=.25in]
		\item Antonucci, A., Mangili, F., Bonesana, C., and \textbf{Adorni, G.} (2022). Intelligent Tutoring Systems by Bayesian Nets with Noisy Gates. The \textit{International FLAIRS Conference Proceedings} \cite{antonucci2021}. 
		
		\item Mangili, F., \textbf{Adorni, G.}, Piatti, A., Bonesana, C., and Antonucci, A. (2022). Modelling Assessment Rubrics through Bayesian Networks: a Pragmatic Approach. In \textit{2022 International Conference on Software, Telecommunications and Computer Networks} (SoftCOM) \cite{softcom}. 
		
		\item \textbf{Adorni, G.}, Mangili, F., Piatti, A., Bonesana, C., and Antonucci, A. (2023a). Rubric-based Learner Modelling via Noisy Gates Bayesian Networks for Computational Thinking Skills Assessment. \textit{Journal of Communications Software and Systems} \cite{adorni2023rubric}. 
		
		
	\end{itemize}
	As an author of these publications, my contribution involved:\\
	\textit{Conceptualisation, Methodology, Validation, Formal analysis, Investigation, Resources, Data curation, Writing -- original draft \& review \& editing, Visualisation.}
\end{custombox}

\section{Summary}

This chapter explores probabilistic models, particularly BNs, for developing IAS, addressing RQ3.
It outlines how BNs with noisy gates can model student skills, comparing these models to Bayesian Knowledge Tracing (BKT). 
The discussion provides insights into the integration of probabilistic methods into IAS, which can help improve the precision and adaptability of assessments across various educational contexts.

Furthermore, the chapter examines how assessment rubrics can be modelled with BNs to evaluate competencies, including the ordering of skills and the inclusion of supplementary competencies to offer a more comprehensive assessment of student performance. This directly supports RQ3 by presenting a probabilistic framework for assessing AT competencies in diverse contexts, enhancing the ability to model complex relationships between skills.

\section[Bayesian Networks (BNs) and learner models]{Bayesian Networks and learner models}\label{sec:bn_learner_model}
BNs can be employed to represent learner skills and observable actions in the context of IASs. 
Key notations used include uppercase letters to denote variables (e.g., $X$, $Y$), lowercase letters for states (e.g., $y_E$), bold letters for sets of variables (e.g., $\mathbf{X}$, $\mathbf{Y}$). 

The structure of a BN over a set of variables is described by a directed acyclic graph $\mathcal{G}$ whose nodes are in one-to-one correspondence with the variables in the set. 
We call parents of a variable $X$, according to $\mathcal{G}$, all the variables are connected directly with $X$ with an arc pointing to it. 
Learner models usually include a set of $n$ latent (i.e., hidden) variables $\mathbf{X}:=(X_1,\ldots, X_n)$, henceforward referred to as \emph{skill nodes}, describing the competence profile of the learner and some $m$ manifest variables $\mathbf{Y}:=(Y_1,\ldots, Y_m)$, hereafter called \emph{answer nodes}, describing the observable actions implemented by the learner to answer each specific task. 

While the orientation of a BN arc may not necessarily reflect a causal interpretation, in practice, graphs that implement an IAS often have a bipartite structure that includes arcs from the skills to the questions but not vice versa. 
This means that each question receives incoming arcs from the relevant skills for answering the question. 
By adopting this bipartite structure, we can model assessment rubrics more suitably. 
This results in a set of simple and interpretable relations that model how the presence or absence of a specific competence directly affects the learner's behaviour when solving tasks that require such competence.
For this purpose, we only consider the case of binary skill nodes that take the value of 1 or the ``true'' state, indicating whether the pupil possesses the skill. 
Additionally, we use binary answer nodes that denote a correct answer or determine whether the pupil has shown the desired behaviour when solving the task.

\begin{figure}[htbp]
	\centerline{
		\begin{tikzpicture}
			\node[nodo] (s1)  at (0,0) {\footnotesize ($X_1$) Scratch};
			\node[nodo] (s2)  at (5,0) {\footnotesize ($X_2$) Python};
			\node[Qnodo] (q1)  at (0.,-2) {\footnotesize ($Y_1$) Build~a maze game};
			\node[Qnodo] (q2)  at (5,-2) {\footnotesize ($Y_2$) Build~a statistical model};
			\draw[arco] (s1) -- (q1);
			\draw[arco] (s2) -- (q1);
			\draw[arco] (s2) -- (q2);
	\end{tikzpicture}}
	\mycaptiontitle{Example of BN-based learner model.}
	\label{fig:bn}
\end{figure}
The relations of a BN-based learner model (skills and questions) can be graphically depicted as in the example of \Cref{fig:bn}. 
The answer nodes describe whether the learner has been able or not to program, for example, a maze game ($Y_1$) or a statistical model ($Y_2$). 
The skill nodes represent the ability to build this program using a block-based programming language such as Scratch ($X_1$) or a text-based programming language such as Python ($X_2$). 
The second skill can be applied to answer both questions and therefore, $X_2$ is a parent node for both answer nodes $Y_1$ and $Y_2$. 
Instead, the first skill can be used to answer just the first question, and therefore, there is no direct arc from $X_1$ to $Y_2$.

Once the graph $\mathcal{G}$ structuring the BN is established, the definition of the BN over the $n+m$ variables of the network $\mathbf{V} := (V_1, V_2,\dots, V_{n+m})$, including both skills ($\mathbf{X}$) and answers ($\mathbf{Y}$), consists in a collection of Conditional Probability Tables (CPTs) giving the probabilities
$P(Y_i = 1|\mathrm{Pa}(Y_i))$ that $Y_i$ takes value one given all possible joint states of its parent nodes $\mathrm{Pa}(Y_i)$. 
Let $\mathbf{V}$ take values in $\Omega_{\mathbf{V}}$, the independence relations imposed from $\mathcal{G}$ by the \emph{Markov condition}, i.e., the fact that each node is assumed to be independent of its non-descendants non-parents given its parents, induce a joint probability mass function over the BN variables that factorises as follows \cite{koller2009probabilistic}:
\begin{equation}\label{eq:joint}
P(\mathbf{V}=\mathbf{v}) = \prod_{v \in \mathbf{v}} P(v|\mathrm{pa}(V))\,,
\end{equation}
where $\mathbf{v}=(v_1, v_2,\dots, v_{n+m})$ represents a given joint state of the variables in $\mathbf{V}$.
BN inference consists of the computation of queries based on \Cref{eq:joint}. In particular, we are interested in \emph{updating} tasks consisting in the computation of the marginal posterior probability mass function for a single skill node $X_q \in \mathbf{X}$ given the observed state $\mathbf{y}_E$ of the answer nodes $\mathbf{Y}_E \subseteq \mathbf{Y}$:
\begin{equation}\label{eq:updating}
P(x_q|\mathbf{y}_E) = \frac{\sum_{\mathbf{v}\in \Omega_{\mathbf{V}|(x_q,\mathbf{y}_E)}} \prod_{v\in \mathbf{v}} P(v|\mathrm{pa}(V))}{\sum_{\mathbf{v}\in \Omega_{\mathbf{V}|\mathbf{y}_E}} \prod_{v\in \mathbf{v}} P(v|\mathrm{pa}(V))}\,,
\end{equation}
where $\Omega_{\mathbf{V}|\mathbf{v}'}:=\{\mathbf{v}: v_i = v'_i ~\forall ~v'_i \in \bf{v}'\}$. 

According to the above model, multiple parent skills may be relevant to the same answer. 
The challenges in the existing model are primarily twofold. 
First, the elicitation process involves an exponential number of parameters due to the potential involvement of multiple parent skills in determining a single answer. Assuming that the answer node $Y_j$ has $n$ parent skills, this results in $2^n$ parameters to be elicited by experts. This high number of parameters to be elicited by experts might discourage practitioners from using these tools in their applications because of a too-demanding elicitation process when many skills are affecting the answer to a question.
Second, the inference task in Bayesian networks, as described in \Cref{eq:updating}, is NP-hard in the general case and highly dependent on the complex graph topology, which is in practice exponential in the graph tree-width \cite{koller2009probabilistic}, making it computationally demanding, especially for models with a high number of parent nodes (high maximum \emph{indegree}), as this involves both large CPTs and tree-width.

To address these issues, in the next section, we propose a solution that introduces noisy gates, specifically the noisy-OR gate \citep{pearl1988probabilistic}, to reduce the number of parameters required for model elicitation, shifting from an exponential to a linear relationship with the number of relevant skills. 
This reduction in parameters streamlines the expert elicitation process and mitigates the potential discouragement of IAs practitioners. Additionally, noisy gates enhance the efficiency of inference tasks, enabling faster computations.
Furthermore, the model recognises that a disjunctive relation among skills may not always be applicable in practice. To accommodate diverse scenarios, using more general logical functions, such as conjunctive relations, is considered, providing flexibility and realism to the learner model while benefiting from reduced parameter complexity in elicitation and inference.


\section{Noisy gates}\label{sec:noisygates}

\subsection{Disjunctive gates}\label{sec:noisyor}
The noisy-OR network induces the following CPT between the $n$ parent skill nodes $\bf{X} = (X_1,\dots,X_n)$ and the observable answer node $Y_j$ \citep{pearl1988probabilistic}:
\begin{equation}\label{eq:noisy}
P(Y_j=0|\bf{X} = (x_1,\ldots,x_n)) = \prod_{i=1}^n (\mathbb{I}_{x_i=0}+\lambda_i\mathbb{I}_{x_i=1})\,,
\end{equation}
where $\mathbb{I}_A$ is the indicator function returning one if $A$ is true and zero otherwise. 
The second term $\lambda_i\mathbb{I}_{x_i=1}$ represents the noise as it introduces the possibility that a skill $X_i$ that the student possesses is not expressed in task $Y_j$ (this phenomenon is also called \textit{slip} elsewhere in this work). The value of $\lambda_i$ implying the biggest uncertainty associated with the task-skill pair ($Y_j$, $X_i$)  is 0.5, whereas the value $\lambda_i = 0$ models the certainty that skill $X_i$, whenever present, will be expressed in solving task $Y_i$ and, vice versa, $\lambda_i = 1$  model the fact that $X_i$ cannot be expressed in task $Y_j$.

\begin{figure}[hbt]
\footnotesize
\centering
\begin{tikzpicture}
\node[n] (xx1)  at (0,1) {${ X_1 }$};
\node[n] (xx2)  at (1.5,1) {${ X_2}$};
\node[] (xx3)  at (3,1) {${\ldots}$};
\node[n] (xx4)  at (4.5,1) {${ X_n}$};
\node[n] (x1)  at (0,0) {${ X_{1,j}'}$};
\node[n] (x2)  at (1.5,0) {${ X_{2,j}'}$};
\node[] (x3)  at (3,0) {${ \ldots}$};
\node[n] (x4)  at (4.5,0) {${ X_{n,j}'}$};
\node[n] (y)  at (2.25,-1) {${ Y_j}$};
\draw[arco] (xx1) -- (x1);
\draw[arco] (xx2) -- (x2);
\draw[arco] (xx4) -- (x4);
\draw[arco] (x1) -- (y);
\draw[arco] (x2) -- (y);
\draw[arco] (x3) -- (y);
\draw[arco] (x4) -- (y);
\end{tikzpicture}
\mycaptiontitle{A noisy gate explicit formulation.}
\label{fig:bn2}
\end{figure}

\Cref{fig:bn2} shows a typical representation of the structure of the noisy-OR network, introducing $n$ auxiliary variables (also called inhibitor nodes), which help clarify \Cref{eq:noisy}. 
To reduce the number of parameters, the structure of this network defines deterministically the state of $Y_j$ as the logical disjunction (OR) of the auxiliary parent nodes.
This first simplification removes the need to specify the answer node CPT given the state of its parent nodes. 
Furthermore, the noisy-OR structure sets the input variable $X_i$ as the unique parent of $X_{i,j}'$ and constraints $X'_{i,j}$ to be zero with probability one when $X_i=0$.
The relationship between skill and answers would be purely logical-deterministic were it not for the noise introduced by the so-called inhibition parameters $\lambda_{i,j} = P(X_{i,j}'=0|X_i=1)$, representing the probability of not expressing skill $i$ in task $j$. This is, thus, the only parameter to be determined.

Auxiliary variables can be interpreted as \emph{inhibitors} of the corresponding skills.
We can regard the auxiliary variable $X_{i,j}'$ as an inhibitor of skill $X_i$ in performing the action described by $Y_j$, since with probability $\lambda_{i,j}$ it makes the skill unavailable to the success of $Y_j$ even if the skill $X_i$ is indeed mastered by the learner. It can be regarded as analogous to the slip probability in BKT models.
In accordance with the above description of the noisy-OR gate, missing skill $i$ implies the inability to apply it to any question $j$, whereas if the learner has the skill, the probability of being able to apply it depends on the specific task and is equal to $1-\lambda_{i,j}$. 
The model parameters should, therefore, be related in some sense to the task's difficulty. 
For instance, they can be in the false state, e.g., $X'_{i,j} = 0$ (with probability $\lambda_{i,j}$), even when the corresponding skill node $X_i$ is in the true state, indicating that, although the learner possesses the skill, it cannot be expressed in task $Y_j$. 
By defining the probability of a failure in expressing a possessed skill in the specific task $j$, the inhibition parameter $\lambda_{i, j}$ provides a measure of the task difficulty.  
If a pair skill-answer has a large inhibition, the state of the answer node tells, in general, little about the state of the skill node. The extreme case of $\lambda_{i,j} = 1$ corresponds to a missing arc in the BN graph between skill $i$ and answer $j$.

\subsection{Leaky models}\label{sec:leaky}
In a noisy-OR gate, when all skills are missing, all auxiliary variables are false; therefore, all answers must be wrong. Such a model excludes the possibility of a lucky guess. 
To avoid this, the noisy gates are made \emph{leaky} by adding a leak node, which represents the possibility of a random guess, i.e., a correct answer or a behaviour given without mastering any required competencies.
The leak is a boolean variable playing the role of an auxiliary skill node $X_{\mathrm{leak}}$, which is set in the observed state $X_{j,\mathrm{leak}}=1$, and added to the parents of all answer nodes for which random guessing is possible.
The chances of guessing answer $Y_j$ at random, i.e., without mastering any of the relevant competencies, is given by parameter $1-\lambda_{j,\mathrm{leak}}$. 
For instance, in a multiple choice question with four options, one of which is correct, one should set $1-\lambda_{j,\mathrm{leak}}=\frac{1}{4}$. $1-\lambda_{j,\mathrm{leak}}$ can therefore be seen as the analogous of the guess probability in BKT \cite{corbett1994knowledge}.

To apply the above model, the domain expert (e.g., the teacher) should first list the parent-less skill nodes (including, eventually, the leak) $X_1, \dots, X_n$, the childless answer nodes $Y_1, \dots, Y_m$ and connect by an arc the skills to all answer nodes in which they can be used. 
Then, the instructor should quantify for each pair of skill-answer nodes, $X_i$ and $Y_j$, connected by an arc, the value of the inhibition $\lambda_{i,j}$. 
This results in a total of at most $n\cdot m$ parameters to be elicited. 
Finally, the expert should state each skill's prior probabilities $\pi_i$.


\subsection[Comparison with Bayesian Knowledge Tracing (BKT)]{Comparison with Bayesian Knowledge Tracing} \label{sec:bkt}
While the BKT, in its standard implementation, traces the evolution of a single skill over time, our approach focuses on fine-grained skills modelling at the specific moment the assessment is performed. However, a parallel can be drawn between the two. BKT models student knowledge at time $t$ as the (binary) latent variable $X(t)$ of a Hidden Markov Model (HMM) \citep{corbett1994knowledge}. Learning is modelled as the transition of $X(t)$ from state zero (lack of knowledge) to state one (knowledge acquired). The model defines four parameters: (i) the \textit{initial} probability, i.e., the probability that the knowledge has been already acquired at the beginning of the activity; (ii) the \textit{learning} probability, that is, the probability of acquiring the probability between $t$ and $t+1$; (iii) the \textit{slip} probability of making a mistake when the knowledge is acquired; (iv) the \textit{guess} probability of doing right in the lack of knowledge. 

In our model, the probability of the \textit{slip} may vary depending on the pair skill $i$ and task $j$, represented by the inhibition $\lambda_{ij}$. The  \textit{guess} probability depends on the task and is equal to $1-\lambda_{\mathrm{leak},j}$. The \textit{initial} probability of a skill $X_i$ is defined by its prior probability $\pi_i$. 
Notice, however, that since our approach, differently from BKT, does not model the learning process, the concept of initial probability here is meant to describe our initial knowledge of the learner competence profile rather than the probability that the skill is initially acquired. For the same reason, no \textit{learning} probability is defined in our model. 

	\section{Assessment rubrics}\label{sec:rubric}
Several possible approaches exist to identify the knowledge components to be included in a learner model. 
We decided to consider only assessment methods based on a task-specific assessment rubric for assessing a given competence through a given task or family of similar tasks \citep{castoldi2009valutare,jonsson2007use}.

A task-specific assessment rubric consists of a two-entry table where each row corresponds to a component of the given competence, described in the light of the given task. In contrast, each column corresponds to a competence level in ascending order of proficiency. 
For each combination of component and level, the rubric provides a qualitative description of the behaviour expected from a person with the given level in the given component.
Identifying a person's competence level consists of matching the learner's behaviours while solving a given task with those described in the assessment rubric. 

For instance, \Cref{tab:examplerubric} shows the task-specific assessment rubric for an example focused on assessing the student's ability to use iterative instructions in algorithms. 
This competence has two levels depending on the tools used by the learner: a visual programming language ($X_1$) or a textual programming language ($X_2$).  
By checking how the learner produced the algorithm, the teacher can see whether he applied any of the methods in the rubrics and assign him the corresponding competence level.
\begin{table}[htb]
\footnotesize
\centering
\mycaption{Example of a task-specific assessment rubric.}{In this rubric, there is only one competence component, the ability to design an algorithm containing loops, and two competence levels, the ability to do it using either a block-based programming language or a text-based programming language.}\label{tab:examplerubric}
\setlength{\tabcolsep}{3mm}
\begin{tabular}{c>{\centering\arraybackslash}m{.8cm}|m{5.1cm}m{5cm}}
\multicolumn{2}{c|}{}   & \multicolumn{2}{c}{{\textbf{Competence level}}} \\
\multicolumn{2}{c|}{}   & {\begin{minipage}{5.1cm}{\begin{tabular}{>{\centering\arraybackslash}m{5.1cm}}\textbf{$X_1$}\\[-1ex]\textbf{$c=1$}\end{tabular}}\end{minipage}}    & {\begin{minipage}{5cm}{\begin{tabular}{>{\centering\arraybackslash}m{5cm}}\textbf{$X_2$}\\[-1ex]\textbf{$c=2$}\end{tabular}}\end{minipage}}  \\
\midrule
{\begin{minipage}{.5cm}\rotatebox{90}{\begin{tabular}{c}\textbf{Competence}\\[-1ex]\textbf{component}\end{tabular}}\end{minipage}} 
& {Loops $r=1$} & Develop an iterative algorithm using a~block-based programming language & Develop an iterative algorithm using a text-based programming language     \\
\end{tabular}
\end{table}

In assessment rubrics, the ordering between competence levels, and sometimes between competence components, plays a fundamental role.
A competence level or component is considered higher than another if the former implies the latter, meaning that a learner with the higher competence can also perform all the tasks that require the lower. 
In practice, the competence level matching the student's behaviours for a given component does not always correspond to the actual learner's state of knowledge. It is also possible that the person possessed a higher level but is underperforming.

To ensure completeness, rubrics must include a basic competence level, describing the observed behaviours of learners who have not yet acquired the competence in question.
This level is typically expressed in a constructive manner, highlighting what the learner is capable of rather than focusing on what they cannot do.
When evaluating a person using a rubric, the final assessment must correspond to one of the rubric's defined levels.

In the case of a task composed of similar sub-tasks., i.e., tasks sharing the same assessment rubric, it is possible to observe behaviours corresponding to different competence levels across various sub-tasks.
As a result, the competence level identified in a given instance does not necessarily reflect the learner's overall state of knowledge, as external factors or temporary difficulties may lead to underperformance.

In the following subsection, we illustrate how this uncertainty can be considered and how an overall assessment based on a full battery of tasks can be produced by modelling the learner competence profile with the BN-based approach described in \Cref{sec:bn_learner_model}. 

\section[Modelling assessment rubrics by BNs]{Modelling assessment rubrics by Bayesian Networks}\label{sec:rubrictobn}
Considering a task-specific assessment rubric, as defined above, it is possible to derive a learner model, as presented in \Cref{sec:bn_learner_model}, hereafter referred to as \emph{baseline model}.
For each cell $(c,r)$ of an assessment rubric with $R$ rows and $C$ columns, we introduce a latent binary competence variable $X_{rc}$, taking value one for a learner mastering the corresponding competence level and zero otherwise. 
Moreover, for each task $t$, in a battery of $T$ similar tasks, and each competence variable $X_{rc}$, we define an observable (manifest) binary variable $Y^t_{rc}$ taking value one if the behaviour described in the assessment rubric's cell $(r,c)$ was applied successfully by the learner in solving task $t$ and zero if he failed using it. 

In addition, we extend this baseline model in two ways. 
Firstly, we explicitly impose the \textit{ordering of competence levels} encoded by the rubric. 
Secondly, we include in the model task-specific \textit{supplementary skills}, which can be combined with each other and with the competencies of the rubric through arbitrary logic functions.

\subsection{Ordering of competences} \label{par:dummies}
In the baseline model, as described in \Cref{sec:bn_learner_model}, it was indirectly accounted for the partial ordering between variables by setting as parents of answer node $Y^t_{rc}$ the skill node $X_{rc}$ and all skill nodes corresponding to higher competence levels. The network was quantified through noisy-OR relations, as described in \Cref{sec:noisyor}.
This structure assumes that an observed behaviour can be explained as the student mastering the corresponding competence level or a higher one if he is underperforming, thus not exploiting his full potential, but cannot be achieved through a lower level. 

As mentioned above, we interpret the (partial) ordering between competencies defined by the assessment rubric as implication constraints, meaning that possessing a particular skill $X_i$ implies that the learner also possesses his inferior competencies. While exploited to design the network structure, this hierarchy of competencies is not strictly imposed by the above baseline model, giving rise to posterior inferences that are usually inconsistent. 

To solve this issue, we enrich the model by adding an auxiliary variable $D_{ik}$ for each relation $X_i \implies X_k$ defined by the rubric. 
A constraint node $D_{ik}$ is always in the observed state one, with $X_i$ and $X_k$ as parent nodes. The desired implication constraint is then implemented by choosing a CPT for $D_{ik}$ such that $P(D_{ik}=1|X_i=1, X_k=0) = 0$. 
The addition to the network of each observed node $D_{ik}$ changes the prior probabilities of $X_i$ and $X_k$, initially set to $\pi_i$ and $\pi_k$. 
Let 
\begin{equation}
\begin{split}
p_{00} & = P(D_{ik}=1|X_i=0, X_k=0) \\
p_{01} & = P(D_{ik}=1|X_i=0, X_k=1) \\
p_{11} & = P(D_{ik}=1|X_i=1, X_k=1),
\end{split}
\end{equation}
be the non-null parameters in the CPT of $D_{ik}$.
After updating with the evidence $D_{ik}=1$, one has 
\begin{equation}\label{eq:priorwithD}
\begin{split}
&P(X_i=1|D_{ik}=1) = \frac{p_{11}\pi_j \pi_k}{K},\\
&P(X_k=1|D_{ik}=1) = \frac{p_{11}+p_{01}\pi_j \pi_k}{K},
\end{split}
\end{equation}
with $K= p_{11}\pi_j \pi_k+p_{01}(1-\pi_j) \pi_k+p_{00}(1-\pi_j) (1-\pi_k )$.

In this work, we simply assume $p_{00} = p_{01} =p_{11}$ and adopt uniform prior probabilities $\pi_i=\pi_k=0.5$. 
Applying them to \Cref{eq:priorwithD} give $P(X_i=1) = \frac{1}{3}$ and $P(X_k=1) = \frac{2}{3}$. 
This result follows from the fact that skill $X_i$ can only be possessed jointly with $X_k$, whereas $X_k$ can also be owned when $X_i=0$. 

Under the assumption $p_{00} = p_{01} =p_{11} = p_{*}$, the prior over the superior skill $X_i$ can be interpreted as the conditional probability of having it given that the learner possesses the inferior skill $X_k$ since 
\begin{equation}
\begin{split}
P(X_i=1|X_k=1, D_{ik}=1)  = &\frac{\pi_i\pi_j p_{*}}{\pi_i\pi_j p_{*} + (1-\pi_i)\pi_j p_{*}} = \pi_i \mbox{.}
\end{split}
\end{equation}

\subsection{Supplementary competencies} \label{par:supp}
While the assessment rubric details the components of the competence of interest and their interactions with the specific task and available tools, it does not necessarily include all the skills required to solve the task successfully.

For instance, considering the assessment rubric proposed in \Cref{tab:examplerubric}, to develop an iterative algorithm with a text-based programming language successfully, the learner might also need knowledge about the different types of statements, e.g., while, repeat, for, do until and so on.
Ignoring such supplementary skills might be misleading in an automatic assessment system, as failures due to the lack of one of them would not be recognised as such and, eventually, be attributed to the absence of the competence components under assessment. Therefore, if not adequately modelled, the lack of unmodelled supplementary skills would translate into an unfairly negative evaluation of the competencies of interest. 

To produce fairer assessments, we extend the model by an additional layer of auxiliary nodes combined with a logic function to allow for the inclusion of a suitable set of supplementary skills.
\begin{figure}[htb]
	\centering
	\includegraphics[width=.55\columnwidth]{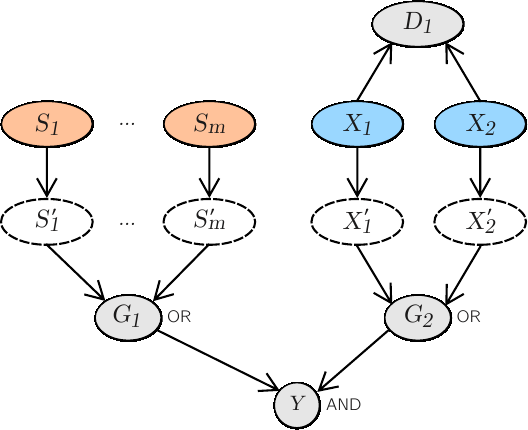}
	\mycaption{Example of BN modelling a task-specific assessment rubric.}{The rubric has two cells, represented by skills $X_1$ and $X_2$ (on the right, in light blue), $m$ supplementary skills grouped in a single set (on the left, in orange), and the constraint $X_2\implies X_1$, represented by the auxiliary variable $D_1$ (on the top right, in light grey).}
	\label{fig:finalbn}
\end{figure}

\Cref{fig:finalbn} shows an example of the structure of the extended network. 
Supplementary skills are described by additional skill nodes $S_1, \dots, S_m$, which are grouped into sets of interchangeable skills (in the case of the example, we have just one set). 
Each of these groups is connected through a noisy-OR to a node in the layer of auxiliary latent nodes, hereafter referred to as group nodes $G_1, \dots, G_l$, representing the success or failure in applying the type of competence described by each group to the specific task $Y$. 
Finally, the group nodes are connected to the answer node through a logic AND or any other logic function suitable for the particular task. 

When supplementary skills can be directly assessed through observing specific learner behaviours or by purposed questions, additional answer nodes can be added to the network as direct children of the relevant supplementary skills.
	
	\part{Assessment instruments}\label{part:assessment_instruments}
\chapter[The unplugged Cross Array Task (CAT)]{The unplugged Cross Array Task} \label{chap:unpluggedCAT}

\begin{custombox}
    {The content of this chapter has been adapted from the following article with permission of all co-authors and publishers:}
    \begin{itemize}[noitemsep,nolistsep,leftmargin=.25in]
        \item Piatti, A., \textbf{Adorni, G.}, El-Hamamsy, L., Negrini, L., Assaf, D., Gambardella, L., and Mondada, F. (2022). The CT-cube: A framework for the design and the assessment of computational thinking activities. \textit{Computers in Human Behavior Reports} \cite{piatti_2022}. 
        
        \item \textbf{Adorni, G.}, Piatti, A., Bumbacher, E., Negrini, L., Mondada, F., Assaf, D., Mangili, F., and Gambardella, L. M. (2025). FADE-CTP: A Framework for the Analysis and Design of Educational Computational Thinking Problems. \textit{Technology,  Knowledge and Learning} \cite{adorni2023ctpframework}.
        \end{itemize}
        As an author of these publications, my contribution involved:\\
        \textit{Conceptualisation, Methodology, Validation, Formal analysis, Investigation, Resources, Data curation, Writing -- original draft \& review \& editing, Visualisation, Supervision.}
\end{custombox}

\section{Summary}

This chapter contributes primarily to RQ2 by focusing on designing and developing an unplugged CT activity called Cross Array Task (CAT), aimed at assessing the progression of algorithmic skills across the entire compulsory education path (K-12) in Switzerland.
It begins by outlining the design of the activity, detailing its objectives and structure, followed by an explanation of how interaction strategies and algorithms can be categorised and how these classifications inform the evaluation metric used to assess AT skills.
By focusing on the development of a practical and scalable assessment tool, the chapter provides insight into how the unplugged CAT can be used effectively across various educational contexts to measure AT.




\section{The cross array}\label{sec:cross_array}

The concept of the \textit{cross array} was developed to engage students in recognising patterns and building AT through a visually stimulating and structured activity.
Visual patterns, such as the cross array, can be effective in engaging learners because they encourage the recognition of regularities, such as repetitions, symmetries, and colour patterns, which are foundational skills in AT. 
This approach draws on cognitive development theories, which suggest that tasks involving visual patterns are well-suited to measure domain-general components of AT, such as pattern recognition, generalisation, and the ability to articulate algorithmic procedures.

The design of the CAT was informed by the principles of situated cognition, which emphasise that learning occurs most effectively in contexts that are meaningful and contextually rich. The task complexity and progression were carefully calibrated to create a balance between accessibility for younger learners and increasing challenges to stimulate growth in AT. The task sequence begins with simpler patterns and gradually increases in complexity, allowing students to build on their existing knowledge and develop new cognitive strategies.

\begin{figure}[htb]
	\centering
	\begin{subfigure}[t]{0.22\linewidth}
		\centering
		\includegraphics[width=\textwidth]{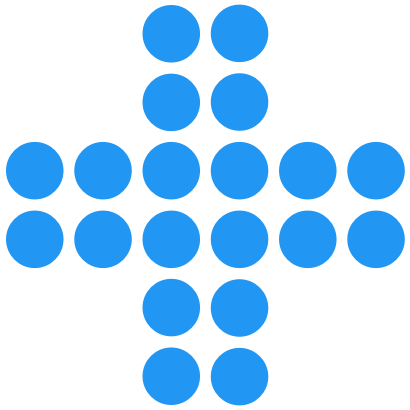}
		\mycaptiontitle{Schema 1.}
		\label{s1}
	\end{subfigure}
	\hfill
	\begin{subfigure}[t]{0.22\linewidth}
		\centering
		\includegraphics[width=\textwidth]{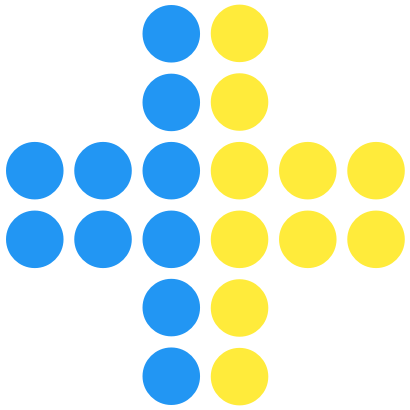}
		\mycaptiontitle{Schema 2.}
		\label{s2}
	\end{subfigure}
	\hfill
	\begin{subfigure}[t]{0.22\linewidth}
		\centering
		\includegraphics[width=\textwidth]{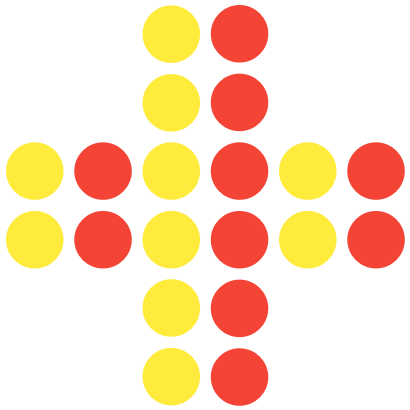}
		\mycaptiontitle{Schema 3.}
		\label{s3}
	\end{subfigure}
	\hfill
	\begin{subfigure}[t]{0.22\linewidth}
		\centering
		\includegraphics[width=\textwidth]{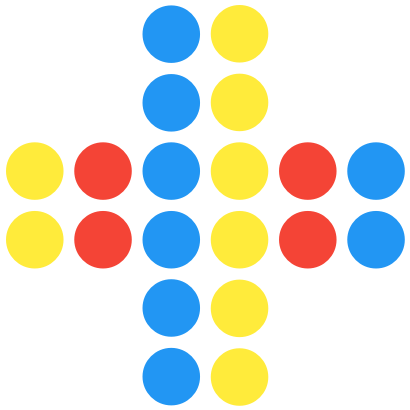}
		\mycaptiontitle{Schema 4.}
	\label{s4}
\end{subfigure}
\begin{subfigure}[t]{0.22\linewidth}
	\centering
	\includegraphics[width=\textwidth]{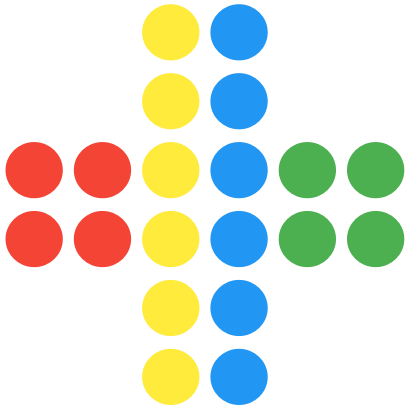}
	\mycaptiontitle{Schema 5.}
\label{s5}
\end{subfigure}
\hfill
\begin{subfigure}[t]{0.22\linewidth}
\centering
\includegraphics[width=\textwidth]{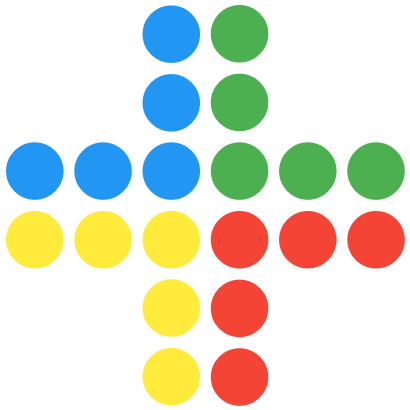}
\mycaptiontitle{Schema 6.}
\label{s6}
\end{subfigure}
\hfill
\begin{subfigure}[t]{0.22\linewidth}
\centering
\includegraphics[width=\textwidth]{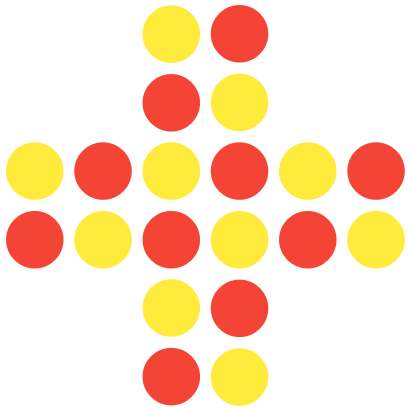}
\mycaptiontitle{Schema 7.}
\label{s7}
\end{subfigure}
\hfill
\begin{subfigure}[t]{0.22\linewidth}
\centering
\includegraphics[width=\textwidth]{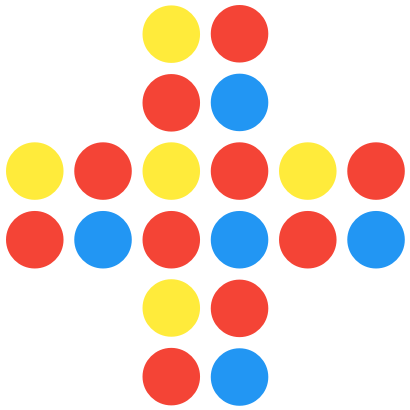}
\mycaptiontitle{Schema 8.}
\label{s8}
\end{subfigure}
\begin{subfigure}[t]{0.22\linewidth}
\centering
\includegraphics[width=\textwidth]{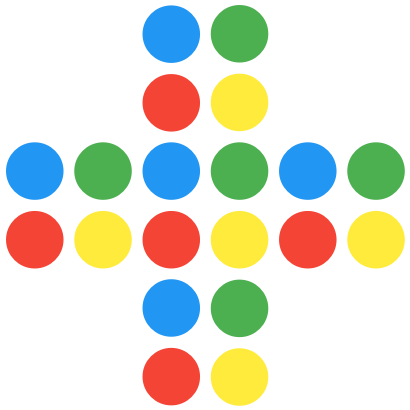}
\mycaptiontitle{Schema 9.}
\label{s9}
\end{subfigure}
\hfill
\begin{subfigure}[t]{0.22\linewidth}
\centering
\includegraphics[width=\textwidth]{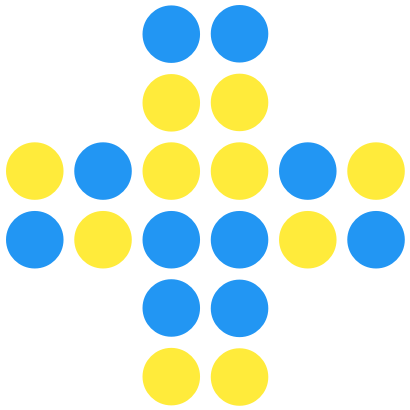}
\mycaptiontitle{Schema 10.}
\label{s10}
\end{subfigure}
\hfill
\begin{subfigure}[t]{0.22\linewidth}
\centering
\includegraphics[width=\textwidth]{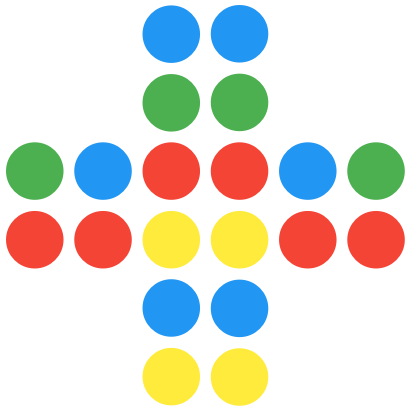}
\mycaptiontitle{Schema 11.}
\label{s11}
\end{subfigure}
\hfill
\begin{subfigure}[t]{0.22\linewidth}
\centering
\includegraphics[width=\textwidth]{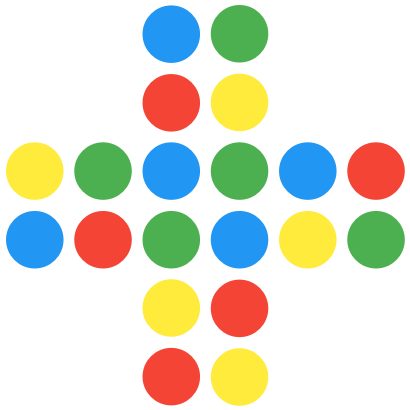}
\mycaptiontitle{Schema 12.}
\label{s12}
\end{subfigure}
\mycaption{Sequence of cross array schemas.}{The figure showcases the 12 schemas proposed in the task, named from Schema 1 to Schema 12, each distinguished by its unique visual regularities and complexities, varying in elements such as colours, symmetries, alternations and other distinctive features..}
\label{fig:sequence}
\end{figure}

Initially, the design was tested with kindergarten students, as it was assumed they would provide insights into how early-stage problem-solving strategies evolve and how they engage with tasks requiring pattern recognition. Based on feedback from these initial tests, the designs were refined, resulting in 12 patterns of varying complexity, presented in  \Cref{fig:sequence}.
Each cross-shaped design, consisting of five $2 \times 2$ square arrays of coloured dots, was selected for its simplicity and ability to introduce different patterns. The final set of 12 schemas provides a progression that supports the development of increasingly sophisticated algorithmic thinking as students advance through the tasks.

The task complexity is designed to maintain engagement and encourage problem-solving strategies that are applicable to the task at hand and transferable to other contexts. Feedback from younger learners and the complexity adjustments made during testing ensured that the tasks offered a meaningful experience, motivating students to engage deeply while refining their AT skills. With these adjustments, we hypothesise that the task will be equally effective for older students, who can engage with the same patterns using more complex strategies. By aligning task complexity with developmental stages, we aim to capture the progression in AT and assess the relationship between task difficulty and students' problem-solving strategies across different age groups.

\section{Activity design}\label{sec:catdesign}
Following the methodological frameworks presented in \Cref{part:methodological_framework}, we designed the unplugged CAT to evaluate AT. 
\begin{figure}[!ht]
	\centering
	\includegraphics[width=.55\linewidth]{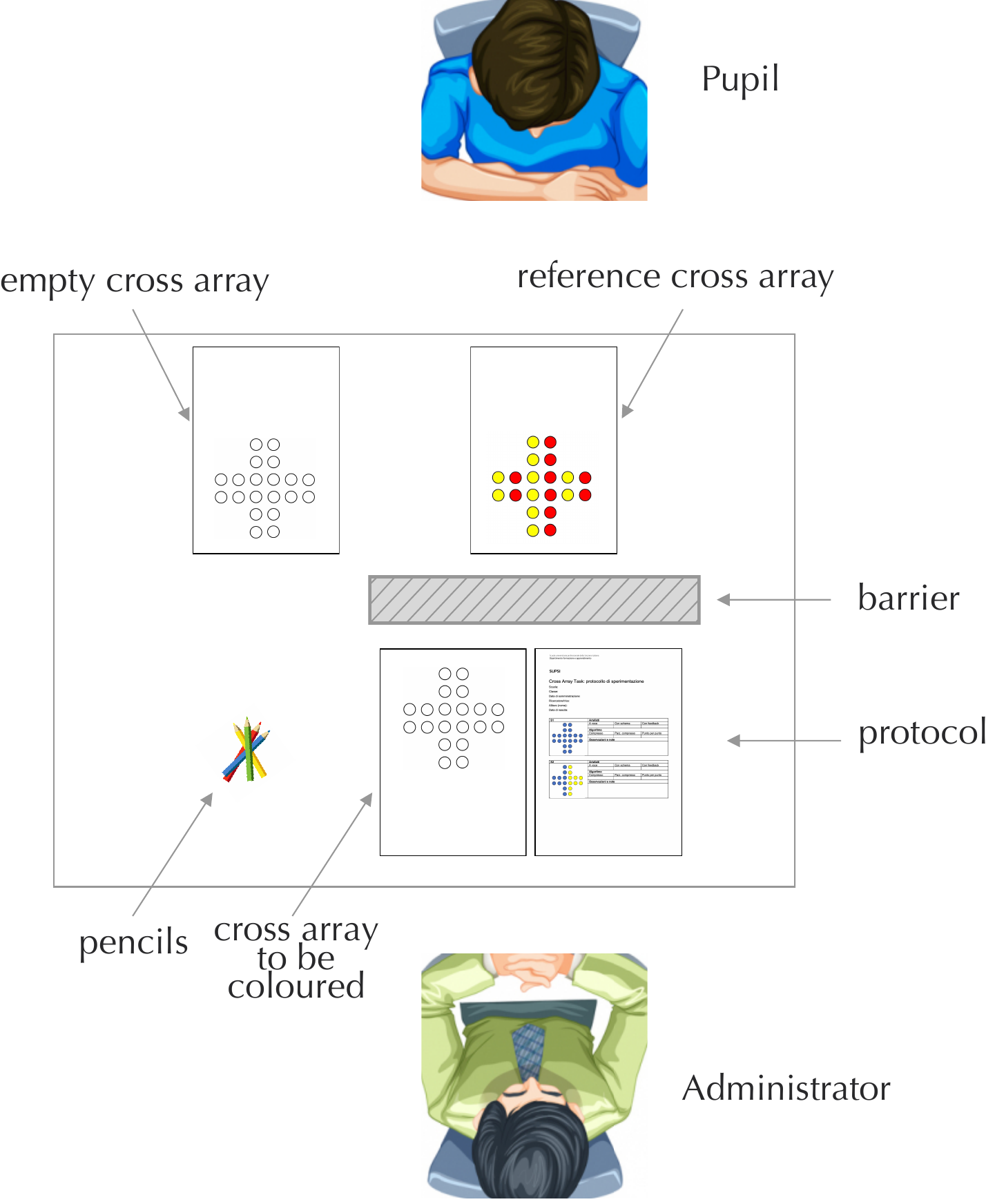}
	\mycaption{Experimental settings (unplugged CAT).} {
		Pupil and administrator are seated at two opposite sides of a table.
		The pupil is tasked with instructing the administrator in recreating a given reference cross array schema. 
		The instructions can be communicated verbally or through gestures on a supporting empty cross array schema.
		Initially, the pupil cannot see the administrator's actions due to a physical barrier preventing visual cues, which can be removed upon request.
		The administrator interprets and records all of the pupil's instructions and algorithms in a protocol.
		}
	\label{setting-cross-array}
\end{figure}


The activity is administered in class (see \Cref{setting-cross-array}), and is characterised by face-to-face interaction between the pupil (problem solver), and the administrator (a human agent).
The task involves the pupil observing a reference cross array and then conceptualising an algorithm to describe it to the administrator, who cannot see the original array. 
The goal is for the administrator to replicate the colouring pattern on a blank cross array based solely on the pupil's verbal instructions.

To begin the activity, the administrator explains the task to the pupil with the following instructions:
\emph{``You have a coloured array in front of you. 
	I have the same array, but uncoloured, in front of me. You should describe your array so I can colour mine the same way. You can try to describe it by voice. 
	You can indicate the dots on the empty array on your right if it is too difficult. 
	If it is still too difficult, you can ask me to remove the screen so that you can look at what I'm colouring.''} 

During the activity, the administrator, in addition, to interpreting students' instructions and using them to colour the schema, records the pupil's algorithmic procedure, the artefactual environment used, and the level of autonomy exhibited during the task in a protocol for later analysis (see \Cref{app:chbr_appE}).
These elements are crucial for assessing the pupil's AT skills and are classified according to specific criteria, explained in the following sections.


\subsection{CT-cube dimensions of the CAT}
During the development of this activity,  we focused on three key aspects of the CT-cube: the cognitive activity performed, the artefactual environment used, and the autonomy of the individuals involved.

\begin{figure}[htb]				
	\centering
	\includegraphics[width=.75\linewidth]{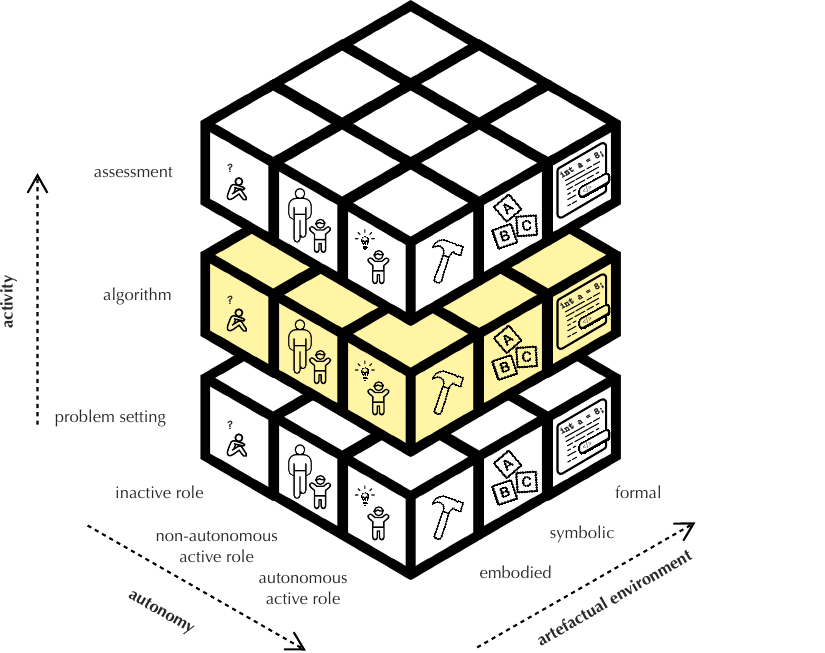}
	\mycaptiontitle{The algorithm activity of the CT-cube.}
	\label{fig:ct-algo}
\end{figure}

Among the possible cognitive activities, the CAT targets the \textit{algorithm} dimension, simplifying both the problem setting and the assessment components to make the task more accessible for young learners, as illustrated in \Cref{fig:ct-algo}.

Regarding the artefactual environment, only two dimensions are explored in the unplugged CAT, the \textit{embodied} and \textit{symbolic} computational worlds. The {formal} computational world is excluded, as it is considered too abstract for the age group targeted by the task \citep{hanna1997,druin1999}.
In particular, pupils can communicate their instructions verbally, describing the process using natural language. This is considered a symbolic artefact, as it relies on words and phrases to represent ideas and concepts.
Alternatively, pupils can enhance their verbal instructions with physical gestures, such as pointing to specific dots on an empty cross array schema. This form of communication, which illustrates the instructions through hand movements, is considered an embodied artefact.

For the autonomy levels, all three levels of autonomy, \textit{inactive}, \textit{active non-autonomous}, and \textit{active autonomous}, which reflect varying degrees of engagement and independence during the activity, are considered.
Pupils are considered inactive if they do not attempt to solve the task or are unable to provide intelligible instructions to the administrator. 
They are classified as non-autonomous if the barrier between them is removed and they can give intelligible instructions to the administrator. 
Finally, they are considered to have an autonomous active role if they provide intelligible instructions to the administrator while the barrier remains in place.

\subsubsection{Activity states}
\begin{figure}[!h]						
	\centering
	\includegraphics[width=.75\linewidth]{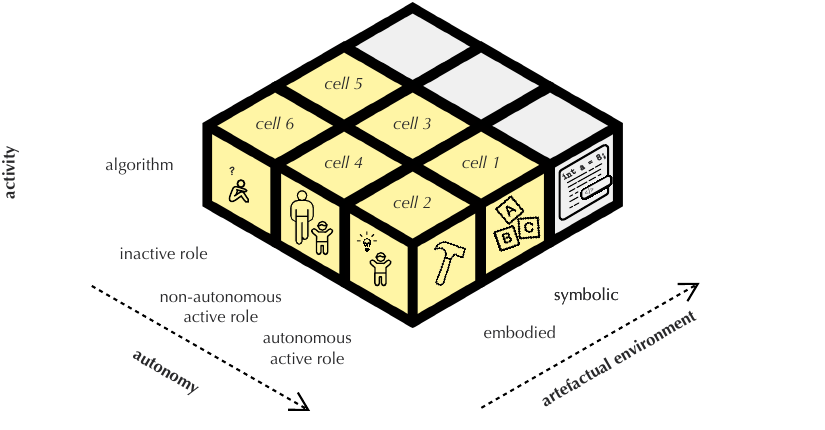}
	\mycaptiontitle{The possible cells in the algorithm activity of the {CT-cube} for the {CAT}.}
	\label{fig:cells}
\end{figure}

\Cref{fig:cells} represents the possible cells of the CT-cube for the algorithmic activity in the CAT, showing the different combinations of activity, autonomy, and artefactual environment.

The activity starts in the CT-cube cell corresponding to \emph{algorithm - autonomous role - symbolic artefact} (cell 1).
If the pupil is inactive or is giving false or incomplete instructions trying to solve the task only by voice, corresponding to \emph{algorithm - inactive role - symbolic artefact} (\textit{cell 5}), the administrator suggests to the pupil to use the empty cross array, corresponding to \emph{algorithm - autonomous role - embodied artefact} (\textit{cell 2}).
If also, in this case, the pupil is inactive or is giving false or incomplete instructions, corresponding to \emph{algorithm - inactive role - embodied artefact} (\textit{cell 6}), the administrator removes the barrier and shows to the pupil what he/she is colouring, corresponding to \emph{algorithm - non-autonomous role - embodied artefact} (\textit{cell 4}).  
The possibility for the pupil to observe what the administrator is doing is considered non-autonomous as it is a particular kind of (indirect) support.
If also, in the latter case, the pupil is inactive or is giving false or incomplete instructions (\textit{cell 5} and \textit{cell 6}), the task is finished and considered \emph{unsuccessful}. 

In each case, the pupil is free to use the empty array on his/her right or to ask to remove the screen at each moment. The task is considered \emph{successful} if the pupil is able to give complete and correct instructions (eventually with corrections during the description, for example, after having removed the screen) to the administrator, independently from the artefacts used by and the active role (non-autonomous or autonomous) of the pupil. In the rest of the paper, we call \emph{algorithm} the entire set of correct instructions given by the pupil.

\FloatBarrier

\section{Activity profile}\label{sec:cat_profile}
\begin{figure*}[ht]
	\centering
	\includegraphics[width=\textwidth]{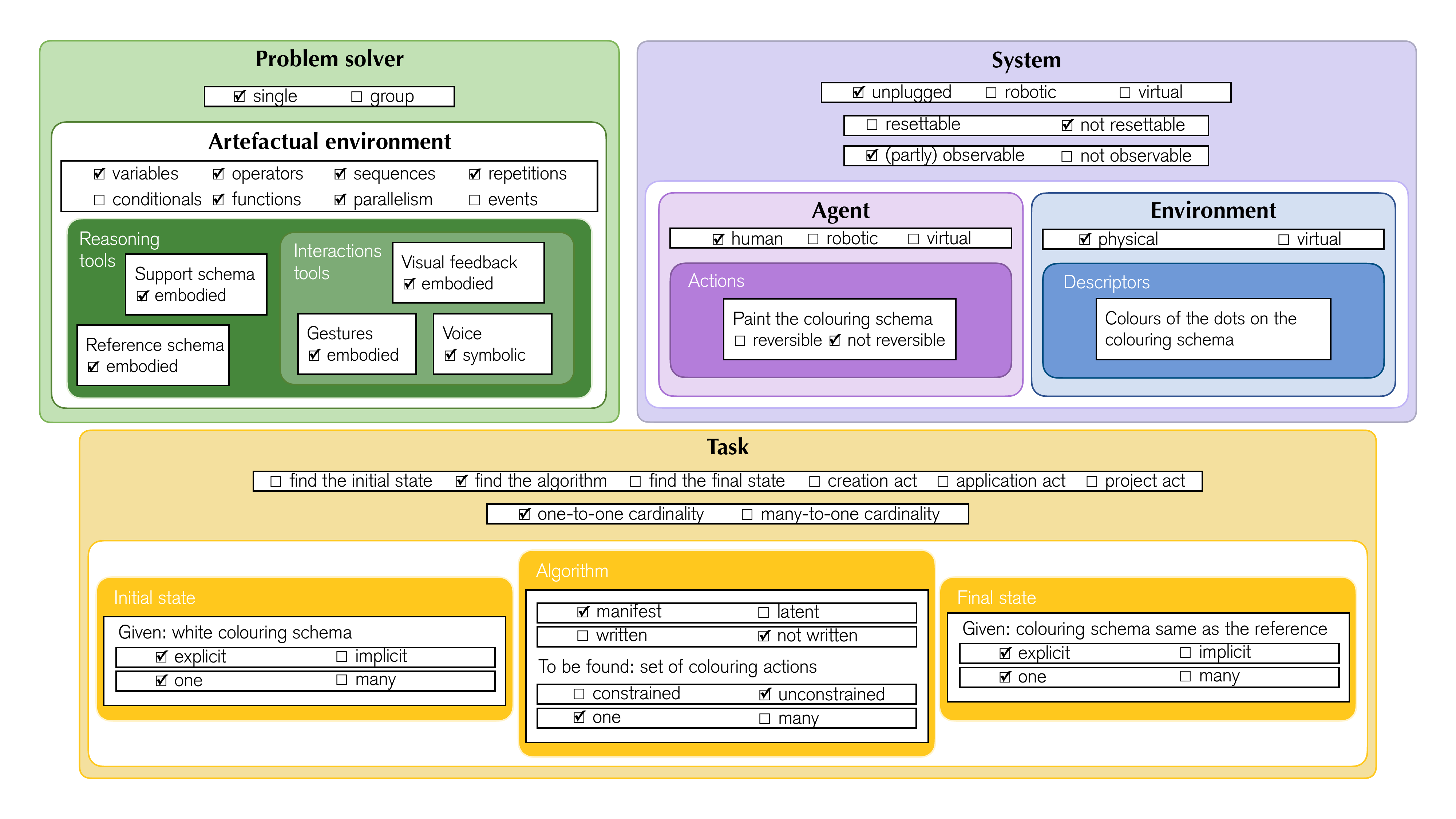}
	\mycaptiontitle{Components and characteristics (unplugged CAT).}
	\label{fig:CAT-features}
\end{figure*}

\begin{figure*}[ht]
	\centering
	\includegraphics[width=\textwidth]{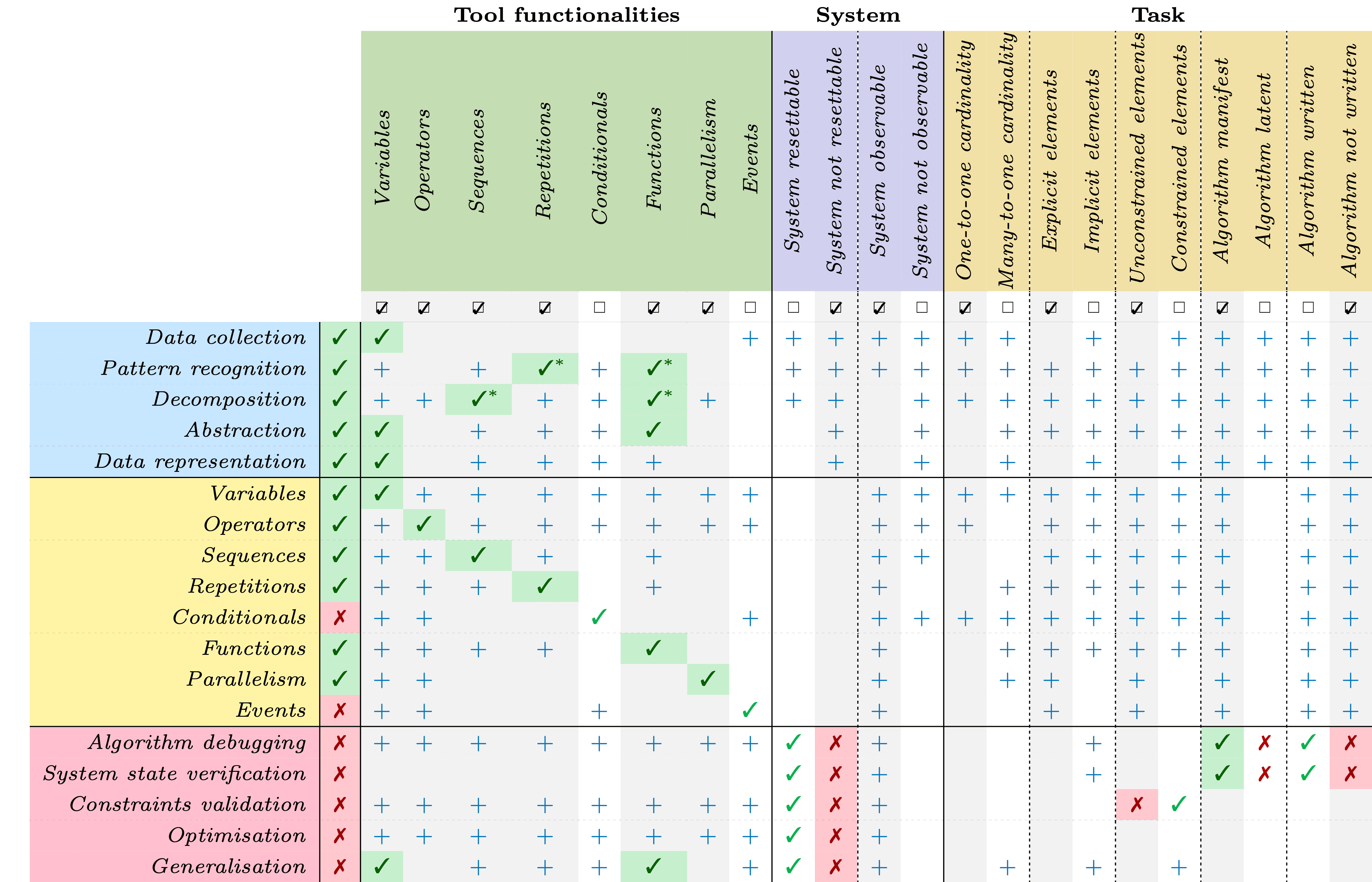}
	\mycaptiontitle{Activity profile (unplugged CAT).}
	\label{tab:CAT-mapping}
\end{figure*}

The components and characteristics of the unplugged CAT are illustrated following our FADE-CTP and graphically represented in \Cref{fig:CAT-features}. 
Additionally, we provide an overview of the overall CT competencies that can be developed through this activity, as well as those that cannot, and finally, we present in \Cref{tab:CAT-mapping} the resulting profile, mapping the relationship between the activity's characteristics and the CT competencies it activates.

\subsection{Components}
\begin{itemize}[noitemsep,nolistsep]
	\item \textit{Problem solver}: the student who has to communicate an algorithm corresponding to the sequence of instructions to reproduce the colouring of the reference schema.
	The artefactual environment comprises cognitive tools such as {support and reference schemas, which are} available to the problem solver to reason about the task.
	Additionally, the problem solver can interact with the system to communicate the algorithm. This can be achieved using a natural language such as the voice (symbolic) or gestures (embodied) on the {colouring schema, an} empty cross array {used from the agent}. 
	Moreover, by removing the screen that separates the problem solver from the agent, he can have visual feedback (embodied) of the cross array being coloured.
	\item \textit{Agent}: the researcher, executor of the problem solver's instructions, responsible for filling the colouring schema according to the problem solver's algorithm. The agent's actions are not resettable.
	\item \textit{Environment}: the cross array to be coloured, whose state is described by the colour of each dot (white, yellow, blue, green, or red). 
	\item \textit{Task}: find the algorithm. The system's state is defined by the colouring cross status, initially white and, at the end, the same as the reference schema. 
	The algorithm is the set of agent instructions to achieve this transformation.
\end{itemize}

\subsection{Characteristics}

\begin{itemize}[noitemsep,nolistsep]
	\item \textit{Tool functionalities}: voice and gestures provide various functionalities associated with algorithmic concepts suitable to design the algorithm, including 
	(i) \textit{variables} can represent different colours of the cross array dots; 
	(ii) \textit{operators} are used to change the colour of the dots performing actions such as colouring a dot, a row, a square and so on;
	(iii) \textit{sequences} determine the order in which the actions should be executed to achieve the desired outcome; 
	(iv) \textit{repetitions} allow for repeating specific sequences of operations, such as colouring the first column in red and repeating it every two columns;
	(v) \textit{functions} consist of operations that perform a specific task and can be applied to different inputs, for example, creating a pattern of alternating red and yellow dots in a square and applying it to different positions of the cross array;
	(vi) \textit{parallelism} involves executing multiple actions simultaneously and can be associated with using symmetries to describe the pattern.
	\item \textit{System resettability}: the system is not resettable since it is impossible to reverse the agent's actions.
	\item \textit{System observability}: the system is partially observable since the cross array being coloured by default is not seen until the end of the task unless the problem solver demands otherwise.
	\item \textit{Task cardinality}: the task has a one-to-one mapping, with given one initial and one final state, and an algorithm to be found.
	\item \textit{Task explicitness}: all elements are given explicitly.
	\item \textit{Task constraints}: the algorithm is unconstrained.    
	\item \textit{Algorithm representation}: the algorithm is represented through voice commands or gestures. It is considered manifest because {it is externalised}, but not written since it is not stored in a permanent format.
\end{itemize}

\subsection{Competencies}

\paragraph{Enabling features for competencies development}
\begin{itemize}[noitemsep,nolistsep]
	\item \textit{Problem setting}: all competencies can be activated thanks to the presence of variables, sequences, repetitions and functions in the tool functionalities.
	The presence of many tool functionalities, the non-resettability of the system and the algorithm representation positively affect and boost problem setting skills.
	The system observability supports data collection and pattern recognition. The one-to-one cardinality, in addition, stimulates decomposition. 
	The explicit and unconstrained definition of the task elements also promotes pattern recognition, decomposition and abstraction. 
	
	\item \textit{Algorithm}: all competencies associated with the algorithmic concepts enabled by the tool functionalities, meaning variables, operators, sequences, repetitions and functions, can be activated and promote one another.
	The form of representation of the algorithm, the system observability, and the explicit and unconstrained definition of the task elements further enhance these. The one-to-one cardinality helps to enhance some of these skills as well.
	
	\item \textit{Assessment}: since the system is not resettable, no assessment skills can be developed.
\end{itemize}

\paragraph{Inhibiting features for competencies development}
\begin{itemize}[noitemsep,nolistsep]
	\item \textit{Conditionals and events}: non-activable as these functionalities are unavailable in the platform. 
	A way to make conditionals available in the tool functionalities would be allowing the problem solver to change a dot colour, for example by communicating instructions such as: ``if the dot is red, then colour it yellow''. By doing this, the problem solver engages with the concept of conditionals and can develop their algorithmic skills.
	The completion of each row in the cross array can be considered an event. The problem solver can specify that they want to fill the cross line by line, and once a line is complete, the researcher will move on to the next line. This allows the problem solver to list only the sequence of colours without repeating the instructions for where to go. The change in the environment (completing a row) triggers the researcher to move to the next row.
	Using conditionals and events can greatly enhance the complexity of the solutions that can be generated and help develop advanced CT skills.
	
	\item \textit{Assessment skills}: the inability to reset the system impairs the development of the student's skills. 
	One possible solution to this issue is enabling the student to reset the colouring schema using a voice command. This would return the schema to its initial blank state, allowing the student to start the task from the beginning and practice their assessment skills.
	To develop system state verification, it is also essential to not reveal the initial or final states. 
	Moreover, constraints should be imposed on the algorithm to develop constraint validation skills, for example, limiting the use of specific operators or the number of times they can be used, allowing the problem solver to develop the ability to think about the constraints and limitations in their algorithms.
\end{itemize}

\section{Competencies assessment}
\begin{figure}[!h]						
	\centering
	\includegraphics[width=.75\linewidth]{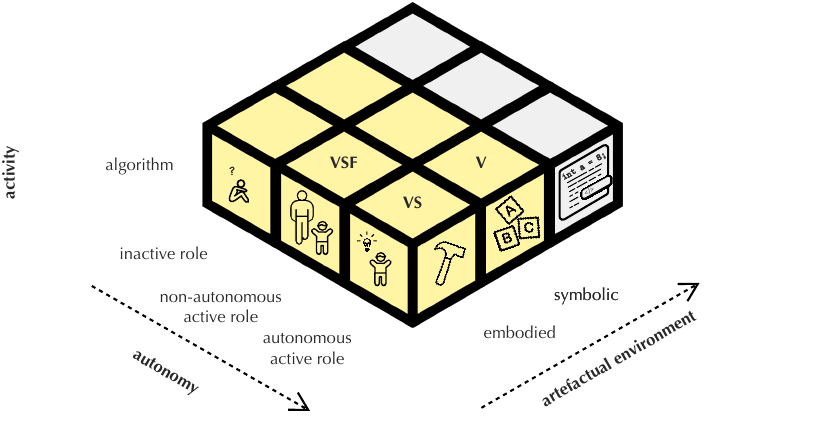}
	\mycaptiontitle{Assessed cells in the algorithm activity of the {CT-cube} for the {CAT}.}
	\label{fig:competencies_assessment_cells_unplugged}
\end{figure}

Although the unplugged CAT activates a range of competencies, our focus is specifically on measuring algorithmic skills, while also considering aspects of situated cognition, specifically the ability to effectively use the available artefacts and the nature of the social dynamics during the task (see \Cref{fig:competencies_assessment_cells_unplugged}).
Additionally, we evaluated participation and success rates, providing further insights into students' engagement and their ability to complete the task successfully.

\subsection{Algorithm dimension} \label{sec:classificationofalgorithms}

The CAT evaluates students' algorithmic skills by examining the complexity of the operations used to describe the cross arrays.
Each operation has an associated level of complexity, which ranges across three distinct levels.
The algorithm dimension, or classification, reflects the highest level of complexity among the operations it contains.
We distinguish between:
\begin{enumerate}[noitemsep,nolistsep]
\item \emph{0D, zero-dimensional algorithms}: this level involves describing the cross array point by point, with each dot coloured individually using a specified colour. These operations are referred to as \texttt{Colour-One-Dot} operations (COD).

\item \emph{1D, one-dimensional algorithms}: this level involves describing the cross array with a series of dots arranged in structures such as rows, columns, diagonals, squares, L-shapes, zig-zags, half-crosses, or entire crosses, all with a specified colour. These operations are referred to as \texttt{Colour-Several-Dots} (CSDs).

\item \emph{2D, two-dimensional algorithms}: this level involves describing the cross array using more advanced operations, such as sequences of CSDs with alternating colours, repetitions or mirroring of COD and/or CSDs operations. 
\end{enumerate}


All 12 cross arrays used in our experimental study could be described using zero-, one- or two-dimensional algorithms. 

An algorithm is \emph{redundant} if one or more dots are described more than once in the algorithm. 
Additionally, we define the \emph{number of operations} used in an algorithm as the total number of CODs and CSDs used in the algorithm, where a COD and/or CSD used inside a loop is considered only once. The maximal number of operations in non-redundant algorithms is 20, the minimal number is 1.

\subsubsection{Example algorithms for describing Schema 3}
Below, we present three different example algorithms for describing Schema 3 in our sequence (see \Cref{s3}). 

\begin{figure}[!ht]
	\centering
	\begin{subfigure}[t]{0.3\linewidth}
		\centering
		\includegraphics[width=\textwidth]{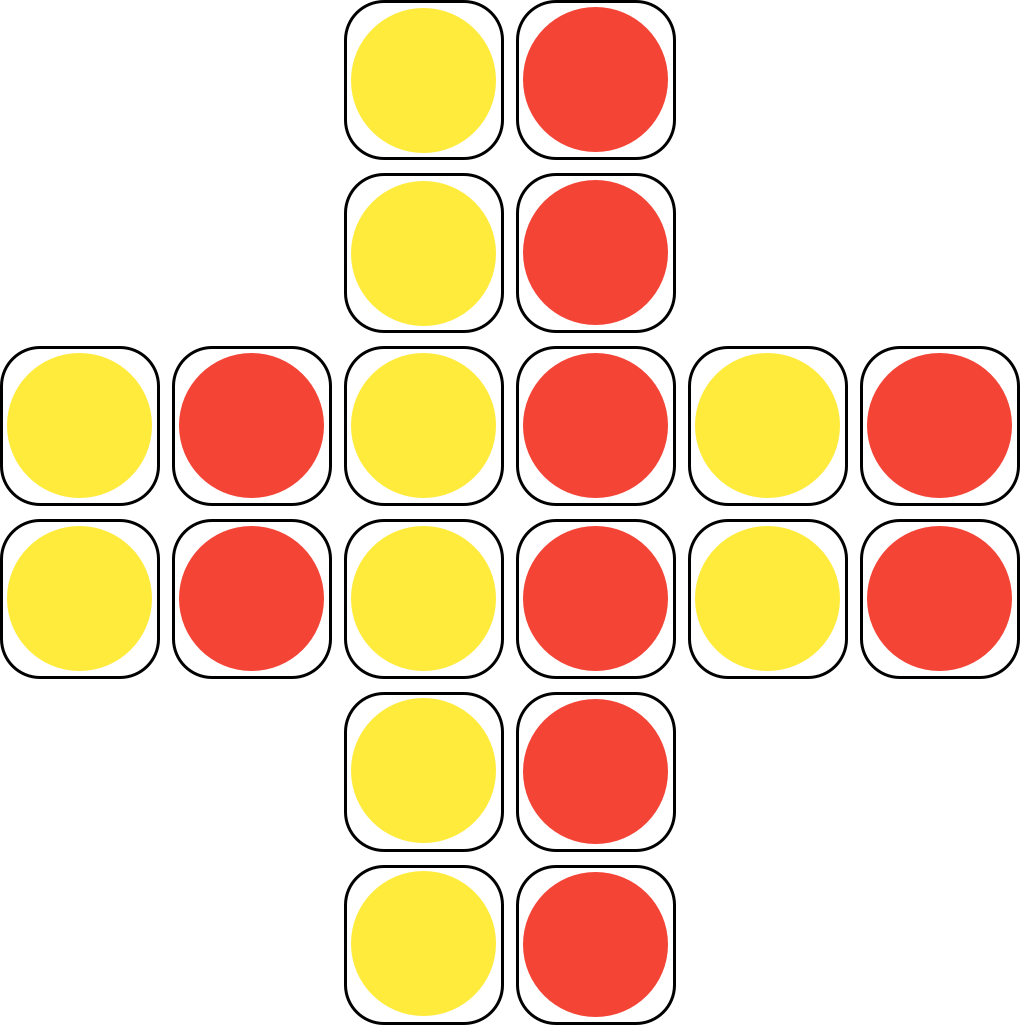}
		\mycaptiontitle{{Schema 3 Algorithm 1.}}
		\label{s3a1}
	\end{subfigure}
	\hfill
	\begin{subfigure}[t]{0.3\linewidth}
		\centering
		\includegraphics[width=\textwidth]{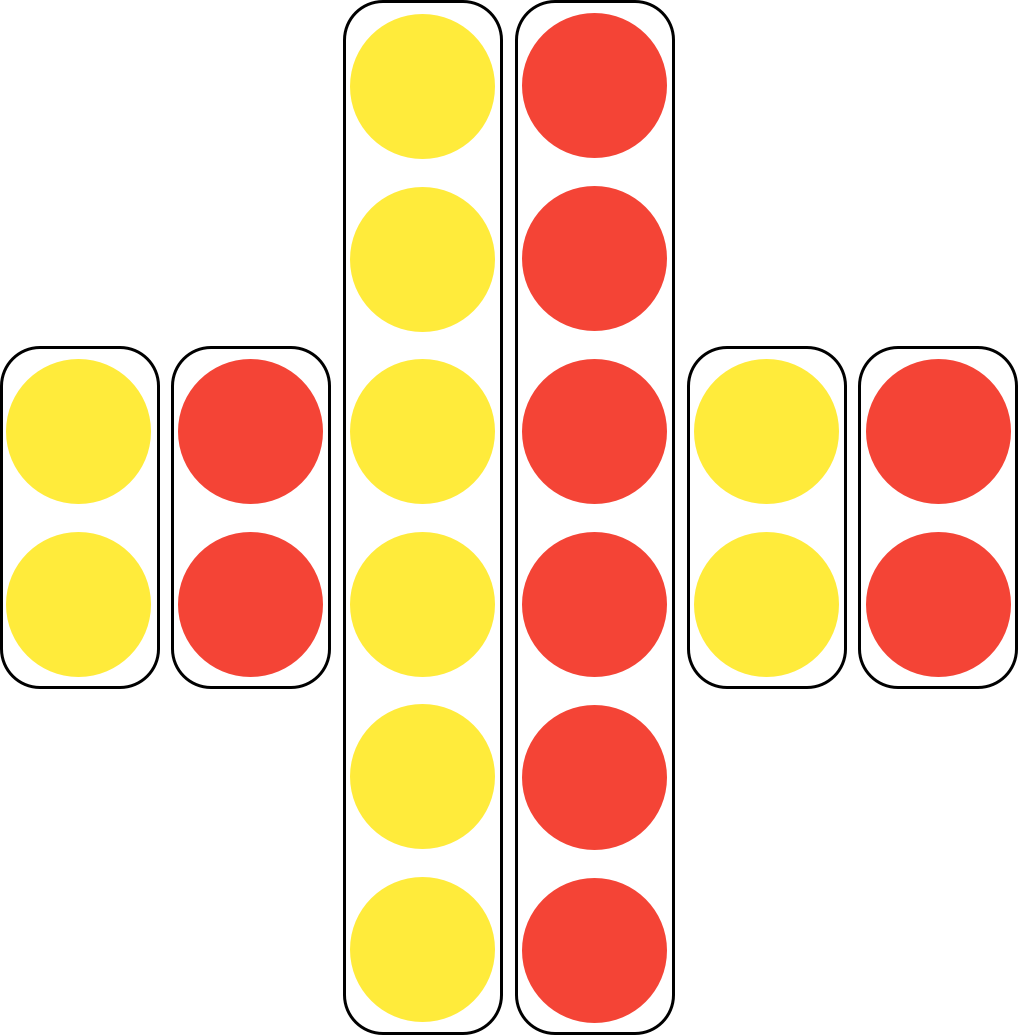}
		\mycaptiontitle{{Schema 3 Algorithm 3.}}
		\label{s3a3}
	\end{subfigure}
	\hfill
	\begin{subfigure}[t]{0.3\linewidth}
		\centering
		\includegraphics[width=\textwidth]{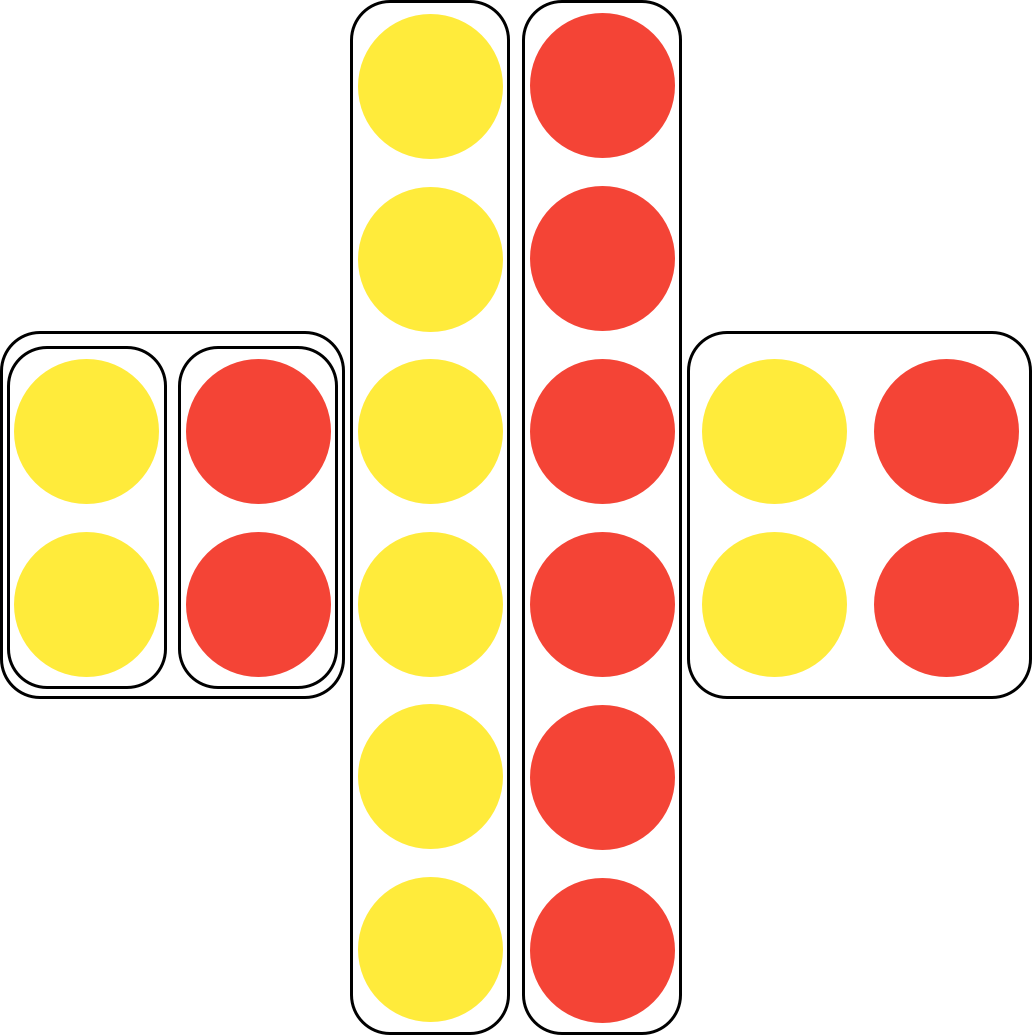}
		\mycaptiontitle{Schema 3 Algorithm 5.}
		\label{s3a5}
	\end{subfigure}
	\mycaptiontitle{Examples of algorithms for Schema 3.}
	\label{fig:s3algorithms}
\end{figure}

If the pupil describes the array point by point without redundancy (see \Cref{s3a1}), he is using a 0D algorithm with 20 CODs (one for each dot), consequently, the number of operations is equal to 20. 

If the pupil describes the array column by column without any loop (see \Cref{s3a3}), the algorithm consists of six CSDs (one for each column) and is 1D. The number of operations corresponds to the number of CSDs (there is no COD) and is equal to 6. 

Finally, we considered an algorithm in which the left square is described through its two columns, and the right square is described as equal to the left one, while the two columns in the middle are described column by column (see \Cref{s3a5}).
\lstset{escapeinside={(*@}{@*)}}
\begin{figure}[ht]
\mycaptionoflisting{Schema 3 Algorithm 5 pseudo-code.}
\label{alg:s3algo5}
  \centering
  \begin{lstlisting}[language=pseudocode, basicstyle=\footnotesize\ttfamily]
# One cycle for each square (e.g., i=1 left square, i=2 right square)
for i from 1 to 2 do(*@\label{lst:line:loop}@*) 
    # Colour the left column of the square in yellow
    CSD(column, yellow)(*@\label{lst:line:csd1}@*)
    # Colour the right column of the square in red
    CSD(column, red)(*@\label{lst:line:csd2}@*)
end for
# Colour the left central column in yellow
CSD(column, yellow)(*@\label{lst:line:csd3}@*) 
# Colour the right central column in red
CSD(column, red)(*@\label{lst:line:csd4}@*) 
\end{lstlisting}
\end{figure}
To generate the right square, it is sufficient to repeat the operations performed to generate the left square and could be described in a pseudo-code using a loop (see Listing~\ref{alg:s3algo5}).
In this case, the algorithm contains a loop on two CSD (see line~\ref{lst:line:loop}), making it 2D. The number of operations is 4, corresponding to the number of CSDs, respectively, on lines \ref{lst:line:csd1}, \ref{lst:line:csd2}, \ref{lst:line:csd3} and \ref{lst:line:csd4} (CSDs inside a loop are considered once). 
    
    

\subsection{Interaction dimension}\label{sec:classificationinteractions}
The CAT, in addition to algorithmic skills, evaluates aspects of situated cognition by assessing how students apply their knowledge in context, including interactions with artefacts and social dynamics.
In particular, the interaction dimension reflects both the complexity of the artefacts used by the students during the task and the level of autonomy demonstrated, determined by the extent to which they asked for visual cues and relied on visual feedback.
We distinguish between:
\begin{enumerate}[noitemsep,nolistsep]
	\item \emph{VSF}: this level involves using voice and hand gestures on an empty cross array, hinging on visual feedback;
	\item \emph{VS}: this level involves using voice and hand gestures on an empty cross array, without hinging on visual feedback, and 
	\item \emph{V}: this level involves using only voice without hand gestures on an empty cross array or visual feedback.
\end{enumerate}

\subsection{CAT score}\label{sec:catscore}

The task is considered successful if the student creates a complete and correct algorithm, regardless of its complexity, the artefactual environment, or the level of autonomy. 
To measure how a pupil's competencies evolve, we define a single metric, the \textit{CAT score}, which quantifies their multi-faceted performance, encompassing both the algorithm and interaction dimensions (see \Cref{tab:unplugged_catscore}).
For more details on the CAT assessment rubric refer to \Cref{sec:modelling-cat-ae}.


\begin{table}[!ht]
	\footnotesize
	\mycaption{CAT score (unplugged CAT).}{Rows represent the algorithm dimensions and columns represent the interaction dimensions. 
		\label{tab:unplugged_catscore}}
	\centering
\begin{tabular}{c>{\centering\arraybackslash}m{.8cm}|>{\centering\arraybackslash}m{.8cm}>{\centering\arraybackslash}m{.8cm}>{\centering\arraybackslash}m{.8cm}}
				\multicolumn{2}{c|}{}   & \multicolumn{3}{c}{{\textbf{Competence level}}} \\
				\multicolumn{2}{c|}{}   & {VSF $c=1$}    & {VS $c=2$}   & {V $c=3$}   \\
				\midrule
				\multirow{3}{*}{\begin{minipage}{.5cm}\rotatebox{90}{\begin{tabular}{c}\textbf{Competence}\\[-1ex]\textbf{component}\end{tabular}}\end{minipage}} 
				& {0D $r=1$} & $0$   & $1$  & $2$             \\
				& {1D $r=2$} & $1$   & $2$  & $3$             \\
				& {2D $r=3$} & $2$   & $3$  & $4$             \\
			\end{tabular}
\end{table}

The algorithm dimension score ranges from 0 (for the simplest level of complexity, 0D) to 2 (for the most complex, 2D). 
The overall algorithm dimension score reflects the most complex operation successfully performed by the student during the assessment. 

The interaction dimension score ranges from 0 (for the simplest level of complexity, VSF) to 2 (for the most complex, V).
The overall interaction dimension score reflects the lowest level of effective complexity of interaction demonstrated by the student during the assessment.
Specifically, the exclusive use of voice (V) is considered more complex and valuable than using voice alongside the empty array (VS). Additionally, a more autonomous role (using voice alone) is valued over a less autonomous one (relying on visual feedback).

We calculate the CAT score for each task completed by the student as the sum of the two scores, ranging from 0 (minimum) to 4 (maximum). 
A higher score is indicative of a student who has navigated the complexities of challenging artefacts, assumed an autonomous role, and/or conceived a higher-dimensional algorithm.

The CAT score alone should not be used to compare students' performance, as different combinations of algorithm complexity and interaction strategies can result in the same total score, despite varying approaches.
For example, a pupil who describes a two-dimensional algorithm using voice and an empty array (2D-VS) receives the same score as a pupil who describes a one-dimensional algorithm exclusively using voice (1D-V), even though the strategies differ in algorithm and interaction complexity. 
However, this does not hinder gaining a deeper understanding of students' algorithmic and interaction skills, as the separate dimensions provide detailed insights into their problem-solving approaches, even when the final score is the same.

%

\subsection{Task metrics}
We also considered metrics to gauge students' proficiency in AT and task execution.

\paragraph{Participation rate} 
The participation rate measures whether students attempted and concluded each task assigned during the CAT assessment, regardless of correctness. Each student is assigned 12 tasks, and the participation rate indicates how many of these tasks. This metric provides an initial overview of students' engagement and persistence in the assessment activities.

\paragraph{Success rate} 
The success rate evaluates the number of tasks that students correctly solved during the CAT assessment. 



 {\tiny }
	\chapter[The virtual Cross Array Task (CAT)]{The virtual Cross Array Task}
\begin{custombox}
    {The content of this chapter has been adapted from the following articles with permission of all co-authors and publishers:}
    \begin{itemize}[noitemsep,nolistsep,leftmargin=.25in]

        \item \textbf{Adorni, G.}, Piatti, S., and Karpenko, V. (2024). Virtual CAT: A multi-interface educational platform for algorithmic thinking assessment. \textit{SoftwareX} \cite{adorni_app_article2023}.
        
        \item \textbf{Adorni, G.} and Piatti, A. (unpublished). Designing the virtual CAT: A digital tool for
        algorithmic thinking assessment in compulsory education \cite{adorni_pilot2023}. 
       
        \item \textbf{Adorni, G.}, Artico, I., Piatti, A., Lutz, E., Gambardella, L. M., Negrini, L., Mon-
        dada, F., and Assaf, D. (2024). Development of algorithmic thinking skills in K-12
        education: A comparative study of unplugged and digital assessment instruments.
        \textit{Computers in Human Behavior Reports} \cite{adorni_2024_largescale}.
        
       \item \textbf{Adorni, G.}, Piatti, A., Bumbacher, E., Negrini, L., Mondada, F., Assaf, D., Mangili, F., and Gambardella, L. M. (2025). FADE-CTP: A Framework for the Analysis and Design of Educational Computational Thinking Problems. \textit{Technology,  Knowledge and Learning} \cite{adorni2023ctpframework}.
        \end{itemize}
        As author of these publications, my contribution involved:\\
        \textit{Conceptualisation, Methodology, Validation, Formal analysis, Investigation, Resources, Data curation, Writing -- original draft \& review \& editing, Visualisation, Supervision.}
\end{custombox}

\section{Summary}

This chapter contributes primarily to RQ2, by focusing on the design and development of the virtual adaptation of the CAT, aimed at enabling automated large-scale assessment of algorithmic skills across the entire compulsory education path (K-12) in Switzerland.
It begins by describing the transition from an unplugged to a virtual format, emphasising the design of the new instrument. 
Next, it outlines the categorisation of the new method of interaction and variations to the algorithm dimension metric, explaining their role in shaping the evaluation metric for assessing AT. 
Finally, a technical overview of the instrument's implementation is provided with a detailed discussion of prototype development, demonstrating how the virtual CAT can be integrated into large-scale assessment systems for use across diverse educational contexts.

\section{Activity design}\label{sec:design_virtualCAT}

Our first experience with the unplugged CAT highlighted several limitations that made it unsuitable for large-scale assessment. 
Designed as a one-on-one activity, it was time-intensive and impractical for simultaneous administration. 
Additionally, reliance on a human administrator introduced potential inconsistencies in interpreting and delivering instructions, and it did not allow for automated assessment, further limiting its scalability and efficiency.
\begin{figure}[hbt]
	\centering
	\includegraphics[width=.55\linewidth]{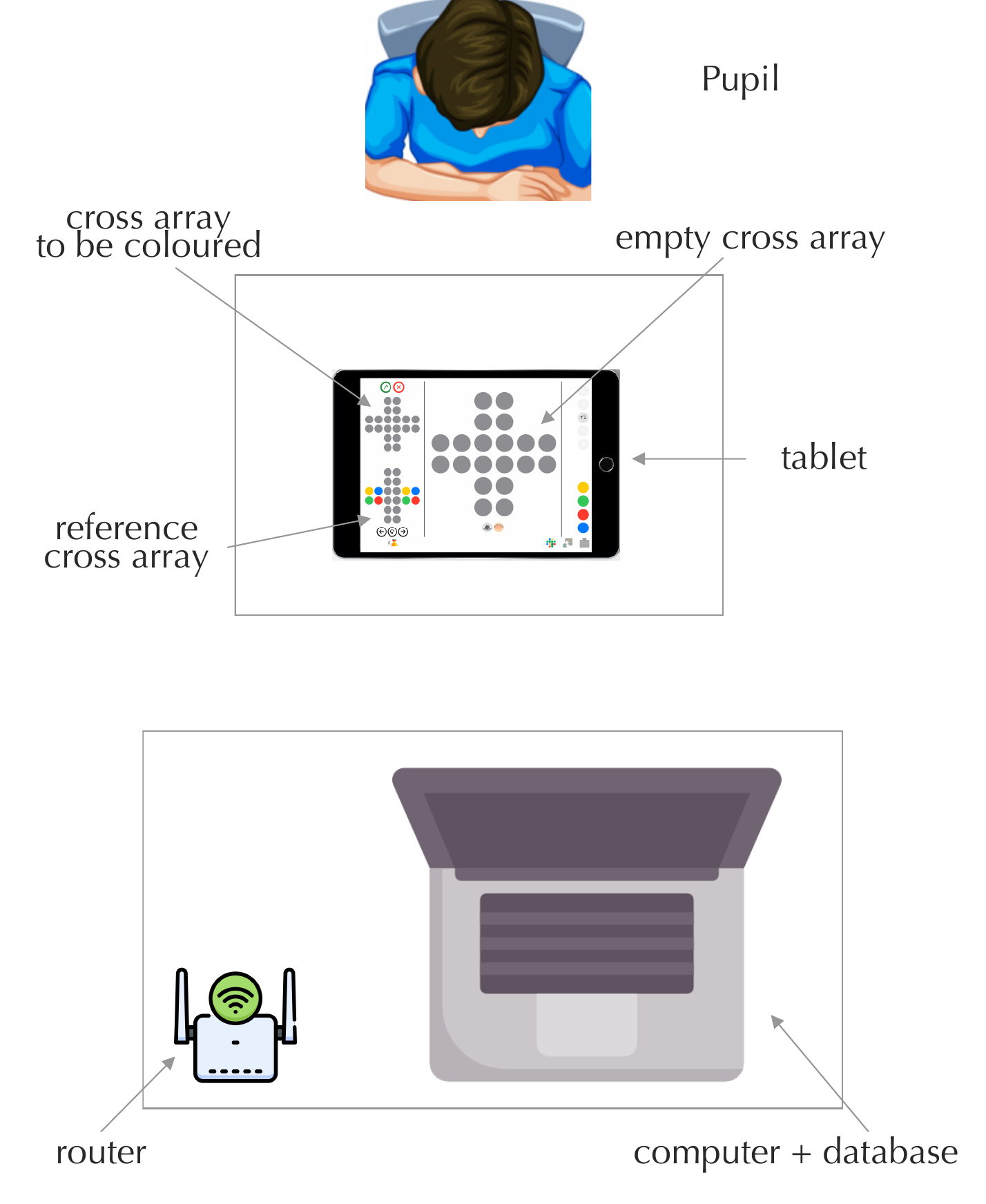}
	\mycaption{Experimental settings (virtual CAT)}{
		The pupil is tasked with recreating a given reference cross array schema using either a gesture-based or visual block-based programming interface.
		Initially, the pupil cannot view the outcome of their instructions due to a feature that blocks visual cues, which can be toggled as needed. 
		The system automatically interprets and logs all of the student's actions and algorithms in an external database.}
	\label{fig:virtualCAT}
\end{figure}
To address these challenges, following the methodological frameworks presented in \Cref{part:methodological_framework}, we developed a digital version of the activity, called virtual CAT \citep{adorni_app_article2023,adorni_pilot2023}.
This adaptation streamlined the assessment process by eliminating the dependency on human administration, ensuring consistency, and allowing for large-scale automated administration.



%
%
%

The virtual CAT retains the core elements of the original activity: 
(i) the \textit{cognitive activity} requires students to devise a set of instructions or an algorithm to replicate coloured patterns on a cross-shaped grid;
(ii) the \textit{artefactual environment} includes a variety of cognitive artefacts based on both embodiment and perception as well as symbolic representation, even though specific artefacts used may vary;
(iii) the students' \textit{autonomy} reflects their level of independence during the activity, ranging from those who do not engage with the task to those who rely on visual feedback up to those who provide clear instructions independently.

The activity is still conducted in class (see \Cref{fig:virtualCAT}), but while the unplugged CAT relied on face-to-face interaction with a human administrator, the virtual CAT relied on a virtual agent that interprets and executes the algorithms devised by students. 
In particular, it acts as a programming language interpreter, translating gesture interactions and visual blocks into a formal programming language that mirrors the operations used in the unplugged activity. Algorithms are automatically recorded, enabling immediate evaluation of students’ strategies and providing feedback along with optional guidance or hints.
This approach eliminates human errors and inconsistencies in instruction interpretation, ensuring a more standardised assessment experience.

Administration via individual devices enables simultaneous participation and assessment by multiple pupils, allowing entire classrooms or larger groups to engage with the activity independently on their devices, thereby overcoming the logistical challenges of one-on-one interactions in the unplugged version without requiring additional human resources.

As for the artefacts available in the virtual environment, two interfaces are provided to accommodate diverse learning styles and preferences.
We did not provide a modality of interaction based on verbal instruction, but we provided an alternative symbolic language, a visual block-based programming interface (CAT-VPI).
This decision was driven by several technical challenges of implementing speech recognition, particularly in a multilingual classroom context with young students \citep{AiminZhou2024,Wilpon1996}.
Most speech recognition systems are trained on adult voices, making them less accurate for children, whose speech differs in pitch and tone \citep{GurunathShivakumar2020,Potamianos2003,Yeung2018,GurunathShivakumar2022}. 
Additionally, there isn't enough data available to improve the accuracy of children’s voices \citep{Claus2013,Fainberg2016}. 
While there are techniques to adjust children's voices to be more like adult voices, they don't work well enough \citep{Ahmad2017,Shahnawazuddin2017,Kathania2018,Kathania2018b}. 
For these reasons, we chose a visual programming interface instead, which is more reliable in this context, and its implementation requires less effort.

The CAT-VPI is designed to make coding accessible to K-12 students, including beginners with no prior programming experience. 
It allows students to construct colouring algorithms using drag-and-drop programming blocks, which mirror the instructions observed in the unplugged version of the activity.
Nevertheless, these blocks are customisable, enabling users to adjust parameters like colour and pattern choices.
This intuitive and flexible approach reduces the likelihood of syntax errors, potentially improving the overall learning experience.
For a visual representation of this interface, refer to \Cref{fig:programming_final}, with further details on its development provided in \Cref{sec:third_prototype}.

The CAT-GI is designed to emulate the hand gestures observed in the unplugged CAT activity, providing a tactile experience similar to interacting with the physical cross array that ensures continuity in the interaction type while leveraging digital capabilities.
Users can build the colouring algorithm by selecting colours, tapping on individual dots, dragging across multiple dots to create patterns, or using icons to perform more advanced actions, such as repeating instructions or mirroring patterns.
For a visual representation of this interface, refer to \Cref{fig:gestures_final}, with further details on its development provided in \Cref{sec:third_prototype}.


The digital version retains the challenge of limited visual feedback present in the unplugged CAT by restricting students' access to the agent's progress unless explicitly enabled, thereby preserving the original task's difficulty and autonomy while offering flexibility in feedback.


Pupils also have the flexibility to choose their preferred interaction mode, navigate between tasks, restart them, confirm completion, or skip them as needed.
Upon completing all tasks, pupils are directed to a results dashboard that comprehensively summarises their performance, including visual representations of attempted tasks, scores, completion status, and time taken.

To ensure accessibility across Switzerland and broader applicability, the virtual CAT supports multiple languages, including Italian, French, and German, reflecting Switzerland's linguistic diversity, and also provides an English version to extend its potential use (see Figure~\ref{fig:language}).

\Cref{tab:unplugged_virtual} summarise the principal differences between the unplugged and virtual versions of the CAT.
\begin{table}[htb]
	\footnotesize
	\mycaptiontitle{Differences between the unplugged and virtual CAT.}\label{tab:unplugged_virtual}
	\setlength{\tabcolsep}{2.5mm}
	\centering
	\begin{tabular}{>{\arraybackslash}m{2.5cm}>{\arraybackslash}m{5cm}>{\arraybackslash}m{5cm}}
		\toprule
		& \textbf{Unplugged CAT} & \textbf{Virtual CAT} \\
		\midrule
		\textbf{Interactions} & Face-to-face \newline(problem solver \& human agent) & Face-to-device \newline(problem solver \& virtual agent) \\[.2cm]
		\textbf{Artefactual \newline environment} & Voice (symbolic) and hand gestures (embodied)& Block-based visual programming interface (symbolic) and gesture interface (embodied)\\[.2cm]
		\textbf{Autonomy} & Removable physical barrier to enable visual feedback & Toggleable button to enable visual feedback \\[.2cm]
		\textbf{Algorithm \newline classification} & A human agent interprets instructions and manually codifies the algorithm & A virtual agent interprets instructions and automatically codifies the algorithm into formal programming language \\[.2cm]
		\textbf{Assessment} & Manual & Automatic \\
		\bottomrule
	\end{tabular}
\end{table}

\FloatBarrier
\section{Activity profile}\label{sec:design_result2}

\begin{figure*}[ht]
	\centering
	\includegraphics[width=\textwidth]{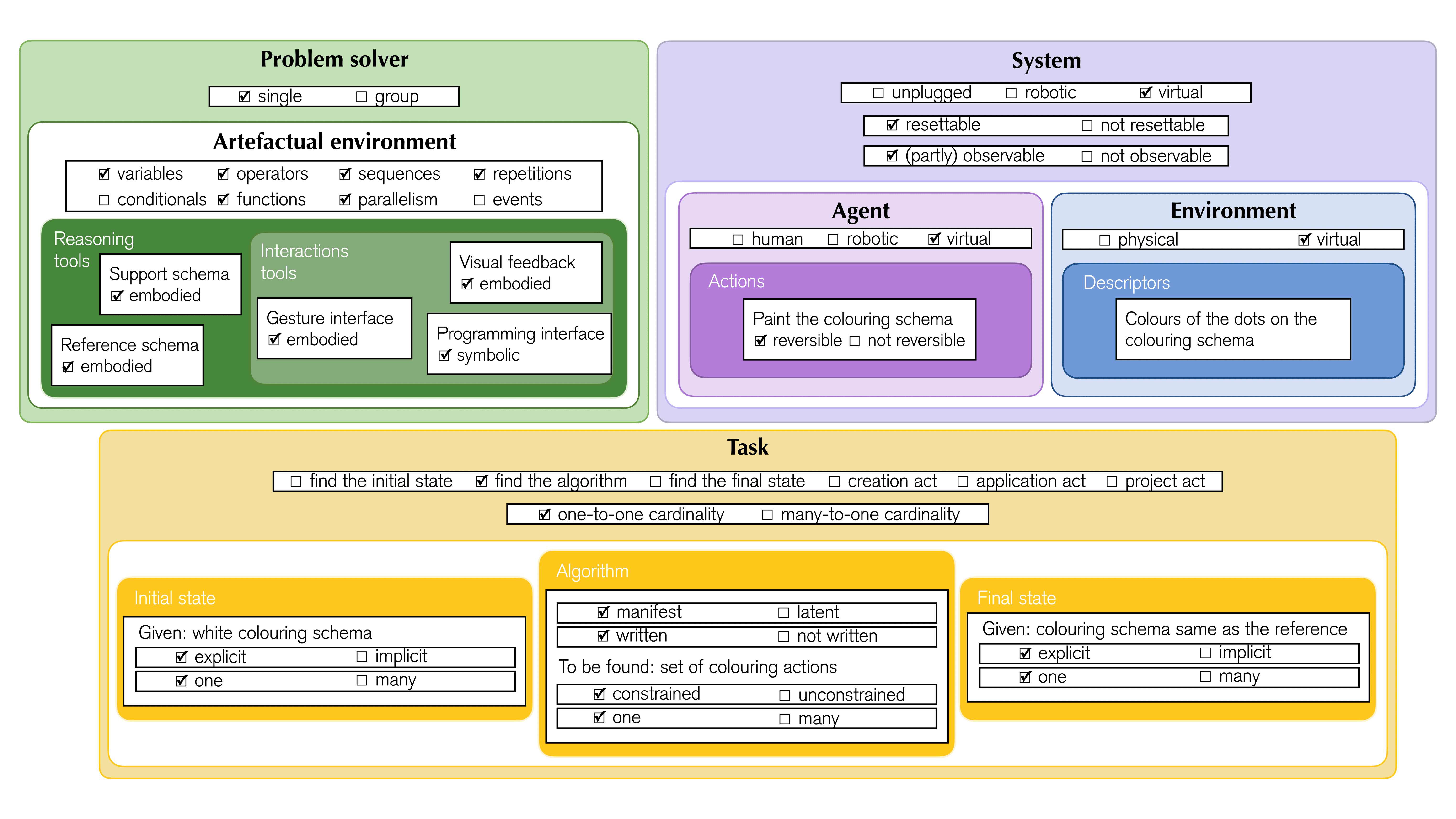}
	\mycaptiontitle{{Components and characteristics (virtual CAT).}}
	\label{fig:virtualCAT-features}
\end{figure*}

\begin{figure*}[h]
	\centering
	\includegraphics[width=\textwidth]{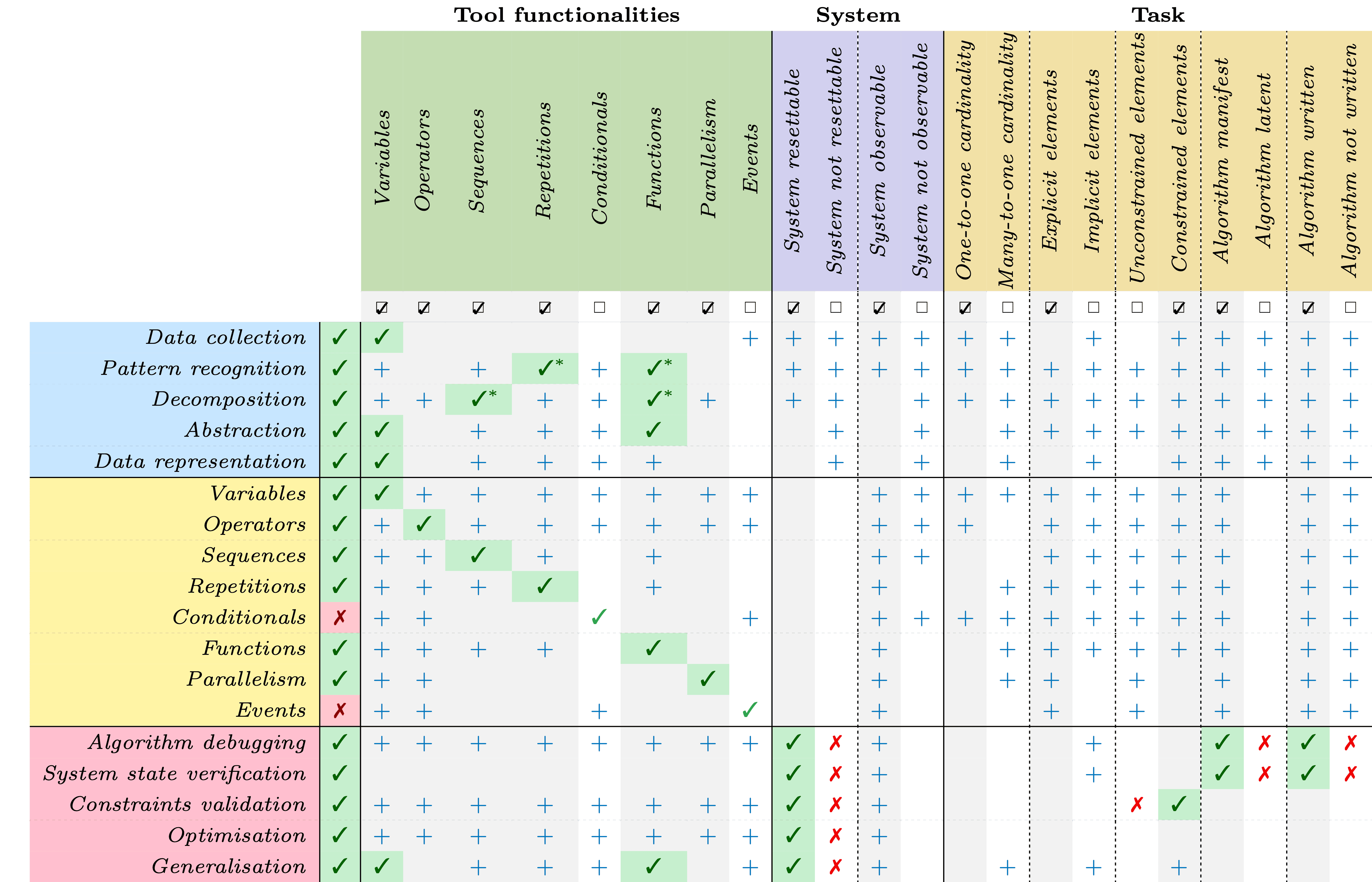}
	\mycaptiontitle{{Activity profile (virtual CAT).}}
	\label{fig:virtualCAT-mapping}
\end{figure*}

As done for the unplugged CAT, the components and characteristics of the virtual CAT are illustrated following our FADE-CTP and graphically represented in \Cref{fig:virtualCAT-features} Additionally, we provide an overview of the overall CT competencies that can be developed through this activity, as well as those that cannot, and finally we present in \Cref{fig:virtualCAT-mapping} the resulting profile, mapping the relationship between the activity’s characteristics and the CT competencies it activates.

\subsection{Components}
\begin{itemize}[noitemsep,nolistsep]
	\item \textit{Problem solver}: the student who has to communicate the algorithm, or the sequence of instructions to reproduce the colouring of the reference schema.
	The artefactual environment comprises cognitive tools such as support and reference schemas, which are available to the problem solver to reason about the task.
	Additionally, the problem solver can interact with the system to communicate the algorithm. This can be achieved using a block-based visual programming interface (symbolic) or a gesture interface (embodied). 
	Moreover, the problem solver can eventually observe the status cross array being coloured, enabling the visual feedback (embodied).
	\item \textit{Agent}: the virtual agent, which automatically interprets and executes the problem solver's instructions. The agent's actions are resettable.
	\item \textit{Environment}: the cross array to be coloured, whose state is described by the colour of each dot (white, yellow, blue, green, or red). 
	\item \textit{Task}: find the algorithm. The system's state is defined by the colouring cross status, initially white and, at the end, the same as the reference schema. 
	The algorithm is the set of agent instructions to achieve this transformation.
\end{itemize}

\subsection{Characteristics}

\begin{itemize}[noitemsep,nolistsep]
	\item \textit{Tool functionalities}: the same core functionalities associated with algorithmic concepts as the unplugged CAT are supported in the virtual CAT through the gesture and programming interfaces.
	\item \textit{System resettability}: the system is instead resettable, since students can reverse the agent's actions by restarting the task or modifying the algorithm.
	\item \textit{System observability}: the system remains partially observable since the cross array being coloured by default is not seen until the end of the task unless the student clicks the button to enable visual feedback.
	\item \textit{Task cardinality}: the task has a one-to-one mapping, with given one initial and one final state, and an algorithm to be found.
	\item \textit{Task explicitness}: all elements are given explicitly.
	\item \textit{Task constraints}: the algorithm is now constrained since pupils can use only the commands made available.
	\item \textit{Algorithm representation}: the algorithm is represented through visual code blocks or gestures.
	It remains manifest, as it is externalised, and in the case of code blocks, it is also written, since it is stored in a permanent format.
\end{itemize}

\Cref{tab:charcateristics_comparison} summarises the key characteristics and differences between the CAT variants. 
\begin{table}[ht]
	\renewcommand{\arraystretch}{1.6}
	\footnotesize
	\centering
	\mycaptiontitle{Differences between the unplugged and virtual CAT characteristics.}
	\label{tab:charcateristics_comparison}
	\begin{tabular}{l|cc}
		& {\textbf{Unplugged CAT}} & {\textbf{Virtual CAT}} \\
		\midrule
		
		\cellcolor[HTML]{C4DDB2}\textit{Tool \newline functionalities} & \multicolumn{2}{>{\centering\arraybackslash}m{5cm}}{Variables, operators, sequences, repetitions, functions and parallelism}\\ \hline 
		\cellcolor[HTML]{D2D0EF}\textit{System \newline resettability} & Not resettable & Resettable \\
		\cellcolor[HTML]{D2D0EF}\textit{System \newline observability} & \multicolumn{2}{c}{Partially observable}\\ \hline 
		\cellcolor[HTML]{F2E1A6}\textit{Task \newline cardinality} & \multicolumn{2}{c}{One-to-one} \\     
		\cellcolor[HTML]{F2E1A6}\textit{Task \newline explicitness} & \multicolumn{2}{c}{Explicit}\\    
		\cellcolor[HTML]{F2E1A6}\textit{Task \newline constraints} &  Unconstrained & Constrained \\
		\cellcolor[HTML]{F2E1A6}\textit{Algorithm \newline representation} & Manifest non-written & Manifest written \\
	\end{tabular}
\end{table}

\subsection{Competencies}

\paragraph{Enabling features for competencies development}
\begin{itemize}[noitemsep,nolistsep]
	\item \textit{Problem setting}: all competencies can be activated, consistently with the unplugged CAT, as the core tool functionalities required to activate these skills remain unchanged.
	
	\item \textit{Algorithm}: the same algorithmic competencies activated in the unplugged CAT are also triggered in the virtual version, as both rely on the same tool functionalities.
	
	\item \textit{Assessment}: unlike the unplugged version, the virtual CAT enables the activation of all assessment skills, thanks to the system’s resettable nature and a manifest, written representation of the algorithm. Additionally, the inclusion of constraints activates constraint validation.
\end{itemize}

\paragraph{Inhibiting features for competencies development}
\begin{itemize}[noitemsep,nolistsep]
	\item \textit{Conditionals and events}: these competencies remain non-activable as these functionalities are unavailable in the platform. 
\end{itemize}

\FloatBarrier

\section{Competencies assessment}
\begin{figure}[!h]						
	\centering
	\includegraphics[width=.75\linewidth]{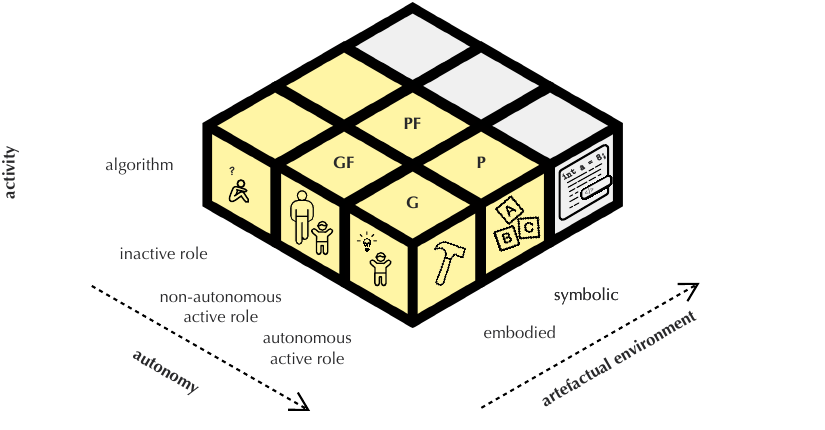}
	\mycaptiontitle{Assessed cells in the algorithm activity of the {CT-cube} for the virtual {CAT}.}
	\label{fig:competencies_assessment_cells_virtual}
\end{figure}

As in the unplugged version, the primary focus of the virtual CAT is on measuring algorithmic skills while considering aspects of situated cognition, participation rate, and success rate (see \Cref{fig:competencies_assessment_cells_virtual}).
However, the digital environment introduces an additional layer of assessment, such as evaluating students' proficiency in task execution. This includes quantifying trial-and-error strategies through the number of restarts and assessing efficiency based on the time taken to solve the tasks.

\subsection{Algorithm dimension}\label{sec:virtual_algorithms}

The algorithm dimensions in the virtual CAT remain consistent with those in the unplugged version, evaluating students' algorithmic skills based on the complexity of the operations used to describe the cross arrays (\textit{0D}, \textit{1D}, and \textit{2D} algorithms).




\subsection{Adjusted algorithm dimension}\label{sec:at_assessment}
Acknowledging the need to assess algorithm efficiency alongside complexity, we have introduced an adapted metric that considers the number of commands used, providing a more nuanced evaluation of students' algorithmic competencies.
It recognises cases where a simpler yet more efficient algorithm may perform better than a complex one with more commands.
The adjusted score, denoted as $\widetilde{\text{AD}}$, is calculated using a formula that balances the highest complexity level achieved by the student against the overall workload: 
\begin{align}
	\widetilde{\text{AD}} = \frac{1 + {P_{\text{max-}d}} + \sum_{{d}} \left( {C_{d}} \cdot P_d \right)}{C_{\text{total}}} \mbox{,}
	\label{eq:AD}
\end{align}
where,
$d$ is the complexity level of the algorithm (i.e., 0, 1, or 2);
$P_{\text{max-}d}$ are the points assigned to the highest complexity level used by the student, computed as the original algorithm dimension score plus one;
$C_d$ is the number of commands at complexity level $d$; 
$P_d$ are the points for the complexity level $d$, computed as the original algorithm dimension score at that complexity level plus one;
$C_{\text{total}}$ is the overall number of commands used across all levels.
The first term in \Cref{eq:AD} gives a score for the most complex algorithm achieved by the student, adjusted for the total commands used, favouring higher-level algorithms but considering the overall workload in terms of the number of commands executed.
The second term calculates a weighted score for each complexity level, factoring in the proportion of commands used at each complexity level relative to the total command count and multiplying it by the points for that level. 


\subsection{Interaction dimension}\label{sec:virtual_interactions}

%


The interaction dimension in the Virtual CAT expands to include four levels of complexity, reflecting both the type of interface used and the level of reliance on visual feedback.
We distinguish between:
\begin{itemize}[noitemsep,nolistsep,align=parleft,leftmargin=1.2cm]
    \item[\textit{GF}:] this level involves using the gesture interface, hinging on visual feedback;
    \item[\textit{G}:] this level involves using the gesture interface, without hinging on visual feedback;
    \item[\textit{PF}:] this level involves using the visual programming interface, hinging on visual feedback;
    \item[\textit{P}:] this level involves using the visual programming interface, without hinging on visual feedback;
\end{itemize}

%


\subsection{CAT score}\label{sec:assessment_variables_virtual_cat}

%
The CAT score for the virtual CAT (see \Cref{tab:catscore_virtual}) is computed following the same approach as the unplugged version, with a key difference in the interaction dimension score, which now ranges from 0 (for the simplest level of complexity, GF) to 3 (for the most complex, P).
The overall CAT score for each task completed by the student is calculated as the sum of the two dimension scores, ranging from 0 (minimum) to 5 (maximum).
For more detail on the development of unplugged CAT assessment rubric refer to \Cref{sec:modelling-cat-ae}.
\begin{table}[htb]
	\footnotesize
	\centering
	\mycaption{CAT score (virtual CAT).}
	{Rows represent the algorithm dimensions and columns represent the interaction dimensions.}
	\label{tab:catscore_virtual}
\begin{tabular}{c>{\centering\arraybackslash}m{.8cm}|>{\centering\arraybackslash}m{.8cm}>{\centering\arraybackslash}m{.8cm}>{\centering\arraybackslash}m{.8cm}>{\centering\arraybackslash}m{.8cm}}
				\multicolumn{2}{c|}{}   & \multicolumn{4}{c}{{\textbf{Competence level}}} \\
				\multicolumn{2}{c|}{}   & {GF $c=1$}    & {G $c=2$}   & {PF $c=3$} & {P $c=4$}   \\
				\midrule
				\multirow{3}{*}{\begin{minipage}{.5cm}\rotatebox{90}{\begin{tabular}{c}\textbf{Competence}\\[-1ex]\textbf{component}\end{tabular}}\end{minipage}} 
				& {0D $r=1$} & $0$   & $1$  & $2$  & $3$             \\
				& {1D $r=2$} & $1$   & $2$  & $3$  & $4$             \\
				& {2D $r=3$} & $2$   & $3$  & $4$  & $5$             \\
			\end{tabular}
\end{table}

\subsection{Task metrics}
Additionally, the CAT assessment instrument evaluates various metrics to gauge students' proficiency in AT and task execution.

\paragraph{Participation rate} 
{The participation rate measures whether students attempted and concluded each task assigned during the CAT assessment, regardless of correctness. Each student is assigned 12 tasks, and the participation rate indicates how many of these tasks. This metric provides an initial overview of students' engagement and persistence in the assessment activities.
}

\paragraph{Success rate} 
{The success rate evaluates the number of tasks that students correctly solved during the CAT assessment, irrespective of efficiency or the number of attempts made.}

\paragraph{Number of restarts} 
{The number of restarts reflects students' approach to problem-solving, particularly their use of trial and error (T\&E) strategies. It counts instances where students choose to restart the tasks, indicating their iterative approach to refining algorithms and achieving desired outcomes.}

\paragraph{Efficiency} 
{Efficiency evaluates how effectively students complete tasks, considering the time taken as a factor.}

\section{Instrument development and implementation}\label{sec:instrument_development_implementation}


In this section, we present our instrument development strategy, guided by the User Experience (UX) design life cycle. This structured approach, involving the systematic collection of data on user behaviours, preferences, and requirements, ensures the development of a user-centred product aligned with their actual needs \citep{hartson2018ux}.
Additionally, we discuss the technical components and architectural choices behind the platform's development, covering its framework, data management, programming language formalization, and interpreter implementation.

\subsection{Development process}
The UX design life cycle is an iterative process that encompasses three main phases: (1) \textit{understand} (U) -- gathering insights into user needs and problem domains; (2) \textit{make} (M) -- designing and prototyping solutions based on the understanding phase; (3) \textit{evaluate} (E) -- testing prototypes and solutions through user feedback and expert analysis \citep{hartson2018ux}.
The cyclic flow between understanding, making, and evaluating emphasises the iterative nature of this process. Insights gained from evaluations often lead to revisiting earlier phases to refine and enhance the design. 
While the specific phases and iterations of the UX design life cycle can vary in the literature, we have adopted the process illustrated in \Cref{fig:design_process} to ensure a structured and user-centred approach.

\begin{figure}[h]
	\centering
	\includegraphics[width=.7\textwidth]{VIRTUAL/design\_process2.pdf}
	\mycaption{UX design life cycle.}{This process encompasses understanding user needs (U), making (M) -- or designing and prototyping solutions --, and evaluating them through user and expert feedback (E). These phases repeat cyclically, with evaluation insights leading to refinements in earlier stages.
	}
	\label{fig:design_process}
\end{figure}

In the context of product development, two types of evaluations are commonly employed to guide and assess design: formative and summative evaluation.
\textit{Formative evaluation} takes place throughout the iterative design process, helping to refine and improve the product before it reaches its final form. It ensures that continuous user feedback informs ongoing refinements of the design, intending to improve usability, functionality, and overall user experience. 
\textit{Summative evaluation}, on the other hand, is conducted once the product has been fully developed, aiming to assess its overall effectiveness and impact. It occurs after the design has been finalised. It focuses primarily on evaluating the effectiveness and impact of the final product, typically through large-scale studies, and does not involve redesign or further iterations \citep{scriven1967methodology,dick2005systematic,williges1984evaluating,CARROLL1992}.

\begin{figure*}[th]
	\centering
	\includegraphics[width=\textwidth]{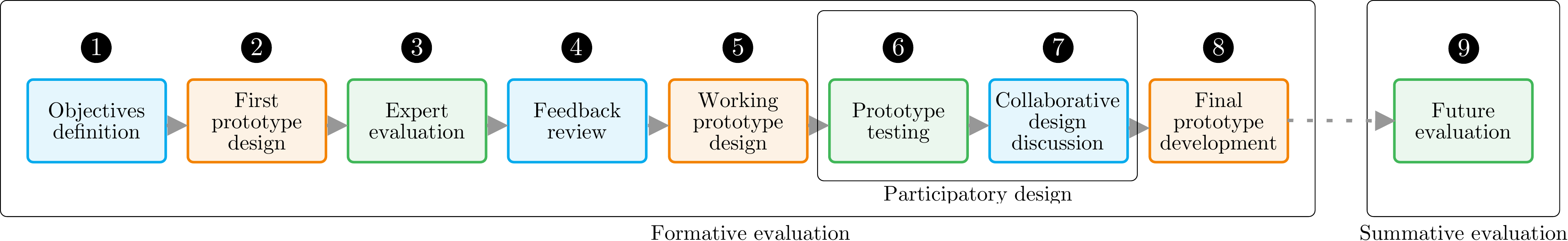}
	\mycaption{Design and evaluation process overview.}{The various stages of the design process, from defining objectives to developing and evaluating prototypes. Different colours represent the phases of the UX design lifecycle: blue for the understand (U) phase, orange for the make (M) phase, and green for the evaluate (E) phase.}
	\label{fig:method}
\end{figure*}

Our design and evaluation process, illustrated in \Cref{fig:method}, followed a structured and iterative procedure aligned with the phases of the UX design cycle of \Cref{fig:design_process}.
The process spanned 17 months, from February 2022 to June 2023.
The first prototype was developed from February to July 2022 (5 months), the second from July 2022 to March 2023 (8 months), and the final version from March to June 2023 (4 months).

\subsubsection{Objectives definition}\label{sec:objectives_definition}
This process began with the objectives definition stage \circled{1}, where we set the instrument's goals and decided how to adapt the unplugged activity to the virtual format.
As discussed in \Cref{sec:design_virtualCAT}, we focused on preserving the pedagogical value of the original task by identifying key components to maintain and determining how to translate them into the digital environment. This step was crucial for ensuring the educational objectives were upheld, laying the foundation for the entire process.

\subsubsection{Initial prototype development}\label{sec:initial_prototype}
The second step in the process focuses on developing the initial digital prototype \circled{2}, prioritising user experience accessibility and usability \citep{hartson2018ux}. 
In educational technology, accessibility centres on crafting solutions to meet users' needs from various backgrounds, regardless of their physical or cognitive abilities \citep{kali2011teaching}.
In contrast, usability focuses on the user experience, aiming at delivering an intuitive and effective learning environment \citep{nikahmad2021usability,gould1985usability}. 
To achieve these objectives, we made several key decisions, including selecting the devices on which the tool would be available, determining supported languages, and defining the layout of the user interfaces, all in accordance with established guidelines and best practices \cite{hartson2018ux}.

The development of the initial prototype was grounded in the architectural and technical decisions outlined in \Cref{sec:implementation_virtualCAT}, ensuring the prototype' functionality and alignment with the educational objectives defined in \Cref{sec:objectives_definition}.
In this phase, we sketched interface layouts, selected appropriate technologies, and built interactive prototypes to simulate user interactions. Throughout this process, we continuously evaluated the user experience, ensuring the prototype was intuitive for students and teachers. Additionally, expert consultations were integrated to validate the design choices, ensuring the prototype met usability and pedagogical standards.

Due to time constraints and limited access to schools and children, we skipped certain prototyping stages, such as producing paper prototypes, and directly developed a functional prototype. This streamlined approach was also necessary because of the age of the children involved in the participatory design, who may struggle with abstract reasoning and therefore require a more accessible prototype \citep{hanna1997,Guran2020,markopoulos2002compare,druin1999}. 
The resulting first prototype is detailed in \Cref{sec:first_prototype}.

\subsubsection{Expert evaluation and prototype redesign}\label{sec:expert_evaluation}

Following, we conducted an expert evaluation \circled{3} to assess the prototype design, usability and accessibility \citep{hartson2018ux}. This step involved the participation of experts in both UX design and educational technology, who examined the prototype and provided detailed feedback on various aspects of the platform, particularly focusing on interface clarity, functionality, and alignment with educational objectives.
The first expert consulted, recruited through our institutional network, is an interaction design teacher-researcher with a background in educational technology and user-centred design.
The expert was provided with a brief description of the platform's intended use before independently exploring the application. 
His feedback was collected during a collaborative session, in which he shared detailed observations after testing the platform. 
In addition, three pedagogical experts with experience in computer science education were invited to review the prototype. These professionals were selected based on recommendations and their known contributions to technology-enhanced learning.
After being introduced to the platform, each expert independently tested the application and subsequently shared their observations during a feedback session. 

During the reflective phase \circled{4}, we carefully analysed and prioritised the changes proposed by the experts, ensuring that the adjustments aligned with both usability principles and pedagogical goals. 

The prototype redesign \circled{5} incorporated these changes and included the development of key technical features, such as a virtual interpreter and the infrastructure necessary for real-time interaction and data processing. 
The feedback received, the modifications decided upon, and the resulting updated prototype are documented in \Cref{sec:second_prototype}.

\subsubsection{Participatory design}\label{sec:pilot_method}

Following the expert evaluations and the creation of the second prototype, the next phase of the design process focused on participatory design and the development of the final application.
This phase began with engaging the target users, students and teachers, in testing the prototype in real-world settings and providing feedback on its usability and effectiveness \circled{6}. 
The goal was to integrate their insights and preferences to refine the platform, ensuring it met their educational needs and user requirements.
Our pilot study was designed as a participatory process involving three key roles: a researcher from our team, students and teachers \citep{schuler1993participatory}.
Details about the selection and participation of students and teachers, as well as the experimental setting and administration procedures, are provided in \Cref{sec:virtual_pilot_participants}.

\paragraph{User feedback elicitation}
During the validation phase, we collected feedback from both students and teachers to evaluate their experience with the tool and identify usability issues, thus refining the tool to meet user needs. 

Pupils were at the heart of the study, and their interactions with the platform were crucial for assessing the tool's usability and identifying new user requirements \citep{druin2002role,kujala2003,valguarnera2023,valguarnera2023challenge,schuler1993participatory}. 
We actively engaged children as informants and evaluators, enabling us to design with their needs and preferences in mind \citep{read2004,read2006,greig2007doing,greig2012doing}.
Their evolving thoughts and reflections, shared during testing activities, provided real-time insights into how they perceived and interacted with the tool \citep{yarosh2011,iversen2013,iversen2012,hourcade2007interaction}.
This participatory approach empowered children to take ownership of the tool's development while fostering critical thinking about its features \citep{iivari2018,kinnula2021,iversen2017,kinnula2017cooperation}. 
It also ensured the process remained enjoyable and rewarding for them, aligning with principles of co-design and participatory research \citep{MullerKuhn1993,MullerWildmanWhite1993,MullerWildmanWhite1994,bogdan1997qualitative}.

Teachers played an essential role by facilitating the study and providing assistance as needed \citep{druin1999,scaife1997designing,kafai1997children,borgers2000,greenbaum2020design,nielsen1994,nielsen1995,baecker2014,holtzblatt1997contextual}.
These teachers were the ones present in the classrooms during the activities and were responsible for ensuring a smooth classroom experience.
Their observations highlighted how students engaged with the tool, identified areas of difficulty, and noted moments of success. 
Teachers' feedback was invaluable in refining the platform to balance educational goals with practical usability and address both pedagogical and logistical challenges in the classroom.

During the study, the administrator from our research team closely monitored pupil progress and interactions collected empirical data on task performance and gathered feedback from students and teachers.
Multiple data elicitation techniques, including think-aloud and observation, have been employed to gain insights into the usability and effectiveness of the design \citep{hanington2019universal,hartson2018ux}.
In particular, students' interactions with the tool have been documented, focusing on their behaviours, verbal feedback, and non-verbal cues. Key observations included moments of confusion, problem-solving strategies, and how students navigated specific features \citep{Guran2020}.
Real-time note-taking captured recurring patterns and usability issues, providing valuable insights into user experience. This structured approach ensured a detailed understanding of the tool’s strengths and areas for improvement, directly informing subsequent design iterations \citep{hanington2019universal,greig2007doing,hartson2018ux}.
These techniques, grounded in Human-Computer Interaction (HCI) and UX design principles, enabled us to triangulate data sources and derive actionable insights for iterative improvements \citep{hanington2019universal,hartson2018ux,druin2000design,lehnert2022Child,hourcade2007interaction,markopoulos2003interaction}.


\paragraph{Collaborative design}

At the end of this process, we conducted a collaborative session with the users \circled{7}, during which design proposals were presented based on the notes and feedback collected by the researcher.
Users were also invited to provide additional input. 
Based on their insights, modifications were proposed and discussed, enabling users to actively contribute to refining the prototype and ensuring the design better aligns with their needs.
In \Cref{sec:third_prototype}, we highlight the feedback received during this collaborative session.

\subsubsection{Final application development}
In the final phase, the prototype is redesigned \circled{8} in response to the feedback and suggestions from the collaborative session, leading to the final working version, which is documented in \Cref{sec:third_prototype}.
The redesign adhered to standard mobile application design principles to enhance usability and accessibility \citep{nayebi2012state,coursaris2011meta,tidwell2010designing,hartson2018ux}.

To create an interface familiar to the user, we incorporated common elements, like a top bar and a left-side menu list.
Legibility and readability were prioritised using large font sizes and ensuring a high contrast between text and background. 
Accessibility considerations were central to the redesign. The interface included a colour-blind mode, high-contrast visuals, and a text-to-speech feature to accommodate users with visual impairments.

Consistency was maintained by using uniform names and labels for similar objects and functions, avoiding synonyms to ensure clarity and reduce cognitive load. 
Frequently used features were placed in easily accessible locations, aligning with common mobile application conventions.
By adhering to these principles, the final application aimed to provide a user-friendly and inclusive experience for a diverse range of users.

The results concerning the formative evaluation are discussed in \Cref{chap:study_unplugged}.
The final summative evaluation \circled{9}, aimed at assessing the effectiveness and impact of the final working prototype on a larger scale, is discussed in \Cref{chap:study_virtual_main}.

\subsection{Implementation}\label{sec:implementation_virtualCAT}



In terms of platform selection, we focused on iPads as the primary target device. This choice was driven by the device's user-friendly touchscreen interface, which aligns with our goal of creating an intuitive, interactive learning experience, particularly for students in K-12 educational settings \citep{scaife1997designing}. 
Additionally, the portability and wide adoption of iPads in educational contexts made them an ideal choice for our application, ensuring the instrument would be accessible to a broad range of students.

The application was developed using the Flutter framework, selected for its robust capabilities and cross-platform support \citep{flutter}. 
This framework enables the creation of a single codebase that operates seamlessly across multiple platforms, including Android, iOS, Linux, macOS, Windows, and web. 
This approach significantly streamlined development efforts and reduced the time required for platform-specific customisation.

Although the application was primarily designed for iPads, its responsive design ensures a consistent and engaging user experience across various devices and screen sizes. This flexibility guarantees that the application functions effectively without compromising on design or usability.

A key feature of Flutter that enhanced our development workflow was its hot reload functionality, which allowed real-time previews of code changes. This feature proved essential in supporting the iterative design process, increasing both efficiency and productivity. 
Furthermore, Flutter’s rich library of pre-built widgets and tools made it easier to develop visually appealing and interactive user interfaces.

The latest version of the virtual CAT application, including its full source code and comprehensive documentation, is openly accessible online \citep{adorni_virtualCATapp_2023}.
For a detailed explanation of the system's data infrastructure, including collection and transfer protocols, application features, development process, and intended use cases, refer to the dedicated software paper \citep{adorni_app_article2023}.


\subsubsection{CAT programming language}\label{sec:commands}

To establish a standardised set of instructions that users could employ within the application interfaces to design the algorithm, we defined the CAT programming language, which codifies and formalises all the commands and actions observed during the original experimental study with the unplugged CAT.
The detailed list of commands available in the formal CAT programming language is provided in \Cref{table:commands}.

\begin{figure}[h]
	\centering
	\includegraphics[height=4.5cm]{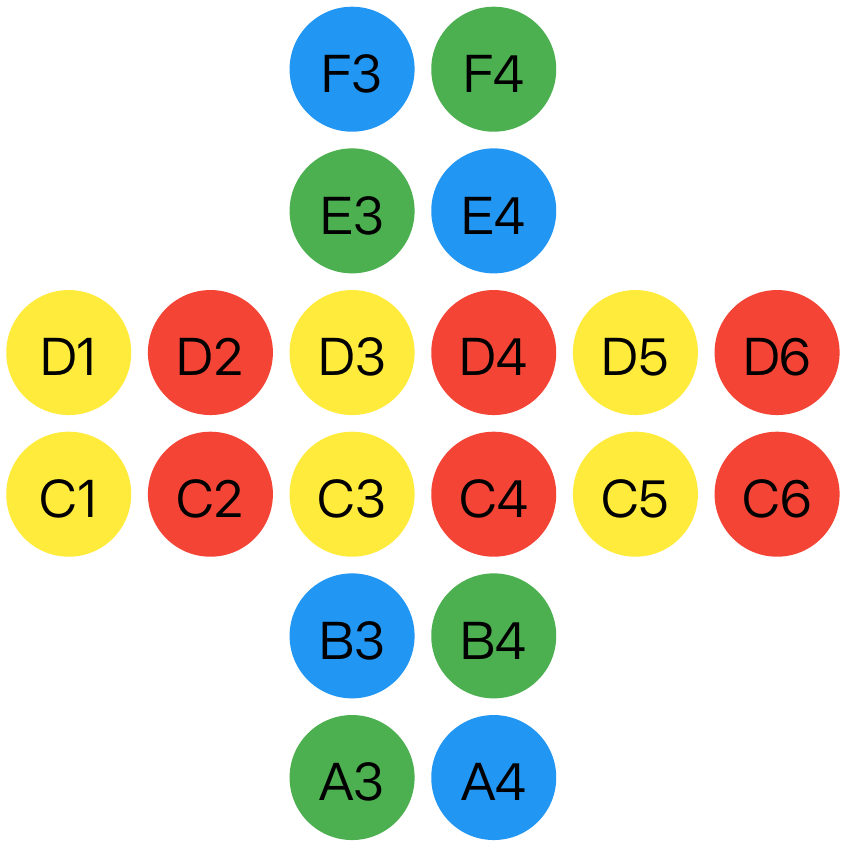}
	\mycaptiontitle{{Example of a cross-board with coordinate labels.}}
	\label{fig:crossboard}
\end{figure}

\paragraph{Cross representation}
The cross-board dots are manipulated and referenced using a coordinate system (see \Cref{fig:crossboard}), where rows are labelled from bottom to top using letters (A-F), and columns are numbered from left to right (1-6). 

\FloatBarrier
\paragraph{Moves}
\begin{figure}[h]
	\centering
	\includegraphics[height=4.5cm]{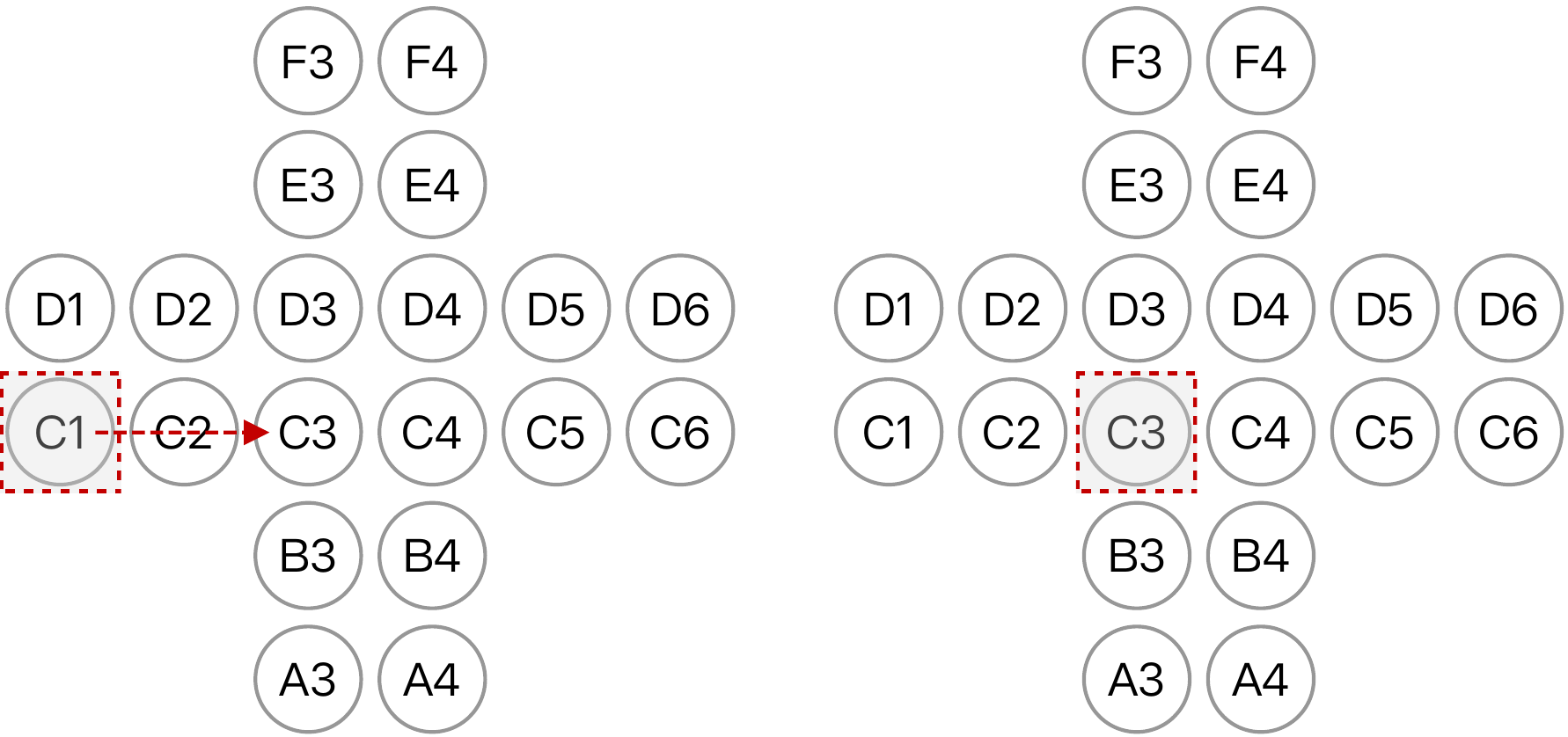}
	\mycaption{Example of movement on the cross-board.}{ 
		Starting from C1, the destination cell C3 can be reached either by using the \texttt{goCell(C3)} command or by traversing two steps to the right using the \texttt{go(right,2)} command.}
	\label{fig:go}
\end{figure}
Moving around the cross-board can be done in two ways (see \Cref{fig:go}):
the \texttt{goCell(cell)} method allows jumping directly to a specific coordinate; the \texttt{go(move, repetitions)} method allows traversing a certain number of dots in one of the eight available directions (cardinal or diagonal) to reach the desired destination.

\begin{figure}[h]
	\centering
	\includegraphics[height=4.5cm]{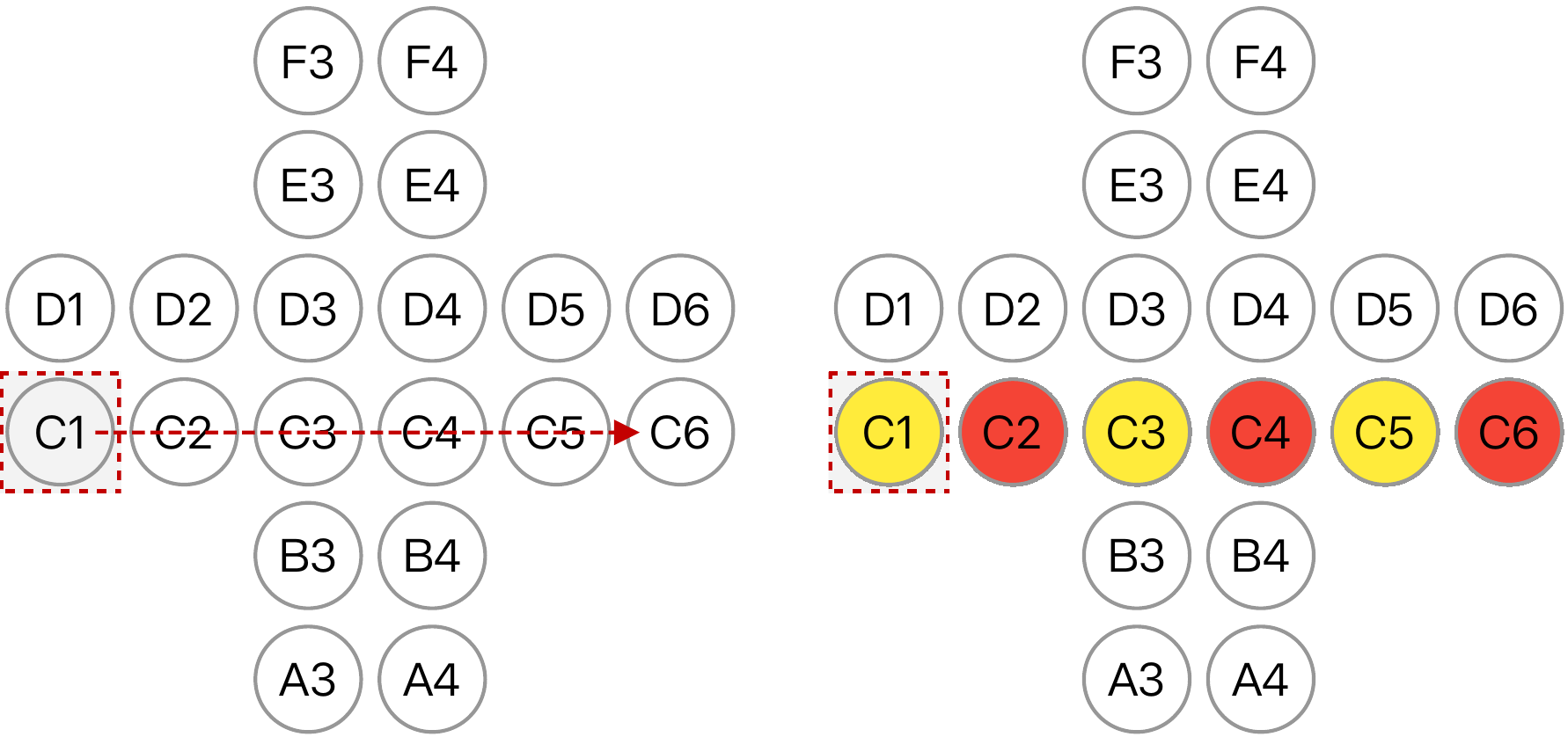}
	\mycaption{Example of colouring a row of six dots.}{
		Starting from C1, the row is coloured alternating yellow and red using either the \texttt{paintPattern(\{yellow, red\}, 6, right)} or the \texttt{paintMultipleCells(\{yellow, red\}, \{C1, C2, C3, C4, C5, C6\})} command.}
	\label{fig:paint}
\end{figure}

\FloatBarrier
 \paragraph{{Basic colouring}}
Colouring the board can be achieved through various methods (see \Cref{fig:paint}). 
The \texttt{paintSingleCell(color)} method allows colouring the dot they are currently positioned on with a single colour. 
The \texttt{paintPattern(colors, repetitions, pattern)} method allows colouring multiple dots according to predefined patterns. 
A sequence of colours can be specified, which will alternate following the selected pattern. 
Additionally, users can choose from five pattern types (cardinal, diagonal, square, L, zigzag), each with various directions. 
The \texttt{paintMultipleCells(colors, cellsPositions)} method enables colouring multiple dots with custom patterns, defined by specifying the coordinates of the cells to be coloured.
The \texttt{fillEmpty(color)} method colours all the uncoloured dots on the board with the same colour. 

\FloatBarrier

\begin{table}[p]
	\footnotesize
	\centering
	\mycaptiontitle{CAT programming language commands index.}
	\label{table:commands}
	\begin{tabular}{>{\arraybackslash}m{1.7cm}|>{\arraybackslash}m{5.8cm}|>{\arraybackslash}m{5cm}}
		\toprule
		\textbf{Category}            & \textbf{Command}                                   & \textbf{Description}  \\\midrule
		
		\multirow{6}*{Movements}  & \texttt{goCell(String cell)}                    
		& Move directly to a specific cell coordinate on the board.  \\
		&&\\
		
		& \texttt{go(String move, int repetitions)}           
		& Moves in one of the eight possible directions, including cardinal and diagonal movements, of a specified number of cells.  \\ \hline

		\multirow{19}*{Colouring}   & \texttt{paintSingleCell(String color)}          
		& Colour the current cells with the specified colour.  \\
		&&\\
		
		& \texttt{paintPattern(List<String> colors, String repetitions, String pattern)} 
		& Colour multiple cells, starting from the current, according to predefined patterns (cardinal, diagonal, square, L, zigzag) with various directions.$^{*}$ \\
		&&\\
		
		& \texttt{paintMultipleCells(List<String> colors, List<String> cellsPositions)} 
		& Colour multiple dots, at specific coordinates, with custom patterns.$^{*}$  \\
		&&\\
		
		& \texttt{fillEmpty(String color)}                
		& Colour all uncoloured dots on the board uniformly with the specified colour.  \\ \hline

		\multirow{5}*{Loops}   & \texttt{repeatCommands(List<String> commands, List<String> positions)} 
		& Repeats a sequence of commands  (e.g., a series of go and paint) at specified coordinates.  \\
		&&\\
		
		& \texttt{copyCells(List<String> origin, List<String> destination)}  
		& Copies colours from one set of coordinates (origin) to another (destination).  \\ \hline
		
		\multirow{8}*{Symmetry}     & \texttt{mirrorBoard(String direction)}          
		& Reflects the coloured dots on the board onto the non-coloured ones, in accordance with the principle of symmetry, along the specified direction (horizontally on the x-axis or vertically on the y-axis). \\
		&&\\
		
		& \texttt{mirrorCells(List<String> cells, String direction)}   
		& Applies symmetry to specific dots across a specified direction.  \\
		&&\\
		
		& \texttt{mirrorCommands(List<String> commands, String direction)} 
		& Reflects a sequence of commands across a specified direction.  \\ 
		\bottomrule
	\end{tabular}
	\\\vspace{0.5em}
	\begin{minipage}{.9\linewidth}
		$^{*}$ If a sequence of colours is specified, they will alternate following the selected pattern.
	\end{minipage}
\end{table}

\begin{figure}[h]
	\centering
	\includegraphics[height=4.5cm]{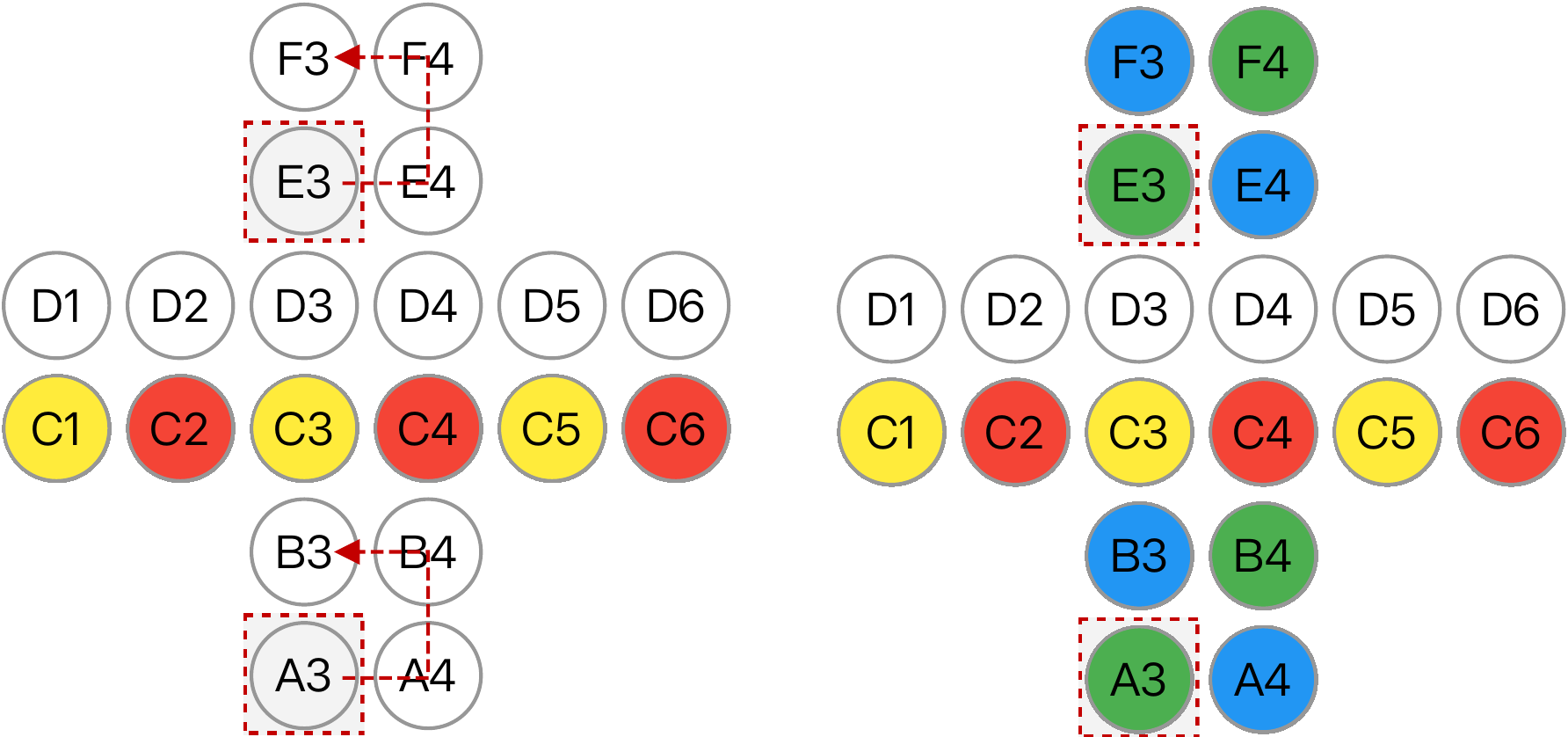}
	\mycaption{Example of repetition of a square pattern.}
		{On cells A3 and E3 is coloured a square pattern with alternating green and blue dots using the \texttt{repeatCommands(\{paintPattern(\{green, blue\}, 4, square\_right\_up\_left)\}, \{A3, E3\})} command.}
	\label{fig:repeat}
\end{figure}

\FloatBarrier
 \paragraph{{Repetition-based colouring}}
Moving beyond the basics, other methods allow for more complex operations, like repetitions (see \Cref{fig:repeat}). 
The \texttt{repeatCommands(commands, positions)} method allows specifying a sequence of commands (e.g., a series of \texttt{go} and \texttt{paint} operations) and applying them to specific coordinates.
The \texttt{copyCells(origin, destination)} method copies the colours from origin coordinates to destination coordinates.

\begin{figure}[h]
 	\centering
 	\includegraphics[height=4.5cm]{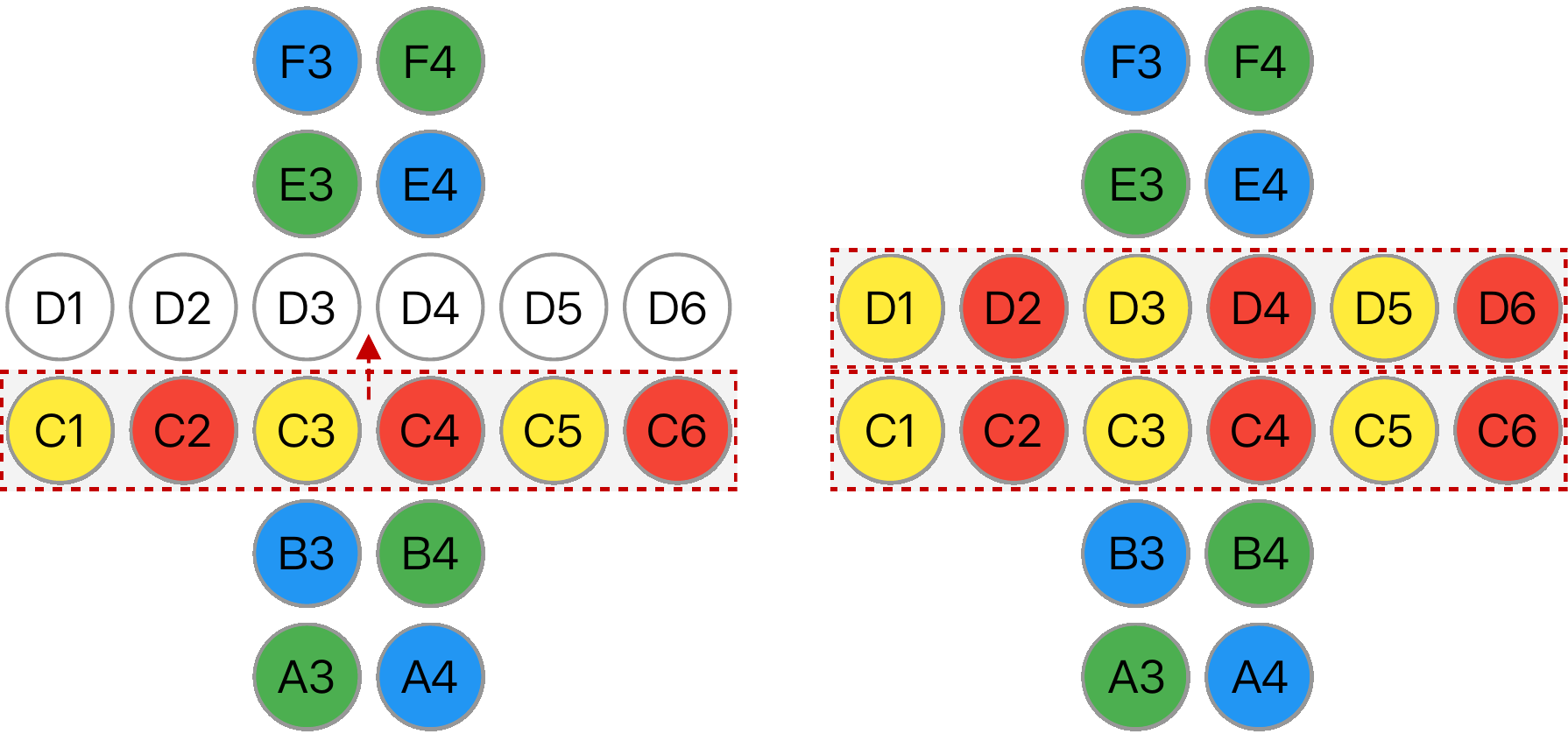}
 	\mycaption{Example of cells mirroring.}
 	{Starting from C1, all dots of the row are mirrored upwards along the horizontal axis using the \texttt{mirrorCells(\{C1, C2, C3, C4, C5, C6\}, horizontal)} command.}
 	\label{fig:mirror}
 \end{figure}

\FloatBarrier
 \paragraph{{Symmetry-based colouring}}
 Finally, symmetrical colouring approaches are available (see \Cref{fig:mirror}). 
The \texttt{mirrorBoard(direction)} method, which reflects the coloured dots on the board onto the non-coloured ones, follows the principle of symmetry. This mirroring can be done horizontally on the x-axis or vertically on the y-axis.
The \texttt{mirrorCells(cells, direction)} method performs similar mirroring operations but on a specified set of dots.
The \texttt{mirrorCommands(commands, direction)} method applies the mirroring to a list of commands.


\FloatBarrier
\subsubsection{CAT programming language interpreter}\label{sec:interpreter}
The virtual CAT programming language interpreter \citep{adorni_virtualCATinterpreter_2023} is a dedicated Dart package that can be integrated into any Flutter project, in our case, the virtual CAT app \citep{adorni_virtualCATapp_2023}.
It translates student actions, including gesture interactions and arranged visual programming blocks, into executable machine-readable instructions.
It analyses the user's input, converting actions into a formal algorithm specified using the CAT programming language.

Each command that composes the algorithm, such as colour selections and other operations, undergoes a validation process to identify and address semantic errors. 
Notably, the interface's design, featuring predefined programming blocks and buttons, obviates the need for syntax checking, as it inherently eliminates the possibility of such errors, significantly streamlining the process.
However, semantic errors can still occur during command execution, for instance, when users attempt to move outside the board boundaries using invalid directions or apply an inappropriate pattern for a colouring command.

Upon validation, the code is executed, and real-time feedback is provided to the user, including the display of current progress on the colouring cross and the CAT score. 
If the interpreter detects errors, it handles them and provides users with error notifications and potential suggestions for correction.

\subsubsection{Data infrastructure}
Considering the often limited availability of secure networks in educational settings, in line with the strict privacy and security demands of Swiss educational environments, we developed a secure data infrastructure that ensures a safe flow of data during the collection and assessment process between the devices used and the central data collection hub \citep{adorni_virtualCATdatainfrastructure_2023}. 

To achieve this, we established a dedicated local network by deploying a router that interconnects all the participant devices, and directs the information to a central computer that acts as the nerve centre for data collection, ensuring that the data remains within a controlled and safeguarded environment.

Gradle is an integral part of this framework, acting as the ignition system for our data process. 
It initiates the server and establishes the necessary network connections, enabling a seamless data exchange from the user devices to our central database. 
This setup allows the data to be synchronised in real-time, ensuring that all information is current and accurately reflected across the system.

To further enhance reliability, in our system's final version, we incorporated offline functionality and automatic progress saving. These features provide flexibility by allowing students to continue their work even during technical issues, such as server disconnections, and offer more flexible time management options during assessments. This ensures that students' progress is preserved and can be resumed without data loss, mitigating disruptions to the learning and testing process.

\begin{figure}[!ht]
    \centering
    \includegraphics[width=.9\linewidth]{VIRTUAL/database\_diagram2.png}
    \mycaption{Database schema.}{The diagram depicts {the logical structure of the database}, including tables such as CANTONS, SCHOOLS, SESSIONS, STUDENTS, ALGORITHMS, RESULTS, and LOGS, {along with their attributes and the relationships between these entities.}}
    \label{fig:database_diagram}
\end{figure}

We leverage the H2 database system for data management and organisation. In particular, Java is used for its database engine execution, and SQL is used for data manipulation and interaction. 

The database schema, illustrated in \Cref{fig:database_diagram}, features interconnected tables designed for specific roles within our educational assessment framework.
Notably, the \texttt{ALGORITHMS} table catalogues algorithms based on complexity levels and specific commands, contributing to the algorithm dimension metric. Simultaneously, to compute the interaction dimension metric, the \texttt{RESULTS} table records information such as the artefact type and the level of autonomy.

Our database comprises several interconnected tables, each serving a specific purpose within our educational assessment framework. 
The \texttt{CANTONS} table lists all 26 Swiss cantons, each uniquely identified by an integer and a name with abbreviations. 
The \texttt{SCHOOLS} table details the institutions participating in the activity, linking each school with its canton. \texttt{SESSIONS} captures each educational session's unique characteristics, including supervisor details, school affiliation, grade level, date, and language of instruction. 
\texttt{STUDENTS} maintains essential demographic (age and gender) and session-related information for individual students. 
The \texttt{ALGORITHMS} table catalogues various algorithms by detailing their complexity levels and the presence of particular algorithmic commands, such as painting in certain patterns or directions and using colours in different ways. 
\texttt{RESULTS} table holds the performance data from user tasks, recording information like the type of interaction interface used, whether the task was completed, and evaluates the accuracy of the tasks by noting the number and correctness of dots coloured. 
Lastly, the \texttt{LOGS} table tracks user activities during tasks in great detail. The data, stored as a JSON object, includes a time-stamped record of users' actions, the commands they use, the interface they interact with, and whether visual feedback was enabled. This information allows us to track the total time users spend on tasks.

\subsection{Prototypes}\label{sec:prototypes}

In this section, we present the three prototypes developed throughout the study: the initial prototype, the version refined after expert evaluation, and the final version of the application following the participatory study.

\subsubsection{First prototype}\label{sec:first_prototype}

\begin{figure*}[p]
	\centering
	\includegraphics[height=9.5cm]{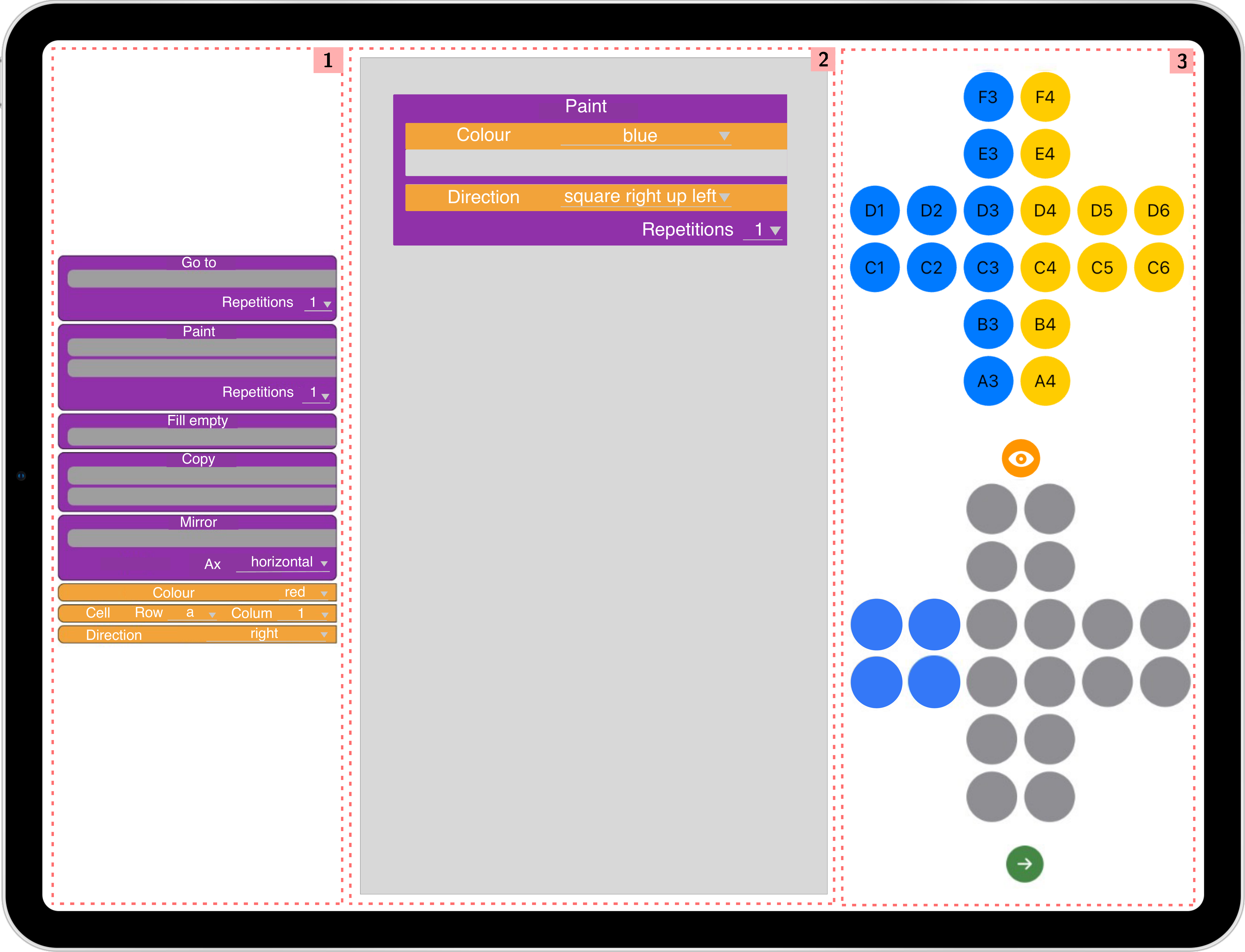}
	\mycaptiontitle{{First prototype of the CAT-VPI.}}
	\label{fig:1st_prototype_CAT-VPI}
	\vspace{10pt}
	\centering
	\includegraphics[height=9.5cm]{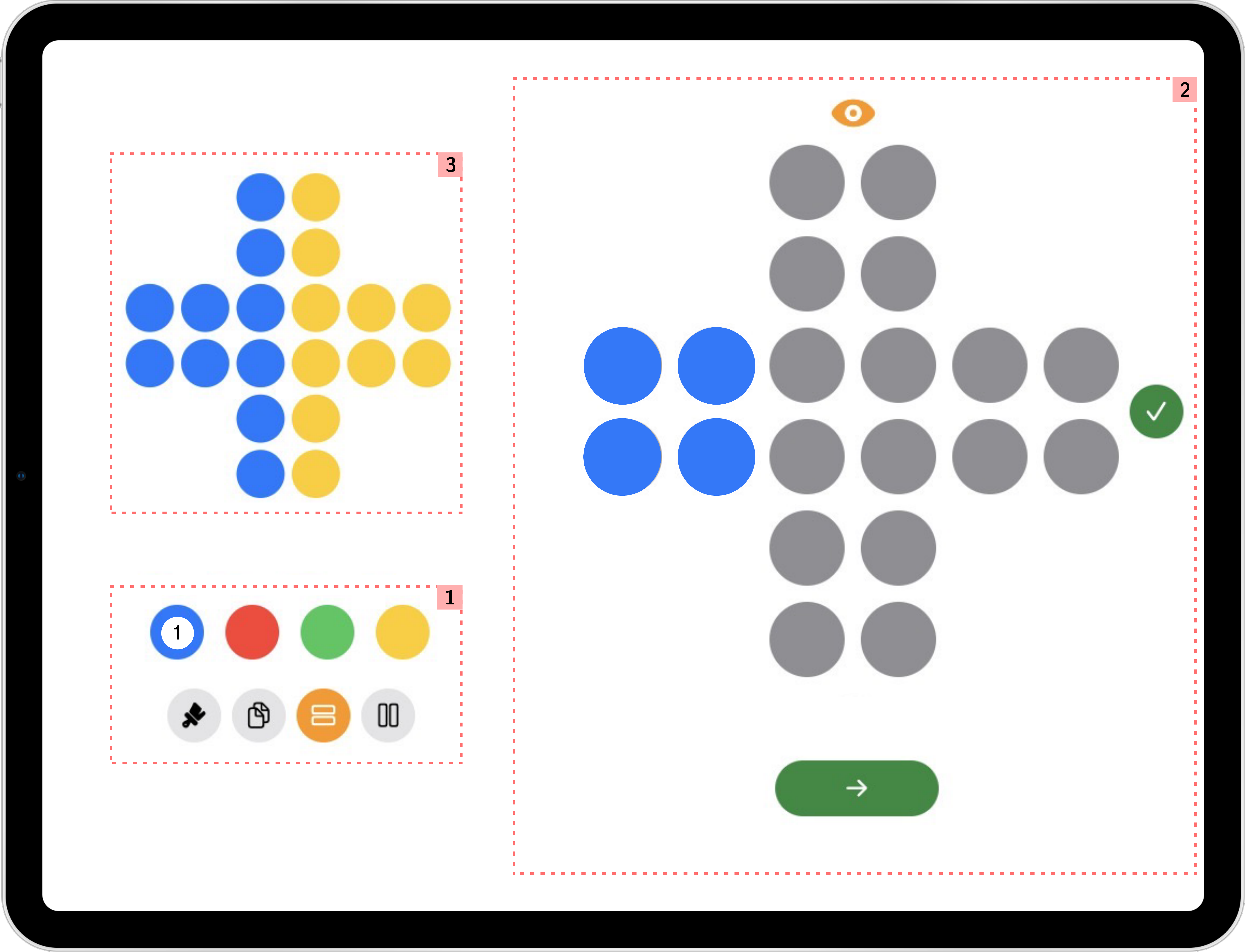}
	\mycaptiontitle{{First prototype of the CAT-GI.}}
	\label{fig:1st_prototype_CAT-GI}
\end{figure*}

The first prototype of the application was developed to explore and test the system's core functionalities, with the goal of creating a foundational version that experts could evaluate to gather feedback for improving its design and usability.

The CAT-VPI, illustrated in \Cref{fig:1st_prototype_CAT-VPI}, features a three-column layout. 
The left column \squared{1} includes predefined code blocks, divided into two types: containers, in purple, are the commands defined in \Cref{sec:commands} (i.e., go to, paint, fill empty, copy, and mirror); and components, in orange, that are the inputs for container blocks, such as the colour to be used, the cell to move to or to colour, or the direction for movement or colouring. 
The central column \squared{2} is the main workspace where users interact with and assemble code blocks.
The right column \squared{3} displays the reference schema to be replicated on top and the colouring schema on the bottom. This section also includes an eye icon to activate visual feedback and a green arrow to proceed to the next schema.

The CAT-GI, illustrated in \Cref{fig:1st_prototype_CAT-GI}, presents a different layout.
The bottom left section \squared{1} contains buttons for interaction, including four selectable colours and the key commands defined in \Cref{sec:commands} (i.e., fill empty, copy, and two types of mirror).
The right section \squared{2} is the main workspace, featuring a large cross array that students interact with after selecting colours and/or commands. To the right of the cross is a green tick to confirm the completion of a colouring action, an eye icon above it to activate visual feedback, and a green arrow at the bottom to proceed to the next task.
Finally, the top left section \squared{3} displays the reference schema to be replicated.


\FloatBarrier
\subsubsection{Second prototype}\label{sec:second_prototype}


Following expert evaluation and detailed feedback from UX and pedagogical experts, a series of modifications were made to improve the initial prototype.


\paragraph{UX expert feedback}
The feedback from the UX expert provided valuable insights into both the strengths and areas for improvement in the prototype.
On the positive side, the expert highlighted the overall clarity of the platform's purpose and its potential to engage users with minimal prior experience. He also appreciated the visual layout, particularly the effectiveness of the workspace design in fostering user engagement.

However, the expert identified two key areas for improvement. 
First, the interface lacked consistency across different screens, which could confuse users. 
He recommended restructuring the interface to create a more uniform and cohesive design across interaction modalities. 
Second, he suggested adopting more intuitive icons to enhance the visual clarity and usability of the interface, particularly for users with limited prior experience in using such tools.

\paragraph{Pedagogical experts feedback}
The feedback from the pedagogical experts provided valuable insights into how the platform aligns with educational goals, highlighting areas for improvement to enhance its pedagogical effectiveness. 
The experts praised the platform's visual engagement and the thoughtful integration of AT concepts, acknowledging the design’s clarity and its potential for supporting learning. 
However, they also identified several areas for refinement to further align the platform with best practices in computer science education.

The experts emphasised the educational benefits of allowing students to experiment, make mistakes, and learn from failures. They believe this iterative ``trial and error'' process fosters deeper learning and understanding. To support this, they recommended incorporating a mechanism that encourages ``trial and error'' preventing students from becoming discouraged by early failures.
While ``trial and error'' can be a valuable strategy, the experts raised concerns about students potentially getting stuck in a loop without making meaningful progress. 
Thus, they also recommended complementing this mechanism with a system to detect when students repeatedly fail or remain inactive for long periods. This would provide targeted hints or guidance to help students reflect on their approach and adjust their strategies, ensuring they continue to move forward without becoming discouraged.
Additionally, one of the recommendations was to ensure the platform is suitable for all ages. In particular, they suggested revisiting the CAT-VPI interface, as it could be too complex for younger students due to the programming blocks and the amount of text to read.

\paragraph{Prototype revision}

\begin{figure*}[!ht]
	\centering
		\includegraphics[height=8.5cm]{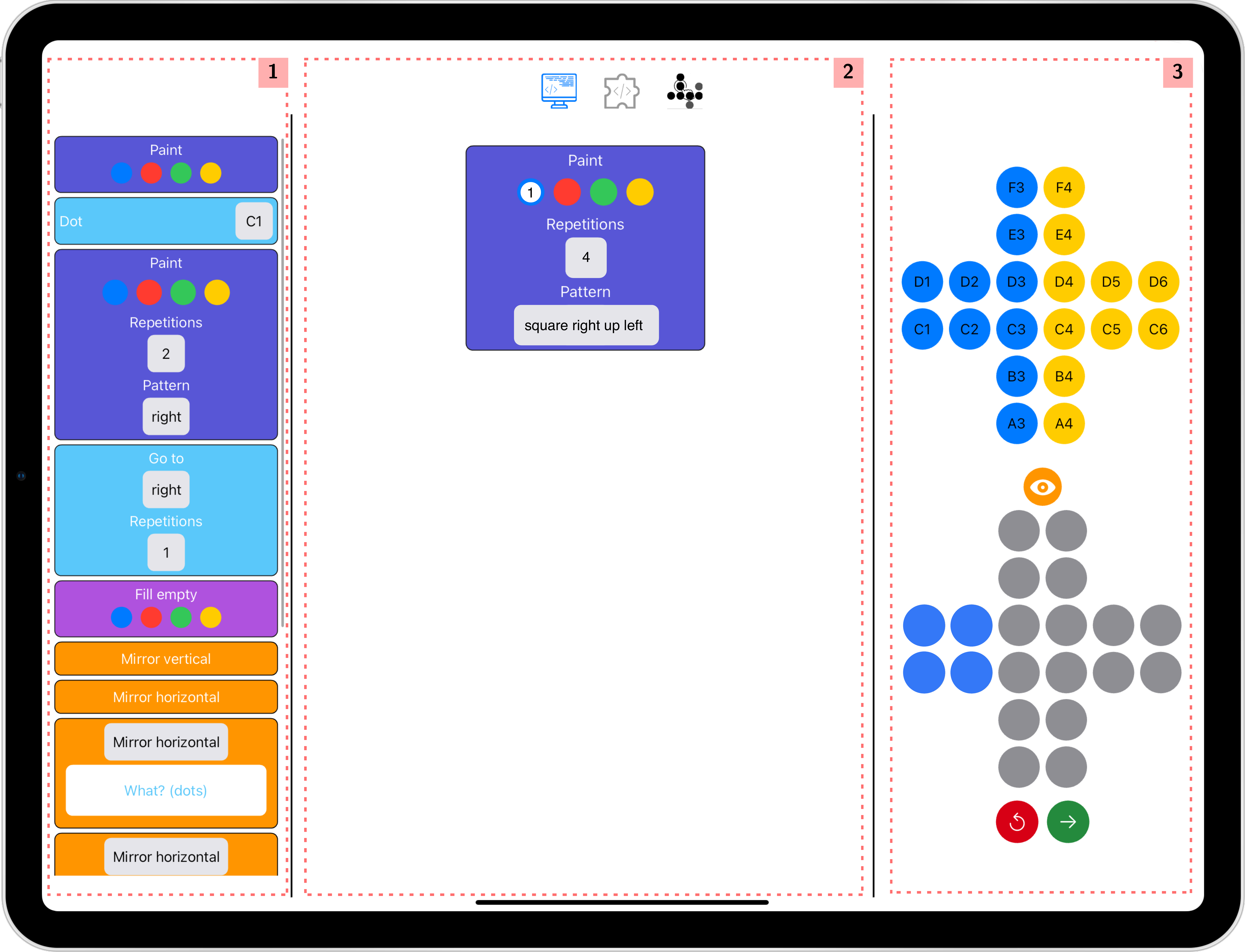}
	\mycaptiontitle{{Second prototype of the CAT-VPI.}}
	\label{fig:2nd_prototype_CAT-VPI}
	
	\vspace{0.5cm}
	
	\centering
	\includegraphics[height=9cm]{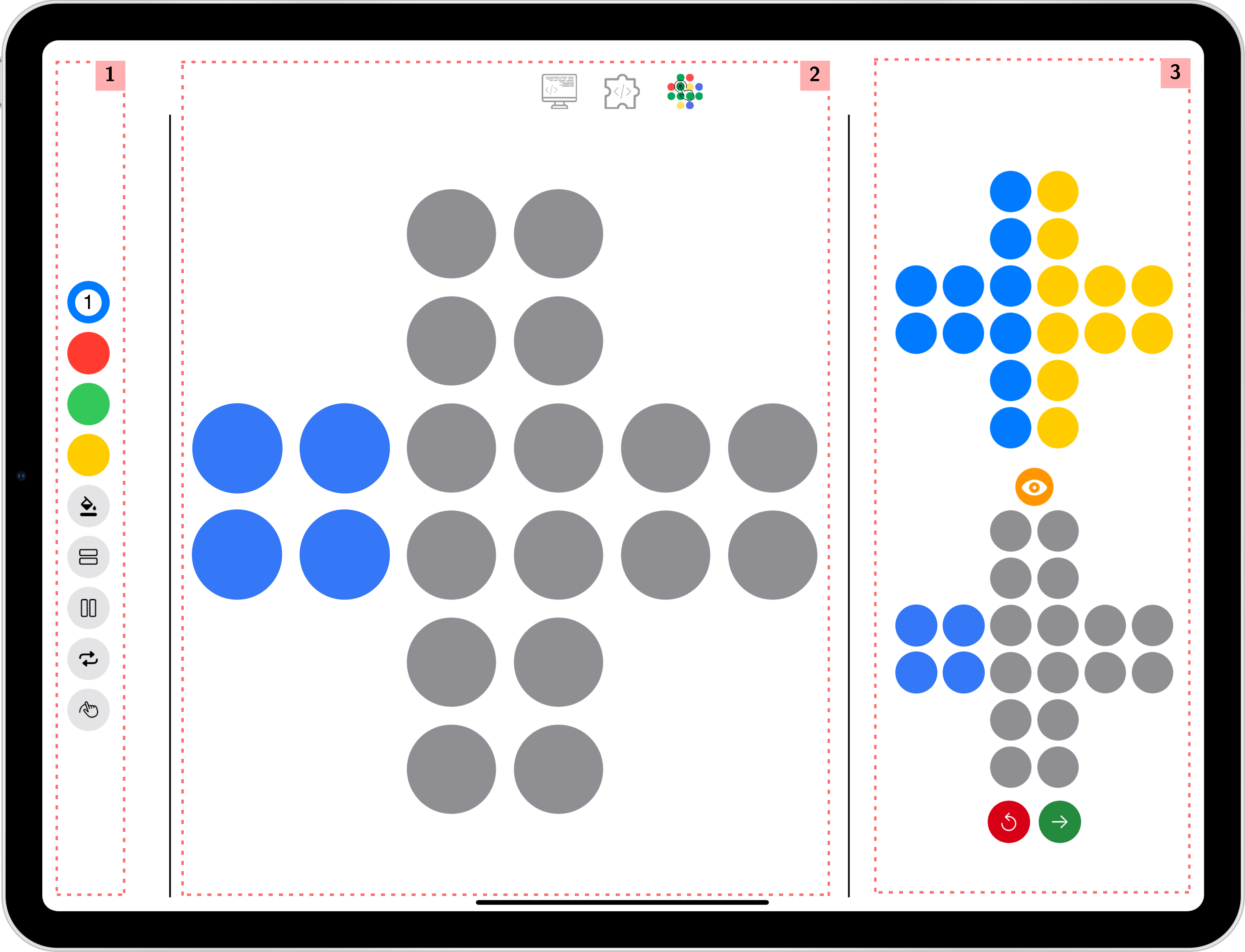}
	\mycaptiontitle{{Second prototype of the CAT-GI.}}
	\label{fig:2nd_prototype_CAT-GI}
\end{figure*}

\Cref{fig:2nd_prototype_CAT-VPI,fig:2nd_prototype_CAT-GI} illustrates the interface layout of the second prototype.
To transition to a working prototype for classroom testing, all interfaces now include three buttons at the top centre of the workspace \squared{2} that allow users to switch between interaction modes.

In response to the feedback from the UX expert, we redesigned the user interfaces to ensure consistency between them by adopting the same three-column layout used in the CAT-VPI, illustrated in \Cref{fig:2nd_prototype_CAT-VPI}. 
The changes were applied solely to the CAT-GI, as shown in \Cref{fig:2nd_prototype_CAT-GI}.
The predefined buttons to select colours and actions are now grouped in the left column \squared{1}.
Previously, the large cross served as both the workspace and the colouring schema. Now, the central section \squared{2} functions as the main workspace, while the right column \squared{3} displays the reference schema at the top and the colouring schema at the bottom, where users can enable visual feedback.
Additionally, new action buttons were added to the CAT-GI, aligning it with the commands available in the CAT-VPI. For example, the ``copy/repeat'' command, absent in the first prototype, has been included, ensuring both interfaces now offer the same set of functionalities.

A major overhaul was conducted to replace the existing icons with more intuitive and universally recognisable symbols.
In the CAT-VPI, the command blocks were simplified for greater clarity. 
The predefined building blocks now use a colour-coding system that groups similar commands together (e.g., indigo for colouring action, orange for the mirror function, etc.).
Most container blocks now come pre-loaded with the necessary components inside, so students don't need to decide which components to include. 
For example, in the case of the paint block, students no longer need to figure out whether to insert the colour or another component, as these are already provided, and they only need to select the component's specific detail, such as the colour. Instructions are provided to guide the student when a component is not pre-loaded.
Other mechanisms were simplified to make the tool more intuitive, streamlining the approach by reducing the steps required for the task and improving the user experience. 
For example, in the previous prototype, for colouring patterns with alternating colours, users had to insert multiple colour components into the paint container and specify the number of repetitions, or cells, to colour.
Now, a dedicated block is available for this operation, where users can select the colours, specify the number of repetitions, and choose a pattern.

In response to the recommendation on accessibility for younger students, we introduced two types of blocks in the CAT-VPI: textual and symbolic.
While only the textual version is shown here, both types are included in \Cref{appendix:appA_pilot} for the final version of the application.
Since most kindergarten pupils cannot yet read, this dual approach accommodates a wider range of users. Textual blocks provide instructions for those who can read, while symbolic blocks use intuitive symbols, offering a language-independent way to interact with the system.
This ensures the platform remains accessible to younger, multilingual, or pre-literate students, making the interface engaging for all learners.

Finally, based on feedback from pedagogical experts regarding the ``trial and error'' process, we implemented a ``retry'' button, represented by a red circular arrow at the bottom of the colouring schema in the right section of the interfaces \squared{3}.
This feature allows students to restart exercises anytime, encouraging them to revisit their mistakes, refine their solutions, and engage in iterative learning.

\FloatBarrier

\subsubsection{Final application}\label{sec:third_prototype}

\begin{figure*}[p]
	\centering
	\includegraphics[height=9.5cm]{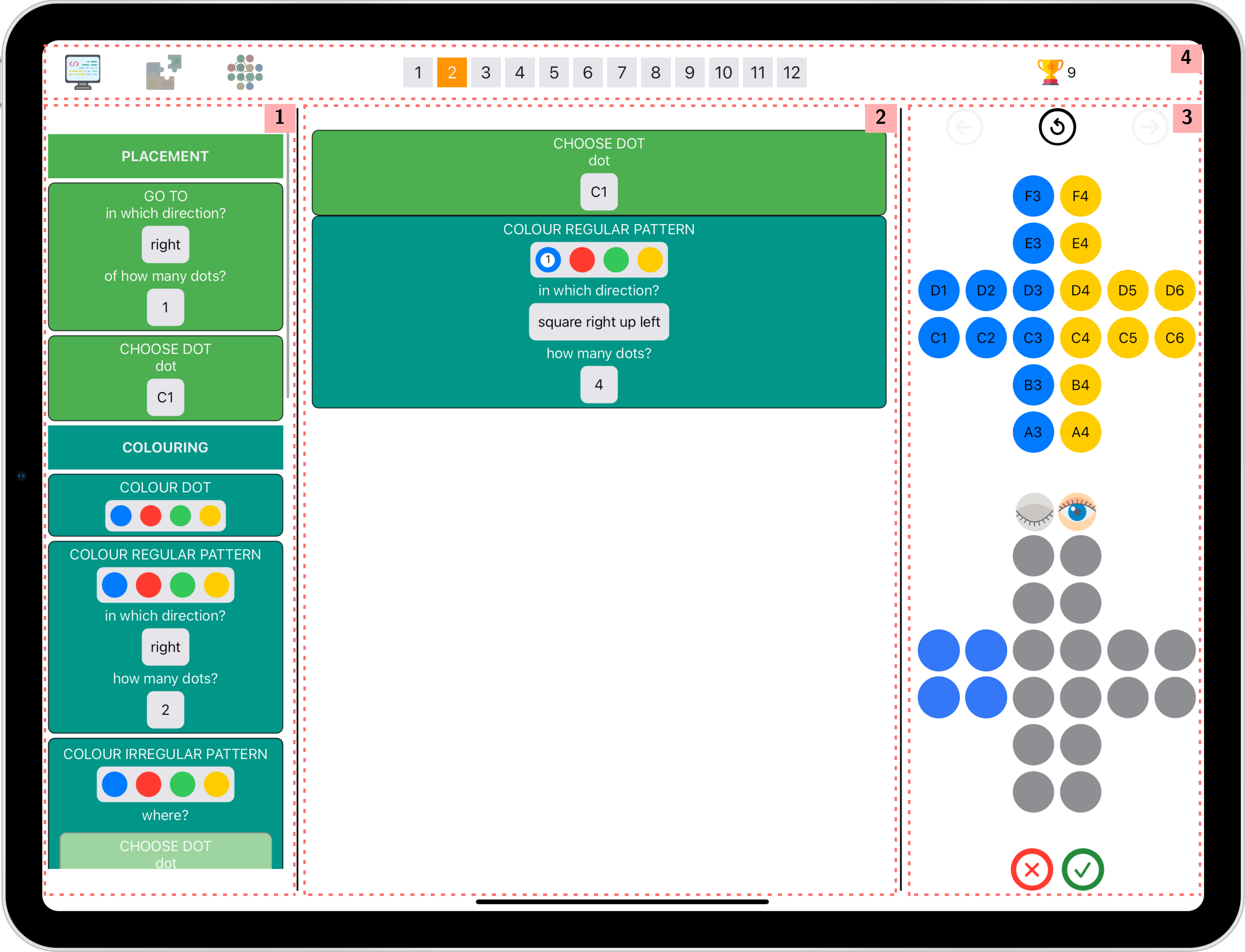}
	\mycaptiontitle{{Final CAT-VPI.}}
	\label{fig:3rd_prototype_CAT-VPI}
	\vspace{10pt}
	\centering
	\includegraphics[height=9.5cm]{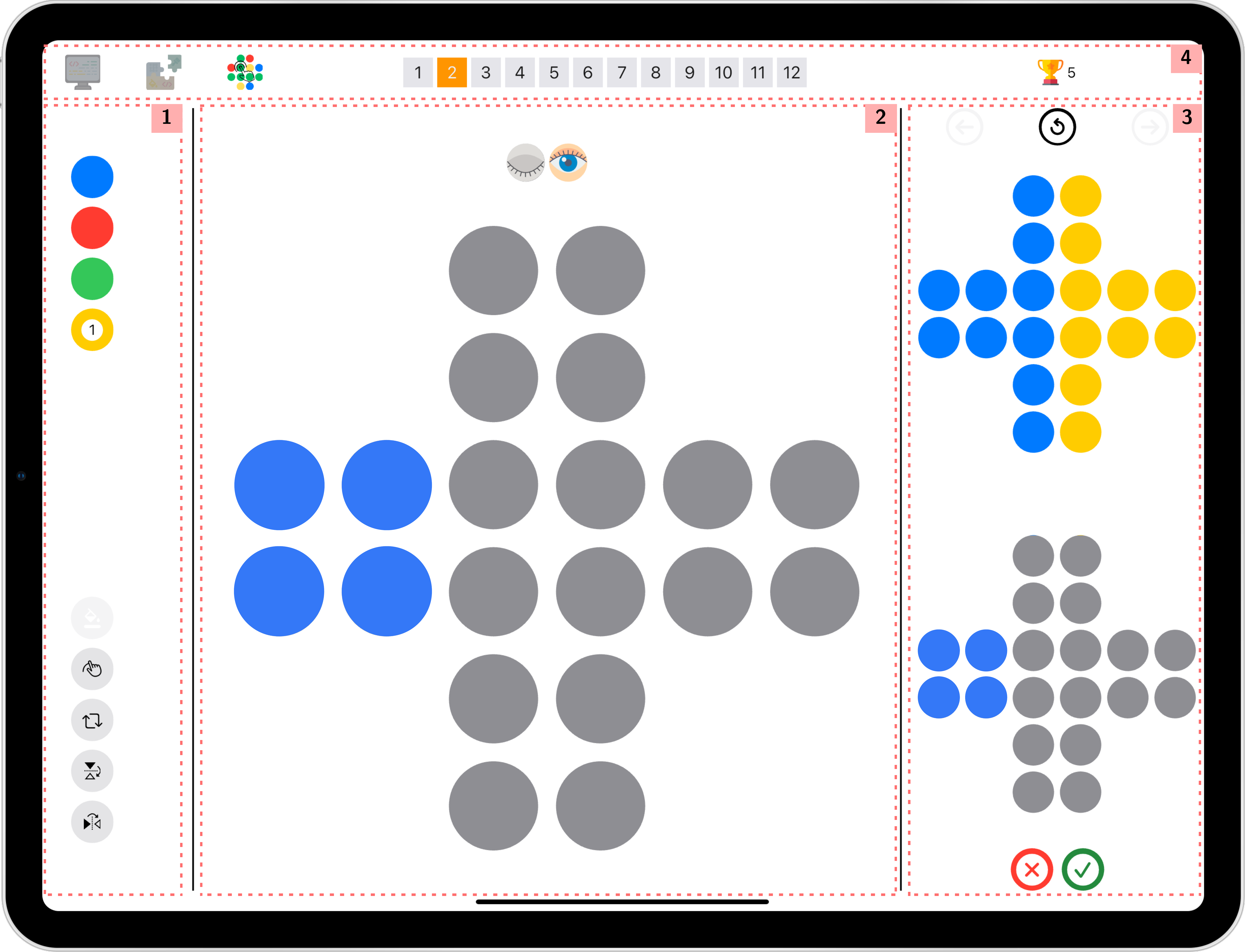}
	\mycaptiontitle{{Final CAT-GI.}}
	\label{fig:3rd_prototype_CAT-GI}
\end{figure*}

The third and final version of the application was developed through active collaboration with teachers and pupils during the pilot session of the participatory study.
\Cref{fig:3rd_prototype_CAT-VPI,fig:3rd_prototype_CAT-GI} illustrate the final versions of the two interfaces.
Screenshots of all application screens are available in \Cref{appendix:appA_pilot}, providing a comprehensive visual reference for the platform's design and functionality.

After active collaboration with teachers and pupils from different age groups and schools during the pilot session of the participatory study. 
Feedback and observations of user interactions guided targeted refinements to the interface and functionality, addressing usability issues, enhancing accessibility, and better aligning the platform with the needs of students and educators.

Collaborating with students yielded invaluable insights that guided several critical changes to the platform. 
Initially, we observed that as time was running out, the need to confirm each schema at the end of the activity individually became cumbersome and unnecessary. 
Moreover, this process led to schemas being incorrectly marked as ``failed'' instead of ``not attempted''. To address this, we introduced a ``surrender'' button at the bottom of the right section of the interfaces \squared{3}, allowing users to skip specific schemas.
This feature also proved useful for students who felt stuck and wanted to move on. 

Another modification stemmed from feedback about the visual feedback button, which some students found unclear. 
In response, we replaced the original button with two new icons: an open eye indicating active visual feedback and a closed eye symbolising that feedback was turned off. 

As students progressed through the activity, some expressed interest in knowing how many schemas remained to be completed. To address this, we added a progress bar at the top of the central column in the interface \squared{4}.
This addition serves a dual purpose: it satisfies students' curiosity about their progress and resolves a limitation observed in the unplugged CAT activity, where the rigid sequencing of tasks restricted students' ability to navigate exercises flexibly. 
Additionally, we included navigation arrows at the top of the right section of the interface \squared{3} to allow students to explore upcoming schemas, providing further flexibility and accommodating those who wish to skip ahead.

Further adjustments were made based on the researcher's observations during the study, which highlighted areas for improvement that were not always evident through feedback alone. 
While observing pupils interacting with the CAT-VPI, it became evident that they were not using all the available commands but only readily visible ones. This was because some commands were not immediately accessible and required scrolling down the column to see them. 
To address this, we grouped related commands into menus in the left column of the interface \squared{1} and revised their colours to improve visibility and accessibility. 

Additionally, we observed that some pupils occasionally forgot to select the colour parameter within the paint blocks.
Thus, we enclosed all customisable parameters within shaded boxes to make it easier for users to identify and adjust them. 

Another observation concerned the use of nested blocks. Despite written instructions, some users struggled to fill these blocks correctly. To improve clarity, we added a transparent representation of the block types that could be inserted within nested blocks and provided more detailed instructions for each label.

Finally, while observing users interact with the CAT-GI, we noticed issues with certain commands, such as the \texttt{fillEmpty} button, which was often used without selecting a colour first. To address this, we implemented conditional activation of buttons, enabling them only when appropriate for the given context.
Additionally, we introduced a visual feedback mechanism, including a shaking effect on incorrect actions and flashing available commands when users deviate from the intended workflow. This feature aims to guide users towards the correct actions, improving the overall user experience.

Following teacher feedback, additional improvements were made to enhance the platform further, focusing on refining the user experience and ensuring the tool met pedagogical and functional needs.
One key suggestion from the teachers was to provide real-time feedback, allowing students to monitor their progress and performance during the activity. 
In response, in the right part of the top bar \squared{4}, we included a display box showing the current score for the ongoing schema. 

Additionally, we introduced a final dashboard that provides a comprehensive summary of student performance across all completed schemas (see Figure~\ref{fig:dashboard}). 
This feature not only aids teachers in quickly assessing student progress but also provides students with a clear overview of their performance, helping them identify areas where they may need to focus more effort.

The initial structure of the training module required a human administrator to guide students through the platform before starting the actual validation session. 
This poses challenges for large-scale implementation, as reliance on this approach could introduce inconsistencies in the explanations given to different student groups, potentially affecting performance.
To address this limitation and enhance the platform's scalability for broader use, we redesigned the training module by integrating standardised in-app video tutorials, enabling users to navigate the platform independently (see Figure~\ref{fig:tutorial}). This ensures consistent instructions for all users, minimises potential biases introduced by varying researcher-led explanations, and supports more efficient large-scale data collection and assessment.

Our vision for future developments involves continuous refinement and expansion of the platform.
To assess user experience, we decided to incorporate a brief survey at the end of the validation module to gather pupils' subjective impressions and insights into their perceptions of the tool (see Figure~\ref{fig:survey}).
This survey aligns with established UX design techniques for data elicitation \citep{hanington2019universal,hartson2018ux} and the Technology Acceptance Model \citep{surendran2012,hartson2018ux}, assessing factors such as ease of use, perceived usefulness, attitude towards use, and behavioural intention to assess users' acceptance of a system. 
It explores various facets of user interaction, from the clarity of app rules and preferred interaction modes to the perceived difficulty of exercises and overall enjoyment. 
Additionally, it prompts participants to reflect on whether they would use the app again in the future.
To accommodate the diverse literacy levels and age groups of our users, the survey features an audio playback option for reading questions aloud, ensuring accessibility even for younger students. Responses are collected using a smiley meter scale (happy, neutral, sad), a child-friendly format shown to be effective in assessing children's attitudes toward interactive technologies \citep{Giannakos2022state,read2006,Guran2020}.

\FloatBarrier

	\chapter[The IAS for the CAT]{The Intelligent Assessment System for the Cross Array Task}\label{sec:casestudy}

\begin{custombox}
	{The content of this chapter has been adapted from the following articles with permission of all co-authors and publishers:}
	\begin{itemize}[noitemsep,nolistsep,leftmargin=.25in]
		\item Antonucci, A., Mangili, F., Bonesana, C., and \textbf{Adorni, G.} (2022). Intelligent Tutoring Systems by Bayesian Nets with Noisy Gates. The \textit{International FLAIRS Conference Proceedings} \cite{antonucci2021}. 
		
		\item Mangili, F., \textbf{Adorni, G.}, Piatti, A., Bonesana, C., and Antonucci, A. (2022). Modelling Assessment Rubrics through Bayesian Networks: a Pragmatic Approach. In \textit{2022 International Conference on Software, Telecommunications and Computer Networks} (SoftCOM) \cite{softcom}. 
		
		\item \textbf{Adorni, G.}, Mangili, F., Piatti, A., Bonesana, C., and Antonucci, A. (2023a). Rubric-based Learner Modelling via Noisy Gates Bayesian Networks for Computational Thinking Skills Assessment. \textit{Journal of Communications Software and Systems} \cite{adorni2023rubric}. 
		
		
	\end{itemize}
	As an author of these publications, my contribution involved:\\
	\textit{Conceptualisation, Methodology, Validation, Formal analysis, Investigation, Resources, Data curation, Writing -- original draft \& review \& editing, Visualisation.}
\end{custombox}

\section{Summary}

This chapter focuses on the definition of the IAS specific to the CAT, contributing to RQ3.
It outlines the modelling of the assessment rubrics for both versions of the CAT, discussing the ordering of competencies, the encoding of answers, and the inclusion of supplementary competencies. 
Additionally, the chapter covers the parameter elicitation process, explaining how the necessary parameters for the model are determined, ensuring the IAS aligns with the intended assessment of AT across different educational contexts. 
This process highlights how a probabilistic IAS can be effectively integrated into the CAT framework to assess AT skills in a scalable and contextually relevant way.

\section{Modelling the CAT assessment rubric}\label{sec:modelling-cat-ae}
As specified in \Cref{sec:rubric}, we defined a task-specific assessment rubric, for both variants of the CAT, unplugged and virtual (see \Cref{tab:catrubric}). 
The instruction sequences conceived by the pupils, called algorithms, are ranked into three categories corresponding to the assessment rubric's competence components (rows). 
Each row represents the pupil's ability to solve a CAT schema using a certain algorithmic dimension. 
The interaction dimension of the pupils, given by their degree of autonomy and the tools used to accomplish the task, have been hierarchically ordered from the highest (right) to lowest (left) and determine the competence levels in the columns of the rubric. 

\begin{table}[ht]
	\footnotesize
	\mycaptiontitle{Definition of the CAT assessment rubric.}\label{tab:catrubric}
	\begin{subtable}{0.49\linewidth}
		\mycaptiontitle{Unplugged CAT. 
			\label{tab:rubric_unplugged}}
		\centering
		\begin{tabular}{c>{\centering\arraybackslash}m{.8cm}|>{\centering\arraybackslash}m{.8cm}>{\centering\arraybackslash}m{.8cm}>{\centering\arraybackslash}m{.8cm}}
			\multicolumn{2}{c|}{}   & \multicolumn{3}{c}{{\textbf{Competence level}}} \\
			\multicolumn{2}{c|}{}   & {VSF $c=1$}    & {VS $c=2$}   & {V $c=3$}   \\
			\midrule
			\multirow{3}{*}{\begin{minipage}{.5cm}\rotatebox{90}{\begin{tabular}{c}\textbf{Competence}\\[-1ex]\textbf{component}\end{tabular}}\end{minipage}} 
			& {0D $r=1$} & $X_{11}$   & $X_{12}$  & $X_{13}$             \\
			& {1D $r=2$} & $X_{21}$   & $X_{22}$  & $X_{23}$             \\
			& {2D $r=3$} & $X_{31}$   & $X_{32}$  & $X_{33}$             \\
		\end{tabular}
	\end{subtable}%
	\hfill
	\begin{subtable}{0.49\linewidth}
		\mycaptiontitle{Virtual CAT. 
			\label{tab:rubric_virtual}}
		\centering
		\begin{tabular}{c>{\centering\arraybackslash}m{.8cm}|>{\centering\arraybackslash}m{.8cm}>{\centering\arraybackslash}m{.8cm}>{\centering\arraybackslash}m{.8cm}>{\centering\arraybackslash}m{.8cm}}
			\multicolumn{2}{c|}{}   & \multicolumn{4}{c}{{\textbf{Competence level}}} \\
			\multicolumn{2}{c|}{}   & {GF $c=1$}    & {G $c=2$}   & {PF $c=3$} & {P $c=4$}   \\
			\midrule
			\multirow{3}{*}{\begin{minipage}{.5cm}\rotatebox{90}{\begin{tabular}{c}\textbf{Competence}\\[-1ex]\textbf{component}\end{tabular}}\end{minipage}} 
			& {0D $r=1$} & $X_{11}$   & $X_{12}$  & $X_{13}$ & $X_{14}$             \\
			& {1D $r=2$} & $X_{21}$   & $X_{22}$  & $X_{23}$ & $X_{24}$             \\
			& {2D $r=3$} & $X_{31}$   & $X_{32}$  & $X_{33}$ & $X_{34}$             \\
		\end{tabular}
	\end{subtable} 
\end{table}

Based on our task-specific assessment rubric, we developed a learner model, as described in \Cref{sec:bn_learner_model}. 
\begin{figure}[htb]
	\centering
	\includegraphics[width=.7\columnwidth]{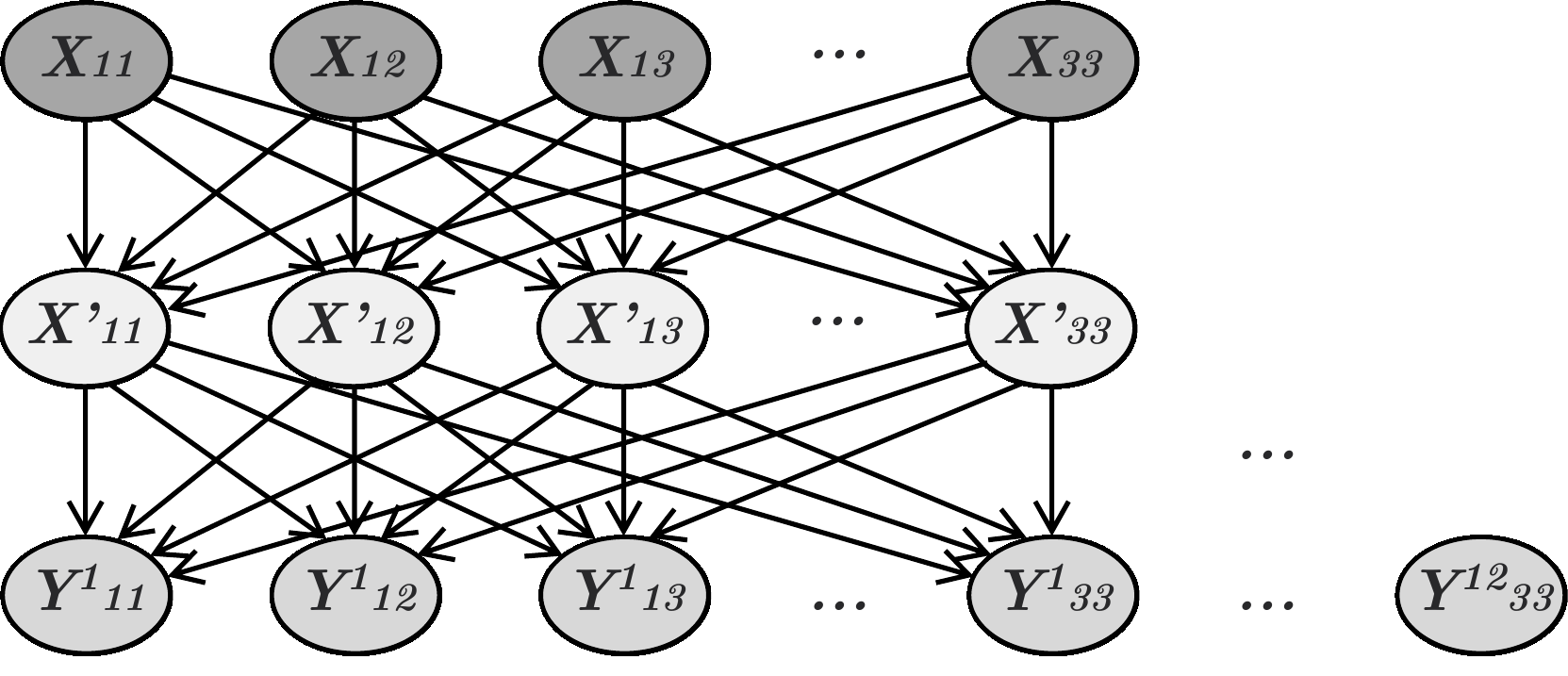}
	\mycaptiontitle{Example of a noisy gates BN modelling the unplugged CAT assessment rubric.}
	\label{fig:unplugged_cat_ar_noisy}
\end{figure}
\Cref{fig:unplugged_cat_ar_noisy} illustrates a simplification unplugged CAT model represented as a BN with noisy gates.
The model includes 9 latent skill nodes ($X_{rc}$) representing the competencies from the rubric, 108 answer nodes ($Y^t_{rc}$) corresponding to manifest behaviours (9 skills $\times$ 12 tasks), and 9 inhibitor nodes ($X'_{rc}$) representing the skill states. 
The network uses noisy gates to model the probability that a skill does not contribute to a specific behaviour or task outcome, providing a probabilistic framework to capture the influence of skills on performance.

\FloatBarrier

\subsection{Ordering of competencies}
As introduced in \Cref{sec:rubrictobn}, the columns of an assessment rubric provide the competence levels in increasing order from left to right. Sometimes, as in this case study, this is true also for the rows, where competence components are ordered from the lower (0D at the top) to the highest (2D at the bottom).  
This follows from the assumption that mastering algorithms of higher complexity implies also mastering simpler ones. The same is valid for the interaction dimension.

Summing up, we can conclude that a competence level $X_{rc}$ is higher than $X_{r'c'}$ whenever $c>c'$ and $r \geq r'$, or $c=c'$ and $r>r'$. When, instead, $c>c'$ but $r<r'$, neither skill can be said to dominate the other. 
From the CAT assessment rubric in \Cref{tab:catrubric}, we define a set of $n$ target skills to be assessed: 9 for the unplugged CAT and 12 for the virtual CAT. 

Accordingly, with the method described in \Cref{sec:rubrictobn}, a latent skill node $X_{rc}$ is included in the BN learner model for each of the $n$ target skills of the rubric. 
The hierarchy of competencies is then modelled by $n$ latent binary variables $D_{rc,r'c'}$, as described in \Cref{sec:rubric}, encoding the implication $X_{rc} \implies X_{r'c'}$ for each pair of consecutive skills in the hierarchy, i.e., such that $(r = r'+1) \land (c= c')$ or $(r = r') \land (c= c'+1)$.

Also, the BN includes an observable answer node $Y^t_{rc}$ for each skill in the rubric and each task $t = 1, \dots, 12$ in the sequence of 12 similar tasks administered during the CAT experiments. 
Observing $Y^t_{rc} = 1$ means that the pupil has solved the $t$-th CAT schema using an algorithm of complexity corresponding to the $c$-th row of the rubric and requesting help in the $r$-th column. 
By way of example, in the unplugged CAT, a student solving the $t$-th schema conceiving a 0D algorithm using voice, empty schema and feedback (0D-VSF) results in the observed node $Y^t_{11}=1$.

\begin{figure}[htb]
	\centering
	\includegraphics[width=.7\columnwidth]{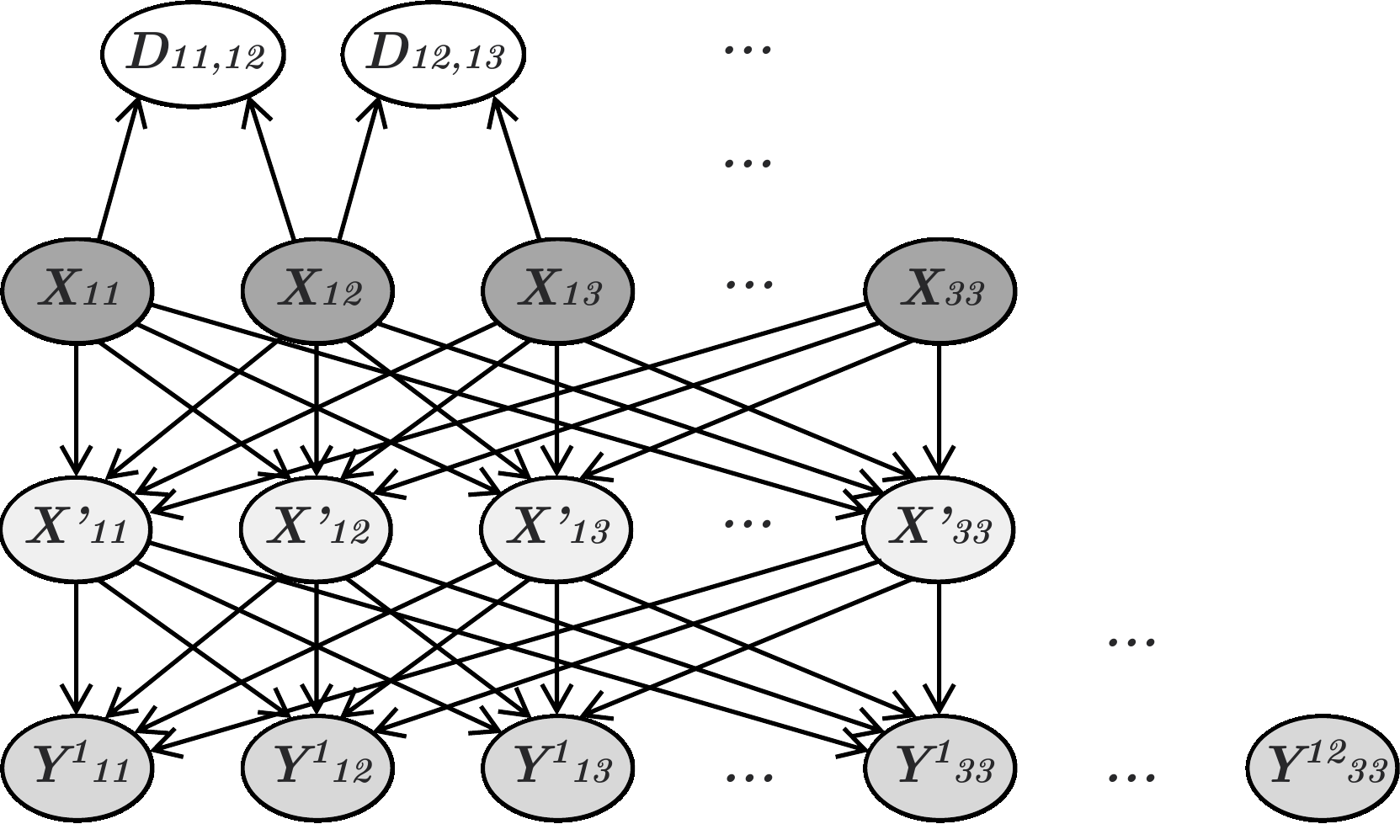}
	\mycaptiontitle{Example of a constrained noisy gates BN modelling the unplugged CAT assessment rubric.}
	\label{fig:unplugged_cat_ar_noisy_constraint}
\end{figure}

\Cref{fig:unplugged_cat_ar_noisy_constraint} illustrates the noisy gates BN for the unplugged CAT case study updated to account for competencies order by including constraints.

\subsection{Answers encoding}
In principle, all answer nodes should be explicitly observed through specific interactions with the pupil. However, this is impossible for the specific activity, as the pupils are free to choose their preferred solving approach during the CAT.
Therefore, to make the answers of the pupils in this activity compatible with our model, we encoded them as follows: a task $t$ solved at level $c^*$ by an algorithm with complexity $r^*$ was translated into $Y^t_{rc} = 1$ for all competence levels $rc$ lower than or equal to $r^*c^*$, thus assuming that, if requested, the pupil would have been able to implement solutions requiring a lower competence level than the one used. Similarly, we set all answer nodes $Y^t_{rc} = 0$ for all higher levels, leaving those not directly comparable unobserved.

As an example, \Cref{tab:unplugged_unconstrained} illustrates the case in which the pupil engaged in the unplugged CAT has generated as a solution for task $t$ a one-dimensional algorithm using only the empty schema and the voice (1D-VS).
This choice also contributed to stressing the ordering of skills. 
\begin{table}[htb]
\footnotesize
\mycaption{Example of answer encoding for 1D-VS in the unplugged CAT.}{Assuming a pupil has generated a 1D-VS solution for the $t$-th schema: $Y^t_{22} = 1$, the two tables illustrated the answer encodings for the unconstrained and constrained cases. The symbol $\varnothing$ indicates that the answer node is not observed.}\label{tab:1dvs}
    \begin{subtable}{0.49\linewidth}
        \mycaptiontitle{{Unconstrained learner model.}\label{tab:unplugged_unconstrained}}
        \centering
        \begin{tabular}{c>{\centering\arraybackslash}m{.8cm}|>{\centering\arraybackslash}m{.8cm}>{\centering\arraybackslash}m{.8cm}>{\centering\arraybackslash}m{.8cm}}
            \multicolumn{2}{c|}{}   & \multicolumn{3}{c}{{\textbf{Competence level}}} \\
            \multicolumn{2}{c|}{}   & {VSF $c=1$}    & {VS $c=2$}   & {V $c=3$}   \\
            \midrule
            \multirow{3}{*}{\begin{minipage}{.5cm}\rotatebox{90}{\begin{tabular}{c}\textbf{Competence}\\[-1ex]\textbf{component}\end{tabular}}\end{minipage}} 
                                    & {0D $r=1$} & \cellcolor[HTML]{D4EFDF}{1}         & \cellcolor[HTML]{D4EFDF}{1}   & $\varnothing$             \\
                                    & {1D $r=2$} & \cellcolor[HTML]{D4EFDF}{1}          & \cellcolor[HTML]{85C1E9}{\textbf{1}}   & \cellcolor[HTML]{FADBD8}{0}            \\
                                    & {2D $r=3$} & $\varnothing$  & \cellcolor[HTML]{FADBD8}{0}  & \cellcolor[HTML]{FADBD8}{0}              \\
        \end{tabular}
    \end{subtable}%
    \hfill
    \begin{subtable}{0.49\linewidth}
        \mycaptiontitle{{Constrained learner model.}\label{tab:unplugged_constrained}}
        \centering
        \begin{tabular}{c>{\centering\arraybackslash}m{.8cm}|>{\centering\arraybackslash}m{.8cm}>{\centering\arraybackslash}m{.8cm}>{\centering\arraybackslash}m{.8cm}}
             \multicolumn{2}{c|}{}   & \multicolumn{3}{c}{{\textbf{Competence level}}} \\
                \multicolumn{2}{c|}{}   & {VSF $c=1$}    & {VS $c=2$}   & {V $c=3$}   \\
                \midrule
                \multirow{3}{*}{\begin{minipage}{.5cm}\rotatebox{90}{\begin{tabular}{c}\textbf{Competence}\\[-1ex]\textbf{component}\end{tabular}}\end{minipage}} 
                                        & {0D $r=1$} & $\varnothing$ & $\varnothing$   & $\varnothing $             \\
                                        & {1D $r=2$} & $\varnothing$  & \cellcolor[HTML]{85C1E9}{\textbf{1}}  & \cellcolor[HTML]{FADBD8}{0}            \\
                                        & {2D $r=3$} & $\varnothing$  & \cellcolor[HTML]{FADBD8}{0}   & $\varnothing$              \\
        \end{tabular}
    \end{subtable} 
\end{table}

However, since in the extended model, the ordering of variables is modelled by explicit constraints imposed through the auxiliary variables $D_{rc, r'c'}$, such a choice would be unnecessary and detrimental, as it would artificially multiply the number of observations. 
Therefore, in the constrained model, a task $t$ solved at level $c^*$ by an algorithm with complexity $r^*$ would be better translated into the single observation $Y^t_{r^*c^*} = 1$.

Since in the experimental setting of the CAT activity, pupils were always allowed to try solving the task with the lowest competence level (0D-VSF for the unplugged task and 0D-GF for the virtual task), a failure could only be observed for that level, with the consequence that only answer nodes $Y^t_{11}$ can be directly observed in the false state $Y^t_{11}=0$. To work around this problem, we set the answer nodes just above the one observed in the true state, i.e., $Y^t_{r^*(c^*+1)}$ and $Y^t_{(r^*+1)c^*}$ to the false state, leaving all other nodes unobserved. 
\Cref{tab:unplugged_constrained} shows how the answer encoding changes in the case of a 1D-VS solution to task $t$ for the constrained model of the unplugged CAT.

\subsection{Supplementary competencies}
Finally, we observed that additional skills beyond those defined in the assessment rubric may be necessary depending on the specific CAT schema, especially for 1D and 2D algorithms. 
Through an analysis of the structures and characteristics of the CAT schemas, we identified three groups of supplementary skills: 10 for the unplugged CAT and 14 for the virtual CAT.
These supplementary skills were added as new nodes to the skill network.
In particular, the first group contains only one skill, represented by the variable $S_1$, which is essential to implement 0D algorithms, the paint single dot operation.
The second group comprises skills required for monochromatic structures, which are associated with 1D algorithms, represented by variables $S_2$ to $S_7$ for the unplugged CAT and $S_2$ to $S_8$ for the virtual CAT.
The third group includes skills necessary for handling polychromatic structures and for operations such as repeating and mirroring a structure associated with 2D algorithms, represented by variables $S_8$ to $S_{10}$ for the unplugged CAT and $S_9$ to $S_{14}$ for the virtual CAT.

From the data collected during the experimental study with both the unplugged and virtual CAT, it was possible to extract direct observations about using each supplementary skill in each task. Consequently, answer nodes $Y^t_{S_i}$ were added to the network for each task $t = 1, \dots, 12$ and each supplementary skill $S_i$, with $i=1, \dots, 10$ fro the unplugged CAT and $i=1, \dots, 14$ for the virtual CAT.
Each schema can be solved using one or more supplementary skills, but using all of them is not always possible.
Answer nodes $Y^t_{S_i}$ take the value one if the pupil has used the $i$-th supplementary skill in the solution of CAT schema $t$, and zero otherwise.

As described in \Cref{sec:rubrictobn}, a noisy-OR combines the variables in the same group into the group auxiliary nodes $G_i$, with $i=1,\dots, 4$, where $G_4$ combines the target skills $X_{rc}$. 
In contrast, the relation between the group nodes and the target skills is conveyed through the logical AND.

\begin{figure}[htb]
	\centering
	\includegraphics[width=\columnwidth]{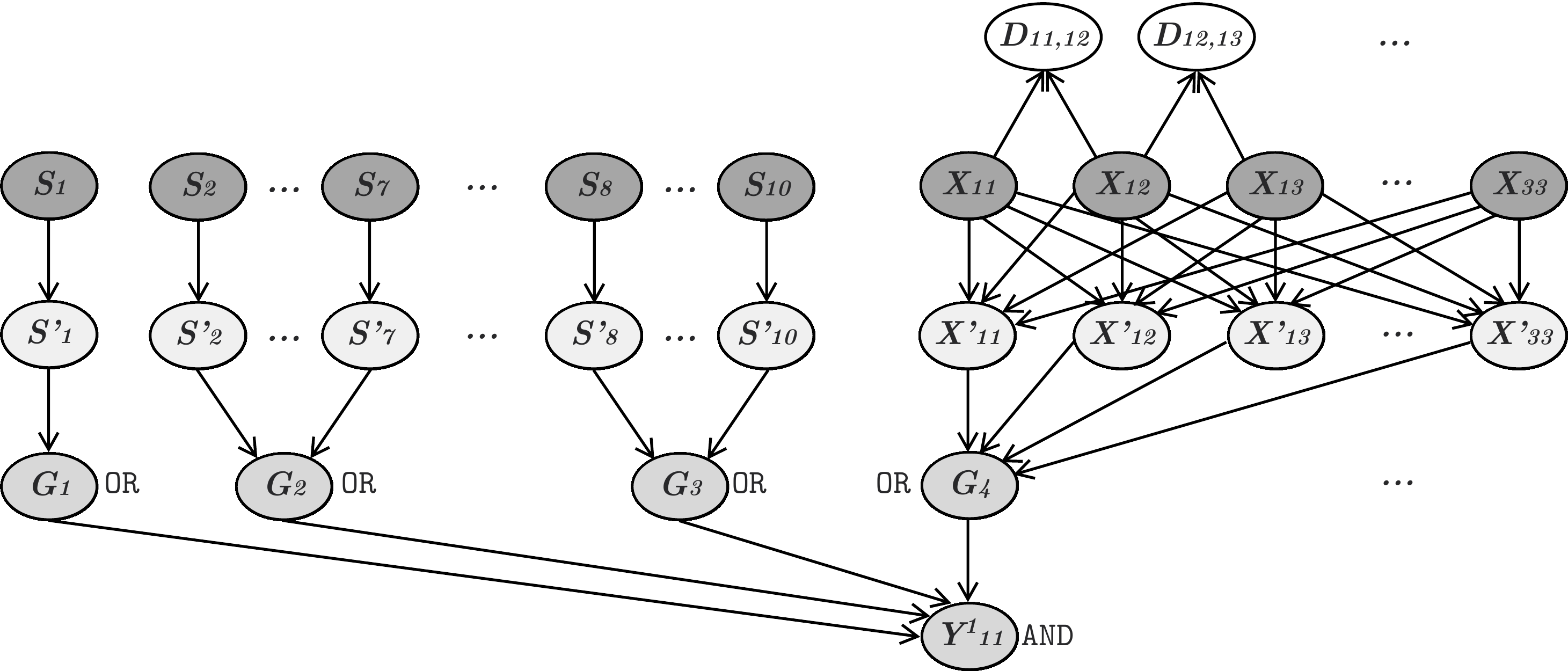}
	\mycaptiontitle{Example of a constrained noisy gates BN modelling the unplugged CAT assessment rubric including supplementary skills.}
	\label{fig:unplugged_cat_ar_noisy_supplementary}
\end{figure}

\Cref{fig:unplugged_cat_ar_noisy_supplementary} illustrates the noisy gates BN for the unplugged CAT case study updated to include supplementary skills.

\section{Parameters' elicitation}

Once the structure of the model is established for the CAT activity, it is necessary to set the values of the prior probabilities $\pi_{*}$, and the 12 inhibitors $\lambda^t_{*}$, $t = 1, \dots, 12$, for both the target and supplementary skills. 
Uniform prior probabilities, i.e., $ \pi_{rc} = 0.50 $, have been assigned to each skill. However, when conditioning given the constraints nodes $D_{rc, r'c'}=1$, their probabilities, before the observation of any answer node, change. For example, for the unplugged CAT, they become $ \pi_{11}=0.95,\, \pi_{12}=0.8,\, \pi_{13}=0.5,\, \pi_{21}=0.8,\, \pi_{22}=0.5,\, \pi_{23}=0.2,\, \pi_{31}=0.5,\, \pi_{32}=0.2,\, \pi_{33}=0.05$.
For the inhibition parameters, we compare two models: the \emph{baseline} model, hereafter referred to as \texttt{Model B}, where all inhibitors are set to the same value, and the \emph{enhanced} model, hereafter referred to as \texttt{Model E}, with parameters elicited by a domain expert.

\texttt{Model B} may look trivial and unrealistic, but it allows one to understand better the effect of the constraints resulting from ordering the skills and supplementary skills on the model inferences. 
The constant value of $\lambda$ was chosen equal to $0.2$, except for the leak node, associated with all answer nodes and modelling a guess probability of $0.1$, resulting in  $\lambda_{\mathrm{leak}} = 0.9$.

\texttt{Model E} builds on the baseline model to address the progressive complexity of the 12 tasks and the challenges students may encounter applying their skills to different schemas. 
The expert elicitation process involved grouping the 12 schemas into eight categories of increasing difficulties based on their characteristics: (i) T1, (ii) T2, (iii) T3, T4, (iv) T5, T6, (v) T7, T8, T9, (vi) T10, (vii) T11, (viii) T12. 
The expert assumed all tasks could be solved with 0D, 1D, and 2D algorithms. Moreover, given a schema $t$ and a manifest variable $Y^t_{rc}$, the same inhibition probability was assumed for all
relevant skills, meaning that all have the same probability of successfully being applied in solving schema $t$ with level $rc$. 
In the proposed method, the inhibitor parameter $\lambda_{rc}$ is used to model the probability of failing a task of a particular difficulty level $rc$, assuming the student has the necessary skills to solve the task. 
When a task is more complex or less help is available to the student, the value of $rc$ increases, which means that the inhibitor parameter also increases. 
This is because when the student possesses the necessary skills to solve a difficult task, the probability of failing is higher than for a simpler task.
Similarly, the inhibitor parameter $\lambda^t_{rc}$ is assigned to a particular schema $t$ and is used to model the difficulty of implementing a solution of level $rc$ for that schema. 
A high value of $\lambda^t_{rc}$ means that it is difficult to implement a solution of level $rc$ for that schema. In other words, the inhibitor parameter $\lambda^t_{rc}$ provides a measure of the difficulty of implementing a particular solution for a given schema at a given level of complexity.

While the students are generally expected to use 2D algorithms to solve the tasks optimally, there may be cases where a simpler 1D solution may be optimal.
Nonetheless, in the current implementation, the first two tasks are designed to serve as starting points for students, introducing them to the activity. They are expected to be solved using simpler 1D algorithms.
However, this particular case was not included in our model. This could have been described by setting high inhibitor values to indicate that certain 2D solutions are more difficult to implement than others, making them less likely to be chosen by students.

\begin{figure}[!h]
	\centering
	\includegraphics[width=\textwidth]{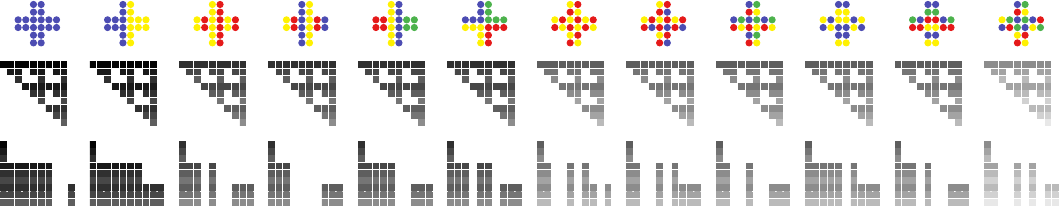}
	\mycaption{The inhibition parameters for the unplugged CAT.}{The 12 CAT schemas $T$ (top); the values of the inhibition parameters $\lambda^t_{rc}$ for the target skill nodes (centre); the value of the inhibition parameters $\lambda^t_{S_i}$ for the supplementary skill nodes (bottom). The inhibition parameters for both the target and supplementary skill are depicted as a matrix of nine rows representing the answers and as many columns as the number of modelled skills. 
		The strength of the skill-answer relation has eleven levels, from $0.1$ to $0.6$, with a step of $0.05$. Darker shades of grey mean lower skill-answer inhibition probabilities, and white squares denote non-relevant skills.}
	\label{fig:its}
	\bigskip
    \centering
    \begin{subfigure}[t]{0.45\linewidth}
        \centering
        \includegraphics[height=4.5cm]{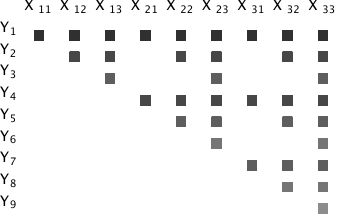}
        \mycaptiontitle{Target skills.}
        \label{fig:inib_target}
    \end{subfigure}
    \hspace{30pt}
    \begin{subfigure}[t]{0.45\linewidth}
        \centering
        \includegraphics[height=4.5cm]{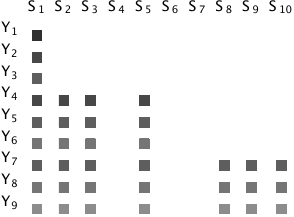}
        \mycaptiontitle{Supplementary skills.}
        \label{fig:inib_supp}
    \end{subfigure}
    \mycaption{The inhibition parameters for the unplugged CAT in schema T3.}{Inhibition parameters $\lambda^{T3}_{*}$ used in the ECS model for schema T3 (Zoom on schema T3 of Fig.~\ref{fig:its}). The parameters are divided into the target skills (top) and the supplementary ones (bottom). 
    The supplementary skills $S_4$ (paint monochromatic squares), $S_6$ (paint monochromatic ls), and $S_7$ (paint monochromatic zigzags) are represented as empty columns because they cannot be used to solve task T3.}
\label{fig:sampleits}
\end{figure}

Our succinct elicitation setup allows for summarising both the BN topology and its parameter values graphically. An example, specifically referring to the unplugged CAT, is shown in the monochromatic rows at the bottom of \Cref{fig:its} and is explained in more detail in \Cref{fig:sampleits} for schema T3. 

The underlying BN has been implemented within the CREMA Java library \citep{huber2020a}, which supports the specifications of noisy gates and inference based on these parametric CPTs. The network size allowed for exact inferences using the Variable Elimination (VE) algorithm \citep{chavira2007compiling}. The model implementation is available on GitHub \cite{Adorni2024noisysoftware}.

\FloatBarrier

	\part{Results}\label{part:results}

\chapter{Experimental study on the unplugged CAT}\label{chap:study_unplugged}
\begin{custombox}
	{The content of this chapter has been adapted from the following article with permission of all co-authors and publishers:}
	\begin{itemize}[noitemsep,nolistsep,leftmargin=.25in]
		\item Piatti, A., \textbf{Adorni, G.}, El-Hamamsy, L., Negrini, L., Assaf, D., Gambardella, L., and Mondada, F. (2022). The CT-cube: A framework for the design and the assessment of computational thinking activities. \textit{Computers in Human Behavior Reports} \cite{piatti_2022}. 
	\end{itemize}
	As an author of this publication, my contribution involved:\\
	\textit{Conceptualisation, Methodology, Validation, Formal analysis, Investigation, Resources, Data curation, Writing -- original draft \& review \& editing, Visualisation.}
\end{custombox}


\section{Summary}

This chapter focuses on an experimental study using the unplugged {CAT} to assess the algorithmic skills of K-12 pupils, contributing to RQ4.
As a preliminary study, our goal was to demonstrate how the CT-cube and the CAT can be used to evaluate students' algorithmic abilities.
We explored how the task works in practice, analysing the algorithms produced by students and their performance according to the CT-cube dimensions, while considering the effects of age and gender. 
%
The chapter includes a detailed description of the study context, participant selection process, data collection methods, and the approach used for data analysis. Finally, the results are presented and discussed to provide insights into the effectiveness of the unplugged CAT for assessing algorithmic skills. 

\section{Methodology}\label{sec:unplugged_method}

\subsection{Study context}\label{sec:unplugged_context}

The experimental study was conducted in Switzerland between March and April 2021.
As the first study in our series, all testing took place in the canton of Ticino, part of Switzerland's Italian-speaking region, chosen for the ease and speed of identifying suitable classes and securing approval from local authorities through our established network.
In our final study, we expanded the sample to other cantons, including regions with different spoken languages, to ensure broader geographical and linguistic representation (see \Cref{chap:study_virtual_main}).

%


\subsection{Participant selection}\label{sec:unplugged_participants} 
To ensure a representative sample across all levels of compulsory education, with a balanced demographic of age, this study included eight classes from three public schools, in particular one preschool class (ages 3-6), three primary school classes (ages 6-11), and four lower secondary school classes (ages 12-16).
It is important to note that the selection of schools and classes was not random. Rather, participating classes were selected through our network of contacts, and the schools were contacted and agreed to take part in the study.

\begin{table}[htb]
	\centering
	\mycaption{Study participants (unplugged CAT).}{Demographics analysis of participants by session, including school ID and type, HG, age category (mean and standard deviation), and gender distribution.}\label{tab:classes}
	\setlength{\tabcolsep}{1.3mm}
	\footnotesize
	\begin{tabular}{clcc@{\hspace{.5\tabcolsep}}c@{\hspace{.8\tabcolsep}}cccc}
		\toprule
		\textbf{Session}   & \textbf{School ID \& type}   &   \textbf{HG} & \multicolumn{3}{c}{\textbf{Age category}}   &   \textbf{Female} &   \textbf{Male} &   \textbf{Total} \\
		\midrule
		1  & A \quad Preschool                 & 0, 1, 2   & 3-6   & yrs& ($\mu$ \, 4.9 $\pm$ 0.9 yrs)   & 8 &     13 &      21 \\ 
		2  & B \quad Primary school            & 3         & 3-6   & yrs& ($\mu$ \, 6.7 $\pm$ 0.5 yrs)   & 4 &      8 &      12 \\
		3  & B \quad Primary school            & 5         & 7-9   & yrs& ($\mu$ \, 8.7 $\pm$ 0.6 yrs)   & 7 &      8 &      15 \\
		4  & B \quad Primary school            & 7         & 10-13 & yrs& ($\mu$ 10.5 $\pm$ 0.6 yrs)     & 8 &     11 &      19 \\
		5  & C \quad Lower secondary school    & 9         & 10-13 & yrs& ($\mu$ 12.5 $\pm$ 0.5 yrs)     & 8 &      7 &      15 \\
		6  & C \quad Lower secondary school    & 10        & 14-16 & yrs& ($\mu$ 13.0 $\pm$ 0.0 yrs)     & 5 &      2 &      7 \\
		7  & C \quad Lower secondary school    & 11        & 14-16 & yrs& ($\mu$ 14.5 $\pm$ 0.7 yrs)     & 9 &      5 &      14 \\
		8  & C \quad Lower secondary school    & 11        & 14-16 & yrs& ($\mu$ 14.5 $\pm$ 0.5 yrs)     & 2 &      4 &      6 \\
		\cmidrule(lr){1-1}\cmidrule(lr){2-2}\cmidrule(lr){3-3}\cmidrule(lr){4-6}\cmidrule(lr){7-7}\cmidrule(lr){8-8}\cmidrule(lr){9-9}
		&   &        &   &&($\mu$ \; 9.9 $\pm$ 3.5 yrs) &51 &     58 &     109 \\
		\bottomrule
	\end{tabular}
\end{table}
The study was conducted in multiple sessions, each assigned a unique ID corresponding to a specific activity or condition, allowing us to track and compare data across conditions and time points.
All pupils in each class who had received parental consent were allowed to voluntarily participate in the study. For ethical considerations regarding participant selection and informed consent procedures, refer to \Cref{sec:ethics}. 
A total of 109 pupils (51 girls and 58 boys) participated in the study.
\Cref{tab:classes} provides a detailed breakdown of participant information across sessions, including school ID, HG, age and gender distribution.

\subsection{Data collection approach and procedures}\label{sec:unplugged_datacollection_procedures}

The study took place during regular class sessions and was administered in Italian.
The procedure was designed to accommodate the participants' age and school environment.
For younger students, task instructions were kept simple and engaging, while older students received more detailed guidance appropriate for their developmental level. The testing environment minimised distractions and ensured that the task did not interfere with the regular school schedule. No time limit was imposed to further reduce stress, allowing students to work at their own pace.
To minimise interference with other students, two pupils at a time were randomly selected from the class and taken to a separate room to complete the task.
Each pupil was randomly assigned to one of the available administrators from our research team, who explained the task and its goals, outlined the rules, and recorded session and participant details, as well as strategies employed by the pupils for each schema of the CAT on a protocol template (see \Cref{app:chbr_appE}).  
For further details on the experimental setting, refer to \Cref{sec:catdesign}.

The information recorded on the templates was later transferred to a database, which included session details, student information, and task performance data. 
Each session was assigned a unique identifier, with specific contextual information such as the date, canton, school ID and type, the HGs level, and the administrator's details. 
Student information was limited to the ID, gender, and date of birth, which was used to calculate their ages.
The task and performance data recorded for each schema completed by the participants included details about the algorithm and interaction dimensions.
Additionally, a separate database was created to store all the unique algorithms generated by schema. This included a description, the sequence of operations followed, the number of operations in the algorithm, the type of patterns used (e.g., point-by-point, rows, squares, etc.), and any redundancy present in the algorithms.


\subsection{Data analysis approach}\label{sec:unplugged_dataanalysis}


The data analysis aimed to provide insights into various aspects of the unplugged CAT assessment, focusing on participation, performance, algorithmic strategies, and the development of competencies across age groups and genders. 
Python was used for data processing, exploratory and statistical analyses, and visualisations of results \citep{pyhton}.

 \subsubsection{Participation and performance}
 
 
We begin by analysing the time-related performance metrics to gain insights into how long students took to complete the tasks.
Next, we examine participation and success rates for each schema, exploring their distributions using descriptive statistics.
Following this, we investigate the distribution of CAT scores across different age groups and performed a chi-square test of independence to assess whether there are significant relationships between age and performance outcomes \citep{Cochran1954,moore_introduction_1989,Yates1934}. 
We employ pairwise comparisons \citep{Yates1934} to assess performance differences between age categories. Specifically, Tukey's Honestly Significant Difference (HSD) test was used to identify which specific age groups showed significant differences in their success rates \citep{Tukey1949}. To control for multiple comparisons and reduce the likelihood of Type I errors (false positives), we applied the Benjamini-Hochberg (BH) correction \citep{Benjamini1995}. This adjustment modifies the p-values to control the false discovery rate, ensuring the reliability and accuracy of our findings. 
Additionally, we conduct a gender-based analysis to examine how performance varies between male and female students, focusing on the distribution of CAT scores. We assessed whether the variations in performance across genders were statistically significant.

\subsubsection{Competencies development}
In the second phase of analysis, we conduct a qualitative analysis on the development of algorithmic and interaction strategies across different age categories, exploring also on how these strategies changed within each schema. 
This analysis aimed to identify patterns in the approaches taken by students at various developmental stages. 
Additionally, we explored how these strategies varied by gender, providing insights into potential gender differences in task approach and competencies development.

%

\subsubsection{Algorithms classification}
Finally, we examined the diversity and frequency of algorithms generated by students for each schema, along with the distribution of algorithmic dimensions. 

%


%



\section{Results}\label{sec:unplugged_results}

\subsection{Participation and performance}

The time required to solve all the 12 schemas in the CAT varied significantly across participants, ranging from a minimum of 10 minutes for older pupils to a maximum of 45 minutes for the younger ones. 
This variation was primarily due to differences in age and cognitive development. Older students tended to solve the tasks more quickly, demonstrating greater efficiency, while younger students required more time to understand and complete the schemas.
The total time spent administering the CAT to all 109 students was approximately 36 hours.

In addition to time-related metrics, we analysed participation and success rates across schemas (see \Cref{tab:performance_by_schema_unplugged}).
Pupils tackled tasks up to schema 7 without intermissions. Beyond that point, some experiments were cut short due to time constraints, although the participation rate remained high, with a minimum of 95\%. These interruptions were primarily due to the end of school hours or the limited attention span of very young preschool pupils \citep{Bennett2006,Mahone2012}.
Regarding success rates, students were generally able to successfully complete all the tasks they were confronted with, demonstrating a strong engagement with the content and an ability to apply their algorithmic skills. 
However, more advanced schemas required additional time or support for younger students.

The high success rate across all schemas suggests that the tasks were accessible and well-aligned with students' cognitive capabilities.
However, this also raises the possibility of a \textit{ceiling effect}, where the tasks may have been too easy for some students, potentially limiting the ability to distinguish variations in skill levels. Future iterations could explore adjusting task complexity or introducing adaptive difficulty to better capture differences in AT.
\begin{table*}[!hb]
	\footnotesize\centering
	\mycaption{Participation and success rates across schemas (unplugged CAT).}{
		Success rate is calculated from the number of students who attempted the schema.}		
	\label{tab:performance_by_schema_unplugged}
	\begin{tabular}{ccccc}
		\toprule
		\rowcolor{white} 
		&
		& \multicolumn{1}{>{\footnotesize\centering\arraybackslash}m{3.5cm}}{{No. pupils participating \linebreak (out of 109)}} 
		& \multicolumn{1}{>{\footnotesize\centering\arraybackslash}m{2cm}}{{\textbf{Participation} \linebreak (\%)}} & \multicolumn{1}{>{\footnotesize\centering\arraybackslash}m{2cm}}{{\textbf{Success} \linebreak (\%)}}\\
		\midrule
		{\multirow{12}{*}{ \rotatebox{90}{\textbf{Schema}}}}
		& 1  & 109 & 100\% & 100\% \\
		& 2  & 109 & 100\% & 100\% \\
		& 3  & 109 & 100\% & 100\% \\
		& 4  & 109 & 100\% & 100\% \\
		& 5  & 109 & 100\% & 100\% \\
		& 6  & 109 & 100\% & 100\% \\
		& 7  & 109 & 100\% & 100\% \\
		& 8  & 107 & \, 98\% & 100\% \\
		& 9  & 105 & \, 96\% & 100\% \\
		& 10 & 105 & \, 96\% & 100\% \\
		& 11 & 105 & \, 96\% & 100\% \\
		& 12 & 104 & \, 95\% & 100\% \\
		\bottomrule
	\end{tabular}
\end{table*}

These results indicate the task suitability for pupils from preschool to secondary school, unlike most formal {CT} and AT assessments, which are limited to a pre-defined age range.
Indeed, the majority of studies focused on evaluating {CT} skills for a specific educational level, more often elementary or middle school. For example, the {CTt} by \citet{roman2017cognitive} is designed for 7th and 8th-grade students (ages 12-14), although it can also be used to measure the {CT} from 5th to 10th grade (ages 10-16). 
The {BCTt} by \citet{zapata2020computational}, which can be considered as an extension of the previous work, is mainly aimed at the first educational stages of primary school from the 1st to 4th grade (ages 5-10) and less adapted for middle school and beyond. 
Appropriate assessments of {CT} for younger children are generally more challenging and require considering developmental appropriateness in both format and content \cite{relkin2020techcheck, relkin2021techcheckk}. 
Being an unplugged assessment, as shown in \Cref{sec:catdesign}, the {CAT} does not require prior programming experience either, making it adequate for young pupils, as the {CTt} \cite{roman2018detected, roman2017cognitive}, the {BCTt} \cite{zapata2020computational} and TechCheck \cite{relkin2020techcheck}. Moreover, avoiding confusing coding with {CT} skills is important.

\begin{figure}[hbt]
	\centering
	\includegraphics[width=.6\linewidth]{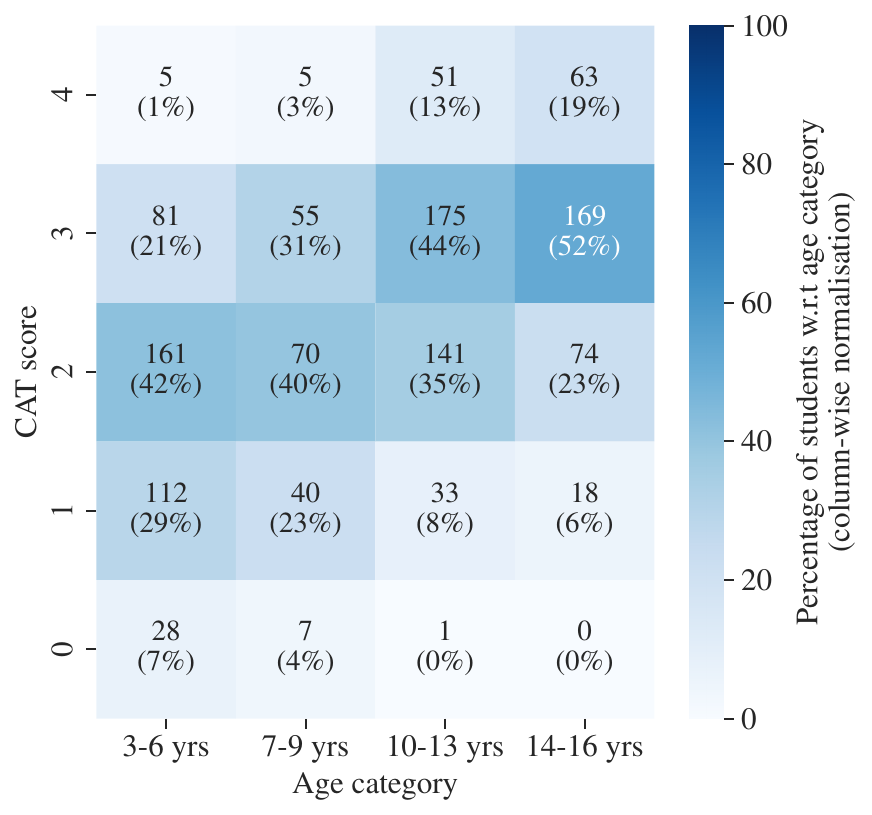}
	\mycaptiontitle{Age-wise distribution of CAT score levels (unplugged CAT).}
	\label{fig:scores_per_age}
\end{figure}

We also investigated how performance differences could be attributed to age. 
\Cref{fig:scores_per_age} shows that the distribution of CAT scores increases with age. This is attributable to the increased autonomy and the capability of using symbolic artefacts to describe the algorithms of older students. 
%
These differences are statistically significant, as confirmed by the chi-square test of independence ($\chi^2(12) = 270.2$, $p<0.0001^{****}$). 

The pairwise comparison in \Cref{tab:scores_per_age_pariwise} reveals the differences in CAT scores are statistically significant between all age categories, except for the two youngest groups (ages 3-6 and 7-9).
This leap can be attributed to the Swiss school system's organisation into educational cycles see \Cref{sec:intro_swiss_schools_organisation}), each with distinct objectives, with a key transition in the curriculum occurring after these two age categories \cite{cantone2015piano,cantone2022piano}. 
\begin{table}[htb]
	\centering
	\mycaptiontitle{Pairwise comparison of CAT scores between age groups (unplugged CAT).}
	\label{tab:scores_per_age_pariwise}
	\setlength{\tabcolsep}{.82cm}
	\footnotesize
	\begin{tabular}{c@{\hspace{.2\tabcolsep}}c>{\centering\arraybackslash}m{2cm}>{\centering\arraybackslash}m{2cm}>{\centering\arraybackslash}m{2cm}}
		\toprule
		& &\multicolumn{1}{c}{\textbf{3-6 yrs}} & \multicolumn{1}{c}{\textbf{7-9 yrs}} & \multicolumn{1}{c}{\textbf{10-13 yrs}}  \\ 
		\midrule 
		\textbf{7-9} &\textbf{yrs} & $\chi^2(4)=$   $5$, $p=0.26$ &  &  \\ 
		\textbf{10-13} &\textbf{yrs} &  $\chi^2(4)=$  $75$, $p<.0001^{****}$    & $\chi^2(4)=$  $73$, $p<.0001^{****}$ & \\  
		\textbf{14-16} &\textbf{yrs} &     $\chi^2(4)=160$, $p<.0001^{****}$ &     $\chi^2(4)=182$, $p<.0001^{****}$ &     $\chi^2(4)=30$, $p<.0001^{****}$    \\
		\bottomrule
	\end{tabular}
	\\\vspace{0.5em}
	\begin{minipage}{.9\linewidth}
		$^{*}$ $p < 0.05$, $^{**}$ $p < 0.01$, $^{***}$ $p < 0.001$, $^{****}$ $p < 0.0001$
	\end{minipage}
\end{table}

\begin{figure*}[!hb]
	\centering
	\includegraphics[width=.92\linewidth]{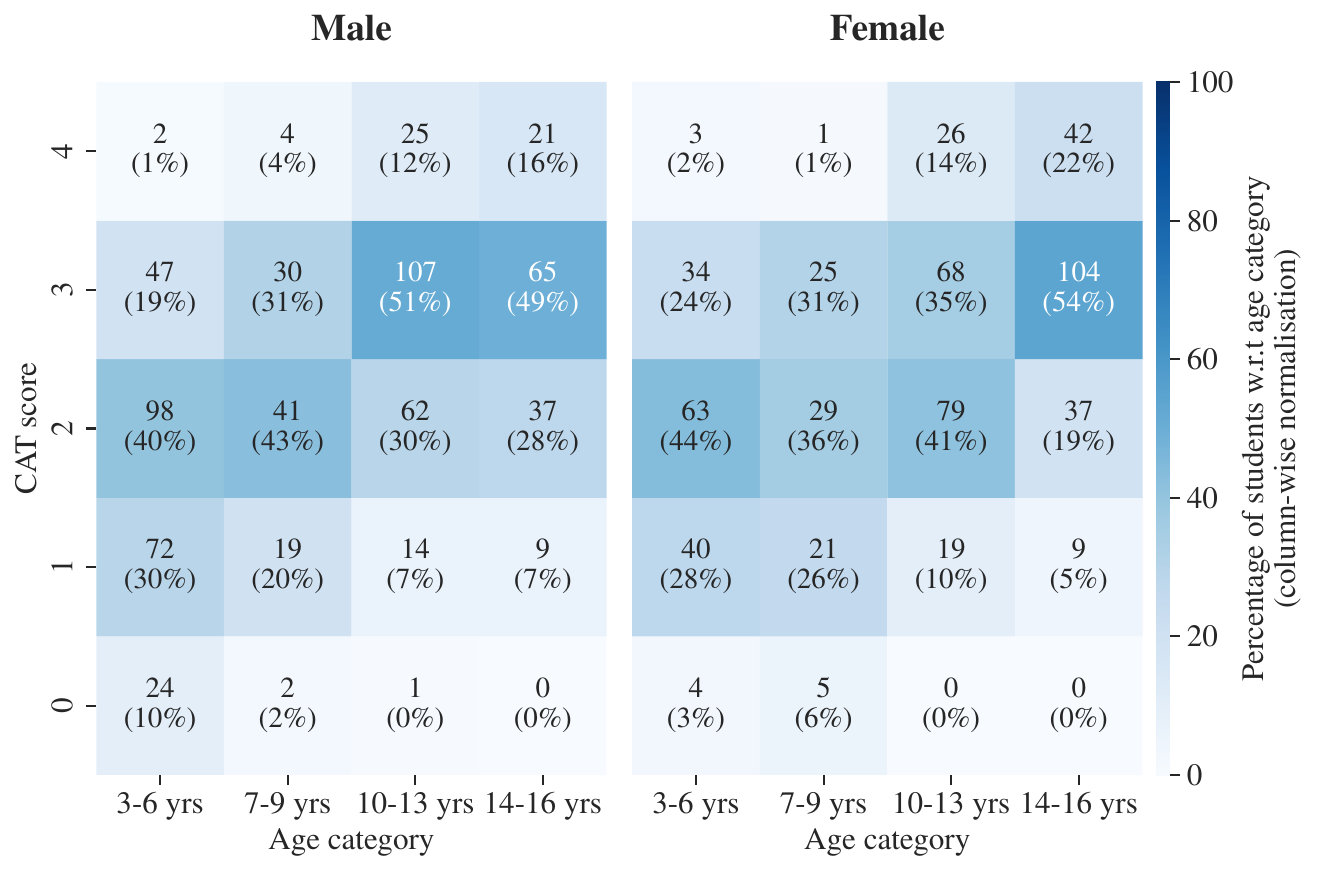}
	\mycaptiontitle{Age-wise distribution of CAT score levels by gender (unplugged CAT).}
	\label{fig:score_gender_age_distribution}
\end{figure*}

Turning our attention to performance variation by gender, \Cref{fig:score_gender_age_distribution} illustrates the distribution of CAT scores across gender and age categories. 
While the overall improvement with age remains evident, no notable differences emerge between genders. 
This finding is supported by the results of the chi-square test of independence, which indicate no significant differences between genders for any age category ($p>0.05$). 

\FloatBarrier

\subsection{Competencies development}\label{sec:unplugged_results_competencies_development}
%

\begin{figure}[ht]
	\includegraphics[width=\linewidth]{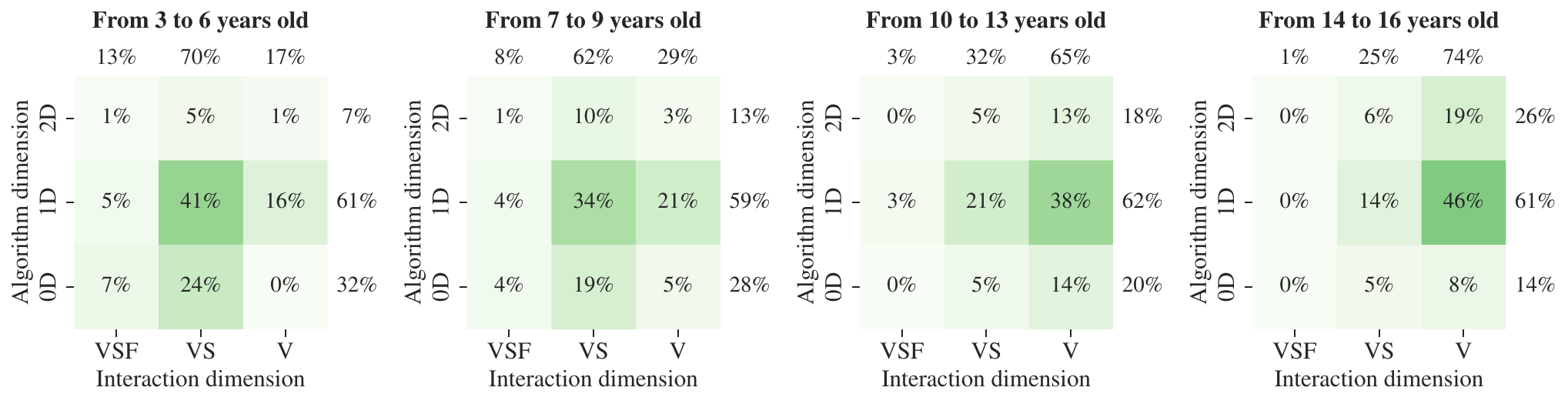}
	\mycaption{Algorithmic and interaction strategies across age (unplugged CAT).}{
		Percentages represent the proportion of each combination within their respective age groups.
		The sum of percentages across rows and columns reveals the aggregate preference or predominance for certain interaction-algorithmic strategies among different age groups.}
	\label{fig:Performance}
\end{figure}

To assess whether these results hold true across both algorithmic and interaction dimensions, we delve deeper into the strategies employed by pupils across different age categories, and subsequently, we analyse these strategies with respect to gender.
\Cref{fig:Performance} illustrates the strategies employed by pupils to solve the {CAT}. Across all age categories, 1D algorithms are the most frequently used, while the proportion of 2D algorithms increases steadily with age. This trend suggests that pupils across all age groups are capable of creating complex algorithms, with their algorithmic skills demonstrating clear growth as they progress in age.
From this visualisation, we can infer that the increase in CAT scores with age is primarily driven by an improved ability to utilise more complex artefacts (e.g., voice alone).
The most frequent interaction dimensions in the first two age categories (ages 3-6 and 7-9) is VS, whereas for the last two age categories (ages 10-13 and 14-16) it is V. This trend suggests that older pupils are increasingly capable of handling more complex artefacts independently.
Interestingly, younger pupils (ages 3-6) displayed non-autonomous behaviour (requiring visual feedback) in only 13\% of cases. This percentage decreases progressively across age groups, reaching just 1\% in the oldest group (ages 14-16).

The ability to conceive more complex algorithms also increases gradually with age.
The two younger age groups (ages 3-6 and 7-9) are capable of conceiving 1D and 2D algorithms in many cases, and in some instances, they can even describe them solely through voice, demonstrating strong algorithmic skills from an early age.
These results suggest that preschool pupils already exhibit algorithmic abilities, which can be elicited through the use of suitable artefacts, particularly embodied or iconic ones.
This evidence is further supported by prior research demonstrating that working on CT skills with very young pupils is not only possible but also effective. Studies have shown that CT skills can emerge and develop rapidly, even in preschool-aged children. 
For instance, \citet{wohl2015teaching} achieved satisfactory learning outcomes with students under 5, concluding that unplugged activities can be used effectively to introduce computational concepts. Similarly, \citet{dietz2019building} explored young children’s developing capacities for problem decomposition and demonstrated that the skills necessary for this type of problem may be initiated to be fostered in preschool years.

A more detailed analysis of these strategies schema by schema (see \Cref{app:unplugged_strategies}) shows that the improvement in interaction dimension is particularly evident from schemas S2 to S6. 
As for the algorithm dimension, the development of the algorithmic skills is more evident from schemas S7 to S9. 

\begin{figure*}[htb]
	\centering
	\includegraphics[width=\linewidth]{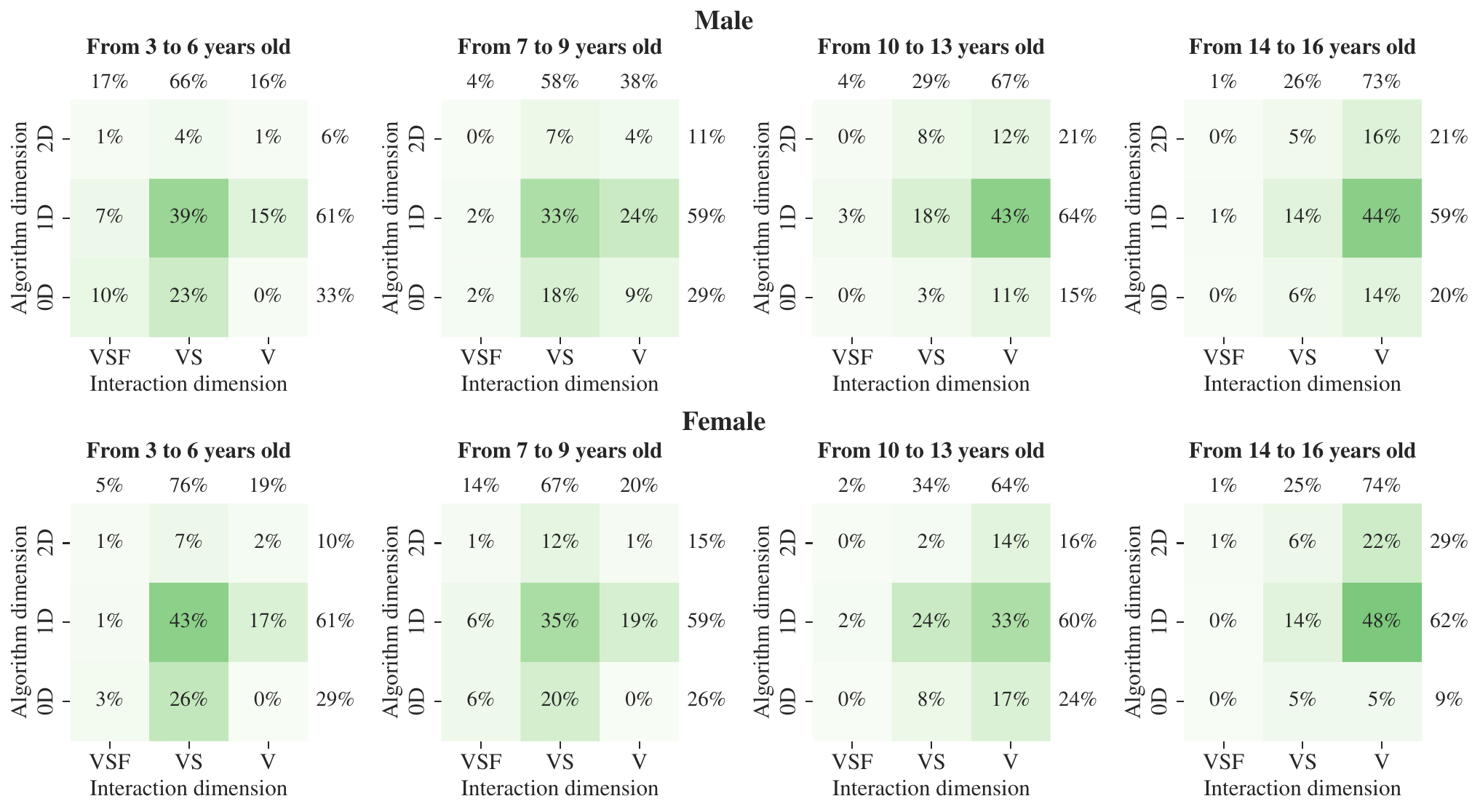}
	\mycaptiontitle{Algorithmic and interaction strategies across age and gender (unplugged CAT).} 
	\label{fig:Performance-gender}
\end{figure*}

Turning again our attention to how these strategies vary by gender, \Cref{fig:Performance-gender} confirms that the observations made for the full sample do not differ according to gender. 
As reported by \citet{tikva_mapping_2021}, gender differences have been examined in many studies, often leading to contradictory results. 
For example, \citet{roman2017cognitive} reported a relationship between the role of gender and the development of {CT}.   
Other studies, such as \citet{relkin2020techcheck}, \citet{delal2020developing}, \citet{atmatzidou2016advancing} and \citet{metin2020activity}, verified gender relationship in their tests and concluded that no significant difference between genders appears when evaluating the students.

\FloatBarrier

\subsection{Algorithms classification}
%

Finally, we focus on analysing the number of different algorithms generated by the students and the distribution of these algorithms according to their dimensions.
\Cref{tab:schemasdimensions} summarises, for each schema of the {CAT}, the number of different algorithms observed and the distribution of algorithms according to their dimension.

Overall, we have observed 137 different algorithms distributed across the 12 schemas of the {CAT}. 
\begin{table}[ht]
	\centering
	\mycaption{Algorithm dimensions distribution across schemas (unplugged CAT).}{
		Values highlighted in bold indicate schemas where the algorithms produced with that algorithmic dimension are at least 20\%.}
	\label{tab:schemasdimensions}
	\setlength{\tabcolsep}{2.15mm}
	\footnotesize
	\begin{tabular}{cc>{\centering\arraybackslash}m{1.7cm}>{\raggedleft\arraybackslash}m{1.1cm}>{\raggedleft\arraybackslash}m{1.1cm}>{\raggedleft\arraybackslash}m{1.1cm}}
		\toprule
		&  &   {{No. unique \linebreak algorithms}} &  \textbf{0D}\linebreak(\%) &  \textbf{1D}\linebreak (\%) &  \textbf{2D}\linebreak(\%) \\
		\midrule 
		{\multirow{12}{*}{ \rotatebox{90}{\textbf{Schema}}}}
		& {1}    &  5 &          2 \%  & \textbf{95 \%} &          4 \% \\
		& {2}    &  7 &          3 \%  & \textbf{95 \%} &          3 \% \\
		& {3}    & 10 &          2 \%  & \textbf{84 \%} & {14 \%} \\
		& {4}    &  8 &          2 \%  & \textbf{89 \%} &          9 \% \\
		& {5}    &  5 &          0 \%  & \textbf{100 \%} &         0 \% \\
		& {6}    &  7 &          5 \%  & \textbf{95 \%} &          0 \% \\
		& {7}    & 19 & \textbf{50 \%} & {18 \%} & \textbf{31 \%} \\
		& {8}   &  21 & \textbf{51 \%} & {12 \%} & \textbf{36 \%} \\
		& {9}   &   9 & \textbf{52 \%} &         0 \%   & \textbf{48 \%} \\
		& {10}  &  25 & {17 \%} & \textbf{58 \%} & \textbf{25 \%} \\
		& {11}  &  15 & {14 \%} & \textbf{61 \%} & \textbf{25 \%} \\
		& {12}  &   6 & \textbf{80 \%} & \textbf{20 \%} &          0 \% \\ 
	\cmidrule(lr){1-2}\cmidrule(lr){3-3}\cmidrule(lr){4-4}\cmidrule(lr){5-5}\cmidrule(lr){6-6}
		&   & 137       &   23 \% & 61 \% & 16 \% \\
		\bottomrule
	\end{tabular}
\end{table}
For most schemas, there is a clear tendency for pupils to resolve the tasks by relying on algorithms of a certain dimension.
Specifically, 1D algorithms are predominantly chosen in schemas S1 to S6, as well as in S10 and S11. 
In schemas S7 to S9, slightly more than half of the pupils opt for 0D algorithms, followed by 2D algorithms. 
In schema S12, the majority of pupils prefer using a 0D algorithm. 
Notably, there are a few schemas (S5, S6, and S12) where 2D algorithms are never selected, one case (S9) where 1D algorithms are never used, and one case (S5) where the point-by-point 0D algorithm is never used, suggesting that pupils preferred more efficient solutions with a higher algorithmic dimension.

\Cref{tab:algorithmsfreq} presents the frequency of algorithm usage across different schemas. 
Despite the variety of algorithms generated for each schema, a few algorithms are consistently recurrent, with two or three being the most commonly used.
As previously mentioned, 1D algorithms are the most frequent, followed by 0D algorithms, while 2D alternatives are never the most frequent. 
A detailed graphical representation of the various algorithms generated for each schema is provided in \Cref{app:chbr_appF}.
\begin{table}[htb]
	\centering
	\mycaption{Algorithms distribution across schemas.}{The percentages represent how often each algorithm (rows) was chosen for that schema (columns). 
	The cell colours indicate the algorithm dimension  (white for 0D, blue for 1D, yellow for 2D).
	Cells highlighted in bold indicate the most frequent algorithm.}
	\label{tab:algorithmsfreq}
	\setlength{\tabcolsep}{2mm}
	\footnotesize
	\begin{tabular}{cccccccccccccc} 
		\toprule
		&\multicolumn{1}{c}{}& \multicolumn{12}{c}{\textbf{Schema}} \\
		& & 1 & 2& 3& 4 & 5 & 6 & 7 & 8 & 9 & 10 & 11 & 12 \\
		\midrule
		\multirow{25}*{\rotatebox[origin=c]{90}{\textbf{Algorithm}}}& {1} 	&	2\%	&	3\% &	2\%	&	2\%	&	0\%	&	5\%	&	\textbf{50\%}	&	\textbf{51\%}	&	\textbf{52\%}	&	18\%	&	14\%	& \textbf{78\%}	\\
		&{2} &	\cellcolor{cyan!25}\textbf{94\%}	&	
		\cellcolor{cyan!25}\textbf{83\%}	&	
		\cellcolor{yellow!25}13\%&	\cellcolor{yellow!25}9\%	& 
		\cellcolor{cyan!25}\textbf{91\%}	&	\cellcolor{cyan!25}\textbf{67\%}	&	
		\cellcolor{yellow!25}11\%	&	\cellcolor{yellow!25}4\%	&	
		\cellcolor{yellow!25}11\%	&	\cellcolor{yellow!25}4\%	&	
		\cellcolor{yellow!25}1\%	&	2\%	\\
		&{3} &	\cellcolor{yellow!25}2\%	&	\cellcolor{yellow!25}3\%	&	
		\cellcolor{cyan!25}\textbf{76\%}	&	\cellcolor{cyan!25}\textbf{82\%}	&	
		\cellcolor{cyan!25}5\%	&	\cellcolor{cyan!25}18\%	&	
		\cellcolor{yellow!25}3\%	&	\cellcolor{yellow!25}4\%	&	
		\cellcolor{yellow!25}2\%	&	\cellcolor{yellow!25}4\%	&		
		\cellcolor{yellow!25}5\%	&	\cellcolor{cyan!25}12\%	\\
		&{4} &	\cellcolor{yellow!25}2\%	&	\cellcolor{cyan!25}9\%	&	
		\cellcolor{cyan!25}1\%	&	\cellcolor{cyan!25}2\%	&	
		\cellcolor{cyan!25}3\%	&	\cellcolor{cyan!25}1\%	&	
		\cellcolor{yellow!25}2\%	&	\cellcolor{yellow!25}2\%	&	
		\cellcolor{yellow!25}2\%	&	\cellcolor{yellow!25}1\%	&		
		\cellcolor{yellow!25}6\%	&	\cellcolor{cyan!25}1\%	\\
		&{5} &	\cellcolor{cyan!25}1\%	&	\cellcolor{cyan!25}1\%	&	
		\cellcolor{yellow!25}1\%	&	\cellcolor{cyan!25}1\%	&	
		\cellcolor{cyan!25}2\%	&	\cellcolor{cyan!25}6\%	&	
		\cellcolor{yellow!25}2\%	&	\cellcolor{yellow!25}1\%	&	
		\cellcolor{yellow!25}3\%	&	\cellcolor{yellow!25}2\%	&		
		\cellcolor{cyan!25}9\%	&	\cellcolor{cyan!25}7\%	\\
		&{6} &	--							&  	\cellcolor{cyan!25}1\%	&	
		\cellcolor{cyan!25}2\%	&	\cellcolor{cyan!25}2\%	&	--							&	
		\cellcolor{cyan!25}1\%	&	\cellcolor{yellow!25}1\%	&	
		\cellcolor{yellow!25}4\%	&	\cellcolor{yellow!25}7\%	&	
		\cellcolor{yellow!25}4\%	&		\cellcolor{yellow!25}2\%	&	
		\cellcolor{cyan!25}1\%	\\					
		&{7} &	--							&  	\cellcolor{cyan!25}1\%	&	
		\cellcolor{cyan!25}2\%	&	\cellcolor{cyan!25}2\%	&	--							&	
		\cellcolor{cyan!25}3\%	&	\cellcolor{yellow!25}1\%	&	
		\cellcolor{yellow!25}2\%	&  \cellcolor{yellow!25}15\%	&	
		\cellcolor{cyan!25}19\%	&		\cellcolor{cyan!25}1\%	&	--		\\					
		&{8} &	--	&  	--	&\cellcolor{cyan!25}2\%	&	\cellcolor{cyan!25}1\%	&	--	& --	&	\cellcolor{yellow!25}6\%	&\cellcolor{yellow!25}4\%&	\cellcolor{yellow!25}7\%	&\cellcolor{cyan!25}2\%	&\cellcolor{cyan!25}4\%	&	--	\\
		&{9} &	--	&  	--	&\cellcolor{cyan!25}1\%	&	--	&	--	& --	&	\cellcolor{cyan!25}1\%	&	\cellcolor{yellow!25}1\% &\cellcolor{yellow!25}1\%	&	\cellcolor{yellow!25}2\%	& \cellcolor{cyan!25}4\%	&	--	\\	
		&{10} &	--	&  	--	&\cellcolor{cyan!25}1\%	&	--	&	--	& --	&	\cellcolor{yellow!25}2\%	&	\cellcolor{cyan!25}6\%& --	&	\cellcolor{yellow!25}2\%	&\cellcolor{yellow!25}1\%	&	--	\\	
		&{11} &	--	&  	--	&	--	& --	&	--	&	--	&\cellcolor{cyan!25}10\%	&	\cellcolor{yellow!25}6\%&	--	&\cellcolor{cyan!25}2\%	&\cellcolor{cyan!25}\textbf{43\%}	&	--	\\
		&{12} &	--	&  	--	&	--	& --	&	--	&	--	&\cellcolor{yellow!25}4\%	&	\cellcolor{yellow!25}4\%&	--	&\cellcolor{yellow!25}1\%	&\cellcolor{yellow!25}2\%	&	--	\\
		&{13} &	--	&  	--	&	--	& --	&	--	&	--	&\cellcolor{cyan!25}1\%	&	\cellcolor{yellow!25}2\%&	--	&\cellcolor{yellow!25}2\%	&\cellcolor{yellow!25}1\%	&	--	\\
		&{14} &	--	&  	--	&	--	& --	&	--	&	--	&	\cellcolor{cyan!25}2\%	&	\cellcolor{yellow!25}1\%	&	--&\cellcolor{cyan!25}2\%	&\cellcolor{cyan!25}1\%	&	--	\\
		&{15} &	--	&  	--	&	--	& --	&	--	&	--	&\cellcolor{yellow!25}1\%	&	\cellcolor{cyan!25}1\%	&	--&\cellcolor{yellow!25}1\%	&\cellcolor{yellow!25}8\%	&	--	\\
		&{16} &	--	&  	--	&	--	& --	&	--	&	--	&	\cellcolor{cyan!25}2\%	&	\cellcolor{cyan!25}3\%	&	--&\cellcolor{yellow!25}1\%	&	--	&	--	\\
		&{17} &	--	&  	--	&	--	& --	&	--	&	--	&\cellcolor{cyan!25}1\%	&	\cellcolor{yellow!25}3\%	&	--&\cellcolor{cyan!25}1\%	&	--&	--	\\
		&{18} &	--	&  	--	&	--	& --	&	--	&	--	&\cellcolor{cyan!25}1\%	&	\cellcolor{yellow!25}1\%	&	--&\cellcolor{cyan!25}\textbf{20\%}	&	--&	--	\\
		&{19} &	--	&  	--	&	--	& --	&	--	&	--	&\cellcolor{cyan!25}1\%	&	\cellcolor{cyan!25}1\%	&	--&\cellcolor{yellow!25}1\%	&	--&	--	\\
		&{20} &	--	&  	--	&	--	& --	&	--	&	--	&	--	&\cellcolor{cyan!25}1\%	&	--&\cellcolor{cyan!25}9\%	&	--&	--	\\
		&{21} &	--	&  	--	&	--	& --	&	--	&	--	&	--	&\cellcolor{cyan!25}1\%	&	--&\cellcolor{yellow!25}1\%	&	--&	--	\\
		&{22} &	--	&  	--	&	--	& --	&	--	&	--	&	--	& --	&	--&\cellcolor{cyan!25}1\%	&	--&	--	\\
		&{23} &	--	&  	--	&	--	& --	&	--	&	--	&	--	& --	&	--&\cellcolor{cyan!25}1\%	&	--&	--	\\
		&{24} &	--	&  	--	&	--	& --	&	--	&	--	&	--	& --	&	--&\cellcolor{cyan!25}1\%	&	--&	--	\\
		&{25} &	--	&  	--	&	--	& --	&	--	&	--	&	--	& --	&	--&\cellcolor{cyan!25}1\%	&	--&	--	\\						
		\bottomrule
	\end{tabular}
\end{table}

\FloatBarrier

\chapter{Experimental study on the virtual CAT (pilot)}\label{chap:study_virtual_pilot}
\begin{custombox}
	{The content of this chapter has been adapted from the following articles with permission of all co-authors and publishers:}
	\begin{itemize}[noitemsep,nolistsep,leftmargin=.25in]
		\item \textbf{Adorni, G.} and Piatti, A. (unpublished). Designing the virtual CAT: A digital tool for algorithmic thinking assessment in compulsory education  \cite{adorni_pilot2023}.
		
		\item \textbf{Adorni, G.}, Piatti, A., Bumbacher, E., Negrini, L., Mondada, F., Assaf, D., Mangili, F., and Gambardella, L. M. (2025). FADE-CTP: A Framework for the Analysis and Design of Educational Computational Thinking Problems. \textit{Technology,  Knowledge and Learning} \cite{adorni2023ctpframework}.
	\end{itemize}
	As an author of these publications, my contribution involved:\\
	\textit{Conceptualisation, Methodology, Validation, Formal analysis, Investigation, Resources, Data curation, Writing -- original draft \& review \& editing, Visualisation, Supervision.}
\end{custombox}

\section{Summary}
This chapter focuses on the pilot experimental study using the virtual CAT to evaluate the usability and suitability of this instrument for large-scale automated assessment of algorithmic skills in K-12 pupils, contributing to RQ4.
By performing a similar analysis to that conducted with the unplugged CAT, we aim to obtain comparable results, demonstrating that the tool is usable and ready for broader data collection. 
The chapter includes a detailed description of the study context, participant selection process, data collection methods, and the approach used for data analysis. 
Finally, the preliminary results are presented and discussed, providing insights into the tool's effectiveness in assessing algorithmic skills and its potential for large-scale implementation.
The code used for the data analysis process is available on GitHub \cite{Adorni_Virtual_CAT_Algorithmic_2024}.


\section{Methodology}\label{sec:virtual_pilot_method}

\subsection{Study context}\label{sec:virtual_pilot_context}

The experimental study was conducted in Switzerland in March 2023. 
As this was a pilot study, during this testing phase, we remained within the canton of Ticino, which is part of the Italian-speaking region of Switzerland.

\subsection{Participant selection}\label{sec:virtual_pilot_participants}
This study aimed to provide a preliminary selection of classes from different stages of compulsory education, focusing on the opposite ends of the system in Switzerland. 
The goal was to demonstrate the platform's effectiveness for diverse school types, with the intention of extending its applicability to the entire range of compulsory education.
Thus, we included three classes from two public schools, in particular one preschool class (ages 3-6) and two secondary school classes (ages 10-13). It is important to note that the selection of schools and classes was not random. Rather, participating classes were selected through our network of contacts, and the schools were contacted and agreed to take part in the study.

As for the unplugged CAT, this study was conducted in multiple sessions. All pupils present in each class who had received parental consent were given the option to voluntarily participate in the study. For ethical considerations regarding participant selection and informed consent procedures, refer to \Cref{sec:ethics}. 
A total of 31 pupils (21 girls and 10 boys) participated in the pilot study. 
\Cref{tab:students_analysis} provides a detailed breakdown of participant information across sessions, including school ID, HG, age and gender distribution.
\begin{table}[htb]
	\footnotesize
	\mycaption{Study participants (virtual CAT -- pilot study).}{Demographics analysis of participants by session, including school ID and type, HG, age category (age category (mean and standard deviation), and gender distribution.}\label{tab:students_analysis}
	\setlength{\tabcolsep}{1.9mm}
	\centering
	\begin{tabular}{clcc@{\hspace{.5\tabcolsep}}c@{\hspace{.8\tabcolsep}}cccc}
		\toprule
		\textbf{Session} & \textbf{School ID \& type}   &   \textbf{HG} & \multicolumn{3}{c}{\textbf{Age category}}   &   \textbf{Female} &   \textbf{Male} &   \textbf{Total} \\
		\midrule
		1   & X \quad Preschool   & 0, 1, 2  & 3-6 & yrs& ($\mu$ \; 5.0 $\pm$ 1.0 yrs)  &  3 &   4 & 7 \\
		2  & Y \quad Secondary school  & 8  & 10-13 & yrs& ($\mu$ 11.2 $\pm$ 0.6 yrs)  & 10 &   0 &   10 \\
		3  & Y \quad Secondary school  & 8  & 10-13 & yrs& ($\mu$ 11.4 $\pm$ 0.5 yrs)  &  8 &   6 &   14 \\
	
		\cmidrule(lr){1-1}\cmidrule(lr){2-2}\cmidrule(lr){3-3}\cmidrule(lr){4-6}\cmidrule(lr){7-7}\cmidrule(lr){8-8}\cmidrule(lr){9-9}
		& & & & &($\mu$ \; 9.9 $\pm$ 2.8 yrs)  & 21 &  10 &   31 \\
		\bottomrule
	\end{tabular}
\end{table}


\subsection{Data collection approach and procedures} \label{sec:virtual_pilot_datacollection_procedures}

The study took place during regular class sessions and was administered in Italian. 
It was structured into two main sequential modules: a training phase followed by a validation phase, both conducted on the same day, one after the other.

Recognised the importance of ensuring pupils' familiarity with the assessment tool, we designed and integrated a training module within the app that serves as a preparatory step for scholars, allowing them to become acquainted with the tool's interface and functionalities (see \Cref{fig:mode}).
During this phase, the administrator led a session guiding students through the app's features using 15 sample cross-array schemas for practice. This enables students to effectively navigate the tool, with teachers supporting the administrator and students as needed.
Training sessions were held in groups based on device availability and typically lasted 30-45 minutes.
The procedure was designed to accommodate the participants’ age and school environment.
For younger students, task instructions were kept simple and engaging, and they were guided to use only the CAT-GI, with simplified instructions and pacing to ensure comfort and understanding \citep{hanna1997}, while older students received more detailed guidance appropriate for their developmental level.
No data collection occurred in this phase. 

After completing the training, students move on to the validation session, during which data collection takes place. This module mirrors the original unplugged activity, where students solved 12 cross-array schemas.
No time limit was imposed to ensure an optimal testing environment, allowing students to work at their own pace. For preschool pupils, to minimise distractions, two pupils at a time were randomly selected and taken to a separate room to complete the task.

To begin the validation process, session and student details are manually input into the app (see \Cref{fig:session,fig:student}). 
Each session was assigned a unique identifier, with specific contextual information such as the date, canton, school ID and type, the HGs level, and the administrator’s details.
Student information was limited to the ID, gender, and date of birth, which was used to calculate their ages.
All information regarding tasks and performance is also collected for the virtual CAT. 
Additionally, this instrument allows to log timestamped actions performed by the students during the tasks, such as adding, confirming, removing, or reordering commands, updating command properties like colours or directions, resetting the algorithm, changing the mode of interaction or visibility, confirming task completion, or surrendering. 
All these data are compiled into a dataset, which has been made available through Zenodo for public access \citep{adorni_virtualCATdatasetpilot_2023} after removing any potentially identifiable information (e.g., school and class) in alignment with open science practices in Switzerland \citep{snfOpenScience}.


\subsection{Data analysis approach}\label{sec:virtual_pilot_dataanalysis}
The data analysis aimed to provide insights into various aspects of the virtual CAT
assessment, focusing on participation, performance, algorithmic strategies, and the development of competencies across age groups.
Python was used for data processing, exploratory analyses, and visualisations of results \citep{pyhton}.

\subsubsection{Participation and performance}
We begin by analysing the time-related performance metrics to gain insights into how long students took to complete the tasks, comparing these across age categories and interaction dimensions.
Next, we examine participation and success rates for each schema, exploring their distributions across age categories using descriptive statistics.
%
%

\subsubsection{Competencies development}
  
In the last phase of analysis, we conduct a preliminary qualitative analysis on the development of algorithmic and interaction strategies across age categories, aimed at identifying patterns in the approaches taken by students at various developmental stages. 

\section{Results}\label{sec:virtual_pilot_results}


\subsection{Participation and performance}\label{sec:partecipation_performance_pilot}

Unlike the unplugged CAT, which required 36 hours for data collection with 109 participants, the virtual CAT enables simultaneous assessment of an entire class, provided each student has access to a device, allowing for assessments across multiple groups. 
The total time for the validation session with 31 students was approximately 4 hours (1 to 1.5 hours per session), a significant reduction compared to the unplugged version.

The time required to solve all 12 schemas shows notable differences across age categories (see \Cref{tab:time_by_age_pilot}).
Younger students (ages 3-6) had slightly longer median times compared to older students (ages 10-13), but the interquartile ranges (Q1-Q3) indicate similar variability in completion times across both groups.
This suggests that while age has some influence on task duration, the consistency of performance is comparable across age categories.
The time students spent using each interaction dimension reveals notable differences (see \Cref{tab:time_by_artefact_pilot}). 
The use of the CAT-VPI (P) consistently required the most time, with a median duration of nearly 13 minutes. 
In contrast, the CAT-GI (G) had the shortest median duration, suggesting either a lower reliance on this dimension or quicker navigation. This aligns with expectations, as G is the less complex interaction dimension, making it more intuitive and easy for students to use.
\begin{table*}[ht]
	\footnotesize\centering
	\mycaption{Activity completion time across age categories (virtual CAT - pilot).}{The time spent by students to complete all 12 schemas, including mean, minimum and maximum values, median, and interquartile ranges (Q1-Q3), grouped by age category. 
	The aggregated statistics at the bottom summarise the overall completion times for the entire dataset.
	}\label{tab:time_by_age_pilot}
	\setlength{\tabcolsep}{3mm}
	\begin{tabular}{cc@{\hspace{.2\tabcolsep}}lcccccc}
		\toprule
		& &  & \textbf{Mean} & \textbf{Min} & \textbf{Q1} & \textbf{Median} & \textbf{Q3} & \textbf{Max} \\
		\midrule
		{\multirow{3}{*}{ \rotatebox{90}{\parbox{1.4cm}{\centering\textbf{Age \\ category}}}}}
		& 3-6    &yrs  & 15m 11s     &09m 08s  &10m 47s &12m 37s     &17m 54s &26m 43s        \\
		& & & & & & & & \\
		& 10-13 & yrs &  12m 48s   &00m 10s  &08m 12s &13m 00s     &17m 16s &27m 48s        \\
%
		\bottomrule
	\end{tabular}
\end{table*}

\begin{table*}[htb]
	\footnotesize\centering
	\mycaption{Time spent using each interaction dimension (virtual CAT - pilot).} 
		{The time students spent using a certain interaction dimension, including mean, minimum and maximum values, median, and interquartile ranges (Q1-Q3).
	}\label{tab:time_by_artefact_pilot}
	\setlength{\tabcolsep}{3.2mm}
	\begin{tabular}{cclcccccc}
		\toprule
		& &  & \textbf{Mean} & \textbf{Min} & \textbf{Q1} & \textbf{Median} & \textbf{Q3} & \textbf{Max} \\
		\midrule
		{\multirow{4}{*}{ \rotatebox{90}{\parbox{1.7cm}{\centering\textbf{Interaction \\ dimension}}}}}
		& \multicolumn{2}{l}{\quad \, GF} & 06m 49s         & 00m 37s        & 03m 42s       & 05m 10s           & 08m 32s       & 26m 14s \\
		& \multicolumn{2}{l}{\quad \, G}  & 02m 24s         & 00m 10s        & 00m 52s       & 01m 13s           & 03m 21s       & 07m 45s        \\
		& \multicolumn{2}{l}{\quad \, PF} & 07m 09s         & 02m 10s        & 03m 25s       & 05m 27s           & 10m 05s       & 15m 58s         \\ 
		& \multicolumn{2}{l}{\quad \, P} & 13m 20s         & 00m 10s        & 09m 13s       & 12m 47s           & 17m 35s       & 27m 48s         \\
		\bottomrule
	\end{tabular}
\end{table*}

The two tables reveal notable outliers in the time spent solving the 12 schemas. 
In \Cref{tab:time_by_age_pilot}, the maximum values significantly deviate from the upper quartiles.
This suggests that while most students complete the tasks within a reasonable time frame, some struggle. This could be attributed to the complexity or unfamiliarity of the interaction dimensions rather than the age of the participants, as similar outliers are observed across both age groups. 
In fact, from \Cref{tab:time_by_artefact_pilot}, outliers are evident in the GF and P interaction dimensions, which represent the less and more complex modes of interaction, respectively. 
For instance, GF exceeds its upper quartile by approximately 18 minutes, which is more than 3 times, while P exceeds its third quartile by around 10 minutes, nearly double.
These extreme values indicate that some students required disproportionately long durations, likely reflecting difficulties in navigating these dimensions or overreliance on them.
While longer times with the P dimension are expected due to its complexity, the outlier in GF is more surprising. We can attribute this to the fact that less proficient students may gravitate toward this interaction mode, possibly because it appears simpler or more intuitive.
In both tables, minimum values present notable outliers, with some students completing tasks in as little as 10 seconds. This suggests that some students might have rushed through the tasks or skipped steps, potentially indicating a lack of engagement.
These outliers may also reflect brief interactions caused by technical issues, such as server disconnections and data loss, which not only impacted the time but also affected task attempts and success rates.
The presence of outliers underscores the need to consider individual differences in task completion times and interaction strategies when interpreting overall performance trends.

\Cref{tab:performance_by_schema_and_age_pilot} provides an overview of participation and success rates across schemas for the two age categories. As expected from the previous analysis of completion times, not all pupils completed every schema. The reasons for this are multiple. 
Some experiments were cut short due to time constraints, as seen in the lower participation rates for the later schemas when students had to interrupt the activity. 
Additionally, there were some technical issues with the server, requiring some students to restart the full task, and some data was lost.
\begin{table}[hbt]
	\footnotesize
	\centering
	\mycaption{Participation and success rates across schemas and age categories (virtual CAT - pilot).}{
		Success rate is calculated from the number of students who attempted the schema.
		Values highlighted in bold indicate schemas with success rate values exceeding 80\%.}
	\label{tab:performance_by_schema_and_age_pilot}
	\setlength{\tabcolsep}{2.8mm}
	\begin{tabular}{ccrrrrrr}
		\toprule
		& &   \multicolumn{3}{c}{\textbf{Participation} \linebreak (\%)} & \multicolumn{3}{c}{\textbf{Success} (\%)} \\
		& &  \multicolumn{1}{>{\footnotesize\centering\arraybackslash}m{1.4cm}}{3-6 yrs \linebreak (out of 7)} &  \multicolumn{1}{>{\footnotesize\centering\arraybackslash}m{1.5cm}}{10-13 yrs \linebreak (out of 24)} & \multicolumn{1}{>{\footnotesize\centering\arraybackslash}m{1.5cm}}{Total \linebreak (out of 31)} &  \multicolumn{1}{>{\footnotesize\centering\arraybackslash}m{1.4cm}}{3-6 yrs} &  \multicolumn{1}{>{\footnotesize\centering\arraybackslash}m{1.4cm}}{10-13 yrs} & \multicolumn{1}{>{\footnotesize\centering\arraybackslash}m{1.5cm}}{Total \linebreak (out of 31)}  \\
		\midrule
		{\multirow{12}{*}{ \rotatebox{90}{\textbf{Schema}}}}
		& 1  & 6 (86 \%)& 24 (100 \%)  & 30 (97 \%)  &  3 \, \, (50\%) &  22 (\textbf{92\%})  & 25 (\textbf{83\%})  \\
		& 2  & 5 (71 \%)& 24 (100 \%) & 29 (94 \%)  &  3 \, \, (60\%) &  21 (\textbf{88\%})  & 24 (\textbf{83\%})  \\
		& 3  & 6 (86 \%)& 23 \, \, (96 \%) & 29 (94 \%)  &  4 \, \, (67\%) &  19 (\textbf{83\%})  & 23 (79\%)  \\
		& 4  & 6 (86 \%)& 20 \, \, (83 \%) & 26 (84 \%)  &  4 \, \, (67\%) &  17 (\textbf{85\%})  & 21 (\textbf{81\%})  \\
		& 5  & 6 (86 \%)& 20 \, \, (83 \%) & 26 (84 \%)  &  6 \,  (\textbf{100\%}) &  16 (\textbf{80\%})  & 22 (\textbf{85\%})  \\
		& 6  & 6 (86 \%)& 22 \, \, (92 \%) & 28 (90 \%)  &  2 \, \, (33\%) &  21 (\textbf{95\%})  & 23 (\textbf{82\%})  \\
		& 7  & 5 (71 \%)& 20 \, \, (83 \%) & 25 (81 \%)  &  1 \, \, (20\%) &  12 (60\%)  & 13 (52\%)  \\
		& 8  & 5 (71 \%)& 21 \, \, (88 \%) & 26 (84 \%)  &  2 \, \, (40\%) &  18 (\textbf{86\%})  & 20 (77\%)  \\
		& 9  & 4 (57 \%)& 21 \, \, (88 \%) & 25 (81 \%)  &  3 \, \, (75\%) &  20 (\textbf{95\%})  & 23 (\textbf{92\%})  \\
		& 10 & 5 (71 \%)& 19 \, \, (79 \%) & 24 (77 \%)  & 3 \, \, (60\%) &  18 (\textbf{95\%})  & 21 (\textbf{88\%})  \\
		& 11 & 4 (57 \%)& 18 \, \, (75 \%) & 22 (71 \%)  & 2 \, \, (50\%) &  14 (78\%)  & 16 (73\%)  \\
		& 12 & 3 (53 \%)& 17 \, \, (71 \%) & 20 (65 \%)  & 1 \, \, (33\%) &  14 (\textbf{82\%})  & 15 (75\%)  \\
		\bottomrule
	\end{tabular}
\end{table}
Regarding success rates, older students generally performed better, while younger pupils showed more variation in their success rates. This could be attributed to differences in cognitive development, familiarity with digital tools, and task complexity, which likely posed more challenges for younger students.
Overall, the success rate across both age groups is around 79\%, indicating that pupils were able to engage effectively with the tasks. The schema with the lowest success rate appears to be S7, likely because it requires more advanced skills. Even the more challenging tasks, designed to be intentionally difficult, such as schemas S9 and S10, were successfully completed by many pupils (with success rates definitely higher than 80\%). This demonstrates the platform's potential to support learners across varying skill levels through its flexibility and diverse interaction modes.

%
%

\FloatBarrier
\subsection{Competencies development}

\begin{figure}[h]
	\centering
	\includegraphics[width=.8\textwidth]{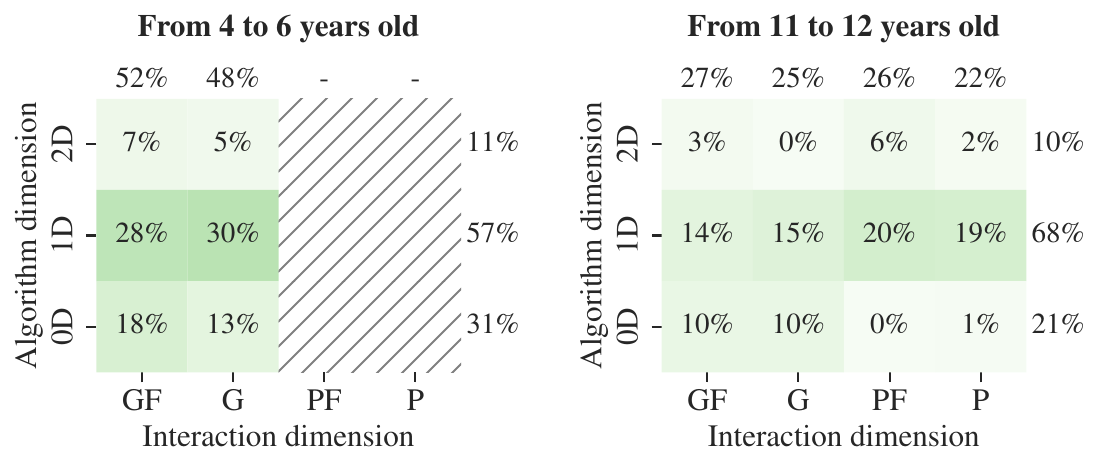}
	\mycaption{Algorithmic and interaction strategies across age (virtual CAT - pilot).}{
		Percentages represent the proportion of each interaction-algorithmic combination within each age group, with aggregated values shown across rows and columns.
	Since the younger age group could not use the CAT-VPI, only data for GF and G interactions are included.
	}
	\label{fig:overall_performance_pilot}
\end{figure}

To assess whether these results hold true across both algorithmic and interaction dimensions, we delve deeper into the strategies employed by pupils across different age categories, and subsequently, we analyse these strategies with respect to gender. \Cref{fig:overall_performance_pilot} illustrates the strategies employed by pupils to solve the CAT, while \Cref{appendix:appB_pilot} offers more detailed illustrations of the strategies schema by schema.
Pupils across different age groups show a balanced use of interaction dimensions, highlighting the application's capacity to cater to various interface preferences and levels of autonomy.
Both groups are proficient in generating algorithms across all three algorithmic dimensions, with 1D algorithms being the most frequently used, consistent with the results of the unplugged CAT (see \Cref{sec:unplugged_results_competencies_development}). 
While younger pupils consistently use all available interfaces without significant variation across algorithmic complexity levels, older pupils design algorithms of varying complexity based on the interaction type selected.
Specifically, simpler algorithms are predominantly constructed using the CAT-GI, which offers an intuitive and cognitively light interaction style. In contrast, the CAT-VPI is favoured for more advanced tasks, as it supports better the creation of complex algorithms. This adaptability among older pupils highlights their strategic use of the platform's features to address tasks of varying difficulty, emphasising the tool’s capability to support diverse skill levels and interaction strategies.

\chapter{Experimental study on the virtual CAT (main)}\label{chap:study_virtual_main}
\begin{custombox}
	{The content of this chapter has been adapted from the following article with permission of all co-authors and publishers:}
	\begin{itemize}[noitemsep,nolistsep,leftmargin=.25in]
		\item \textbf{Adorni, G.}, Artico, I., Piatti, A., Lutz, E., Gambardella, L. M., Negrini, L., Mon-
		dada, F., and Assaf, D. (2024). Development of algorithmic thinking skills in K-12
		education: A comparative study of unplugged and digital assessment instruments.
		\textit{Computers in Human Behavior Reports} \cite{adorni_2024_largescale}. 
	\end{itemize}
	As an author of this publication, my contribution involved:\\
	\textit{Conceptualisation, Methodology, Validation, Formal analysis, Investigation, Resources, Data curation, Writing -- original draft \& review \& editing, Visualisation, Supervision.}
\end{custombox}

\section{Summary}

This chapter focuses on a main experimental study using the virtual CAT to assess the
algorithmic skills of K-12 pupils in Switzerland, contributing to RQ4.
We extended the preliminary analysis conducted with the unplugged CAT and the pilot study with the virtual CAT to provide more generalisable results from a larger sample.
The chapter includes a detailed description of the study context, participant selection process, data collection methods, and the approach used for data analysis. 
Finally, the results are presented and discussed, providing insights into baseline competencies in AT in compulsory education, how these competencies develop across school grades, and how this development is influenced by instrumental factors such as the artefactual environment, as well as contextual factors like gender, educational environment (e.g., school level and grade), regional factors (e.g., canton of the school), and their interactions.
The code used for the data analysis process is available on GitHub \cite{Adorni_Virtual_CAT_Algorithmic_2024}.

\section{Methodology}\label{sec:virtual_main_method}

\subsection{Study context}\label{sec:virtual_main_context}
The experimental study was conducted between May and June 2023, specifically within the Solothurn and Ticino cantons, which are part of the German- and Italian-speaking regions of Switzerland, respectively.

\subsection{Participant selection}\label{sec:virtual_main_participants}
To ensure a representative sample across all levels of compulsory education, with a balanced selection of age, gender and demographic origin, this study included nine classes across five public schools across Ticino and Solothurn cantons, in particular two preschool classes (ages 3-6), two primary school classes (ages 6-11), and five lower secondary school classes (ages 12-16). 
Participating classes were selected through our network of contacts, and the schools were contacted and agreed to take part in the study. All ethical considerations regarding participant selection and informed consent procedures are detailed in \Cref{sec:ethics}.

\begin{table*}[hbt]
	\footnotesize\centering
	\mycaption{Study participants (virtual CAT - main).}{ 
		Demographics analysis of participants by session, including canton (TI for Ticino and, SO for Solothurn), school ID and type, HG, age category (mean and standard deviation), and gender distribution.
		} 
	\label{tab:students_analysis_by_school_virtual}
	\setlength{\tabcolsep}{.6mm}
	\begin{tabular}{cclcc@{\hspace{.5\tabcolsep}}c@{\hspace{.8\tabcolsep}}cccc}
		
		\toprule
		\textbf{Session} & \textbf{Canton} & \textbf{School ID \& type} & \textbf{HG} & \multicolumn{3}{c}{\textbf{Age category}} & \textbf{Female} & \textbf{Male} & \textbf{Total} \\
		\midrule
		1V & TI & A \, \,  Preschool & 0, 1, 2 & 3-6 & yrs& ($\mu$ \; 5.0 $\pm$ 0.8 yrs) & 6 & 7 & 13 \\
		2V & SO & D \, \,  Preschool & 2 & 3-6 & yrs& ($\mu$ \; 5.9 $\pm$ 0.3 yrs) & 8 & 6 & 14 \\
		3V & TI & E \, \,  Primary school & 4 & 7-9 & yrs& ($\mu$ \; 7.7 $\pm$ 0.6 yrs) & 7 & 8 & 15 \\
		4V & SO & D \, \,  Primary school & 6 & 7-9 & yrs& ($\mu$ \; 9.9 $\pm$ 0.3 yrs) & 8 & 10 & 18 \\
		5V & TI & F \, \,  Lower secondary school & 8 & 10-13 & yrs& ($\mu$ 11.6 $\pm$ 0.5 yrs) &11 & 9 & 20 \\
		6V & TI & F \, \,  Lower secondary school & 10 & 14-16 & yrs& ($\mu$ 13.9 $\pm$ 0.8 yrs) & 8 & 5 & 13 \\
		7V & TI & G \, \,  Lower secondary school & 10 & 14-16 & yrs& ($\mu$ 13.6 $\pm$ 0.6 yrs) & 7 & 7 & 14 \\
		8V & TI & G \, \,  Lower secondary school & 11 & 14-16 & yrs& ($\mu$ 14.7 $\pm$ 0.5 yrs) & 6 & 5 & 11 \\
		9V & SO & D \, \,  Lower secondary school & 11 & {14-16} & yrs& ($\mu$ {15.5} $\pm$ 0.5 yrs) & 4 & 7 & 11 \\
		
		\cmidrule(lr){1-1}\cmidrule(lr){2-2}\cmidrule(lr){3-3}\cmidrule(lr){4-4}\cmidrule(lr){5-7}\cmidrule(lr){8-8}\cmidrule(lr){9-9}\cmidrule(lr){10-10}
		& & & & &&($\mu$ 10.7 $\pm$ 3.6 yrs)&65 & 64 & 129 \\
		\bottomrule
	\end{tabular}
\end{table*}
A total of 129 pupils (65 girls and 64 boys) participated in the study. 
\Cref{tab:students_analysis_by_school_virtual}  provides a detailed breakdown of participant information across sessions, including canton, school ID, HG, age and gender distribution.
While a balanced distribution is evident across factors like school type, HG, age category, and gender, there is a slight imbalance in geographic representation, with fewer students from Solothurn compared to Ticino, but overall, these demographic analyses offer valuable insights into the diverse characteristics of participants in both studies.

\subsection{Data collection approach and procedures}
\label{sec:virtual_main_datacollection_procedures}

The data collection procedures for this study closely mirrored those used in the virtual CAT pilot study (see \Cref{sec:virtual_pilot_datacollection_procedures}).
One notable difference in the main study was the variability in administering the activity across linguistic regions: one person managed sessions in the Italian-speaking region, while another conducted those in the German-speaking region. To ensure consistency, the Italian-speaking administrator supervised the later sessions in the German-speaking region.
Differences in how the tutorials were administered could have impacted the delivery of instructional materials and the CAT’s administration, potentially affecting the study’s outcomes.

The validation session followed the same data collection procedure as the pilot study, with identical data being recorded.
After removing any potentially identifiable information (e.g., school and class), the final dataset has been made publicly available through Zenodo \citep{adorni_virtualCATdatasetmain_2024}, in alignment with open science practices in Switzerland \citep{snfOpenScience}.

\subsection{Data analysis approach}\label{sec:virtual_main_dataanalysis}

The data analysis aimed to provide insights into various aspects of the virtual CAT assessment, focusing on participation, performance, algorithmic strategies, and the development of competencies across age groups and genders. 
Python was used for data processing, exploratory and visualisations of results  \citep{pyhton}, while R was used for statistical analyses \citep{R}.

\subsubsection{Participation and performance}
We begin by analysing the time-related performance metrics, first comparing task completion times across age categories, followed by an examination of these metrics across interaction dimensions.
As part of this analysis, we examine the frequency of interaction dimensions to uncover preferences and tendencies in artefact use and autonomy among different age groups, as we hypothesise that a relationship exists between age categories and interaction dimensions.
Next, we examine participation and success rates for each schema, exploring their distributions across age categories. 
%
%
Following, we investigated the distribution of CAT scores across different age groups and performed a chi-square test of independence to assess whether there are significant relationships between age and performance outcomes \citep{Cochran1954,moore_introduction_1989,Yates1934}. 
Pairwise comparisons between age categories were conducted using Tukey's HSD test with BH correction \citep{Yates1934,Tukey1949,Benjamini1995}.


\subsubsection{Competencies development}

In the second phase of analysis, we aimed to understand the influence of various factors on the algorithm dimension. This phase builds on observations and hypotheses made during the unplugged CAT experimental study and the pilot study with the virtual CAT, where potential correlations were identified but not statistically tested. 
To validate and extend these findings, we performed a series of statistical tests to assess the relationship between the complexity of the algorithms produced and factors such as age, interaction type and Trial and Error (T\&E) approaches, as well as exploring the joint influence between some of these factors on AT.

To explore the direct relationship of algorithmic dimension with age and interaction dimension, we conducted an Analysis of Variance (ANOVA) to determine whether these factors significantly predicted the complexity of the algorithms produced \citep{chambers1992,Hastie2001,James2013,Cox1979,Cochran1954}.
Post-hoc analyses were performed using Tukey's HSD test with BH adjustments \citep{Yates1934,Tukey1949,Benjamini1995} and pairwise t-tests with Bonferroni correction \citep{Bland1995,Perneger1998,Sedgwick2014} to explore these differences further.
Finally, to assess the distribution of the higher algorithmic dimensions (2D) across different interaction dimensions and age categories, we performed a chi-squared test of proportions \citep{Yates1934,Cochran1954}.
Additionally, to further explore the relationship between interaction and algorithm dimensions, t-tests were conducted  \citep{bartlett1937properties,moore_introduction_1989,Wilson1927}.
%

We also investigated students' T\&E behaviours, specifically focusing on the predictors of task restarts.
We used Ordinary Least Squares (OLS) regression to examine the relationship between restart frequency and factors such as schemas, age, gender, and interaction type \citep{Zdaniuk2014,Hastie2001,James2013,Stone1990}. 
Subsequently, we analysed the impact of T\&E behaviours on the algorithmic dimension to understand how this approach might influence the complexity of the algorithms produced and the achieved performance.

Finally, we explored the combined effect of factors such as interaction type, age, gender, and schemas on algorithmic dimension using ANOVA \citep{chambers1992,Hastie2001,James2013,Cox1979,Cochran1954} and linear regression models \citep{Hastie2001,Seber1984,Fisk1982,chambers1992,martin1979multivariate}. 
We further investigated the relationship between interaction dimension and age by conducting an Estimated Marginal Means (EMMs) analysis to examine their combined effect on algorithm complexity \citep{emmeans}. 
Additionally, we used chi-squared tests of proportions to explore differences in the distribution of higher algorithm dimensions (2D) across various age categories and interaction types \citep{Yates1934,Cochran1954}.

\section{Results}\label{sec:virtual_main_results}

%

\subsection{Participation and performance}\label{sec:participation_performance_main}


Examining task completion times across age categories (see \Cref{tab:time_by_age}), we found that older students do not necessarily complete tasks faster, contrary to expectations. While median completion times generally increase with age, the second group deviates from this trend, suggesting a complex relationship between age and completion time. Interaction methods may also play a crucial role in this relationship. For example, older students might take longer to resolve tasks because they tend to use more advanced artefacts with higher autonomy, which are inherently more complex and could contribute to longer completion times.

\begin{table*}[htb]
	\footnotesize\centering
	\mycaption{Activity completion time across age categories (virtual CAT - main).} 
		{The time spent by students to complete all 12 schemas, including mean, minimum and maximum values, median, and interquartile ranges (Q1-Q3), grouped by age category.
			The overall summary statistics at the bottom provide an overview of completion times for the entire dataset.
	}\label{tab:time_by_age}
	\setlength{\tabcolsep}{3mm}
	\begin{tabular}{cc@{\hspace{.2\tabcolsep}}lcccccc}
		\toprule
		& &  & \textbf{Mean} & \textbf{Min} & \textbf{Q1} & \textbf{Median} & \textbf{Q3} & \textbf{Max} \\
		\midrule
		{\multirow{4}{*}{ \rotatebox{90}{\parbox{1.65cm}{\centering\textbf{Age \\ category}}}}}
		& 3-6 &yrs & 20m 53s & 03m 41s & 15m 48s & 21m 50s & 27m 14s & 35m 51s\\
		& 7-9 &yrs & 13m 13s & 05m 27s & 09m 52s & 12m 24s & 13m 54s & 29m 39s\\
		& 10-13 & yrs & 26m 41s & 05m 19s & 17m 04s & 25m 04s & 41m 51s & 52m 26s\\
		& 14-16 & yrs & 29m 34s & 02m 51s & 18m 58s & 28m 08s & 40m 29s & 79m 36s\\
		\cmidrule(lr){1-3}\cmidrule(lr){4-4}\cmidrule(lr){5-5}\cmidrule(lr){6-6}\cmidrule(lr){7-7}\cmidrule(lr){8-8}\cmidrule(lr){9-9}
		
		&&& 23m 07s & 02m 51s  & 12m 14s & 20m 51s & 30m 03s & 79m 36s  \\
		\bottomrule
	\end{tabular}
\end{table*}

\begin{figure}[htb]
	\centering
	\includegraphics[width=.7\linewidth]{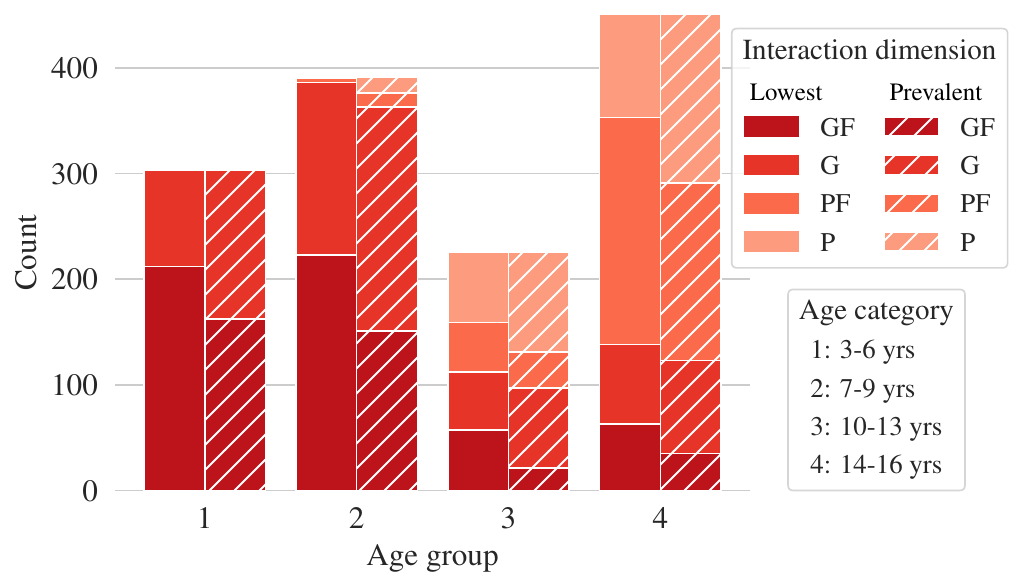}
	\mycaption{Age-wise distribution of interaction dimensions (virtual CAT - main).}{
		The y-axis displays counts of the lowest and most used interaction dimensions, represented by solid and striped bars, across four age categories on the x-axis.
		Since the younger age group could not use the CAT-VPI, only data for GF and G interactions are included.
	}
	\label{fig:artefact_by_age_group_main}
\end{figure}

Analysing students' interaction preferences among age categories, focusing on the least complex interaction dimension they achieved during the task and the most commonly used one, we aim to discern variations in usage patterns to contextualise our previous findings.
The trend in \Cref{fig:artefact_by_age_group_main} reveals an evolution in interaction preferences as pupils grow older.
Among the youngest group (ages 3-6), due to their limited exposure to technology, we restricted interactions to the CAT-GI interface.
While these restrictions were lifted for the second group of students (ages 7-9), the CAT-GI remained the more popular assignment, even if some pupils began exploring the CAT-VPI, signalling an emerging interest in more complex interactions.
Students in the third age group (ages 10-13) showed a balanced use of all four interaction dimensions, reflecting growing versatility and adaptability.
Finally, older students (ages 14-16) predominantly opted for the most complex interaction dimensions (PF and P), demonstrating a preference for advanced and sophisticated methods.
Our previously observed variations in task completion times across age groups can be explained by their interaction preferences. The second group (ages 7-9) showed the shortest completion times, likely due to their continued reliance on simpler methods, like CAT-GI. In contrast, older students (ages 14-16) predominantly employed more advanced interactions, like CAT-VPI, which demand higher autonomy and cognitive effort, likely contributing to the longer completion times and supporting our previous hypothesis.

The analysis of time spent on specific interaction dimensions shows clear patterns (see \Cref{tab:time_by_artefact_main}).
Users spend less time on the CAT-GI interface, aligning with pilot study findings (see \Cref{sec:partecipation_performance_pilot}), suggesting quicker navigation within this modality, while more time is spent on the CAT-VPI, with wide interquartile ranges (Q1-Q3) indicating variability in user engagement.
Regardless of the interface used, users take longer to complete tasks when relying on visual feedback, likely due to the time needed for continuous monitoring and adjustment.
\begin{table*}[htb]
	\footnotesize\centering
	\mycaption{Time spent using each interaction dimension (virtual CAT - main).} 
		{The time students spent using a certain interaction dimension, including mean, minimum and maximum values, median, and interquartile ranges (Q1-Q3).
	}\label{tab:time_by_artefact_main}
	\setlength{\tabcolsep}{3.2mm}
	\begin{tabular}{cclcccccc}
		\toprule
		& &  & \textbf{Mean} & \textbf{Min} & \textbf{Q1} & \textbf{Median} & \textbf{Q3} & \textbf{Max} \\
		\midrule
		{\multirow{4}{*}{ \rotatebox{90}{\parbox{1.65cm}{\centering\textbf{Interaction \\ dimension}}}}}
		& \multicolumn{2}{l}{\quad \, GF} & 11m 38s & 00m 29s  & 05m 21s & 10m 00s & 16m 59s & 35m 36s \\
		& \multicolumn{2}{l}{\quad \, G}  & 02m 49s & 00m 04s  & 00m 45s & 01m 49s & 04m 30s & 11m 52s \\
		& \multicolumn{2}{l}{\quad \, PF} & 21m 13s & 02m 31s  & 08m 32s & 18m 20s & 30m 05s & 79m 36s \\ 
		& \multicolumn{2}{l}{\quad \, P}  & 12m 15s & 00m 01s  & 00m 29s & 07m 18s & 19m 33s & 45m 20s \\
		\bottomrule
	\end{tabular}
\end{table*}

\FloatBarrier

\Cref{tab:performance_by_schema_main} provides a detailed overview of student participation and success on individual schemas.
Students were now able to skip tasks, an option unavailable in the unplugged CAT or the virtual CAT pilot. Despite this change, the participation rate remained notably high, reflecting strong determination and intrinsic interest, although a slight decline was observed in the later tasks.
%
%
Regarding success rates, in the unplugged CAT, all students successfully completed each schema they attempted, as they were guided to correct any errors. However, success rates for the virtual CAT varied across schemas. 
Schema 1 had the highest success rate, reflecting its role as an introductory task, while Schema 12 showed lower success rates, indicating increased complexity. The drop in success rates for Schemas 2 and 8, despite a similar number of attempts as Schema 1, may suggest heightened task complexity or a mismatch between students' skills and schema demands.
The non-linear decline in success rates as schema numbers increase suggests that students perceive varying levels of difficulty, which may not always align with educators' intended progression.
\begin{table*}[hb]
\footnotesize\centering
\mycaption{Participation and success rates across schemas (virtual CAT - main).}{
Success rate is calculated from the number of students who attempted the schema. Values highlighted in bold indicate schemas with success rate values exceeding 80\%.}
\label{tab:performance_by_schema_main}
\begin{tabular}{cccccc}
	\toprule
	\rowcolor{white} 
	&
	& \multicolumn{1}{>{\footnotesize\centering\arraybackslash}m{3.5cm}}{{No. pupils participating \linebreak (out of 129)}} 
	& \multicolumn{1}{>{\footnotesize\centering\arraybackslash}m{2cm}}{{\textbf{Participation} \linebreak (\%)}} 
	& \multicolumn{1}{>{\footnotesize\centering\arraybackslash}m{3.5cm}}{{No. pupils succeeding \linebreak \mbox{~}}} 
	& \multicolumn{1}{>{\footnotesize\centering\arraybackslash}m{2cm}}{{\textbf{Success} \linebreak (\%)}} \\
	\midrule
	{\multirow{12}{*}{ \rotatebox{90}{\textbf{Schema}}}}
	& 1 & 126&  \, 98\%  & 119& \textbf{94\%} \\
	& 2 & 127&  \, 98\%  &  \, 93 & 73\%  \\
	& 3 & 127&  \, 98\%  & 100& 79\% \\
	& 4 & 129& 100\% & 106 & \textbf{82\%} \\
	& 5 & 128&  \, 99\%  & 109& \textbf{85\%} \\
	& 6 & 127&  \, 98\%  & 112& \textbf{88\%} \\
	& 7 & 125&  \, 97\%  &  \, 98 & 78\%  \\
	& 8 & 126&  \, 98\%  &  \, 92 & 73\%  \\
	& 9 & 121&  \, 94\%  &  \, 98 & \textbf{81\%}  \\
	& 10& 118 &  \, 91\%  &  \, 91 & 77\%  \\
	& 11& 110&  \, 85\%  &  \, 82 & 75\%  \\
	& 12& 110&  \, 85\%  &  \, 78 & 71\%  \\
	\bottomrule
\end{tabular}

\end{table*}

\Cref{tab:performance_by_age_main} illustrates trends in student participation and success across age categories, specifically showing the percentage of students who completed the entire sequence of tasks or solved the entire sequence correctly.
This type of participation rate is particularly beneficial for assessing whether the addition of the option to skip tasks has significantly affected participation rates for specific age groups. Overall, participation rates remain relatively high, with the second age group (ages 7-9) showing the highest participation.
Regarding success rates, while we would expect an increase with age, this is not entirely the case, as the youngest pupils (ages 3-6) achieved the highest success rate. This can likely be attributed to the relationship between age categories and the choice of interaction method, which may have influenced the outcomes.
\begin{table*}[hbt]
	\footnotesize\centering
	\mycaption{Participation and success rates across age categories (virtual CAT - main).}{
			The number and percentage of students who attempted and successfully completed all 12 schemas, grouped by age category, along with the median and interquartile range (IQR). Success rate is calculated from the number of students participating.
	}\label{tab:performance_by_age_main}
	\setlength{\tabcolsep}{.4mm}
	\begin{tabular}{cc@{\hspace{.2\tabcolsep}}lcccccc}
		\toprule
		&  &
		& \multicolumn{1}{>{\footnotesize\centering\arraybackslash}m{2.3cm}}{{No. pupils \linebreak participating \linebreak in all schemas}} 
		& \multicolumn{1}{>{\footnotesize\centering\arraybackslash}m{2cm}}{{\textbf{Participation} \linebreak (\%)}}
		& \multicolumn{1}{>{\footnotesize\centering\arraybackslash}m{1.7cm}}{{Median schemas participated (Q1-Q3)}}
		& \multicolumn{1}{>{\footnotesize\centering\arraybackslash}m{2.3cm}}{{No. pupils \linebreak succeeding \linebreak in all schemas}} 
		& \multicolumn{1}{>{\footnotesize\centering\arraybackslash}m{1.3cm}}{{\textbf{Success} \linebreak (\%)}} 
		& \multicolumn{1}{>{\footnotesize\centering\arraybackslash}m{1.7cm}}{{Median schemas succeeded (Q1-Q3) }}\\
		
		\midrule
		{\multirow{4}{*}{ \rotatebox{90}{\parbox{1.65cm}{\centering\textbf{Age \\ category}}}}}
		& 3-6 & yrs   & 18/27 & 67\% & 12 (11-12)  & \, 6 & 33\% & 10 (8-11) \\
		& 7-9 & yrs   & 30/33 & 91\% & 12 (12-12)  & \, 4 & 13\% & 10 (9-11) \\
		& 10-13 & yrs & 13/20 & 65\% & 12 (11-12)  & \, 3 & 23\% & 10 (7-11) \\
		& 14-16 & yrs & 38/49 & 78\% & 12 (12-12)  &  11  & 29\% & 10 (8-11) \\
		\bottomrule
	\end{tabular}
\end{table*}

\FloatBarrier

\begin{figure*}[!hb]
	\centering
	\includegraphics[width=.6\linewidth]{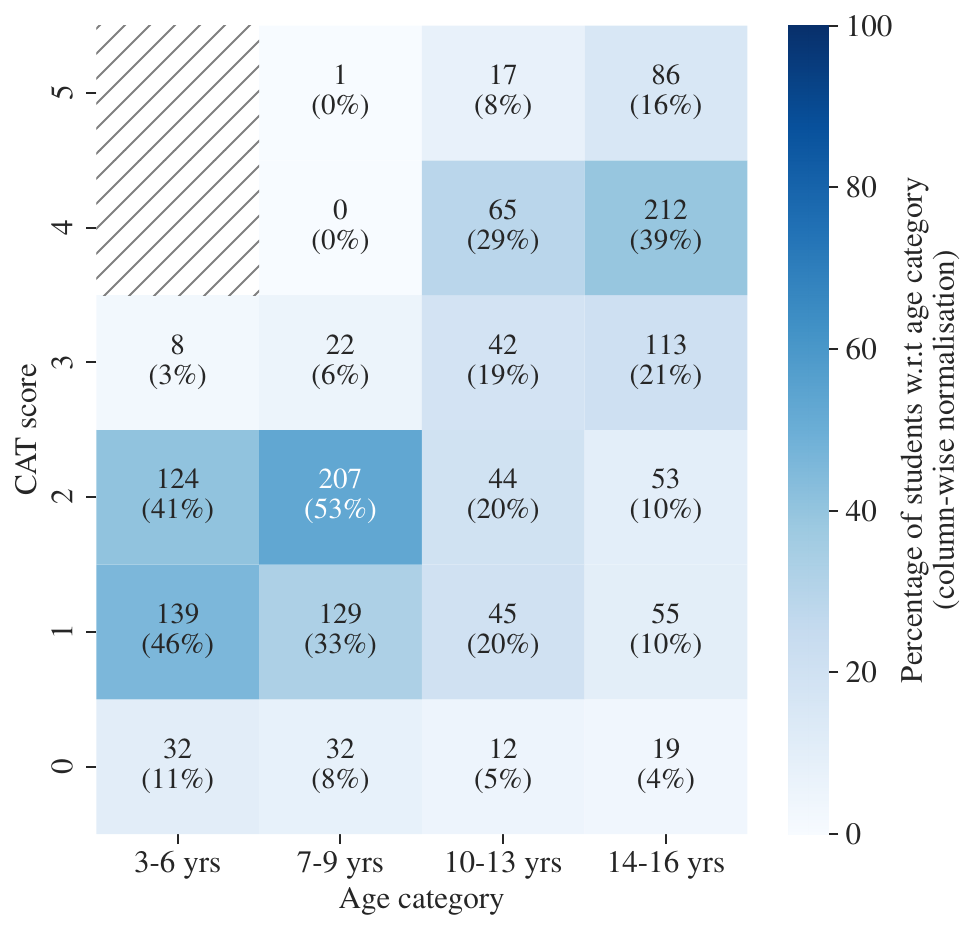}
	\mycaption{Age-wise distribution of CAT score levels (virtual CAT - main).}{
		Since the younger age group could not use the CAT-VPI, only data for GF and G interactions are included.}
	\label{fig:score_by_age_category_main}
\end{figure*}

Examining the distribution of CAT scores across age groups, as seen for the unplugged CAT, we found that performance improves with age (see \Cref{fig:score_by_age_category_main}), and these differences are statistically significant ($\chi^2=735.73$, $p < 1e-15^{****}$). 
Again, pairwise comparisons in \Cref{tab:pairwise_comparison_main} show that this significance holds for all age categories except the two youngest groups (ages 3-6 and 7-9). 

\begin{table}[ht]
	\centering
	\mycaptiontitle{Pairwise comparison of CAT scores between age groups (virtual CAT - main).}
	\label{tab:pairwise_comparison_main}
	\setlength{\tabcolsep}{.82cm}
	\footnotesize
	\begin{tabular}{c@{\hspace{.2\tabcolsep}}c>{\centering\arraybackslash}m{2cm}>{\centering\arraybackslash}m{2cm}>{\centering\arraybackslash}m{2cm}}
		\toprule
		& &\multicolumn{1}{c}{\textbf{3-6 yrs}} & \multicolumn{1}{c}{\textbf{7-9 yrs}} & \multicolumn{1}{c}{\textbf{10-13 yrs}}  \\ 
		\midrule 
		\textbf{7-9} &\textbf{yrs} & $p=0542$ &  &  \\ 
		\textbf{10-13} &\textbf{yrs} &  $p<.0001^{****}$    & $p<.0001^{****}$ & \\  
		\textbf{14-16} &\textbf{yrs} &    $p<.0001^{****}$ &    $p<.0001^{****}$ &    $p<.0001^{****}$    \\
		\bottomrule
	\end{tabular}
	\\\vspace{0.5em}
	\begin{minipage}{.9\linewidth}
		$^{*}$ $p < 0.05$, $^{**}$ $p < 0.01$, $^{***}$ $p < 0.001$, $^{****}$ $p < 0.0001$
	\end{minipage}
\end{table}

\FloatBarrier

\subsection{Competencies development}

\subsubsection{Impact of age on algorithmic skills}
To understand the influence of age on the development of AT, we performed a series of statistical tests based on observations and hypotheses made during previous studies.
ANOVA results show a positive correlation between age and algorithmic dimension ($p < 1e-6^{****}$), indicating that as age increases, there is a corresponding increase in the complexity of the algorithm produced.  
Specifically, the older age category (ages 14-16) showed a higher algorithmic dimension and positive coefficients (0.23200, $1e-06^{***}$). 
The chi-squared tests of proportions reveal significant age-related variations ($p < 0.0001^{****}$) in the higher algorithm dimensions (2D), with proportions increasing with age, reaching 36\% for the older age category. 

\FloatBarrier
\subsubsection{Impact of interaction strategies on algorithmic skills}

\begin{figure*}[!ht]
	\centering
	\includegraphics[width=\linewidth]{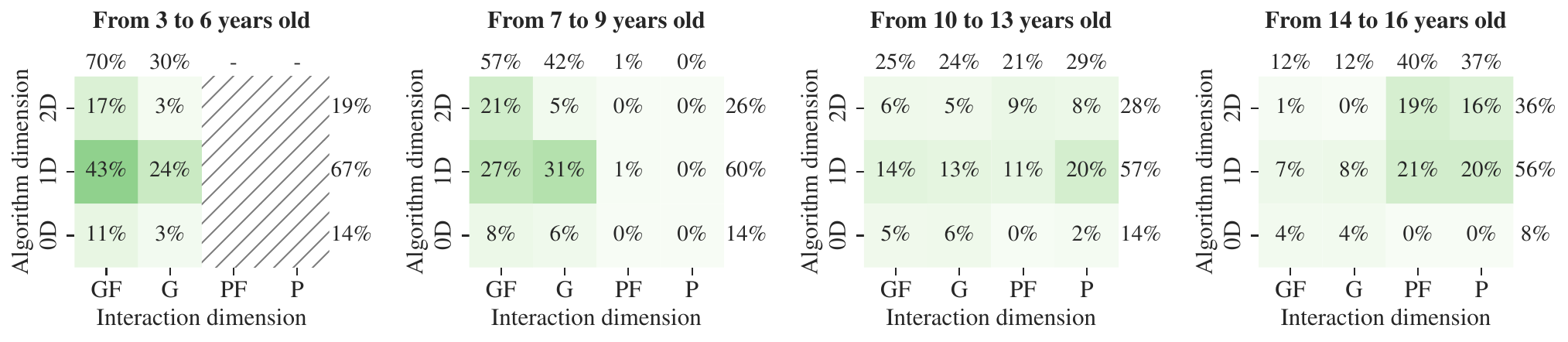}
	\mycaption{Algorithmic and interaction strategies across age (virtual CAT - main).}{
		Percentages represent the proportion of each interaction-algorithmic combination within each age group, with aggregated values shown across rows and columns.
		Since the younger age group could not use the CAT-VPI, only data for GF and G interactions are included.
	}  \label{fig:strategies_main}
\end{figure*}

Delving deeper into the strategies employed by pupils across different age categories, \Cref{fig:strategies_main} confirm that in the virtual domain, as for the unplugged, there is a developmental tendency towards more complex interactions, as shown in \Cref{fig:artefact_by_age_group_main}.
%
Regardless of age, there is a common tendency to create 1D algorithms, with a developmental progression towards more complex 2D algorithms, consistent with the unplugged study findings. 
The youngest age category (ages 3-6) demonstrated higher proficiency in conceiving complex 2D algorithms, surpassing simpler 0D ones, highlighting that virtual artefacts seemingly facilitate producing more complex algorithms at an earlier age.





We proceeded by conducting statistical analysis to determine whether the interaction dimension is a predictor of the algorithm dimension and identify if certain interaction strategies are more effective in producing complex algorithms than others.
ANOVA tests showed that artefact dimension is a significant predictor of algorithmic dimension ($p < 1e-15^{****}$).
Further analysis using Tukey's HSD test with BH adjustment and pairwise t-tests with Bonferroni adjustment highlighted the substantial impact of the PF interaction on algorithm dimension when compared to other artefacts ($p < 0.001^{***}$).
The chi-squared test of proportions confirmed these findings, showing significant variations in the production of higher algorithm dimensions (2D) across artefact categories ($\chi^2= 140.38$, $p < 1e-15^{****}$). The virtual PF artefact had the highest proportion of higher algorithm dimensions (46\%), followed by P (39\%), highlighting its effectiveness in promoting more complex algorithms.
Finally, we performed T-tests to compare the algorithmic capabilities across the unplugged and virtual domains. The results reinforced our findings, revealing a significant difference in favour of the virtual interactions, which consistently led to the production of more complex algorithms ($t = -10.25365$, $p < 1e-23$).

\FloatBarrier
\subsubsection{Impact of trial and error strategies on algorithmic skills}
\begin{figure*}[!h]
	\centering
	\includegraphics[width=.49\linewidth]{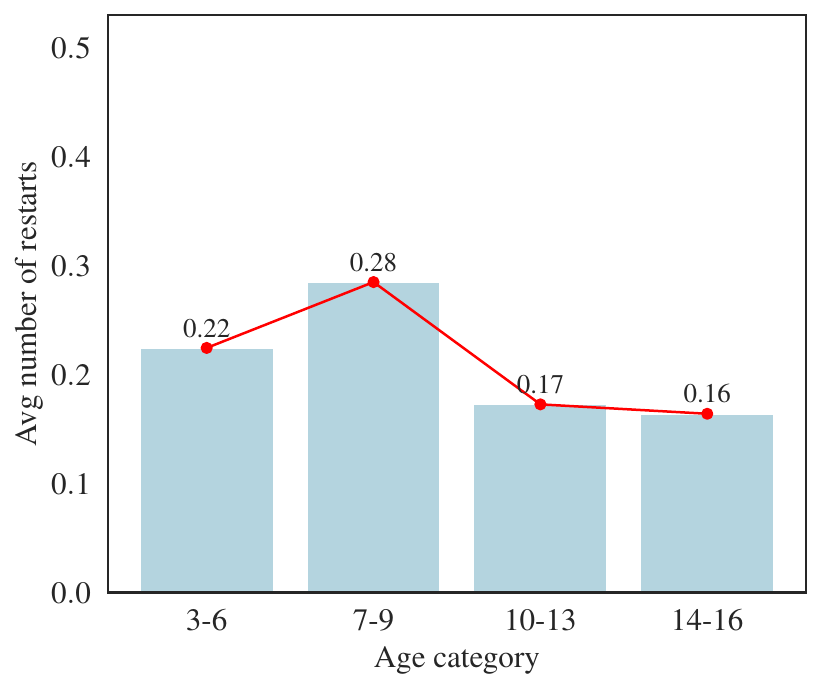}
	\mycaptiontitle{Restarts distribution per age.}
\label{fig:restart_distribution_Age category}
\end{figure*}

In this supplementary analysis, we aim to identify the predictors of students' use of the restart feature, a key aspect of T\&E strategies, and assess its impact on the complexity of the algorithms produced.
Our OLS regression analysis indicates that neither the characteristics of schemas (-0.0098, $p = 0.161$) nor gender (-0.0059, $p = 0.902$) significantly affect restart behaviour.
Instead, a negative correlation was found between interaction dimension and restart frequency, indicating that students working with increasingly complex artefacts tend to restart their tasks less frequently.
This is supported by \Cref{fig:restart_distribution_Assigned_interaction_dimension}, which shows that, for the simplest artefacts, the average number of restarts decreases as artefact complexity increases (-0.0283, $p = 0.180$). 
Additionally, \Cref{fig:restart_distribution_Prevalent_interaction_dimension} highlights a non-linear relationship between prevalent artefact complexity and restarts (-0.0168, $p = 0.427$). 
The slight increase in restarts from non-autonomous (GF and PF) to autonomous (G and P) suggests that visual feedback is important in supporting students' task participation and reducing restarts, possibly indicating that uncertainty arises when such feedback is absent.
Also, age plays a significant role in restart behaviour, as reflected in a statistically significant inverse relationship (-0.0167, $p = 0.012^{*}$), indicating that older students are less inclined to restart tasks.
\Cref{fig:restart_distribution_Age category} illustrates this decrease in the average number of restarts with increasing age categories.
A peak in restarts among the second age group (ages 7-9) suggests increased exploration or developing problem-solving efficiency, while the decline in restarts among older groups (ages 10-13 and 14-16) may signal improved problem-solving skills and a greater ability to integrate past experiences, indicating a shift towards more advanced problem-solving approaches as students mature.
\begin{figure*}[ht]
	\centering
	\begin{subfigure}[t]{.49\linewidth}
		\centering
		\includegraphics[width=\linewidth]{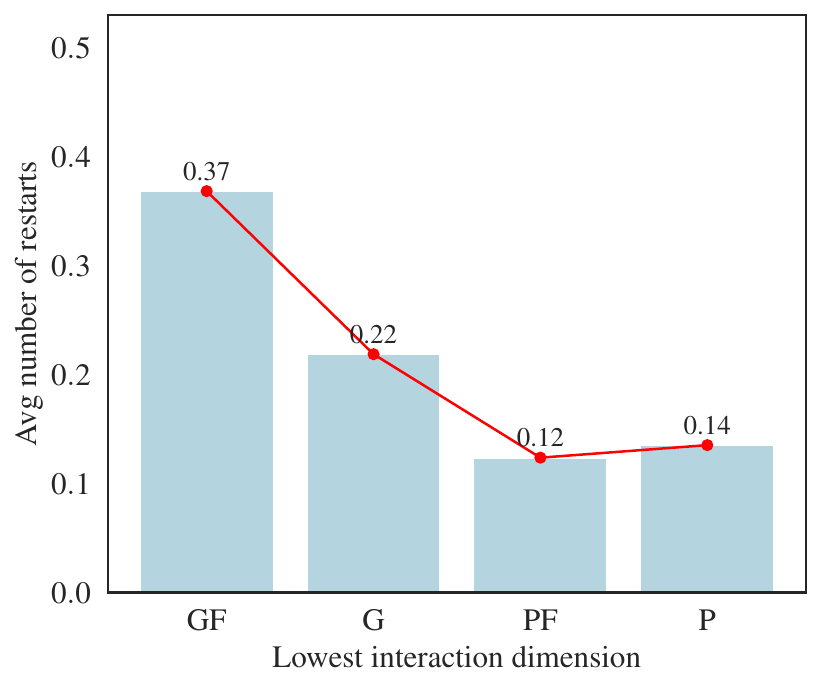}
		\caption{\centering}
		\label{fig:restart_distribution_Assigned interaction dimension}
	\end{subfigure}
	\hfill
	\begin{subfigure}[t]{.49\linewidth}
		\centering
		\includegraphics[width=\linewidth]{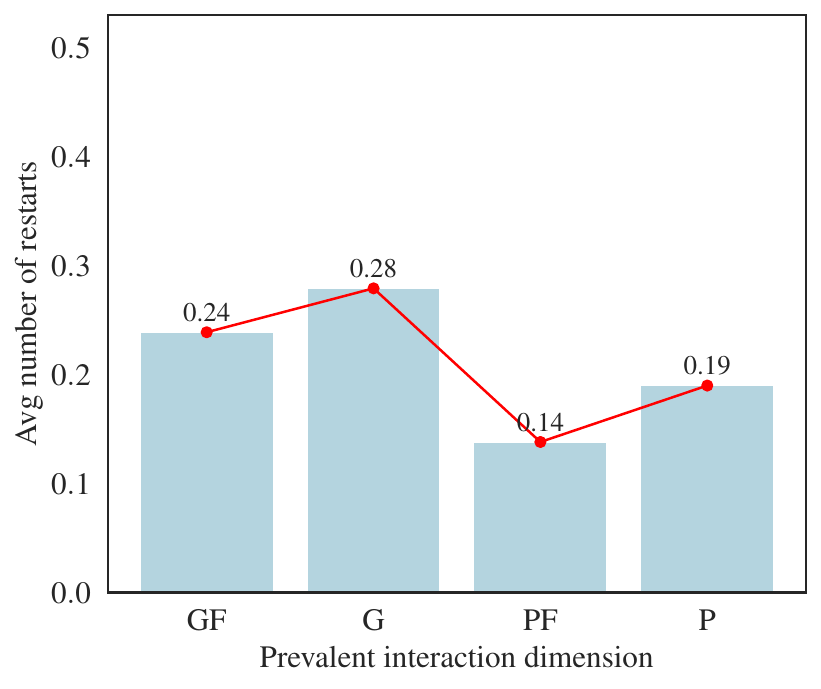}
		\caption{\centering}
		\label{fig:restart_distribution_Prevalent interaction dimension}
	\end{subfigure}
	\mycaptiontitle{Restarts distribution per interaction dimension.}
	\label{fig:restart_distribution_Interaction dimension}
\end{figure*}

Finally, our regression analysis examined the relationship between restart behaviour and the algorithm dimension. 
No significant association was found (-0.0221, $p= 0.232$), and testing for non-linear effects using polynomial terms also revealed no significant patterns ($p=0.200, 0.407$, and $0.497$).
This suggests that restart frequency does not predict algorithmic performance, either linearly or non-linearly.
Regarding CAT scores, we obtained different results. While the direct link between restart frequency and CAT scores was not statistically significant (-0.0677, $p = 0.110$), further polynomial terms investigation indicated a more complex relationship. 
Initially, higher restart frequency was associated with lower CAT scores (-0.4509, $p = 0.001^{**}$), likely reflecting difficulty or confusion. 
However, the addition of a quadratic term showed a positive effect (0.0895, $p = 0.031^{*}$), suggesting a non-linear pattern where, beyond a certain threshold, more restarts led to improved scores, suggesting a learning or optimisation process.
The cubic term, however, was not significant (-0.0037, $p = 0.151$), indicating no added explanatory value from further complexity.

This analysis reveals an intriguing dynamic between the variables. 
Age and interaction dimensions significantly influence restart frequency, indicating that restarts are not random but reflect user characteristics such as experience (linked to age) and the complexity of the interactions they navigate. 
Interestingly, restart frequency does not directly impact algorithm dimension, suggesting that the frequency of restarts is not a reliable indicator of algorithmic performance or the ability to handle complex problems. 
However, restart frequency does influence the CAT score, a composite measure of algorithm and interaction dimensions, through a non-linear relationship, highlighting the critical role of interaction dimension, which directly impacts the CAT score and indirectly shapes it through restart frequency, emphasising the need to consider interaction complexity when interpreting user performance.


\FloatBarrier
\subsubsection{Joint impact of interaction strategies and age on algorithmic skills }

Examining the combined effect of interaction strategies and age on algorithmic skills, we found this relationship to be statistically significant ($p = 0.000106^{***}$). 
The linear regression model, EMMs analysis, and the chi-squared test of proportions revealed notable differences in how various interactions influence the algorithm dimension across age groups.
%
%
The results show significant variations in interaction-algorithmic combinations for all age groups ($p < 0.01^{*}$). 
Younger children (ages 3-6) predominantly benefit from GF interactions (24\%, EMM = 1.090, SE = 0.0394), while older age groups display more diverse patterns, with PF (EMM = 1.465, SE = 0.0391) and P (EMM = 1.431, SE = 0.0409) interactions becoming more common. 
prevalent, particularly for pupils aged 14-16, where PF interactions account for 47\% and P interactions for 44\%. 
This is especially evident in the older age group (ages 14-16), where the most pronounced disparities are observed ($\chi^2 = 81.434$, $p < 1e-15^{****}$), with PF interactions accounting for 47\% and P interactions for 44\%. 



\FloatBarrier
\subsubsection{Joint impact of interaction strategies and gender on algorithmic skills}

Examining the combined effect of interaction strategies and gender on algorithmic skills, ANOVA revealed a marginally statistically significant relationship ($p = 0.0521$), while no significance was found in the unplugged domain ($p = 0.2655$).
The linear regression analysis further confirms these findings, highlighting a significant relationship between the simplest virtual artefact G and gender ($0.21394$, $p = 0.00642^{**}$), indicating a smaller negative impact of the G interaction on algorithmic dimension for males than females.


\FloatBarrier
\subsubsection{Joint impact of interaction strategies and schemas on algorithmic skills}

Examining the combined effect of interaction strategies and schemas characteristics on algorithmic skills, ANOVA revealed a significant interaction between these factors in both domains, with a stronger effect in the virtual domain ($p=1e-13^{****}$) compared to the unplugged one (($p=1e-9^{****}$). 
For the virtual CAT, linear regression analysis identifies significant positive interactions between artefacts and specific schemas. Notably, artefact P combined with schemas 2, 3, 8, 9, 10, 11, and 12, and artefact PF with schemas 9, 10, 11, and 12, lead to a higher algorithmic dimension.
In contrast, for the unplugged CAT, the effects, while still significant, are primarily negative. For instance, artefact VS combined with schemas 3, 4, 7, 8, and 12, as well as artefact PF with schemas 9, 10, 11, and 12, significantly reduce algorithmic complexity.


\FloatBarrier
\subsubsection{Joint impact of age and gender on algorithmic skills}

Examining the combined effect of age and gender on algorithmic skills, ANOVA revealed a significant interaction between these factors for the virtual domain ($p = 0.000115^{***}$) and less pronounced for the unplugged domain ($p = 0.00364^{**}$).
In the virtual domain, linear regression analysis shows that male participants aged 10 to 13 years exhibit lower algorithmic skills compared to females ($-0.29045$, $p = 0.00641$), reflecting a reversal of the pattern observed in the unplugged environment, where males in the same age range show higher algorithmic skills ($0.22245$, $p = 0.011^{*}$).

\chapter{Factors influencing CAT performance}\label{chap:performance_across_assessements}
\begin{custombox}
	{The content of this chapter has been adapted from the following article with permission of all co-authors and publishers:}
	\begin{itemize}[noitemsep,nolistsep,leftmargin=.25in]
		\item \textbf{Adorni, G.}, Artico, I., Piatti, A., Lutz, E., Gambardella, L. M., Negrini, L., Mon-
		dada, F., and Assaf, D. (2024). Development of algorithmic thinking skills in K-12
		education: A comparative study of unplugged and digital assessment instruments.
		\textit{Computers in Human Behavior Reports} \cite{adorni_2024_largescale}. 
	\end{itemize}
	As an author of this publication, my contribution involved:\\
	\textit{Conceptualisation, Methodology, Validation, Formal analysis, Investigation, Resources, Data curation, Writing -- original draft \& review \& editing, Visualisation, Supervision.}
\end{custombox}
\section{Summary}
This chapter focuses on analysing the factors affecting K-12 pupils' performance in both the unplugged and virtual CAT assessments using hierarchical statistical models, contributing to RQ4.
It examines how fixed and random effects explain performance differences and discusses the model selection process and the significance of predictors across both domains.
The code used for the data analysis process is available on GitHub \cite{Adorni_Virtual_CAT_Algorithmic_2024}.

\section{Methodology}

We employed Linear Mixed Models (LMMs) to examine factors influencing student performance for both the unplugged and virtual CAT assessments. This approach is suitable for hierarchical data, as students are nested within sessions, schools, and cantons \citep{hox2017multilevel,raudenbush2002hierarchical}.
LMMs allowed us to account for both fixed effects (e.g., gender) and random effects (e.g., variability across sessions), providing insights into how different predictors influence CAT performance while appropriately handling the hierarchical correlations in data.

The model selection and refinement process was carried out using the virtual CAT dataset. 
We started by constructing a baseline model using the Restricted Maximum Likelihood (REML) approach, with Satterthwaite's approximation for degrees of freedom to account for unequal variances and sample sizes \citep{bartlett1937properties}. This adjustment ensures a more accurate test of significance for the predictors. The model was implemented using the \texttt{lmer} function from the \texttt{lmerTest} package in \texttt{R} \citep{Kuznetsova2017, R}.

After defining the initial model, the model selection process was carried out using the Likelihood Ratio Test (LRT) via the \texttt{anova} method in the \texttt{lmreTest} package in R \citep{bartlett1937properties,chambers1992,Kuznetsova2017,R}. This method allowed us to compare alternative model specifications to assess the contribution of each predictor to explaining student performance.

Following model selection, we applied the chosen model to the virtual and unplugged datasets and conducted a Type III ANOVA with Satterthwaite's method to evaluate the significance of different terms in explaining the variability in CAT scores across both domains. 
For the virtual dataset, we examined the effect of task completion time on performance, exploring its relationship with variables such as age and interaction types. This analysis was performed solely for the virtual domain due to the lack of task completion data in the unplugged dataset. Task completion time was treated as a random effect to account for individual differences in completion pace, task complexity, and concentration, thereby avoiding potential biases that could arise if it were treated as a fixed effect.

Lastly, we combined the two datasets and introduced new fixed-effect predictors to differentiate between the virtual and unplugged domains. We then used LRT to assess the impact of gender on performance while controlling for school-related variability.

\section{Results}


\definecolor{Gray}{gray}{0.9}
\subsection{Model selection and refinement}
Our baseline model (M0), defined in \Cref{eq:m0}, includes various components to account for factors influencing student performance.
The outcome variable $\texttt{CAT\_SCORE}$ represents the performance score. 
As fixed effects, we included $\texttt{CANTON}$ due to the limited number of cantons (only two cantons) and $\texttt{GENDER}$, a binary predictor for gender. 
As random effects, we included $\texttt{STUDENT}$, accounting for natural heterogeneity among students, acknowledging unique factors like abilities, prior knowledge, and unobserved characteristics inherent to each student; $\texttt{SESSION\_GRADE}$ captures variations related to HGs and testing sessions, such as time of day, classroom conditions, and differences introduced by various teachers, as well as disparities across educational levels, acknowledging each grade's unique curricular and teaching aspects; $\texttt{SCHOOL}$ represents variability among different schools, encompassing their unique environments and resources; and $\texttt{SCHEMA}$ accounts for variability among the 12 distinct tasks, isolating the task-specific characteristics.
$\beta_0$, $\beta_1$, and $\beta_2$ are the coefficients for the fixed effects, while $\epsilon$ is the error term representing the unexplained variability.
\begin{align}
	\begin{split}
		\texttt{CAT\_SCORE} = & \beta_0 + \beta_1 \cdot \texttt{CANTON} + \beta_2 \cdot \texttt{GENDER} + \\ 
		& + u_{\texttt{STUDENT}} + u_{\texttt{SESSION\_GRADE}} +  \\
		& + u_{\texttt{SCHOOL}} + u_{\texttt{SCHEMA}} + \epsilon \mbox{.}
	\end{split}
	\label{eq:m0}
\end{align}

The baseline model has been analysed to assess the significance of the predictors, and as shown in \Cref{tab:baseline_summary}, no statistically significant differences in CAT scores were found between students from Ticino and Solothurn cantons ($p = 0.719$).
Similarly, there were non-significant variations in CAT scores between male and female students, with a negative coefficient suggesting slightly lower performance in males ($p = 0.497$). 
Variance estimates reveal variation in student performance across different grouping levels, with notable differences observed at the school and student levels.

\begin{table}[!ht]
	\centering
	\footnotesize
	\mycaption{Baseline model (M0) summary.}{
		The REML criterion at convergence is 4044.3 for the baseline linear mixed-effects model.}
	\label{tab:baseline_summary}
	\begin{subtable}{1\linewidth}
		\mycaptiontitle{{Scaled residuals.}\label{tab:baseline_residuals}}
		\setlength{\tabcolsep}{4.4mm}
		\centering
		\begin{tabular}{ccccc}
			\toprule
			\textbf{Min}      & \textbf{Q1}     & \textbf{Median}  & \textbf{Q3}     & \textbf{Max}     \\
			\midrule
			$-3.785$  & $-0.582$ & $0.050$  & $0.611$ & $3.569$  \\
			\bottomrule
		\end{tabular}
	\end{subtable}
	\\\vspace{.5cm}
	\begin{subtable}{1\linewidth}
		\mycaptiontitle{Random effects. \label{tab:baseline_random}}
		\centering
		\setlength{\tabcolsep}{4.6mm}
		\begin{tabular}{lccc}
			\toprule
			\textbf{Groups$^*$}        & \textbf{Name}        &\textbf{Variance} & \textbf{\textit{SD}}$^a$ \\
			\midrule
			Student       & (Intercept) & $0.490$ & $0.700$  \\
			Schema        & (Intercept) & $0.057$ & $0.238$  \\
			Session-Grade & (Intercept) & $0.054$ & $0.232$  \\
			School        & (Intercept) & $0.931$ & $0.965$  \\
			Residual      &             & $0.756$ & $0.870$  \\
			\bottomrule
		\end{tabular}
		\\\vspace{0.5em}
		\begin{minipage}{.6\linewidth}
			$^a$ Standard deviation\\
			$^*$ Student (129), Schema (12), Session-Grade (9), School (5);\\
			\hspace*{.5em} Number of observations: 1457
		\end{minipage}
	\end{subtable}
	\\\vspace{.5cm}
	\begin{subtable}{1\linewidth}
		\setlength{\tabcolsep}{2.1mm}
		\mycaptiontitle{Fixed effects.} 
		\label{tab:baseline_fixed}
		\centering
		\begin{tabular}{lrlrrc}
			\toprule
			& \multicolumn{1}{c}{\textbf{Estimate}} & \multicolumn{1}{c}{\textbf{\textit{SE}}$^b$} & \multicolumn{1}{c}{\textbf{df}} & \multicolumn{1}{c}{$\boldsymbol{t}^c$} & \multicolumn{1}{c}{$\boldsymbol{p}^d$}\\
			\midrule
			(Intercept) & $2.169$   & $0.986$ & $2.826  $ & $2.200 $ & $0.121$ \\
			Gender$^1$     & $-0.090$  & $0.132$ & $120.082$ & $-0.681$ & $0.497$ \\
			Canton$^2$  & $0.439$   & $1.101$ & $2.815  $ & $0.398 $ & $0.719$ \\
			\bottomrule
		\end{tabular}
		\\\vspace{0.5em}
		\begin{minipage}{.6\linewidth}
			$^b$ Standard error
			
			$^c$ $t$ value
			
			$^d$ $p$ value $= Pr(>{|t|})$
			
			$^1$ Gender: Male
			
			$^2$ Canton: Ticino
		\end{minipage}
	\end{subtable}
\end{table}

To further refine our analysis, we performed an LRT to evaluate the contribution of canton and gender as predictors.
The results in \Cref{tab:lrt-canton} indicate that including the canton predictor doesn't significantly improve the model's fit ($p = 0.5998$), likely due to the limited representation of Swiss cantons in the data (only 2 out of 26 sampled). Therefore, we opt for the simpler reduced model (M1).

\begin{table}[htb]
\footnotesize
    \centering
    \mycaption{LRT to evaluate the inclusion of canton as a predictor.}{
    Comparison between the reduced model (M1) without the canton predictor and the baseline model (M0) including it. 
    }
    \label{tab:lrt-canton}
    \setlength{\tabcolsep}{6.6mm}
    \begin{tabular}{lcrl}
        \toprule
        \textbf{Model} & \textbf{AIC}$^a$ & \multicolumn{1}{c}{{$\boldsymbol{{\chi}^2}$}} & \multicolumn{1}{c}{{$\boldsymbol{p}$}$^b$} \\
        \midrule
        M1  & $4058.1$  &        \\
        M0  & $4059.8$   & $0.275$ & $0.5998$  \\
        \bottomrule
    \end{tabular}
    \\\vspace{0.5em}
        \begin{minipage}{.6\linewidth}
        $^a$ Akaike Information Criterion for the model evaluated as \\
        \hspace*{.5em} $-2\cdot(\text{logLik} - \text{npar})$. Smaller is better.

        $^b$ $p$ value $= Pr(>{\chi}^2)$
        \end{minipage}
\end{table}

To assess the impact of gender, whose effect was not statistically significant in the baseline model (M0), we compared three models, each considering different combinations of predictors: the reduced model (M1) without the canton predictor, another reduced model (M2) without both canton and gender and an improved model (M3) without canton but with gender as a random slope within schools. 
The assumption in model M3 stems from observations in our experimental studies, which suggest that contextual factors, such as school environment and resources, may influence gender dynamics.
The LRT results in \Cref{tab:lrt-gender} show that including gender as a fixed effect (M2) does not significantly enhance model fit ($p = 0.4881$), suggesting that gender alone is not a substantial predictor of CAT scores. 
However, the improved model (M3) showed a significant improvement in fit ($p < 1e-3^{***}$), indicating that gender-related differences may vary across schools.
\begin{table}[htb]
\footnotesize
    \centering
    \mycaption{LRT to evaluate the inclusion of gender as a predictor.} 
    {Comparison between the reduced model (M1) without the canton predictor, another reduced model (M2) without canton and gender, and an improved model (M3) without canton, but that considers gender as a random slope within schools.
    }
    \label{tab:lrt-gender}
    \setlength{\tabcolsep}{5.8mm}
    \begin{tabular}{lcrl}
        \toprule
        \textbf{Model} & \textbf{AIC}$^a$ & \multicolumn{1}{c}{{$\boldsymbol{{\chi}^2}$}} & \multicolumn{1}{c}{{$\boldsymbol{p}$}$^b$} \\
        \midrule
        M2       & $4056.6$ & & \\
        M1       & $4058.1$ &  $0.481$ &  $0.4881$ \\
        \cellcolor{Gray}M3       & \cellcolor{Gray}$4048.3$ & \cellcolor{Gray}$11.785$ & \cellcolor{Gray}{$0.0006^{***}$} \\
        \bottomrule
    \end{tabular}
    \\\vspace{0.5em}
        \begin{minipage}{.6\linewidth}
        $^a$ Akaike Information Criterion for the model evaluated as \\
        \hspace*{.5em} $-2\cdot(\text{logLik} - \text{npar})$. Smaller is better.

        $^b$ $p$ value $= Pr(>{\chi}^2)$

        $^{*}$ $p < 0.05$, $^{**}$ $p < 0.01$, $^{***}$ $p < 0.001$, $^{****}$ $p < 0.0001$
        \hspace*{.5em} Grey shadings indicate statistically significant fixed effects
        \end{minipage}
\end{table}

\FloatBarrier

\subsection{Model application and predictor significance}

Given these results, we consider model M3, defined in \Cref{eq:m3}. 
This model comprises the intercept coefficient $\beta_0$,  and the coefficients for the effect of gender within schools $\beta_1$ and $\beta_2$.
The random effects include $\texttt{STUDENT}$, $\texttt{SESSION\_GRADE}$, and $\texttt{SCHEMA}$.
Finally, $\epsilon_{\text{GENDER}}$ represents the error term for the interaction between gender and school, and $\epsilon$ denotes the unexplained variability in the model.
\begin{align}
	\begin{split}
		\texttt{CAT\_SCORE} = & \beta_0 + (\beta_1 + \beta_2 \cdot \texttt{GENDER} + \epsilon_{\texttt{GENDER}}|\texttt{SCHOOL}) + \\
		& + u_{\texttt{STUDENT}} + u_{\texttt{SESSION\_GRADE}} + \\ 
		& + u_{\texttt{SCHEMA}} + \epsilon \mbox{.}
	\end{split}
	\label{eq:m3}
\end{align}

Once the model was defined, it was applied to both the unplugged and virtual CAT data.
The model summary in \Cref{tab:model_summary} shows a significant intercept in the fixed effects for both domains (U: $p = 0.014^{*}$, V: $p = 0.00286^{**}$), suggesting that the average 
\begin{table}[p]
\centering
\footnotesize
\mycaption{Model (M3) summary.} 
{The REML criterion at convergence is 3377.3 for the unplugged domain (U) and 4032.5 for the virtual domain (V) in the linear mixed-effects models.}
\label{tab:model_summary}
\begin{subtable}{\linewidth}
\setlength{\tabcolsep}{4mm}
  \mycaptiontitle{Scaled residuals.}\label{tab:model_residuals}
  \centering
 \begin{tabular}{cccccc}
 \toprule
 & \textbf{Min}  & \textbf{Q1} & \textbf{Median}  & \textbf{Q3} & \textbf{Max} \\
 \midrule
 U & $-3.574$  & $-0.656$ & $-0.048$  & $0.563$ & $2.855$  \\
 \midrule
 V & $-3.797$  & $-0.578$ & \, $0.048$  & $0.616$ & $3.537$  \\
 \bottomrule
 \end{tabular}
\end{subtable}
\\\vspace{.5cm}
\begin{subtable}{\linewidth}
\setlength{\tabcolsep}{2.7mm}
  \mycaptiontitle{Random effects.} 
  \label{tab:model_random}
  \centering
 \begin{tabular}{cllccc}
 \toprule
 &\textbf{Groups$^*$} & \multicolumn{1}{c}{\textbf{Name}} &\textbf{Variance} & \textbf{\textit{SD}}$^a$& \textbf{Corr} \\
 \midrule
 \multirow{6}*{U} & Student & (Intercept) & $0.235$ & $0.485$ & \\
 & Schema                   & (Intercept) & $0.197$ & $0.443$ & \\
 & Session-Grade            & (Intercept) & $0.024$ & $0.154$ & \\
 & School                   & (Intercept) & $0.479$ & $0.692$ & \\
 &                          & Gender$^1$     & $0.027$ & $0.164$ & $0.14$ \\
 & Residual                 &             & $0.671$ & $0.819$ & \\
 \midrule
 \multirow{6}*{V} & Student & (Intercept) & $0.407$ & $0.638$ & \\
 & Schema                   & (Intercept) & $0.057$ & $0.237$ & \\
 & Session-Grade            & (Intercept) & $0.059$ & $0.243$ & \\
 & School                   & (Intercept) & $0.873$ & $0.935$ & \\
 &                          & Gender$^1$     & $0.253$ & $0.503$ & $-0.49$ \\
 & Residual                 &             & $0.756$ & $0.870$ & \\
 \bottomrule
 \end{tabular}
 \\\vspace{0.5em}
 \begin{minipage}{.65\linewidth}
 $^a$ Standard deviation
 
 $^1$ Gender: Male 

 $^*$ U: Student (109), Schema (12), Session-Grade (8), School (3); \\
 \hspace*{.5em} Number of observations: 1280\\
 \hspace*{.5em} V: Student (129), Schema (12), Session-Grade (9), School (5);\\
 \hspace*{.5em} Number of observations: 1457\\
 
 \end{minipage}
\end{subtable}
\\\vspace{.5cm}
\begin{subtable}{\linewidth}
\setlength{\tabcolsep}{2.4mm}
  \mycaptiontitle{Fixed effects.}
  \label{tab:model_fixed}
  \centering
 \begin{tabular}{clclrrl}
 \toprule
  & & \textbf{Estimate} & \multicolumn{1}{c}{\textbf{\textit{SE}}$^b$} & \multicolumn{1}{c}{\textbf{df}} & \multicolumn{1}{c}{$\boldsymbol{t}^c$} & \multicolumn{1}{c}{$\boldsymbol{p}^d$}\\
 \midrule
 \cellcolor{Gray}U & \cellcolor{Gray}(Intercept) & \cellcolor{Gray}$2.827$  & \cellcolor{Gray}$0.432$ & \cellcolor{Gray}$2.397$  & \cellcolor{Gray}$6.54 $ & \cellcolor{Gray}{$0.014^{*}$}  \\\midrule
 \cellcolor{Gray}V & \cellcolor{Gray}(Intercept) & \cellcolor{Gray}$2.429$  & \cellcolor{Gray}$0.389$ & \cellcolor{Gray}$4.194$  & \cellcolor{Gray}$6.244$  & \cellcolor{Gray}{$0.003^{**}$} \\
 \bottomrule
 \end{tabular}
 \\\vspace{0.5em}
 \begin{minipage}{.65\linewidth}
 $^b$ Standard error

 $^c$ $t$ value
 
 $^d$ $p$ value $= Pr(>{|t|})$

 $^{*}$ $p < 0.05$, $^{**}$ $p < 0.01$, $^{***}$ $p < 0.001$, $^{****}$ $p < 0.0001$\\
 \hspace*{.5em} Grey shadings indicate statistically significant fixed effects
 \end{minipage}
\end{subtable}
\end{table}
CAT score significantly differs from zero when accounting for random effects, indicating a baseline proficiency level within the student population.
Exploring random effects, school-level differences account for the largest variance in both domains (U = 0.47939, V = 0.87332), suggesting that school-specific factors strongly influence CAT scores.
Gender variance is similar across domains, with a lower positive correlation in the unplugged domain (0.14) and a negative one in the virtual domain (-0.49), suggesting different gender dynamics: in the virtual domain, schools with higher overall performance tend to have lower scores for male students, and vice versa.
%
The virtual domain has higher student variance (U = 0.23476, V = 0.40687), indicating greater individual variability, possibly influenced also by unmeasured factors.
%
Session-grade variance is smaller for unplugged (U = 0.02378, V = 0.05883), indicating more consistent performance, likely due to stable session-grade factors like age progression, curriculum complexity, teaching methods, or cohort effects.
%
Lastly, the unplugged domain shows a larger variance in schemas compared to the virtual domain (U = 0.19660, V = 0.05605), suggesting that factors such as the nature of the activity, the way information is presented, and features like the ability to skip or solve schemas in a preferred order contribute to this variation.

\begin{table}[htb]
\centering
\footnotesize
\mycaption{Type III ANOVA, with Satterthwaite's method, on model (M3).} 
{Rows shaded in grey indicate statistically significant variables.
}
\label{tab:model_ranova}
\setlength{\tabcolsep}{3.1mm}
 \begin{tabular}{clcrr}
 \toprule
 && \textbf{AIC}$^a$ & \multicolumn{1}{c}{\textbf{LRT}$^b$}  & \multicolumn{1}{c}{{$\boldsymbol{p}$}$^c$} \\
 \midrule
 \multirow{5}*{U} & & $3393.3$ &  & \\
 & \cellcolor{Gray}Schema  & \cellcolor{Gray}$3653.2$ &  \cellcolor{Gray}$261.903$ &  \cellcolor{Gray}{$<1e-15^{****}$}\\
 & Gender$^d$                  & $3389.7$ &  $0.438$ &  $0.803$  \; \; \; \\
 & Session-Grade            & $3393.1$ &  $1.780$ &  $0.182$  \; \; \; \\
 & \cellcolor{Gray}Student & \cellcolor{Gray}$3584.7$ &  \cellcolor{Gray}$193.418$ &  \cellcolor{Gray}{$<1e-15^{****}$}\\
 \midrule
 \multirow{5}*{V} & & $4048.5$  &  & \\
 & \cellcolor{Gray}Schema  & \cellcolor{Gray}$4112.5$  &  \cellcolor{Gray}$ 65.75$  &  \cellcolor{Gray}{$<1e-15^{****}$} \! \\
 & \cellcolor{Gray}Gender$^d$  & \cellcolor{Gray}$4056.7$ &  \cellcolor{Gray}$ 12.20$  & \cellcolor{Gray}{$0.002^{**}$} \, \,  \\
 & Session-Grade & $4049.6$ & $3.11$  & $0.078$ \; \; \; \\
 & \cellcolor{Gray}Student & \cellcolor{Gray}$4397.0$  & \cellcolor{Gray} $350.52$  &  \cellcolor{Gray}{$<1e-15^{****}$} \! \\
 \bottomrule
 \end{tabular}
 \\\vspace{0.5em}
 \begin{minipage}{.6\linewidth}
 $^a$ Akaike Information Criterion for the model evaluated as\\
 \hspace*{.5em} $-2\cdot(\text{logLik} - \text{npar})$. Smaller is better.

 $^b$ LRT statistic; twice the difference in log-likelihood, \\
 \hspace*{.5em} which is asymptotically chi-square distributed.

 $^c$ $p$ value $= Pr(>{\chi}^2)$

 $^d$ Gender in (1 + Gender | School)

 $^{*}$ $p < 0.05$, $^{**}$ $p < 0.01$, $^{***}$ $p < 0.001$, $^{****}$ $p < 0.0001$\\
 \hspace*{.5em} Grey shadings indicate statistically significant fixed effects
 \end{minipage}
\end{table}
The ANOVA results in \Cref{tab:model_ranova} further confirm the significance of the random effects, particularly for schema and student ($p < 1e-15^{****}$), showing that task type and individual student differences significantly affect CAT scores.
%
Interestingly, the impact of gender on CAT scores varies significantly across schools in the virtual domain ($p = 0.002^{**}$) but not in the unplugged one ($p = 0.803$), suggesting that gender-related factors have a greater effect in virtual environments.

\FloatBarrier
\subsubsection{Gender influence within schools on performance}

\begin{figure*}[h]
	\centering
	\begin{subfigure}[t]{.49\linewidth}
		\centering
		\includegraphics[width=\linewidth]{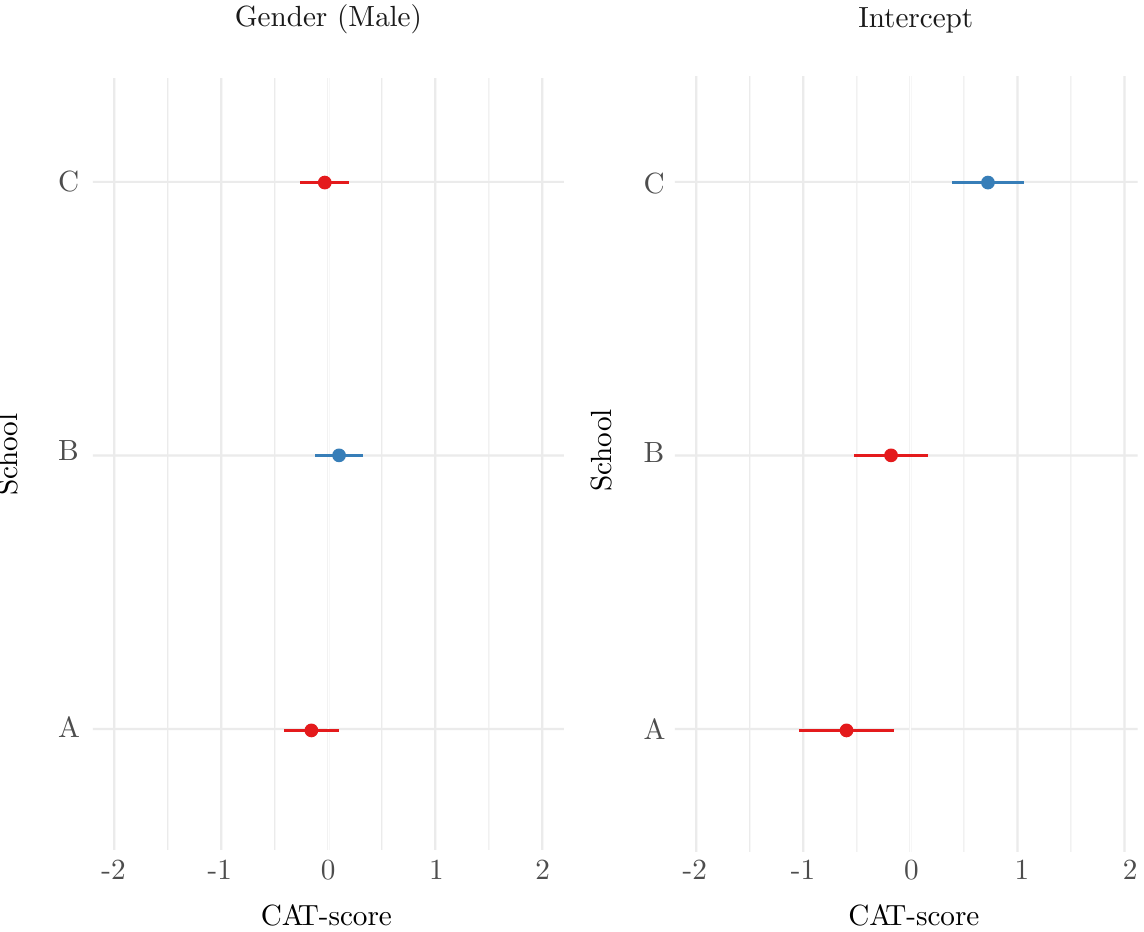}
		\mycaptiontitle{{Unplugged CAT.}
		}
		\label{fig:unplugged_re_school}
	\end{subfigure}
	\hfill
	\begin{subfigure}[t]{.49\linewidth}
		\centering
		\includegraphics[width=\linewidth]{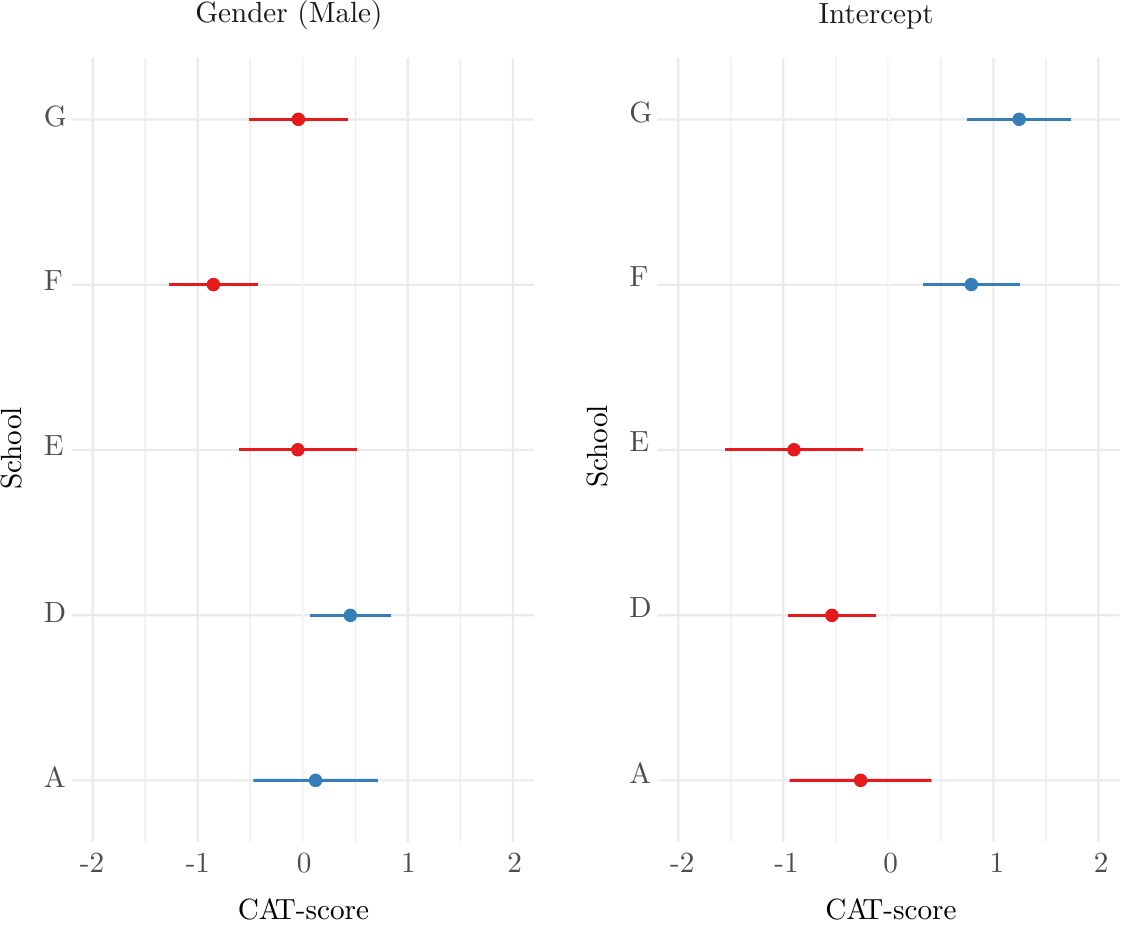}
		\mycaptiontitle{{Virtual CAT.}  
		}
		\label{fig:virtual_re_school}
	\end{subfigure}
	\mycaption{Gender-related school performance variations.} 
		{The plot on the left captures the variability in CAT scores between schools for male students compared to female students, illustrating how scores differ across schools based on gender. 
			On the right, the plot shows the intercept, representing the average variability between schools, exclusively focusing on female students.
			Blue points represent scores above average, while reds those below. 
			Horizontal lines represent the estimates' confidence intervals.
		}
		\label{fig:re_school}
	\end{figure*}

Comparing gender-related school performance in the two domains, \Cref{fig:re_school} shows no significant effects for the unplugged domain, while the virtual domain exhibits variability in CAT scores between male and female students across schools. 
Certain schools (e.g., D) show higher CAT scores for male students than others (e.g., F), challenging the idea of a uniform gender effect.

Focusing on average performances across schools, we discern differences in the baseline performance for both domains. 
In the unplugged case, A performs below average, B slightly below, and C notably above. 
In the virtual case, schools exhibit varied impacts on female students, with some (A, D, and E) showing decreased CAT scores and others (F and G) demonstrating increased scores.

\FloatBarrier
\subsubsection{Individual student variability in performance}

\begin{figure*}[htb]
	\centering
	\begin{subfigure}[t]{.49\linewidth}
		\centering
		\includegraphics[width=\linewidth]{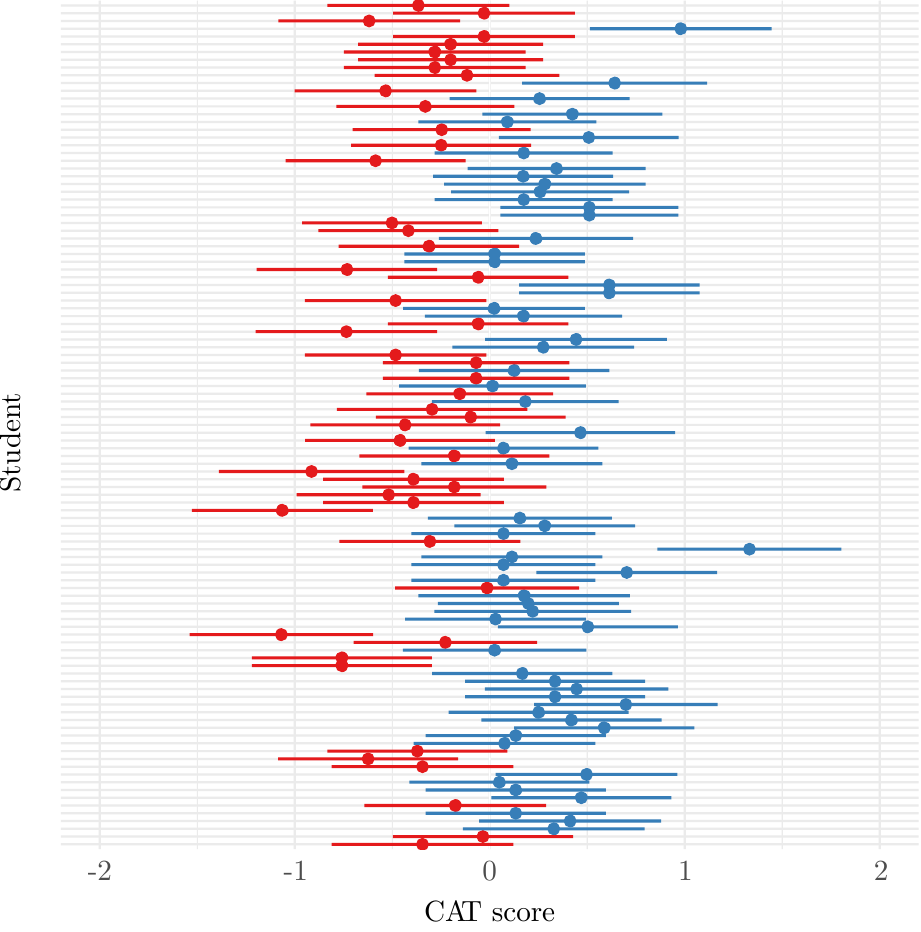}
		\mycaptiontitle{{Unplugged CAT.}}
		\label{fig:unplugged_re_student}
	\end{subfigure}
	\begin{subfigure}[t]{.49\linewidth}
		\centering
		\includegraphics[width=\linewidth]{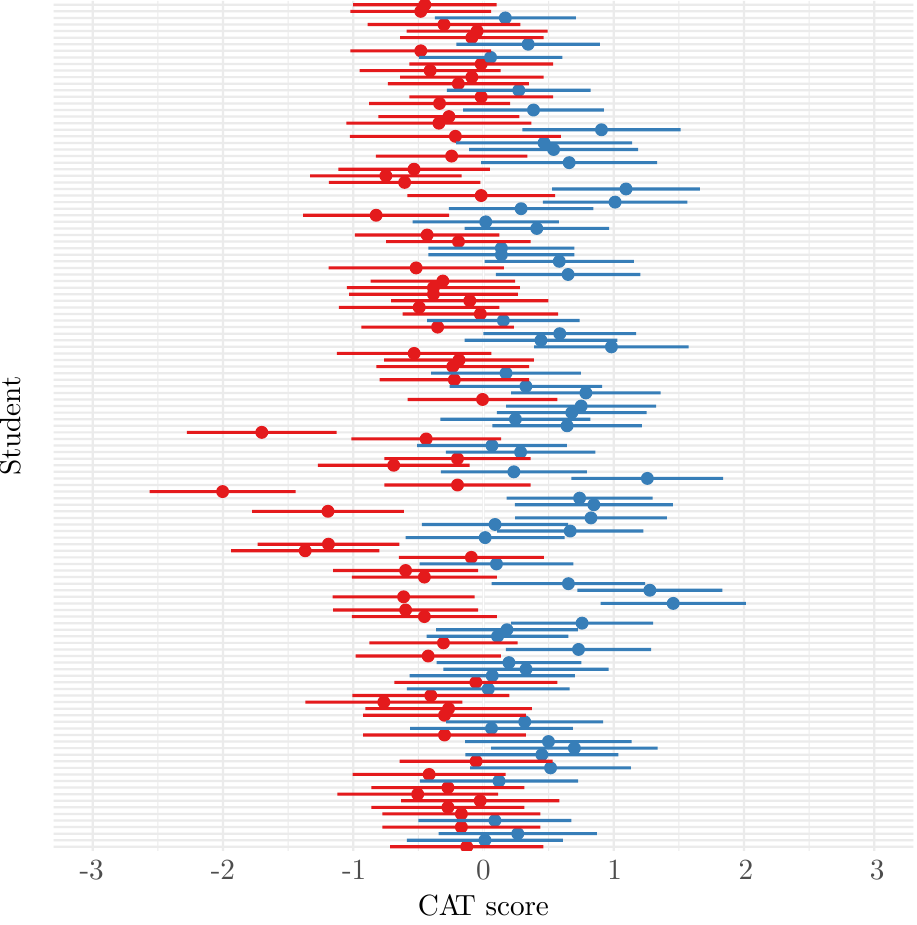}
		\mycaptiontitle{{Virtual CAT.}}
		\label{fig:virtual_re_student}
	\end{subfigure}
	\mycaption{Individual student performance variations.} 
	{Each point represents the student deviation from the average CAT score, with blue indicating scores above average and red below. 
		Horizontal lines represent the estimates' confidence intervals.}
	\label{fig:re_student}
\end{figure*}

The analysis of student performance in \Cref{fig:re_student} shows significant individual variability in both domains.
The presence of high achievers (blue dots to the right) and those facing challenges (red dots to the left) is consistent across both datasets, highlighting a diverse range of performances. This indicates a substantial amount of unexplained variability not accounted for by other factors considered in the study.

Despite accounting for school-level differences in the model, unexplained variability in CAT scores persists among students, indicating significant differences in performance residuals across various schools. This suggests that factors associated with the distinct educational environments of each school might contribute to the observed variance. Statistical analysis, specifically Levene's test, supports this observation ($p < 1e-15^{****}$).

\FloatBarrier
\subsubsection{Session grade impact on performance}

\begin{figure*}[htb]
	\centering
	\begin{subfigure}[t]{.49\linewidth}
		\centering
		\includegraphics[width=\linewidth]{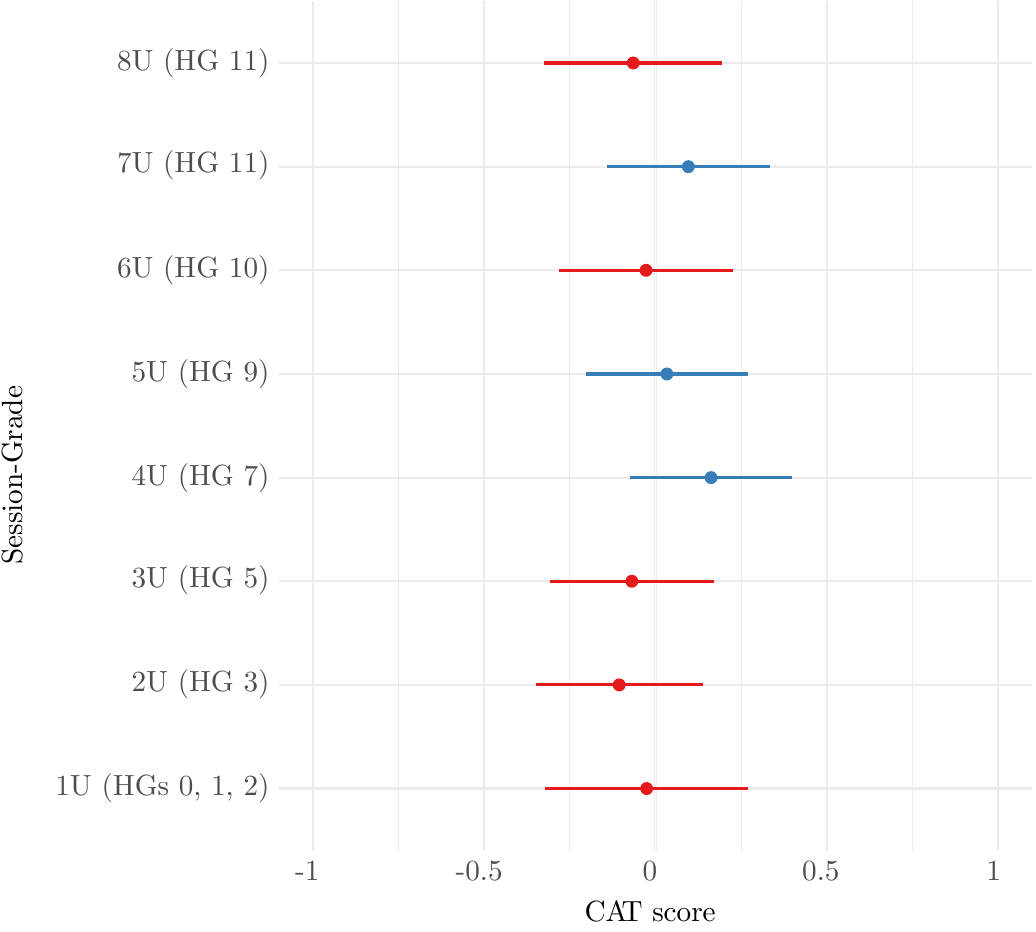}
		\mycaptiontitle{{Unplugged CAT.}}
		\label{fig:unplugged_re_sessiongrade}
	\end{subfigure}
	\begin{subfigure}[t]{.49\linewidth}
		\centering
		\includegraphics[width=\linewidth]{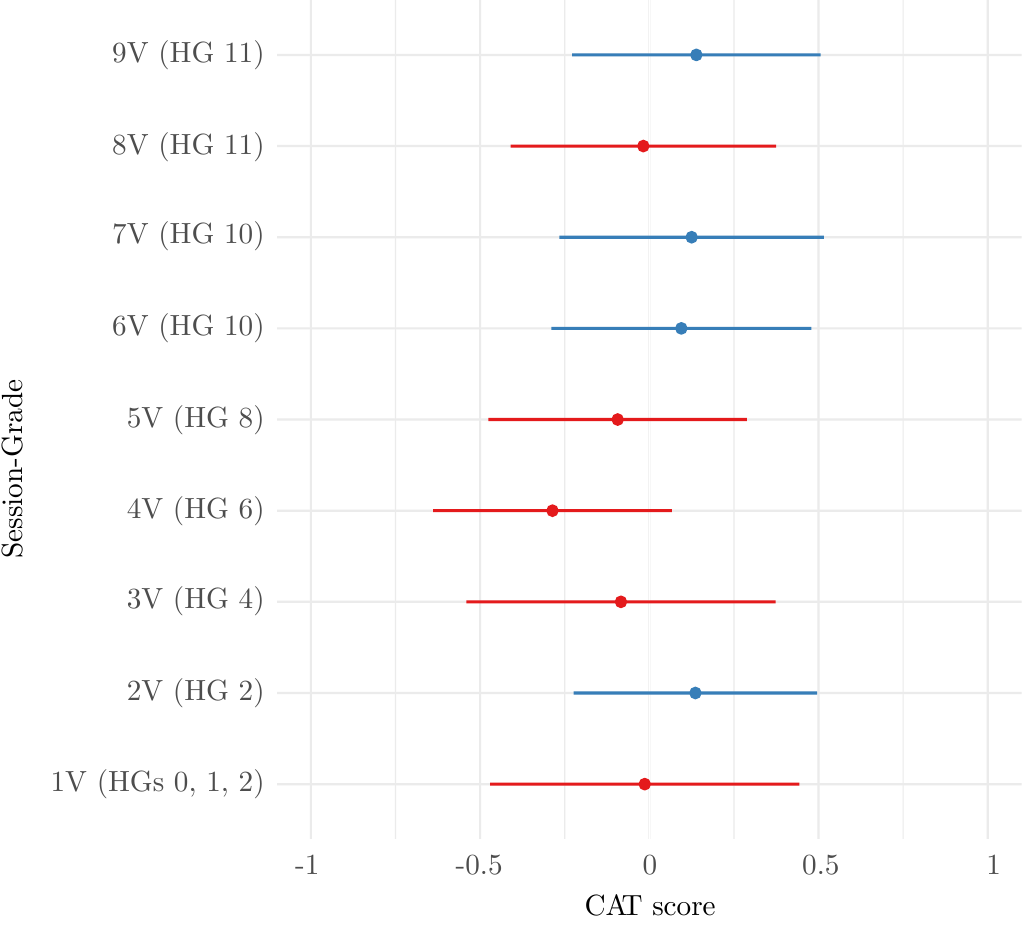}
		\mycaptiontitle{{Virtual CAT.}}
		\label{fig:virtual_re_sessiongrade}
	\end{subfigure}
	\mycaption{Session-Grade performance variations.}  
	{Each point represents the session deviation from the average CAT score, with blue indicating scores above average and red below. 
		Horizontal lines represent the estimates' confidence intervals. 
	}
	\label{fig:re_sessiongrade}
\end{figure*}

Examining the impact of session and grade on scores in \Cref{fig:re_sessiongrade}, no statistical differences in performance across sessions and grades are observed in both the unplugged and virtual domains. 
The pattern of fluctuations implies a complex relationship between sessions, grades, and CAT scores. Notably, lower performance is observed from the initial to the middle sessions, coinciding with lower HarmoS grades (HGs). Positive deviations in higher sessions suggest that older students generally perform better. This consistency implies that advanced cognitive skills and better adaptation to educational demands may contribute to improved performance among older students.

\FloatBarrier
\subsubsection{Schema-based differences in performance}\label{sec:schema-reml}
\begin{figure*}[ht]
	\centering
	\begin{subfigure}[t]{.49\linewidth}
		\centering
		\includegraphics[width=\linewidth]{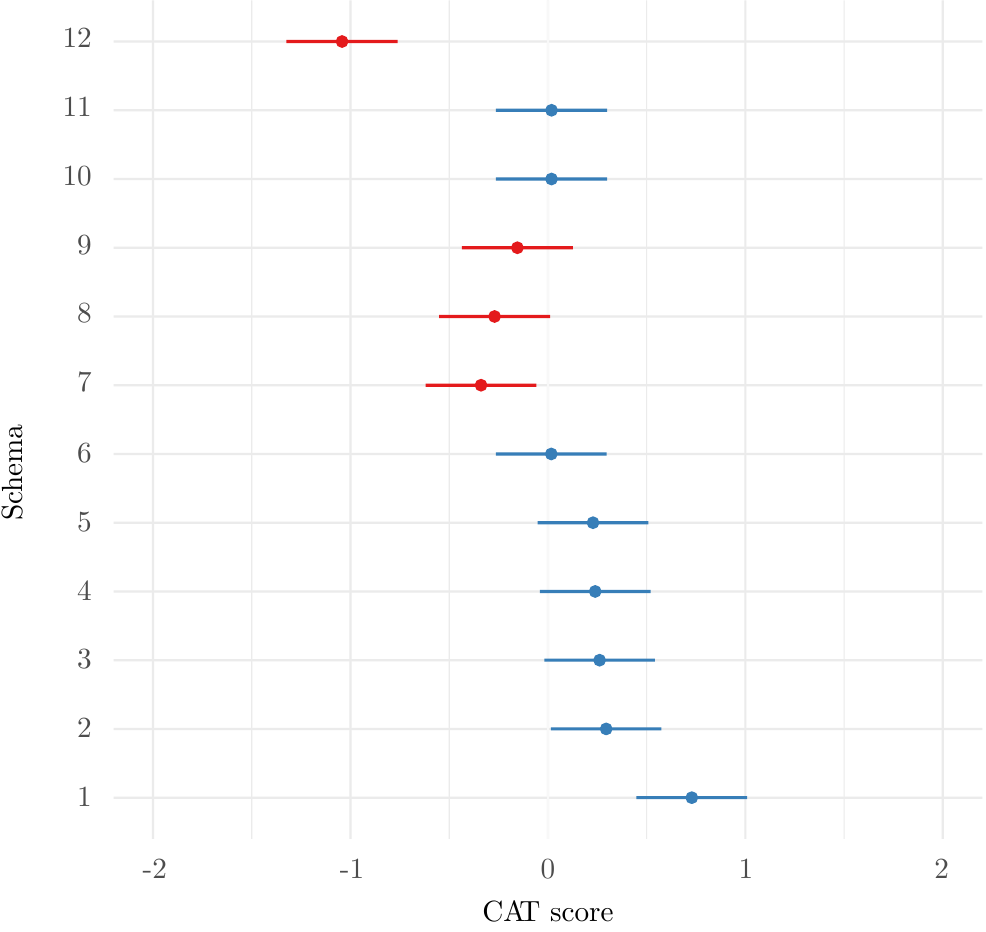}
		\mycaptiontitle{Unplugged CAT.}
		\label{fig:unplugged_re_schema}
	\end{subfigure}
	\begin{subfigure}[t]{.49\linewidth}
		\centering
		\includegraphics[width=\linewidth]{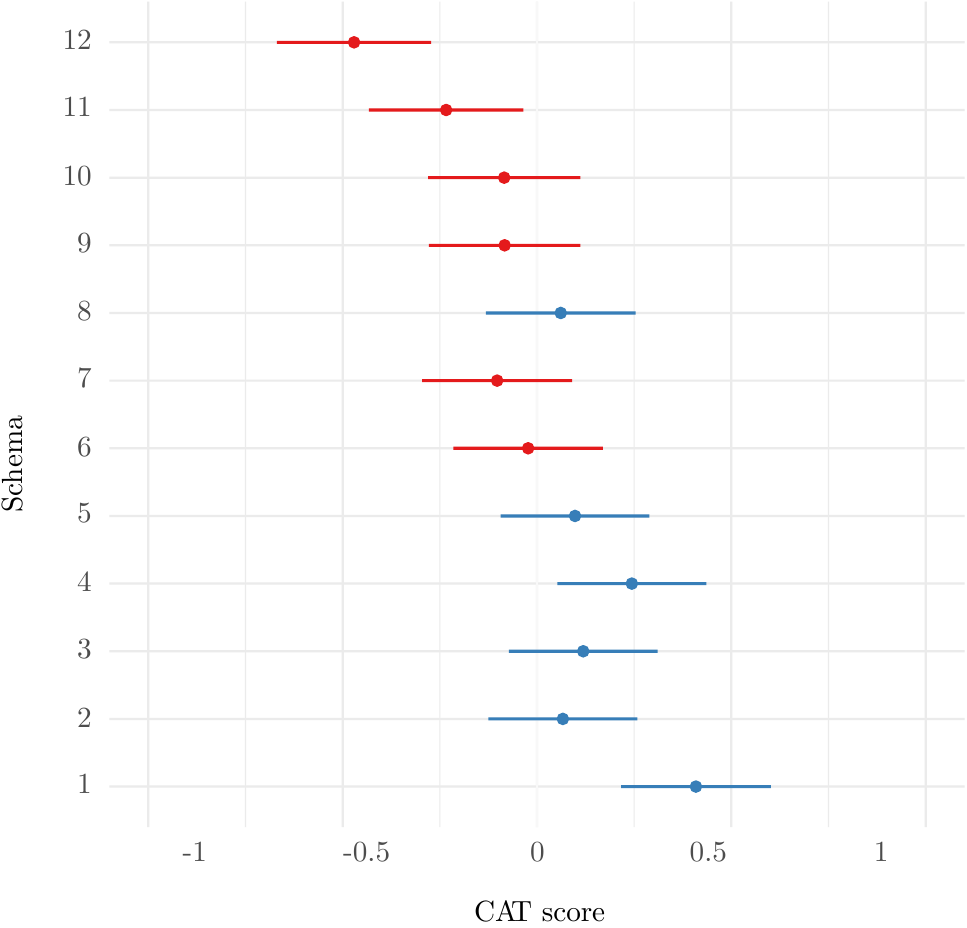}
		\mycaptiontitle{Virtual CAT.} 
		\label{fig:virtual_re_schema}
	\end{subfigure}
	\mycaption{Schema-based performance variations.} 
	{Each point represents the schema deviation from the average CAT score, with blue indicating scores above average and red below. Horizontal lines represent the estimates' confidence intervals. 
	}
	\label{fig:re_schema}
\end{figure*}

Task performance varies across different schemas in both the unplugged and virtual domains, as depicted in \Cref{fig:re_schema}.
For the unplugged dataset, initial schemas (1 to 6) generally yield good performance, although there's a decline as the schema number increases, hinting at rising task difficulty. 
Schemas 7 to 9 show below-benchmark scores but with improving trends, suggesting student adaptation or better task alignment.
Scores rise in schemas 10 and 11 but drop significantly in schema 12, possibly due to task difficulty or misalignment with student abilities.
A consistent decreasing trend is observed in the virtual dataset, although with some irregularities. Performance is above the benchmark for less difficult tasks (1 to 5) and declines below the benchmark (6 to 12) with increasing task difficulty. 
Notably, the mean CAT score for schema 8 is above the benchmark, suggesting better-than-expected performance on average. 
The non-linear decline in performance as schema numbers increase suggests that students perceive varying levels of difficulty, which may not align with the intended task progression.

\begin{figure*}[htb]
	\centering
	\begin{subfigure}[t]{.49\linewidth}
		\centering
		\includegraphics[width=\linewidth]{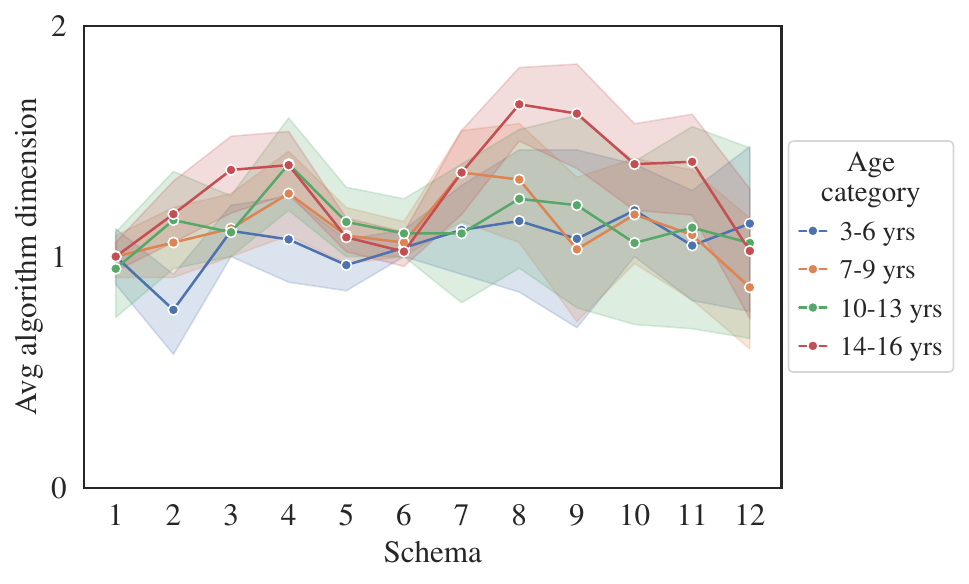}
		\mycaptiontitle{{Standard algorithm dimension metric}.}
		\label{fig:line_ALGORITHM_DIMENSION_by_task_and_AGE_CATEGORY}
	\end{subfigure}
	\begin{subfigure}[t]{.49\linewidth}
		\centering
		\includegraphics[width=\linewidth]{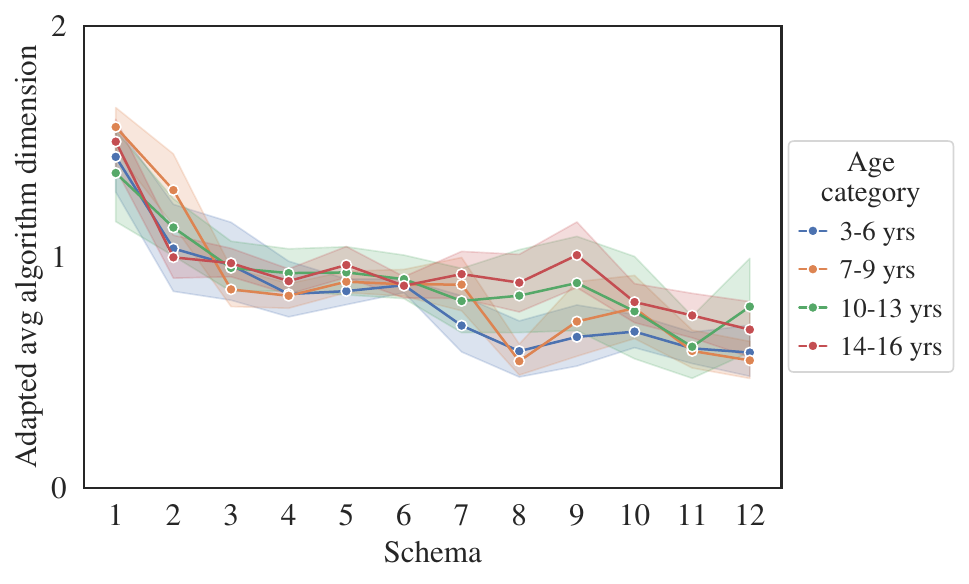}
		\mycaptiontitle{Adapted algorithm dimension metric.}
		\label{fig:line_WEIGHTED_ALGORITHM_DIMENSION_by_task_and_AGE_CATEGORY}
	\end{subfigure}
	
	\mycaption{Algorithm dimension variations across age categories at schema level.} 
	{The y-axis represents the average variations in algorithm dimension for each age category, plotted against different schemas on the x-axis.
	}
	\label{fig:line_ALGORITHM_DIMENSION}
\end{figure*}

To explore performance trends and irregularities across different tasks, we specifically looked at the algorithm dimension instead of overall performance. 
This examination pertains specifically to the virtual CAT, where we have precise and comprehensive information on all the commands students use in crafting their algorithms.
\Cref{fig:line_ALGORITHM_DIMENSION_by_task_and_AGE_CATEGORY} indicates that the algorithm dimension varies across tasks, suggesting that students adapt their problem-solving strategies to each task rather than following a linear regression of algorithm complexity. Notably, for schemas 1, 2, 5, 6, and 12, students often use 1D dimensional algorithms driven by practical considerations. 

\begin{figure*}[htb]
	\centering
	\begin{subfigure}[t]{.49\linewidth}
		\centering
		\includegraphics[width=\linewidth]{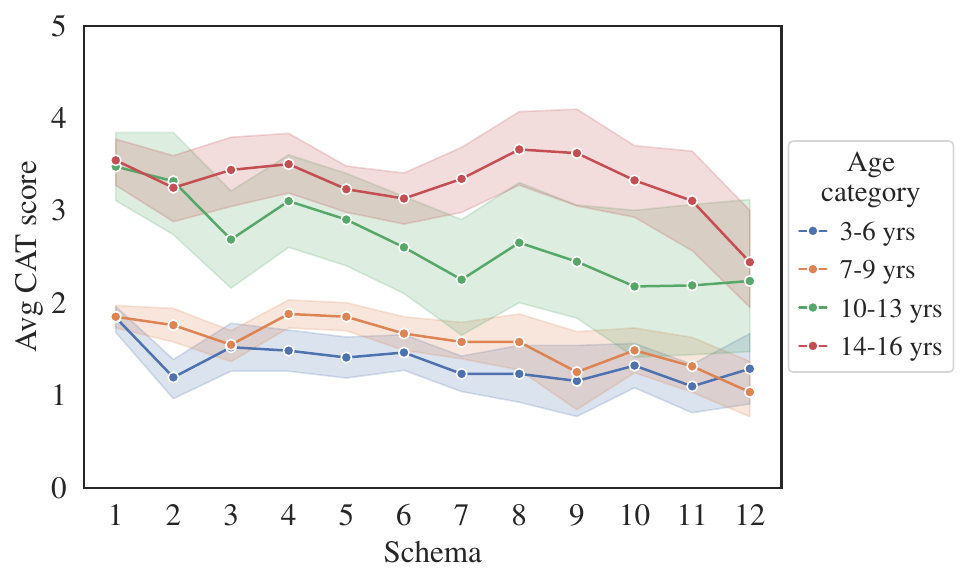}
		\mycaptiontitle{{Standard CAT score metric.}}
		\label{fig:line_CAT_SCORE_by_task_and_AGE_CATEGORY}
	\end{subfigure}
	\begin{subfigure}[t]{.49\linewidth}
		\centering
		\includegraphics[width=\linewidth]{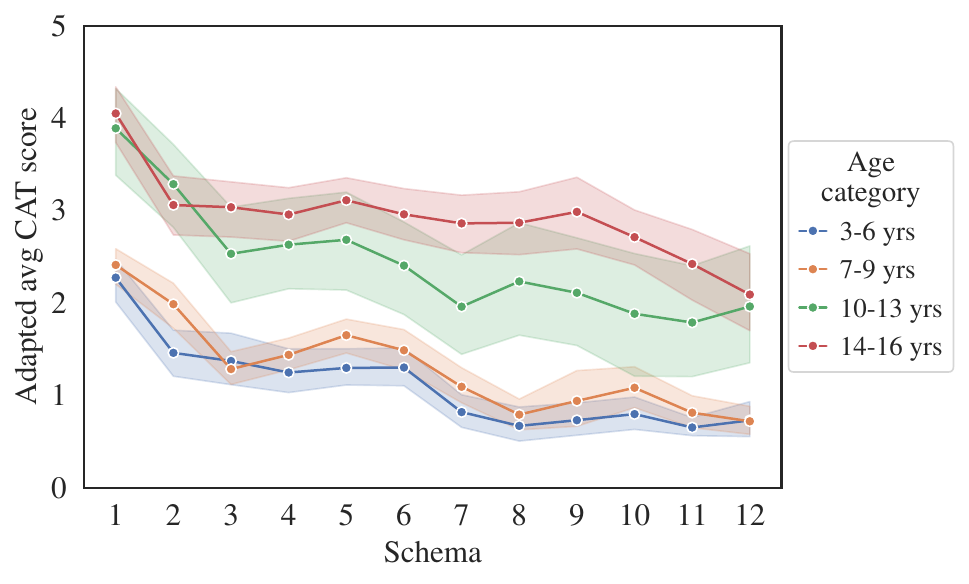}
		\mycaptiontitle{{Adapted CAT score metric.}}
		\label{fig:line_WEIGHTED_CAT_SCORE_by_task_and_AGE_CATEGORY}
	\end{subfigure}
	\mycaption{Performance variations across age categories at schema level.} 
	{The y-axis represents the average variations in CAT score for each age category, plotted against different schemas on the x-axis.}
	\label{fig:line_CAT_SCORE}
\end{figure*}

Sometimes, a simpler, less complex algorithm with fewer moves is more effective than a more intricate one. This preference for efficiency doesn't imply lower performance but reflects a pragmatic approach to problem-solving, as argued in \Cref{sec:at_assessment}.
To assess student performance, we introduced an alternative method considering both algorithm complexity and efficiency. The adapted algorithm dimension metric, presented in \Cref{fig:line_WEIGHTED_ALGORITHM_DIMENSION_by_task_and_AGE_CATEGORY}, demonstrates a more linear decrease in average algorithm dimensions.
\Cref{fig:line_WEIGHTED_CAT_SCORE_by_task_and_AGE_CATEGORY} shows the original and updated distribution of performance across schemas using the new metric.

\FloatBarrier 
\subsubsection{Task completion time effects}

\begin{figure}[htb]
	\centering
	\includegraphics[width=.5\linewidth]{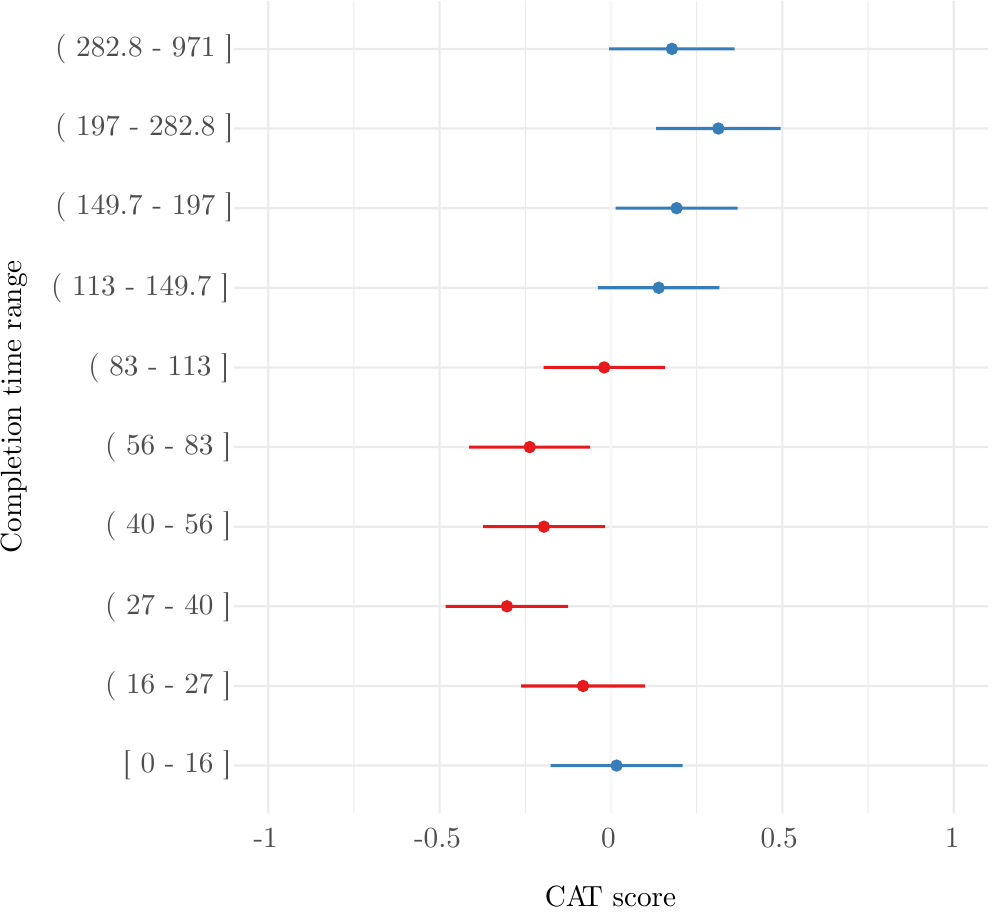}
	\mycaption{Task completion time and performance variations.}
	{Each point represents the deviation of task completion time intervals from the average CAT score, with blue indicating scores above average and red below. 
		Horizontal lines represent the estimates' confidence intervals.}
	\label{fig:re_time}
\end{figure}

In the concluding phase of our analysis, we incorporated task completion time as a random effect into our existing model. This specific examination was exclusive to the virtual CAT, benefiting from detailed records of task completion times. We aimed to uncover the correlation between the time students spent on tasks and their ensuing performance levels.
\Cref{fig:re_time} shows a non-linear relationship between task completion time and performance. Both extremely brief and significantly extended durations appear beneficial, resulting in higher CAT scores. This observation suggests that rapid responses may be driven by strong intuition or familiarity, while longer times may reflect a more analytical approach, likely enhancing performance.
On the other hand, intermediate completion times do not seem to capitalise on the strengths of either approach, potentially explaining the observed dip in scores and the negative impact on performance associated with moderate haste.

\FloatBarrier


\subsection{Domain-specific analysis}

To understand the factors influencing student performance in both unplugged and virtual settings, we combined the datasets to formulate the final model (M4), defined in \Cref{eq:m4}, aimed at assessing how the domain impacts CAT scores, along with other contributing factors.
The difference from M3 is the inclusion of the variable domain as a predictor of the CAT score.
\begin{align}
	\begin{split}
		\texttt{CAT\_SCORE} = & \beta_0 + \beta_1 \cdot \texttt{DOMAIN} + \\
		& + (\beta_2 + \beta_3 \cdot \texttt{GENDER} + \epsilon_{\texttt{GENDER}}|\texttt{SCHOOL}) + \\
		& + u_{\texttt{STUDENT}} + u_{\texttt{SESSION\_GRADE}} + \\ 
		& + u_{\texttt{SCHEMA}} + \epsilon \mbox{.}
	\end{split}
	\label{eq:m4}
\end{align}

 \begin{table}[!t]
	 \centering
	 \footnotesize
	 \mycaption{Model (M4) summary.} {The REML criterion at convergence is 7445.5 in the linear mixed-effects model.}
	 \label{tab:model_summary_full}
	 \begin{subtable}{\linewidth}
		     \mycaptiontitle{Scaled residuals.}\label{tab:model_residuals_full}
		     \centering
		     \setlength{\tabcolsep}{5mm}
		        \begin{tabular}{ccccc}
		        	\toprule
		        	\textbf{Min}  & \textbf{Q1} & \textbf{Median}  & \textbf{Q3} & \textbf{Max} \\
		        	\midrule
			         -4.005  & -0.628 & 0.005  & 0.618 & 3.640  \\
			         \bottomrule
			         \end{tabular}
		 \end{subtable}
	 \\\vspace{.5cm}
	 \begin{subtable}{1\linewidth}
		 \setlength{\tabcolsep}{3mm}
		     \mycaptiontitle{Random effects.} \label{tab:model_random_full}
		     \centering
		         \begin{tabular}{llccc}
			         \toprule
			         \textbf{Groups$^*$} & \multicolumn{1}{c}{\textbf{Name}} &\textbf{Variance} & \textbf{\textit{SD}}$^a$& \textbf{Corr} \\
			         \midrule
			         Student       & (Intercept)   & 0.328 & 0.573 & \\
			         Session-Grade & (Intercept)   & 0.039 & 0.196 & \\
			         Schema        & (Intercept)   & 0.110 & 0.331 & \\
			         School        & (Intercept)   & 0.758 & 0.871 & \\
			                       & Gender$^1$    & 0.167 & 0.409 & -0.39 \\
			         Residual      &               & 0.730 & 0.854 & \\
			         \bottomrule
			         \end{tabular}
			         \\\vspace{0.5em}
			         \begin{minipage}{.65\linewidth}
			         	$^a$ Standard deviation
			         	
			         	$^1$ Gender: Male 
			         	
			         	$^*$ Student (238), Session-Grade (17), Schema (12), School (7); \\
			         	\hspace*{.5em} Number of observations: 2746\\
			         	
			         \end{minipage}
        \end{subtable}
	 \\\vspace{.5cm}
	 \begin{subtable}{1\linewidth}
		     \mycaptiontitle{Fixed effects.} 
			     \label{tab:model_fixed_full}
		     \centering
		     \setlength{\tabcolsep}{2.4mm}
		         \begin{tabular}{lcrrrr}
			         \toprule
			                         & {\textbf{Estimate}} & \multicolumn{1}{c}{\textbf{\textit{SE}}$^b$} & \multicolumn{1}{c}{\textbf{df}} & \multicolumn{1}{c}{$\bm{t}^c$} & \multicolumn{1}{c}{$\bm{p}^d$}\\
			         \midrule
			         \cellcolor{Gray}(Intercept) & \cellcolor{Gray}2.638     & \cellcolor{Gray}0.382 & \cellcolor{Gray}10.088   & \cellcolor{Gray}6.905  & \cellcolor{Gray}{$<$1e-04$^{****}$} \\
			         Domain$^2$  & -0.040\;  & 0.317 & 12.219   & -0.127  & 0.901  \: \: \: \\
			         \bottomrule
			         \end{tabular}
			          \\\vspace{0.5em}
			         \begin{minipage}{.65\linewidth}
			         	$^b$ Standard error
			         	
			         	$^c$ $t$ value
			         	
			         	$^d$ $p$ value $= Pr(>{|t|})$
			         	
			         	$^2$ Domain: Virtual
			         	
			         	$^{*}$ $p < 0.05$, $^{**}$ $p < 0.01$, $^{***}$ $p < 0.001$, $^{****}$ $p < 0.0001$\\
			         	\hspace*{.5em} Grey shadings indicate statistically significant fixed effects
			         \end{minipage}
			         \end{subtable}
	 \end{table}
	
From \Cref{tab:model_summary_full}, we observe that variations in student performance across different groups, including individual students, session grades, schemas, and gender across schools, align with patterns identified in the model (M3) on individual datasets.
Nevertheless, the model revealed that the domain effect on CAT scores lacked statistical significance ($p = 0.901$), thereby strengthening the coherence of these results across various settings. This underscores the robustness of the conclusion, highlighting the significance of domain-independent factors in shaping CAT scores.

The ANOVA results in \Cref{tab:model_ranova_full} highlight that all factors significantly influence CAT scores in both virtual and unplugged settings. 

\begin{table}[htb]
\centering
\footnotesize
\mycaptiontitle{Type III ANOVA, with Satterthwaite's method, on model (M4).} 
\label{tab:model_ranova_full}
\setlength{\tabcolsep}{4.3mm}
        \begin{tabular}{lcrl}
        
        \toprule
        & \textbf{AIC}$^a$     & \multicolumn{1}{c}{\textbf{LRT}$^b$}  & \multicolumn{1}{c}{$\bm{p}$$^c$}     \\
        \midrule
                                          &                 $7461.0$ &         &                \\
        \cellcolor{Gray}Schema            & \cellcolor{Gray}$7761.2$ &  \cellcolor{Gray} $302.16$ &  \cellcolor{Gray}{$<1e-15^{****}$}\\
        \cellcolor{Gray}Gender$^d$           & \cellcolor{Gray}$7470.7$ &  \cellcolor{Gray} $ 13.64$ &  \: \, \, \, \cellcolor{Gray}{$0.001^{**}$} \\
        \cellcolor{Gray}Session-Grade     & \cellcolor{Gray}$7463.0$ &  \cellcolor{Gray} $  3.96$ &  \: \, \, \, \cellcolor{Gray}{$0.046^{*}$} \\
        \cellcolor{Gray}Student           & \cellcolor{Gray}$8003.2$ &  \cellcolor{Gray} $544.16$ &  \cellcolor{Gray}{$<1e-15^{****}$}\\
        \bottomrule
        \end{tabular}
        \\\vspace{0.5em}
        \begin{minipage}{.6\linewidth}
        $^a$ Akaike Information Criterion for the model evaluated as\\ 
        \hspace*{.5em} $-2\cdot(\text{logLik} - \text{npar})$. Smaller is better.

        $^b$ LRT statistic; twice the difference in log-likelihood, \\
        \hspace*{.5em} which is asymptotically chi-square distributed.

        $^c$ $p$ value $= Pr(>{\chi}^2)$
        
        $^d$ Gender in (1 + Gender | School)

        $^{*}$ $p < 0.05$, $^{**}$ $p < 0.01$, $^{***}$ $p < 0.001$, $^{****}$ $p < 0.0001$\\
        \hspace*{.5em} Grey shadings indicate statistically significant fixed effects

        \end{minipage}
\end{table}

To assess the impact of gender on CAT scores, we compared three models, each considering different combinations of predictors: the model with gender as a random slope within schools (M4), a reduced model without gender as predictor (M5), and a model with gender as a fixed effect (M6).
The LRT results in \Cref{tab:lrt-gender_full} show that introducing gender as a fixed effect (M6) does not significantly improve the model compared to the reduced model (M5) ($p=0.4270$). However, the inclusion of gender as a random slope within schools (M4) significantly enhances the model fit ($p = 1e-3^{***}$), emphasising that the impact of gender on CAT scores varies across different school environments. This underscores the importance of considering the interaction between gender and the school context when assessing its effect on educational outcomes.
\begin{table}[htb]
	\footnotesize
	\centering
	\mycaption{LRT to evaluate the global effect of Gender on the full dataset.} 
	{Comparison between the reduced model (M5) without the gender predictor, the model (M6) that considers gender a fixed effect, and the initial model (M4) that considers gender as a random slope within schools.
	}
	\label{tab:lrt-gender_full}
	\setlength{\tabcolsep}{5.8mm}
	\begin{tabular}{lcrl}
		\toprule
		\textbf{Model} & \textbf{AIC}$^a$ &  \multicolumn{1}{c}{\textbf{${\chi}^2$}} & \multicolumn{1}{c}{{$\bm{p}$}$^b$} \\
		\midrule
		M5       & $7472.2$  &   &  \\
		M6       & $7473.6$  &  $0.631$ & $0.4270$ \\
		\cellcolor{Gray}M4       & \cellcolor{Gray}$7462.5$  & \cellcolor{Gray}$13.063$ & \cellcolor{Gray}{$0.0003^{***}$} \\
		\bottomrule
	\end{tabular}
	\\\vspace{0.5em}
	\begin{minipage}{.6\linewidth}
		$^a$ Akaike Information Criterion for the model evaluated as\\
		\hspace*{.5em} $-2\cdot(\text{logLik} - \text{npar})$. Smaller is better.
		
		$^b$ $p$ value $= Pr(>{\chi}^2)$
		
		$^{*}$ $p < 0.05$, $^{**}$ $p < 0.01$, $^{***}$ $p < 0.001$, $^{****}$ $p < 0.0001$\\
		\hspace*{.5em} Grey shadings indicate statistically significant fixed effects
		
	\end{minipage}
\end{table}

	\chapter{Competencies assessment with IAS}

\begin{custombox}
	{The content of this chapter has been adapted from the following articles with permission of all co-authors and publishers:}
	\begin{itemize}[noitemsep,nolistsep,leftmargin=.25in]
		
		\item Mangili, F., \textbf{Adorni, G.}, Piatti, A., Bonesana, C., and Antonucci, A. (2022). Modelling Assessment Rubrics through Bayesian Networks: a Pragmatic Approach. In \textit{2022 International Conference on Software, Telecommunications and Computer Networks} (SoftCOM) \cite{softcom}. 
		
		\item \textbf{Adorni, G.}, Mangili, F., Piatti, A., Bonesana, C., and Antonucci, A. (2023a). Rubric-based Learner Modelling via Noisy Gates Bayesian Networks for Computational Thinking Skills Assessment. \textit{Journal of Communications Software and Systems} \cite{adorni2023rubric}. 
		
		
	\end{itemize}
	As an author of these publications, my contribution involved:\\
	\textit{Conceptualisation, Methodology, Validation, Formal analysis, Investigation, Resources, Data curation, Writing -- original draft \& review \& editing, Visualisation.}
\end{custombox}

\section{Summary}

This chapter focuses on using our IAS to provide a more holistic evaluation of students' competencies, contributing primarily to RQ3.
Specifically, we apply the model to the data from the CAT unplugged, while for the virtual CAT, we integrate it into our tool for real-time assessment. We present the results and compare them with the task-specific assessment provided by the standard CAT score. Additionally, we compare the findings between the unplugged and virtual CAT assessments.

\section{Methodology}

The standard assessment, reflected by the CAT score (see \Cref{tab:unplugged_catscore,tab:catscore_virtual}), is task-specific and evaluates the pupil’s competence in completing individual tasks within a specific schema. While it provides insight into task performance, it does not offer a holistic view of the pupil's overall skills.
In contrast, the assessment provided by the IAS delivers a more comprehensive evaluation of the pupil's performance. 
The BN computes probabilistic outputs, or posterior probabilities, for each target skill, as well as any relevant supplementary skills.
These probabilities represent the likelihood that the pupil has mastered each skill based on their responses throughout the tasks.
The BN-based CAT score is derived from these posterior probabilities and provides a global evaluation of the pupil's skill level. It is calculated by summing the marginal posterior probabilities of the target skill nodes, offering an estimate of how many competence levels the pupil has mastered. This summary score reflects a broader, more nuanced assessment of the pupil's abilities, aggregating the inferences drawn from all 12 tasks and accounting for both task performance and the model's estimation of the pupil's skill mastery.

To compare the standard assessment with the one obtained from the IAS, we analyse the correlation between the \textit{BN-based CAT score} obtained with the IAS and the \textit{average CAT score}, a baseline measure calculated as the mean of the individual task-specific CAT scores, for both the unplugged and virtual CAT versions. 

Additionally, we present the inference times for the various models used in both the unplugged and virtual CAT versions. This information illustrates the computational efficiency of the BN-based evaluation process when comparing the different models.

To further investigate the models' effectiveness, we focus on a set of competence profiles selected from a group of students with interesting or notable performance patterns, comparing their average CAT score with the BN-based CAT scores derived from each model. This comparison highlights how the BN-based evaluation provides a more holistic and detailed understanding of pupil competencies.
Furthermore, we also present the posterior probabilities for each skill based on the student responses. These probabilities represent the models' estimation of the pupil’s proficiency, offering an overall estimate of skill mastery. 


\section{Evaluation of the unplugged CAT data}

\begin{figure*}[!ht]
 \centering
 \includegraphics[height=8cm]{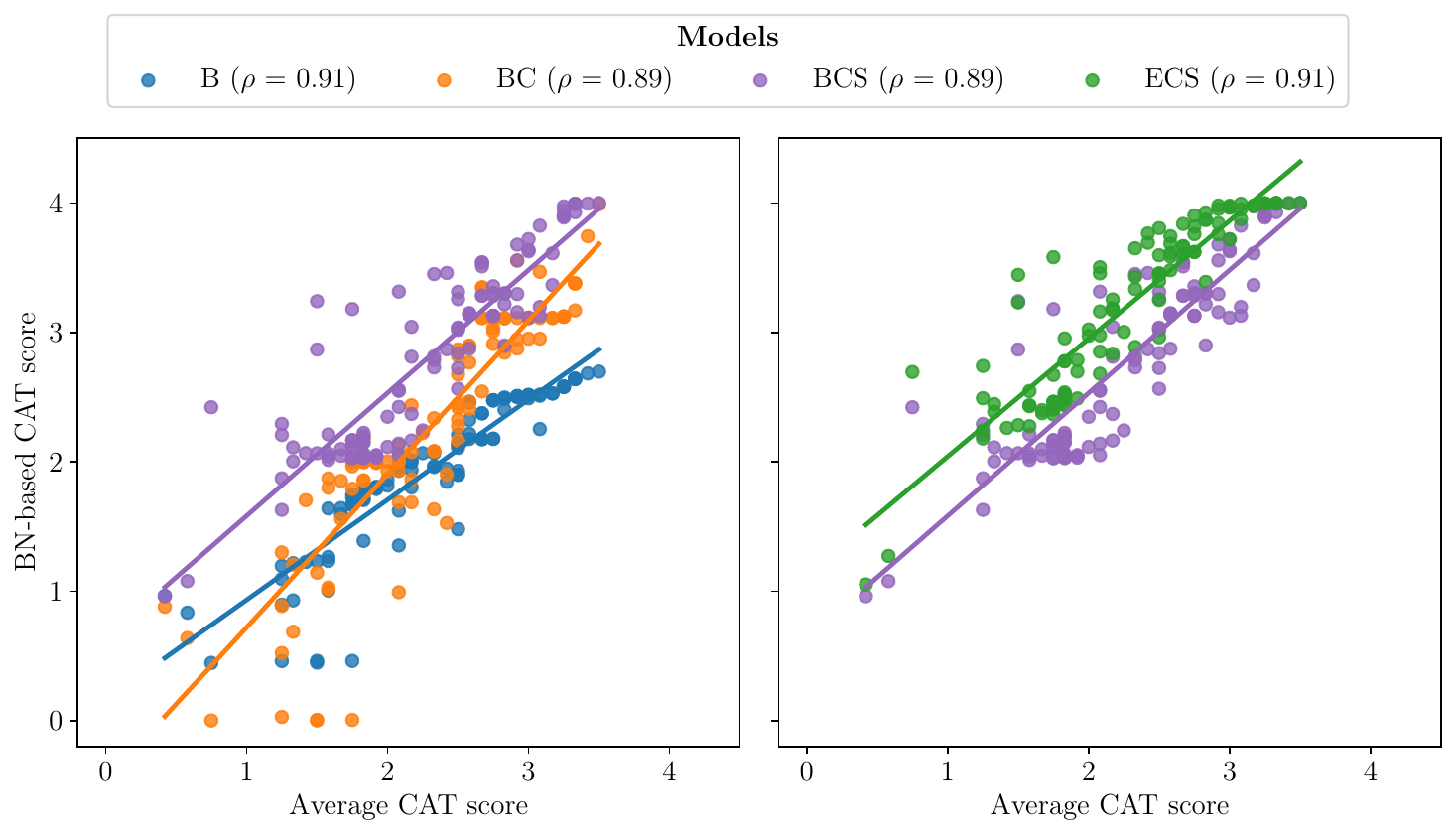}
  \mycaption{Comparison of BN-based and average CAT scores (unplugged CAT).}{
  	Scatterplots showing Pearson correlation coefficients $\rho$ between BN-based CAT scores of the four models and the average CAT score. On the left, the three baseline models are compared, and on the right, the baseline model with constraints and supplementary skills is compared to the enhanced model.
  	}
 \label{fig:scatter}
\end{figure*}

To evaluate model reliability and consistency with expert-based evaluation, we processed the responses of the 109 pupils from the unplugged CAT experimental study, calculating the posterior probabilities for the 9 target skills and, depending on the model, for the 10 supplementary skills.
\Cref{fig:scatter} illustrates the correlation between the average CAT score and the BN-based CAT scores for the following models: the baseline model (\texttt{Model B}), the baseline model with constraints (\texttt{Model BC}), the one which also includes the supplementary skills (\texttt{Model BCS}) and finally the enhanced model including both constraints and supplementary skills (\texttt{Model ECS}). 
The BN-based CAT score, originally in the $[0, 9]$ range, has been rescaled in the $[0, 4]$ range to align with the CAT score, defined on this scale, for easier and more direct comparison.
In all cases, the Pearson correlation coefficient ($\rho$) is very high, indicating a strong consistency between the BN-based and the expert-based assessment.


The inference times in \Cref{tab:times} reveal significant variation across the four models.
The models that assess only standard skills (\texttt{Model B} and \texttt{Model BC}) have relatively short inference times, with the total time for these models remaining under 30 seconds. In contrast, the models that incorporate supplementary skills (\texttt{Model BCS} and \texttt{Model ECS}) experience considerably longer inference times, exceeding 5 minutes for the total assessment.
This increase in time is expected, as the addition of supplementary skills leads to a more complex model with a larger parameter space.
\begin{table}[ht]
	\footnotesize
	\centering
	\mycaption{Comparison of inference times across models (unplugged CAT).}{For each model are reported the total inference time, the inference time per student (calculated by dividing the total inference time by the number of students), and the inference time per task (calculated by dividing the inference time per student by the number of tasks). \label{tab:times}}
	\setlength{\tabcolsep}{7mm}
	\begin{tabular}{l>{\centering\arraybackslash}m{2.5cm}>{\centering\arraybackslash}m{2.5cm}>{\centering\arraybackslash}m{2.5cm}}
		\toprule
		\multicolumn{1}{l}{{\textbf{Model}}} & \textbf{Total \linebreak inference time} & \textbf{Inference time per student}  & \textbf{Inference time per task} \\
		\midrule
		\texttt{B}            & \, 29.615s  & 0.272s & 0.023s \\
		\texttt{BC}           & \, 28.940s & 0.266s & 0.022s  \\
		\texttt{\texttt{BCS}} & 316.555s & 2.904s & 0.242s \\
		\texttt{\texttt{ECS}} & 306.517s  & 2.812s  & 0.234s \\          
		\bottomrule
	\end{tabular}
\end{table}
Despite the differences in total inference times, it is important to highlight that the average inference time per student remains relatively short across all models, consistently under 3 seconds.
These results suggest that all models are computationally efficient enough for real-time application, providing timely assessments for individual pupils without compromising the quality of evaluation.


\begin{table}[p]
\footnotesize
\centering
\mycaption{Tasks answers for a representative set of pupils (unplugged CAT).}{For each task (T1-T12) are reported the target and supplementary skills applied. Supplementary skills correspond to: $S_1$ - paint dot, $S_2$ - fill empty dots, $S_3$ - paint monochromatic rows or columns, $S_4$ - paint monochromatic squares, $S_5$ - paint monochromatic diagonals, $S_6$ - paint monochromatic L-shaped patterns, $S_7$ - paint monochromatic zig-zags, $S_8$ - paint polychromatic rows or columns, $S_9$ - paint polychromatic diagonals or zig-zags, and $S_{10}$ - repetition of a pattern. \label{tab:answers}}
\setlength{\tabcolsep}{3.1mm}
    \begin{subtable}{\linewidth}
    \centering
        \begin{tabular}{c>{\centering\arraybackslash}m{1.5cm}>{\centering\arraybackslash}m{1.5cm}>{\centering\arraybackslash}m{1.5cm}>{\centering\arraybackslash}m{1.5cm}>{\centering\arraybackslash}m{1.5cm}>{\centering\arraybackslash}m{1.5cm}}
        \toprule
        {\textbf{Pupil}}& \multicolumn{1}{c}{{\textbf{T1}}} & \multicolumn{1}{c}{{\textbf{T2}}} & \multicolumn{1}{c}{{\textbf{T3}}} & \multicolumn{1}{c}{{\textbf{T4}}} & \multicolumn{1}{c}{{\textbf{T5}}} & \multicolumn{1}{c}{{\textbf{T6}}} \\
        \midrule
        \multirow{2}{*}{21}  &   1D-V	& 1D-V	  & 2D-V	 & 1D-V	     & 1D-V	     & 1D-V	  \\
        &S$_{2}$	&S$_{2}$; S$_{6}$	&S$_{3}$; S$_{10}$	&S$_{3}$	&S$_{3}$; S$_{4}$	&S$_{6}$	\\ \midrule
        \multirow{2}{*}{33}  &   1D-V	& 1D-VS	  & 1D-VS	 & 1D-VSF	 & 1D-VS	 & 1D-VS    \\
                             &   S$_2$	& S$_2$; S$_{6}$	& S$_{3}$	& S$_{3}$	& S$_{3}$	 &  S$_1$; S$_{3}$  \\\midrule
        \multirow{2}{*}{81}  &   1D-V	& 1D-V	  & 1D-V	 & 1D-VS	 & 1D-V	     & 1D-V	  \\
                             &   S$_{2}$	& S$_{2}$; S$_{6}$	& S$_{3}$	& S$_{3}$	& S$_{3}$; S$_{4}$ & 	S$_{6}$	    \\ \midrule
        \multirow{2}{*}{92}  &   1D-V	& 1D-V	  & 1D-V	 & 1D-V	     & 1D-V	     & 1D-V	    \\
         &S$_{2}$ & S$_{2}$; S$_{6}$ & S$_{3}$ & S$_{3}$ & S$_{3}$; S$_{4}$ & S$_{6}$ \\
        \bottomrule
        \end{tabular}
    \end{subtable} 
    \\\vspace{.5cm}
    \begin{subtable}{\linewidth}
    \centering
        \begin{tabular}{c>{\centering\arraybackslash}m{1.5cm}>{\centering\arraybackslash}m{1.5cm}>{\centering\arraybackslash}m{1.5cm}>{\centering\arraybackslash}m{1.5cm}>{\centering\arraybackslash}m{1.5cm}>{\centering\arraybackslash}m{1.5cm}}
        \toprule
        {\textbf{Pupil}}& \multicolumn{1}{c}{{\textbf{T7}}} & \multicolumn{1}{c}{{\textbf{T8}}} & \multicolumn{1}{c}{{\textbf{T9}}} & \multicolumn{1}{c}{{\textbf{T10}}}& \multicolumn{1}{c}{{\textbf{T11}}}& \multicolumn{1}{c}{{\textbf{T12}}}\\
        \midrule
        \multirow{2}{*}{21}     & 2D-V	& 2D-V	& 2D-V	  & 1D-V	& 1D-V	& 1D-V \\
        &S$_{8}$; S$_{10}$	&S$_{1}$; S$_{5}$; S$_{10}$	&S$_{1}$; S$_{10}$	&S$_{1}$; S$_{4}$	&S$_{1}$	&S$_{1}$; S$_{5}$\\ \midrule
        \multirow{2}{*}{33}    & 1D-VS	& \multicolumn{1}{c}{\multirow{2}{*}{fail}}	& \multicolumn{1}{c}{\multirow{2}{*}{fail}}	  & \multicolumn{1}{c}{\multirow{2}{*}{fail}}	& \multicolumn{1}{c}{\multirow{2}{*}{fail}}	& \multicolumn{1}{c}{\multirow{2}{*}{fail}} \\
                             &  S$_{5}$	        & 	& & 	& 	&  \\\midrule
        \multirow{2}{*}{81}  &  2D-VSF	& 0D-VS	& 2D-V	  & \multicolumn{1}{c}{\multirow{2}{*}{fail}}	& \multicolumn{1}{c}{\multirow{2}{*}{fail}}	& \multicolumn{1}{c}{\multirow{2}{*}{fail}} \\
                             &  S$_1$; S$_{5}$; S$_{10}$	& S$_{1}$	& S$_{1}$; S$_{10}$& 	& 	&  \\ \midrule
        \multirow{2}{*}{92}  &  0D-V	& 0D-V	& 0D-VSF  & 1D-VS	& 2D-V	& 0D-V \\
         &S$_{1}$ & S$_{1}$ & S$_{1}$ &S$_{4}$; S$_{5}$ & S$_{1}$; S$_{10}$ & 	S$_{1}$\\
        \bottomrule
        \end{tabular}
    \end{subtable} 
\end{table}

When examining the posterior probabilities for individual students, we can gain deeper insights into their competence profiles, showcasing the interpretability of the model. By comparing the competence profiles produced by the four models considered, we can highlight significant differences in their assessments.
\Cref{tab:answers} reports the answers provided by four representative pupils, allowing for a more detailed understanding of their performance.
\Cref{tab:scores} compares the BN-based CAT scores to the traditional CAT scores, while \Cref{tab:posteriors,tab:posteriors_supp} display the corresponding posterior probabilities inferred by the models for the target skills and the supplementary skills, respectively. 

\begin{table}[ht]
	\footnotesize
	\centering
	\mycaptiontitle{Comparison of the average CAT score and the BN-based CAT scores across models for a representative subset of pupils (unplugged CAT).}\label{tab:scores}
	\begin{tabular}{ccl}
		\toprule
		{{\textbf{Pupil}}} & \textbf{Average CAT score} & \textbf{BN-based CAT score}  \\ 
		\midrule
		\multirow{4}{*}{{21}} &\multirow{4}{*}{3.30} &  2.23 (\texttt{Model B}) \\
		&                     & 1.65 (\texttt{Model BC})\\
		&                     & 1.98 (\texttt{Model BCS}) \\
		&                     & 2.00 (\texttt{Model ECS})\\ \midrule
		\multirow{4}{*}{{33}} &\multirow{4}{*}{0.75} & 2.00 (\texttt{Model B}) \\
		&                      & 0.00 (\texttt{Model BC})\\
		&                      & 1.33 (\texttt{Model BCS}) \\
		&                      & 1.47 (\texttt{Model ECS})\\ \midrule
		\multirow{4}{*}{{81}} &\multirow{4}{*}{1.75} & 2.90 (\texttt{Model B}) \\
		&                      & 0.07 (\texttt{Model BC})\\
		&                      & 1.62 (\texttt{Model BCS}) \\
		&                      & 1.82 (\texttt{Model ECS})\\ \midrule
		\multirow{4}{*}{{92}} &\multirow{4}{*}{2.50}  & 1.77 (\texttt{Model B}) \\
		&                      & 1.42 (\texttt{Model BC})\\
		&                      & 1.59 (\texttt{Model BCS}) \\
		&                      & 1.79 (\texttt{Model ECS})\\
		\bottomrule
	\end{tabular}
\end{table}

\newcolumntype{C}{>{\centering\arraybackslash}m{.8cm}}

\begin{table}[p]
\footnotesize
\centering
\mycaptiontitle{Posterior probabilities for target skills across models for a representative subset of pupils (unplugged CAT).}
  \label{tab:posteriors}
\setlength{\tabcolsep}{2.35mm}
\begin{tabular}{clCCCCCCCCC}
\toprule
\textbf{Pupil} &{\textbf{Model}}& { $X_{11} $}&{ $X_{12} $}&{ $X_{13} $}&{ $X_{21} $}&{ $X_{22}$}&{$X_{23}$}&{ $X_{31}$}&{ $X_{32}$}&{ $X_{33}$}\\ 
\midrule
\multirow{4}{*}{{21}} & \texttt{B}  & \gradient{0.50}	&	\gradient{0.51}	&	\gradient{0.67}	&	\gradient{0.51}	&	\gradient{0.57}	&	\gradient{0.96}	&	\gradient{0.59}	&	\gradient{0.83}	&	\gradient{0.80}\\
                              & \texttt{BC} & \gradient{1.00}	&	\gradient{1.00}	&	\gradient{1.00}	&	\gradient{1.00}	&	\gradient{1.00}	&	\gradient{1.00}	&	\gradient{0.69}	&	\gradient{0.38}	&	\gradient{0.07}\\
                              & \texttt{BCS} & \gradient{1.00}	&	\gradient{1.00}	&	\gradient{1.00}	&	\gradient{1.00}	&	\gradient{1.00}	&	\gradient{1.00}	&	\gradient{0.97}	&	\gradient{0.95}	&	\gradient{0.92}\\
                              & \texttt{ECS} & \gradient{1.00}	&	\gradient{1.00}	&	\gradient{1.00}	&	\gradient{1.00}	&	\gradient{1.00}	&	\gradient{1.00}	&	\gradient{1.00}	&	\gradient{1.00}	&	\gradient{1.00}\\ \midrule
\multirow{4}{*}{{33}} & \texttt{B}  & \gradient{0.00}	&	\gradient{0.00}	&	\gradient{0.00}	&	\gradient{0.00}	&	\gradient{1.00}	&	\gradient{0.00}	&	\gradient{0.00}	&	\gradient{0.00}	&	\gradient{0.00}\\
                              & \texttt{BC} & \gradient{0.00}	&	\gradient{0.00}	&	\gradient{0.00}	&	\gradient{0.00}	&	\gradient{0.00}	&	\gradient{0.00}	&	\gradient{0.00}	&	\gradient{0.00}	&	\gradient{0.00}\\
                              & \texttt{BCS} & \gradient{1.00}	&	\gradient{1.00}	&	\gradient{0.52}	&	\gradient{1.00}	&	\gradient{1.00}	&	\gradient{0.05}	&	\gradient{0.59}	&	\gradient{0.30}	&	\gradient{0.00}\\
                              & \texttt{ECS} & \gradient{1.00}	&	\gradient{1.00}	&	\gradient{0.69}	&	\gradient{1.00}	&	\gradient{1.00}	&	\gradient{0.39}	&	\gradient{0.63}	&	\gradient{0.33}	&	\gradient{0.03}\\ \midrule
\multirow{4}{*}{{81}} & \texttt{B}  & \gradient{0.03}	&	\gradient{0.00}	&	\gradient{0.00}	&	\gradient{0.00}	&	\gradient{0.00}	&	\gradient{1.00}	&	\gradient{0.00}	&	\gradient{0.00}	&	\gradient{0.00}\\
                              & \texttt{BC} & \gradient{0.01}	&	\gradient{0.00}	&	\gradient{0.00}	&	\gradient{0.00}	&	\gradient{0.00}	&	\gradient{0.00}	&	\gradient{0.00}	&	\gradient{0.00}	&	\gradient{0.00}\\
                              & \texttt{BCS} & \gradient{1.00}	&	\gradient{1.00}	&	\gradient{1.00}	&	\gradient{1.00}	&	\gradient{1.00}	&	\gradient{1.00}	&	\gradient{0.91}	&	\gradient{0.21}	&\gradient{0.03}\\
                              & \texttt{ECS} & \gradient{1.00}	&	\gradient{1.00}	&	\gradient{1.00}	&	\gradient{1.00}	&	\gradient{1.00}	&	\gradient{1.00}	&	\gradient{0.95}	&	\gradient{0.67}	&	\gradient{0.44}\\ \midrule
\multirow{4}{*}{{92}} & \texttt{B}  & \gradient{0.55}	&	\gradient{0.40}	&	\gradient{0.41}	&	\gradient{0.33}	&	\gradient{0.13}	&	\gradient{1.00}	&	\gradient{0.46}	&	\gradient{0.05}	&	\gradient{0.00}\\
                              & \texttt{BC} & \gradient{1.00}	&	\gradient{0.99}	&	\gradient{0.99}	&	\gradient{0.76}	&	\gradient{0.70}	&	\gradient{0.68}	&	\gradient{0.13}	&	\gradient{0.00}	&	\gradient{0.00}\\
                              & \texttt{BCS} & \gradient{1.00}	&	\gradient{1.00}	&	\gradient{1.00}	&	\gradient{1.00}	&	\gradient{1.00}	&	\gradient{1.00}	&	\gradient{0.59}	&	\gradient{0.19}	&	\gradient{0.01}\\
                              & \texttt{ECS} & \gradient{1.00}	&	\gradient{1.00}	&	\gradient{1.00}	&	\gradient{1.00}	&	\gradient{1.00}	&	\gradient{1.00}	&	\gradient{0.79}	&	\gradient{0.58}	&	\gradient{0.41}\\ 
                              \bottomrule
\end{tabular}
	\\\vspace{.5cm}
\footnotesize
\centering
\mycaptiontitle{Posterior probabilities for supplementary skills across models for a representative subset of pupils (unplugged CAT).}
 \label{tab:posteriors_supp}
\setlength{\tabcolsep}{1.8mm}
\begin{tabular}{ccCCCCCCCCCC}
\toprule
{\textbf{Pupil}}& {\textbf{Model}}& {S$_{1}$}&{S$_{2}$}&{S$_{3}$}&{S$_{4}$}&{S$_{5}$}&{S$_{6}$}&{S$_{7}$}&{S$_{8}$}&{S$_{9}$}&{S$_{10}$} \\ 
\midrule
\multirow{2}{*}{{21}} & \texttt{BCS} & \gradient{1.00} &	\gradient{1.00}	& \gradient{1.00} &	\gradient{1.00} &	\gradient{1.00} &	\gradient{1.00} &	\gradient{0.40} &	\gradient{1.00} &	\gradient{0.26} &	\gradient{1.00}\\
                       & \texttt{ECS} & \gradient{1.00} &	\gradient{1.00}	& \gradient{1.00} &	\gradient{1.00} &	\gradient{1.00} &	\gradient{1.00} &	\gradient{0.52} &	\gradient{1.00} &	\gradient{0.38} &	\gradient{1.00}\\ \midrule
\multirow{2}{*}{{33}} & \texttt{BCS} & \gradient{1.00} & \gradient{1.00} & \gradient{1.00} & \gradient{0.42} & \gradient{1.00} & \gradient{1.00} & \gradient{0.38} & \gradient{0.15} & \gradient{0.16} & \gradient{0.13} \\
                       & \texttt{ECS} & \gradient{1.00} & \gradient{1.00} & \gradient{1.00} & \gradient{0.43} & \gradient{1.00} & \gradient{1.00} & \gradient{0.42} & \gradient{0.21} & \gradient{0.22} & \gradient{0.19} \\ \midrule       
\multirow{2}{*}{{81}} & \texttt{BCS} & \gradient{1.00} & \gradient{1.00} & \gradient{1.00} & \gradient{1.00} & \gradient{1.00} & \gradient{1.00} & \gradient{0.36} & \gradient{0.34} & \gradient{0.31} & \gradient{1.00} \\
                       & \texttt{ECS} & \gradient{1.00} & \gradient{1.00} & \gradient{1.00} & \gradient{1.00} & \gradient{1.00} & \gradient{1.00} & \gradient{0.39} & \gradient{0.41} & \gradient{0.35} & \gradient{1.00} \\ \midrule
\multirow{2}{*}{{92}} & \texttt{BCS} & \gradient{1.00}	& \gradient{1.00}	& \gradient{1.00}	& \gradient{1.00}	& \gradient{1.00}	& \gradient{1.00}	& \gradient{0.34}	& \gradient{0.31}	& \gradient{0.31}	& \gradient{1.00}\\
                       & \texttt{ECS} & \gradient{1.00}	& \gradient{1.00}	& \gradient{1.00}	& \gradient{1.00}	& \gradient{1.00}	& \gradient{1.00}	& \gradient{0.36}	& \gradient{0.35}	& \gradient{0.35}	& \gradient{1.00}\\
                       \bottomrule
\end{tabular}
\end{table}

Pupil 21 is a high-performing student, in terms of CAT and BN-based scores,  who consistently used complex interactions for all CAT schemas and primarily used 2D algorithms.

Pupils 33 and 81 cannot be considered high-performing since they failed to complete some of the CAT schemas.
Pupil 33 solved only the first seven schemas, always using 1D algorithms and almost always relying on the VS artefact. In comparison, student 81 was successful in the first nine schemas, where he applied different algorithms and artefacts, but mostly the 1D-V. 
For both students 33 and 81, the BN-based CAT scores predicted by the four models vary significantly, indicating that the models may be producing different predictions of their abilities. The difference between the original and BN-based CAT scores is inconsistent across the models. For both students, the largest difference between the original and BN-based CAT scores is observed in \texttt{Model B}, which predicts a much higher score for both students.
On the other hand, \texttt{Model BC} predicts a meagre BN-based CAT score close to 0 for both students, indicating that this model may not be the most accurate for these particular students. This suggests that other models may be better suited for predicting their performance on the CAT.

Pupil 92's performance was strong, as he successfully completed all 12 tasks using different skill levels. 
He solved the first six schemas with the 1D-V skill, reduced the algorithm's complexity in the following ones, changed artefact for some of the more complex tasks, and applied the highest level skill, 2D-V, in a tricky schema.
Regarding the BN-based scores, all four models predicted a lower BN-based CAT score for student 92 than the original CAT score, although the differences were not as large as those observed for students 33 and 81. This suggests that student 92 is a relatively strong performer overall, but there is potential for improvement in his performance.

For all students, the baseline model assigns posterior probabilities equal or very close to one to the most used skill levels but fails to recognise that they also possess lower level skills, to which rather small probabilities, eventually equal to zero, as for the worst performing students 33 and 81, are assigned. 
On the one hand, when the ordering between skills is explicitly imposed, this problem is solved: the probabilities of lower skills increase, and those of higher skills decrease. 
This may lead to an excessive penalisation of higher skills, as in the case of pupils 33 and 81, where, as a consequence of the repeated failures in applying even the lowest competence level, \texttt{Model BC} decides for the total absence of the competencies under examination, returning a posterior probability of zero, even for the skills successfully used by the students in several schemas. These inferences look too severe for these situations,  where an expert would rather attribute the errors to the specific difficulties of the failed tasks rather than the total lack of algorithmic skills. 
On the other hand, when the supplementary skills are included in the assessment (\texttt{Model BCS} and \texttt{Model ECS}), this issue is solved, and the result of the posterior inference is consistent with the hierarchy of competencies defined by the rubric and the observations collected. 
In this case, the model understands that the failure follows from a lack of the supplementary skills necessary to solve specific schemas with more complex algorithms and not from a lack of target skills.

For instance, according to \texttt{Model BCS} and \texttt{Model ECS}, pupil 21 is likely to miss monochromatic zigzags (S$_7$) and polychromatic diagonals and zigzags (S$_9$), justifying the failure in applying the possessed 2D competence in schemas related to these supplementary skills.
Finally, employing more elaborate models, such as \texttt{Model ECS}, may, in some cases, reward the ability to apply high-level skills in more complex tasks, i.e., those assigned with higher inhibition probabilities, such as for pupil 92 who managed to solve schema T11, a difficult one according to the parameters' elicitation in \Cref{fig:its}, using a 2D-V skill and thus 2D algorithms are given a much higher probability by \texttt{Model ECS} than by \texttt{Model BCS}.

\section{Real-time evaluation of the virtual CAT data}

\begin{figure*}[!ht]
	\centering
	\includegraphics[height=8cm]{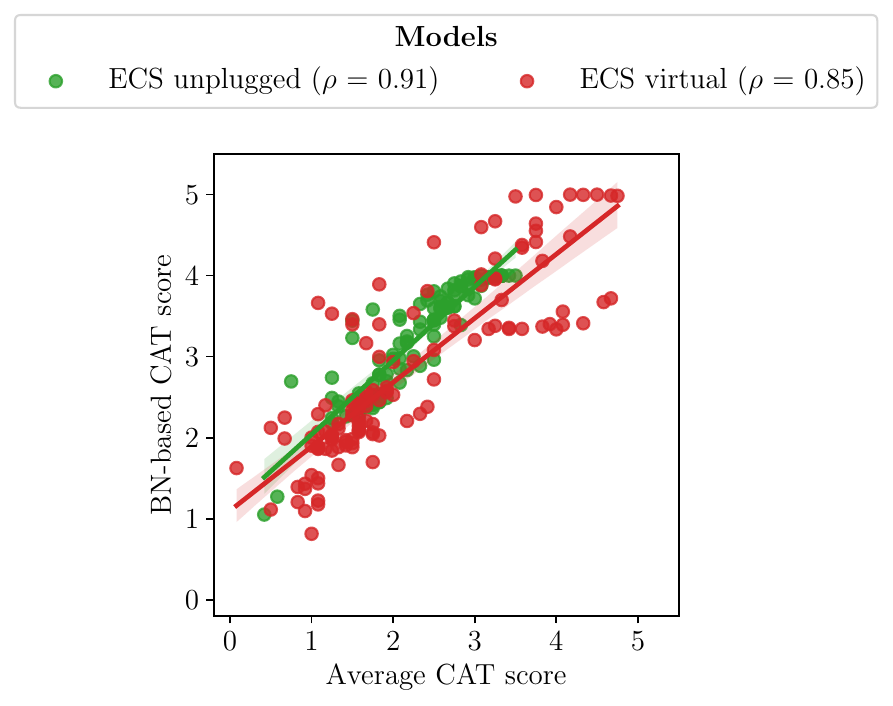}
	  \mycaption{Comparison of BN-based and average CAT scores (unplugged and virtual CAT).}{
		Scatterplot showing Pearson correlation coefficients $\rho$ between BN-based CAT scores of the enhanced model, for both unplugged and virtual CAT, and the average CAT score.}
	\label{fig:scatter_comparison}
\end{figure*}
For the virtual CAT, we considered only the enhanced model, including both constraints and supplementary skills (\texttt{Model ECS}).
To assess its reliability and consistency with expert-based evaluation, we processed the responses of the 129 pupils from the virtual CAT experimental study, calculating the posterior probabilities for the 12 target skills and the 14 supplementary skills.
\Cref{fig:scatter_comparison} illustrates the correlation between the average CAT score and the BN-based CAT scores for the enhanced model in both the unplugged and virtual settings.
In this case, the BN-based CAT score, originally ranging from $[0, 12]$, was rescaled to the $[0, 5]$ range to match the CAT score.
For the virtual CAT, the Pearson correlation coefficient ($\rho$) is slightly lower at 0.85 but still demonstrates a strong alignment with the expert-based assessment.  This suggests that \texttt{Model ECS} is reliable and consistent with expert evaluation in both contexts. A lower correlation does not necessarily imply a poorer model but rather indicates that the model remains consistent, albeit with some minor differences in the alignment between the two evaluations.

\begin{table}[!hb]
	\footnotesize
	\centering
		\mycaption{Comparison of inference times across models (unplugged vs virtual CAT).}{For the two enhanced models are reported the total inference time, the inference time per student (calculated by dividing the total inference time by the number of students), and the inference time per task (calculated by dividing the inference time per student by the number of tasks). \label{tab:times_comparison}}
	\setlength{\tabcolsep}{6mm}
	\begin{tabular}{l>{\centering\arraybackslash}m{2.5cm}>{\centering\arraybackslash}m{2.5cm}>{\centering\arraybackslash}m{2.5cm}}
		\toprule
		\multicolumn{1}{l}{{\textbf{Model}}} & \textbf{Total \linebreak inference time} & \textbf{Inference time per student}  & \textbf{Inference time per task} \\
		\midrule
		{\texttt{ECS} unplugged} & \, 306.517s  & \, 2.812s  & 0.234s \\          
		{\texttt{ECS} virtual} & 3229.390s  & 25.034s  & 2.086s \\       
		\bottomrule
	\end{tabular}
\end{table}

The comparison of inference times in \Cref{tab:times_comparison} reveals significant variation across the two domains.
For the unplugged CAT, the computational cost of the inference is relatively low, whereas, for the virtual CAT, the total inference time increases by a factor of approximately 10 (from 5 minutes to 53 minutes), with an inference time per student of 25 seconds and an inference time per task of 2 seconds.
The higher inference time for the virtual CAT is primarily due to the increased complexity of the model, which includes a larger set of target skills (12 in total) and additional supplementary skills (14). This increase in the number of variables requires more computational resources and time to calculate the posterior probabilities for each student’s performance. 
Although the inference times for the virtual CAT model appear significantly higher, it's important to note that the IAS does not perform post-hoc inference on the data as in the unplugged CAT. Instead, being integrated within the virtual CAT app, inferences are made in real time as students complete each task.
This means they do not have to wait for the entire estimated time per student for the assessment to be processed, only for the estimated inference time per task.

We continue by analysing the model's results on a group of students to gain insights into their competence profiles.
\Cref{tab:answers_virtual} reports the answers provided by four representative pupils,
\Cref{tab:scores_virtual} compares their average CAT scores to the BN-based CAT scores, and \Cref{tab:posteriors_virtual,tab:posteriors_supp_virtual} display the corresponding posterior probabilities inferred by the models for the target skills and the supplementary skills, respectively. 
\begin{table}[ht]
	\footnotesize
	\centering
\mycaptiontitle{Comparison of the average CAT score and the BN-based CAT score for the enhanced model for a representative subset of pupils (virtual CAT).}\label{tab:scores_virtual}
\begin{tabular}{ccc}
	\toprule
	{{\textbf{Pupil}}} & \textbf{Average CAT score} & \textbf{BN-based CAT score}  \\ 
		\midrule
		  {{5}} & {1.50} &  2.34 \\ \midrule
		{{89}} & {0.08} & 1.63 \\ \midrule
		{{39}} & {2.42} & 3.81 \\ \midrule
		{{65}} & {4.50}  & 5.00\\
		\bottomrule
	\end{tabular}
\end{table}

\begin{table}[p]
	\footnotesize
	\centering
	\mycaption{Tasks answers for a representative set of pupils (virtual CAT).}{For each task (T1-T12) are reported the target and supplementary skills applied. Supplementary skills correspond to: $S_1$ - paint dot, $S_2$ - fill empty dots, $S_3$ - paint monochromatic custom patterns, $S_4$ - paint monochromatic rows or columns, $S_5$ - paint monochromatic squares, $S_6$ - paint monochromatic diagonals, $S_7$ - paint monochromatic L-shaped patterns, $S_8$ - paint monochromatic zig-zags, $S_9$ - paint polychromatic custom patterns, $S_{10}$ - paint polychromatic rows or columns, $S_{11}$ - paint polychromatic squares or L-shaped patterns, $S_{12}$ - paint polychromatic diagonals or zig-zags, and $S_{13}$ - repetition or copy of patterns, $S_{14}$ - mirror patterns. \label{tab:answers_virtual}}
	\setlength{\tabcolsep}{4.2mm}
	\begin{subtable}{\linewidth}
		\centering
		\begin{tabular}{ccccccc}
			\toprule
			{\textbf{Pupil}}& \multicolumn{1}{c}{{\textbf{T1}}} & \multicolumn{1}{c}{{\textbf{T2}}} & \multicolumn{1}{c}{{\textbf{T3}}} & \multicolumn{1}{c}{{\textbf{T4}}} & \multicolumn{1}{c}{{\textbf{T5}}} & \multicolumn{1}{c}{{\textbf{T6}}} \\
			\midrule
			\multirow{2}{*}{5}  &   1D-G	& 1D-G	  & 2D-GF	 & 1D-G	     & 1D-G	     & 1D-GF	  \\
			&S$_{2}$	&S$_{3}$ & S$_{4}$	&S$_{4}$ &S$_{3}$; S$_{4}$; S$_{5}$	&S$_{3}$; S$_{4}$; S$_{7}$	\\ \midrule
			\multirow{2}{*}{89}  &   1D-G	& 1D-GF	  & 1D-GF	 & 1D-GF	 & 1D-GF	 & \multicolumn{1}{c}{\multirow{2}{*}{fail}}    \\
			&S$_{2}$&S$_{2}$;S$_{3}$&S$_{2}$;S$_{3}$&S$_{2}$;S$_{3}$&S$_{2}$;S$_{4}$;S$_{5}$&\\ \midrule
			\multirow{2}{*}{39}  &   1D-PF	& 0D-P	  & 1D-GF	 & 1D-PF	 & 1D-GF	     & 1D-G	  \\
			& S$_{1}$;S$_{2}$&S$_{1}$&S$_{1}$;S$_{4}$&S$_{4}$&S$_{4}$;S$_{5}$&S$_{3}$;S$_{7}$\\    \midrule
			\multirow{2}{*}{65}  &   1D-P	& 2D-P	  & 2D-P	 & 2D-P	     & 1D-P	     & 1D-P	    \\
			&S$_{2}$&S$_{2}$;S$_{4}$;S$_{13}$&S$_{2}$;S$_{4}$;S$_{13}$&S$_{2}$;S$_{4}$;S$_{13}$&S$_{4}$;S$_{5}$&S$_{7}$\\
			\bottomrule
		\end{tabular}
	\end{subtable} 
	\\\vspace{.5cm}
	\setlength{\tabcolsep}{2.5mm}
	\begin{subtable}{\linewidth}
		\centering
		\begin{tabular}{ccccccc}
			\toprule
			{\textbf{Pupil}}& \multicolumn{1}{c}{{\textbf{T7}}} & \multicolumn{1}{c}{{\textbf{T8}}} & \multicolumn{1}{c}{{\textbf{T9}}} & \multicolumn{1}{c}{{\textbf{T10}}}& \multicolumn{1}{c}{{\textbf{T11}}}& \multicolumn{1}{c}{{\textbf{T12}}}\\
			\midrule
			\multirow{2}{*}{5}     & 0D-G	& 2D-GF	& 2D-GF	  & 1D-GF	& 1D-GF	& 1D-GF \\
			&S$_{1}$ &S$_{1}$; S$_{6}$; S$_{8}$; S$_{12}$ 	&S$_{1}$; S$_{4}$; S$_{10}$	&S$_{3}$ &S$_{1}$; S$_{3}$; S$_{4}$	&S$_{1}$; S$_{6}$; S$_{8}$\\ \midrule
			\multirow{2}{*}{89}    & \multicolumn{1}{c}{\multirow{2}{*}{fail}}	& \multicolumn{1}{c}{\multirow{2}{*}{fail}}	& \multicolumn{1}{c}{\multirow{2}{*}{fail}}	  &  1D-GF	& \multicolumn{1}{c}{\multirow{2}{*}{fail}}	& \multicolumn{1}{c}{\multirow{2}{*}{fail}} \\
			&        & 	& & S$_{3}$;S$_{4}$;S$_{5}$	& 	&  \\\midrule
			\multirow{2}{*}{39}  &  2D-PF	& 2D-PF	& 2D-PF	  & 1D-G	& 2D-GF	& 0D-GF \\
			&S$_{10}$&S$_{10}$&S$_{10}$&S$_{4}$;S$_{5}$;S$_{6}$;S$_{7}$&S$_{1}$;S$_{4}$;S$_{7}$;S$_{10}$&  S$_{1}$\\ \midrule
			\multirow{2}{*}{65}  &  2D-PF	& 2D-P	& 2D-P  & 2D-PF	& 2D-P	& 2D-PF \\
			&S$_{10}$;S$_{11}$&S$_{10}$&S$_{10}$&S$_{1}$;S$_{11}$;S$_{12}$;S$_{13}$&S$_{4}$;S$_{10}$;S$_{14}$&S$_{8}$;S$_{11}$\\
			\bottomrule
		\end{tabular}
	\end{subtable} 
\end{table}

Pupil 5 can be considered an average performer, with a CAT score of 1.5/5 assigned by the expert and a slightly higher score according to the IAS. This score suggests some level of competence, though Pupil 5 primarily relied on simpler 1D algorithms. The pupil only engaged with 2D twice and used a 0D algorithm once. Despite these limitations, he was able to successfully complete all tasks, relying on the gesture-based interface, often accompanied by visual feedback.
This approach, while useful, could indicate a preference for more intuitive, immediate methods of interaction rather than a deeper engagement with different strategies or more abstract problem-solving techniques. 

In contrast, pupil 89's scores are definitely lower, indicating a significant struggle with the tasks. This pupil failed multiple tasks and, when successful, consistently relied on basic 1D-GF strategies.

Pupil 39, on the other hand, achieved higher scores, indicating a high level of competence and application of skills. He effectively applied all interaction methods and produced algorithms of varying complexity, even creating some complex 2D algorithms for the most difficult tasks, demonstrating a diverse use of supplementary skills.

Finally, Pupil 65 achieved the highest score, indicating exceptional performance. This pupil demonstrated mastery across all tasks, applying a wide range of supplementary skills, mostly producing 2D algorithms using the most complex interaction methods.

\begin{table}[htb]
	\footnotesize
	\centering
	\mycaptiontitle{Posterior probabilities for target skills in \texttt{Model ECS} for a representative subset of pupils (virtual CAT).}
	\label{tab:posteriors_virtual}
	\setlength{\tabcolsep}{2.35mm}
	\begin{tabular}{cCCCCCC}
		\toprule
		\textbf{Pupil} & { $X_{11} $}&{ $X_{12} $}&{ $X_{13} $}&{ $X_{14} $}&{ $X_{21} $}&{ $X_{22}$}\\ 
		\midrule
		5   & \gradient{1.00}& \gradient{1.00}	&	\gradient{0.53}	&	\gradient{0.19}	&	\gradient{1.00}	&	\gradient{1.00}	\\ \midrule
		{{89}}   & \gradient{1.00}& \gradient{0.71}	&	\gradient{0.45}	&	\gradient{0.22}	&	\gradient{1.00}	&	\gradient{0.28}	\\ \midrule
		{{39}}   & \gradient{1.00}& \gradient{1.00}	&	\gradient{0.99}	&	\gradient{0.85}	&	\gradient{1.00}	&	\gradient{1.00}	\\ \midrule
		{{65}}   & \gradient{1.00}& \gradient{1.00}& \gradient{1.00}& \gradient{1.00}& \gradient{1.00}& \gradient{1.00}\\
		\bottomrule
	\end{tabular}
		\\\vspace{.5cm}
		
			\begin{tabular}{cCCCCCC}
			\toprule
			\textbf{Pupil} &{$X_{23}$}&{$X_{24}$}&{ $X_{31}$}&{ $X_{32}$}&{ $X_{33}$}&{ $X_{34}$}\\ 
			\midrule
			5   &	\gradient{0.00}	&	\gradient{0.00}	&	\gradient{0.87}	&	\gradient{0.02}&	\gradient{0.00}&	\gradient{0.00}\\ \midrule
			{{89}}   & 	\gradient{0.00}	&	\gradient{0.00}	&	\gradient{0.24}	&	\gradient{0.01}&	\gradient{0.00}&	\gradient{0.00}\\ \midrule
			{{39}}  &	\gradient{0.95}	&	\gradient{0.02}	&	\gradient{0.90}	&	\gradient{0.74}&	\gradient{0.70}&	\gradient{0.00}\\ \midrule
			{{65}}   & \gradient{1.00}& \gradient{1.00}& \gradient{1.00}& \gradient{1.00}& \gradient{1.00}& \gradient{1.00}\\
			\bottomrule
		\end{tabular}
\end{table}

\begin{table}[htb]
	\footnotesize
	\centering
		\mycaptiontitle{Posterior probabilities for supplementary skills in \texttt{Model ECS} for a representative subset of pupils (virtual CAT).}
\label{tab:posteriors_supp_virtual}
	\setlength{\tabcolsep}{1.8mm}
	\begin{tabular}{cCCCCCCC}
		\toprule
		{\textbf{Pupil}}&  {S$_{1}$}&{S$_{2}$}&{S$_{3}$}&{S$_{4}$}&{S$_{5}$}&{S$_{6}$}&{S$_{7}$}\\ 
		\midrule
		{{5}} & \gradient{1.00} &	\gradient{1.00}	& \gradient{1.00} &	\gradient{1.00} &	\gradient{1.00} &	\gradient{1.00} &	\gradient{1.00} \\ \midrule
		{{89}} & \gradient{1.00} &\gradient{1.00}  &\gradient{1.00} &\gradient{1.00} &\gradient{1.00} &\gradient{0.32} &\gradient{0.44} \\ \midrule       
	    {{39}} &  \gradient{1.00} &\gradient{1.00}  &\gradient{1.00} &\gradient{1.00} &\gradient{1.00} &\gradient{1.00} &\gradient{1.00} \\ \midrule       
		{{65}} & \gradient{1.00} &\gradient{1.00}  &\gradient{0.51} &\gradient{1.00} &\gradient{1.00} &\gradient{0.49} &\gradient{1.00}  \\
		\bottomrule
	\end{tabular}
		\\\vspace{.5cm}
	\begin{tabular}{cCCCCCCC}
		\toprule
		{\textbf{Pupil}}& 
		{S$_{8}$}&{S$_{9}$}&{S$_{10}$} &{S$_{11}$} &{S$_{12}$} &{S$_{13}$} &{S$_{14}$} \\ 
		\midrule
		{{5}} &	\gradient{1.00} &	\gradient{0.32} &	\gradient{1.00} & \gradient{0.32}	& \gradient{1.00}	& \gradient{0.29}	& \gradient{0.29} \\ \midrule
		{{89}} &\gradient{0.40} &\gradient{0.27} &\gradient{0.28} &\gradient{0.27} &\gradient{0.28} &\gradient{0.25} &\gradient{0.25} \\ \midrule       
		{{39}} &\gradient{0.42} &\gradient{0.35} &\gradient{1.00} &\gradient{0.35} &\gradient{0.31} &\gradient{0.28} &\gradient{0.28} \\ \midrule       
		{{65}} &\gradient{1.00} &\gradient{0.38} &\gradient{1.00} &\gradient{1.00} &\gradient{0.39} &\gradient{1.00} &\gradient{1.00} \\
		\bottomrule
	\end{tabular}
\end{table}

Based on posterior probabilities, given the observed data, we analyse the likelihood of each skill being present for each student.
Pupil 6 demonstrates strong competence in 0D and 1D algorithmic tasks using the gesture interface. However, their ability to perform more complex 2D tasks is limited. They show minimal exploration of complex polychromatic patterns and lack proficiency in most 2D patterns, such as copying and mirroring. Successful application of 2D skills appears to be possible only when heavily relying on visual feedback.

Pupil 89 primarily uses the simplest interface, demonstrating strong performance in 0D algorithms, weak performance in 1D, and nearly no proficiency in 2D, with a clear preference for simple patterns and a sharp decline in competence with more complex patterns.

Pupil 39 demonstrates well-rounded competence, effectively using gesture-based interfaces to create algorithms of varying complexity, with high proficiency in nearly all 0D and 1D skills and good performance in 2D tasks.
However, he is less skilled with the programming interface and can only create simple algorithms when supported by visual feedback. While he shows strong mastery of simple patterns, except for zigzag, his performance declines significantly for complex patterns and 2D-related skills.

Finally, pupil 65 demonstrates exceptional competence, with maximum probabilities across all skills, showcasing a comprehensive and versatile approach that includes advanced 2D patterns, with the exception of diagonal and custom patterns and interactions, highlighting his excellence in exploring complex and abstract strategies.

	
	\part{Discussion and Conclusion}\label{part:dicussion}
	
	\chapter{Summary and interpretation of findings}\label{chap:summary-interpretation}
This chapter outlines the key findings of the study, with each subsection focusing on a research question, summarising results and comparing them to related work in the field.


\section[Developing an age-based competence model for CT]{Developing an age-based competence model for Computational Thinking}

To answer the first research question, ``How can a competence model for CT be defined to assess skills across different age groups and educational contexts?'', we developed two complementary frameworks. 

To begin with, we adopted an alternative approach to classical CT models, which often emphasise individual cognitive abilities while overlooking critical factors such as the social and environmental influences and the developmental progression that shape cognitive processes during learning activities \citep{shute2017demystifying,tikva_mapping_2021,roth2013situated,heersmink2013taxonomy,RomnGonzlez2024}.
Based on the theory of situated cognition, we developed the CT-cube, a theoretical framework that extends existing CT models by addressing both the developmental and context-specific nature of CT. It integrates the cognitive processes essential for problem-solving with social and environmental factors, thus facilitating the design of CT activities that capture the multidimensional nature of CT and assessment of CT skills across different developmental stages and contexts.
To validate its applicability, we applied this framework to design the CAT, an unplugged activity aimed at assessing AT in compulsory school pupils. 
The results indicate that the CT-cube effectively supports the assessment of CT skills in authentic classroom scenarios, highlighting its potential as a competence model for creating activities that reliably assess abilities across varied educational contexts and developmental stages.

Following this, we developed a second framework that builds upon the principles established in the CT-cube, aiming to address the lack of comprehensive competence models for CT from a different perspective. 
While many existing models focus on defining CT skills and competencies, they often fail to provide sufficient guidance on designing CTPs that effectively foster and assess these skills \citep{brennan2012new,grover2017computational,weintrop2016defining,lafuentemartinez2022assessing,saxena2020designing,Relkin2019}. 
Furthermore, many theoretical models are overly complex, incomplete, or overlap with one another \citep{tikva_mapping_2021,shute2017demystifying}. 
This second framework, FADE-CTP, focuses on identifying and analysing the components and characteristics of CTPs, recognising that the structure and context of these problems are crucial for CT skill development \citep{heersmink2013taxonomy,roth2013situated}. 
To address gaps and overlaps in existing competence models, we created a structured catalogue of CT competencies that consolidates and organises the various models in the literature.
This catalogue is linked to the components and characteristics of CTPs, facilitating the analysis of existing CTPs by identifying which competencies they can develop or assess based on their inherent characteristics. Additionally, it guides the design of new CTPs targeted at specific CT skills by outlining the necessary characteristics to activate their development.

In this way, both frameworks provide complementary approaches for defining and assessing CT competencies across different age groups and educational contexts, contributing to the development of more targeted and effective tools for fostering CT in educational settings.

\section[Developing a large-scale assessment instrument for AT]{Developing a large-scale assessment instrument for Algorithmic Thinking}

To answer the second research question, ``How can an activity and related instruments be developed to assess AT competencies on a large scale across different age groups and educational contexts, and what characteristics should they have to ensure their effectiveness and validity?'', building on the CT-cube and the FADE-CTP frameworks, we developed the CAT, an unplugged CT activity designed to assess algorithmic skills in compulsory school pupils.
We tested the CAT in an experimental study and demonstrated its suitability for reliably measuring AT skills and providing insights into the progression of these skills across different developmental stages. 

To enable large-scale assessment, we adapted the unplugged CAT into a digital format, creating the virtual CAT. 
This adaptation preserved the core problem-solving tasks and educational goals while allowing students to complete activities independently, reducing the need for administrator supervision and addressing scalability challenges.
While in this form, the virtual CAT does not yet fully integrate IASs, its automated evaluation significantly improves efficiency and scalability compared to the unplugged version, maintaining the integrity and quality of the assessment.
This approach aligns with research highlighting the potential of technology-enhanced assessments to generate rich insights, support formative practices, and adapt to diverse educational settings \citep{Sweeney2017,Chiu2020}.
The virtual CAT was tested in two experimental studies, confirming its usability, accessibility, and versatility in catering to students across various developmental stages, backgrounds, and educational contexts.
The platform encouraged active participation, with many students successfully completing their tasks.
Moreover, the results demonstrated its efficiency in supporting large-scale assessments, meeting time and resource demands, and its potential for future integration with IASs to enhance scalability.



\section[eloping an IAS]{Developing an Intelligent Assessment System}

To answer the third research question, ``How can a probabilistic IAS be designed and integrated into the instrument for assessing AT skills across different age groups and educational contexts?'', we developed an IAS by translating a task-specific assessment rubric into a BN with noisy gates.
BNs are widely recognised in recent literature as an effective method for modelling student knowledge.
By leveraging the assessment rubric, our approach ensures a structured definition of relationships and parameters within the BN, enhancing its interpretability and applicability.
Our implementation exploits the noisy gates mechanism to simplify parameter elicitation, making the system more efficient while preserving accuracy in assessing AT skills.
Unlike conventional methods that assign a single score per student-task, our approach uses posterior probabilities to construct a comprehensive learner model that provides a more detailed understanding of students' competence profiles, highlighting their proficiency across various skill levels \citep{xing2021automatic,wu2020student,rodriguez2021bayesian,mousavinasab2021intelligent}.


Specifically, we designed four BN-based models with increasing sophistication, starting with a simple baseline where all inhibitors share the same value. We then added constraints to model the ordering of competencies, incorporated supplementary competencies, and finally developed an enhanced model combining both features with expert-elicited parameters reflecting the intrinsic difficulty of tasks and competencies.
The models were evaluated by comparing their assessments to expert evaluations, showing a high correlation and confirming the consistency of the BN-based assessments with expert judgments. Additionally, differences in posterior probabilities among the models highlighted the impact of improvements introduced during development, validating the iterative refinement process.
Even the baseline model produced inferences closely aligned with expert assessments, suggesting that even a minimal parametrisation can serve as an effective starting point for further enhancements. 
Also, the interpretability of the models was demonstrated through the analysis and comparison of competence profiles generated by the four models.

\section[Examining AT competencies in Swiss educational settings]{Examining Algorithmic Thinking competencies in Swiss educational settings}

To answer the last research question, ``What are the key AT competencies in the Swiss educational landscape, how do they develop across age groups, and what demographic or contextual factors are associated with variations in these competencies?'', we tested the unplugged CAT and virtual CAT activities in real-world classroom settings.

Our investigation of K-12 pupils' algorithmic skills using the unplugged CAT revealed three key findings.
(i) Algorithmic skills improve with age, particularly in autonomy and the ability to use more complex artefacts to describe algorithms. The most significant increase occurs between lower and upper primary school pupils, aligning with prior studies. For instance, \citet{dietz2019building} demonstrate that older children become more successful and efficient at completing tasks, according to various measures, confirming the relationship between age and success rate. Similarly, \citet{klahr1981formal} observe improved problem-solving abilities and planning processes among older preschool children.
(ii) Very young learners, such as those in preschool and lower primary school, are already capable of conceiving and describing complex algorithms, supporting the literature indicating the rapid development of AT skills in preschool-aged children \citep{dietz2019building,vujicic2021development,nikolopoulou2023stem,wahyuningsih2020steam,Voronina2016}. This underscores the notion that complex problem-solving abilities can emerge earlier than previously thought \citep{sarama2009early,kanaki2022age,georgiou2021developing}.
(iii) There is no significant difference between genders with respect to the algorithmic skills of K-12 pupils, which contrasts with much of the existing literature. 
This finding could be specific to the context of our study, where no global gender differences were observed, possibly due to the interaction with other variables not measured.
School-specific factors such as pedagogical methods, institutional culture, and student cohort dynamics may significantly influence performance variations across genders \citep{Rachmatullah2022, Wang2017}.
The quality of instruction, classroom management, and local educational practices also play a key role \citep{ElHamamsy2023stem, Wang2019}.
These results highlight the importance of considering local contexts, as academic achievements can vary across genders and regions \citep{Wang2019}.

Our investigation of K-12 pupils' algorithmic skills using the virtual CAT revealed four key findings.
(i) Algorithmic skills develop progressively with age, consistent with findings from the unplugged CAT.
Younger students tend to use T\&E strategies, especially with new tasks, but as they mature, they adopt more sophisticated problem-solving techniques and rely less on T\&E.
These results align with existing research on the developmental progression of AT and problem-solving \citep{Vlachogianni2021, ElHamamsy2023stem, Kong2022, del2020computational, roman2017cognitive}. Like previous studies, our results confirm that younger students rely more on T\&E when solutions are unclear, and their problem-solving methods become more advanced with age \citep{kanaki2022age, Tönnsen_2021, Chevalier2020}.
Our study further explores how T\&E behaviour affects performance outcomes, noting that while excessive reliance on T\&E may initially hinder performance, iterative attempts can promote learning, but relying solely on T\&E without reflective thinking may limit deeper understanding and algorithmic competence \citep{Chevalier2020, shute2017demystifying}.
(ii) Our examination found no global gender effects on AT performance, which is consistent with the results from the unplugged CAT. 
However, we observed nuanced differences influenced by various factors.
In the virtual CAT, simpler artefacts had less impact on algorithmic complexity for males than for females, while the unplugged CAT showed no significant gender differences.
Age also moderated these differences, with males aged 10 to 13 outperforming females in unplugged settings but lagging in virtual ones. 
Furthermore, school performance data showed variability across institutions, with some showing higher performance for males and others for females. 
These findings suggest that gender’s impact on AT performance is shaped by multiple factors, consistent with existing research \citep{Ardito2020,Mouza2020,Sun2022,Plante2013,Kong2022}.
The literature highlights the importance of early exposure to AT and effective teacher preparation to reduce gender gaps and promote equity \citep{ElHamamsy2023stem} while emphasising the need for targeted interventions and supportive educational environments to address gender differences in early childhood \citep{Master2021}.
These nuanced differences raise important questions about the factors contributing to variability and highlight the need to explore how some schools can better support all students, regardless of gender.
(iii) The wide range of performances highlights the individual differences influenced by personal abilities, learning preferences, and external circumstances. 
This diversity underscores the need for equitable learning environments that accommodate various needs and learning styles, recognising that a one-size-fits-all approach may not be effective for all students. Addressing these differences is essential for ensuring that every student has the opportunity to succeed and develop their AT skills.
This perspective aligns with research suggesting that tailored, personalised and adaptive educational approaches that address individual needs and characteristics can enhance learning experiences \citep{millan2000adaptive,soofi2019systematic,Vomlel2004,desmarais2012review,mousavinasab2021intelligent,Hooshyar2016}. 
By adapting educational practices to address diverse learning preferences and abilities, educators can create more inclusive and supportive environments that foster success for all students.
(iv) Our analysis highlights the significant impact of different interaction modalities on the development of AT skills, revealing variations across age groups.
Younger students predominantly use simpler artefacts, while older students shift to more complex artefacts, indicating a developmental progression toward more sophisticated problem-solving techniques. Interestingly, younger learners can also engage effectively with complex artefacts, suggesting that exposure to such tools can foster advanced algorithmic skills at an earlier age than traditionally assumed.
Interestingly, students using the virtual CAT generally demonstrated greater proficiency in advanced AT skills compared to those engaging with the unplugged CAT.
The effectiveness can be attributed to the interactive and stimulating nature of digital environments, which provide a richer learning experience, aligning with theories emphasising the role of immersive learning environments in cognitive development \citep{Lui2023,makransky2021cognitive,wohl2015teaching}.



\chapter{Practical implications} \label{chap:practical-implications}
%

The CT-cube and FADE-CTP frameworks offer practical tools for assessing and developing CT skills in different educational contexts. 
The CT-cube helps design activities that account for both cognitive development and the context in which learning occurs, making it useful for evaluating CT skills in real classroom settings.
The FADE-CTP framework focuses on identifying the key components of CTPs, helping to design tasks that target specific CT skills. It also allows for the evaluation of existing CTPs to determine which competencies they address.
These frameworks provide guidance for educators and curriculum designers to create developmentally appropriate assessments and activities that foster CT skills across various age groups and learning environments.

The development of the unplugged CAT and its digital counterpart, the virtual CAT, offers practical benefits for educators and policymakers looking for scalable, efficient tools to assess AT skills across diverse student populations. 
The integration of a probabilistic IAS based on BNs enhances the precision and flexibility of assessments, supporting adaptive testing in educational contexts.
The detailed evaluations provided by the system can guide instructional decisions, helping educators focus on competencies that need further development, especially in heterogeneous classrooms where students progress at different rates.
Additionally, translating assessment rubrics into flexible mathematical models makes this approach accessible to educators with limited technical expertise, promoting the widespread adoption of IASs for real-time learner interaction.
The effort required to create and refine the model is minimal, making this approach both scalable and easily adaptable to various educational environments, with the potential for further customisation with little additional effort.

The findings of this research have several practical implications for educators and policymakers in Switzerland and beyond.
The developmental progression of AT skills observed in this study emphasises the importance of providing age-appropriate learning experiences that nurture these competencies. Introducing complex problem-solving tasks earlier than traditionally thought could benefit younger students, who have demonstrated an ability to engage effectively with such challenges.
The variability in performance, particularly regarding gender and contextual factors like school environments and artefacts used, highlights the necessity of tailored educational strategies. 
While global gender differences were not evident, the nuanced patterns observed suggest that addressing specific contextual variables is essential to promoting equity and inclusivity in AT education. Creating adaptive and inclusive learning environments that cater to diverse needs and learning styles is critical for supporting all students effectively.
The use of digital tools, such as the virtual CAT, has shown promise in fostering cognitive development and algorithmic understanding, especially as students progress through different developmental stages. However, integrating these tools must be done thoughtfully, balancing their potential to enhance learning with the risks associated with excessive screen time. Research has shown that prolonged screen use can negatively affect cognitive, social-emotional, and physical development \citep{Muppalla2023,Ponti2023,SwiderCios2023}. To mitigate these risks, educators should set reasonable limits on screen time and complement digital activities with other developmental approaches.
By adopting a balanced, personalised, and inclusive approach, educational strategies can harness the benefits of interactive technologies while ensuring the overall well-being and success of learners.

\chapter{Limitations and Future works} \label{chap:limitations-future-work}

This chapter discusses the limitations of this study and outlines potential directions for future research to address these challenges and build upon the findings.

\section{Extending and validating the competence models}
Despite the benefits and insights offered by our proposed competence models, further research is needed to validate and extend their applicability. 

Specifically, the CT-cube framework, while promising, requires additional investigation to confirm its utility as a robust tool for the design and assessment of CT activities. Future studies should explore its application not only to algorithmic capabilities but also to problem setting tasks and more complex assessment scenarios. Moreover, expanding the research to other domains, such as educational robotics, and exploring more extensive regions, or even the entirety, of the CT-cube, for example, including formal artefactual environments, would be instrumental in assessing its full potential and versatility.

Although the FADE-CTP framework provides valuable guidance on the competencies that can be nurtured through specific CTPs, it does not yet address the levels of abstraction at which these competencies can be cultivated. Future work could focus on delineating whether these competencies emerge at foundational levels, such as recognising or understanding algorithmic concepts, or at more advanced stages, such as applying or synthesising them \citep{bloom1956taxonomy,CTF}. 
Furthermore, broadening the framework to include more competencies, for example, those related to creativity, would greatly enhance its applicability. Such developments would support the design of more holistic CT activities and contribute to the advancement of educational strategies in this field.

\section{Instrument potential for learning and teaching}



The CAT holds promise not only as an assessment tool but also as an instrument for fostering the development of AT across various educational contexts. While traditionally employed for measuring students' algorithmic capabilities, CAT’s structure allows it to be effectively repurposed for teaching and learning, thereby supporting both skill development and knowledge acquisition.
As an instructional tool, CAT offers a versatile platform to engage students in active problem-solving activities that help them build foundational algorithmic concepts. Through iterative challenges and guided problem exploration, students can deepen their understanding of algorithmic structures and logical reasoning. The iterative nature of the tasks provides opportunities for students to experiment with different approaches, reflect on their solutions, and refine their strategies over time, promoting an active learning process.
Moreover, integrating CAT into learning environments can facilitate personalised learning pathways. Its adaptable framework enables teachers to tailor tasks based on individual or group needs, helping to scaffold students' learning as they progress from basic concepts to more complex AT. The use of the tool in the classroom could further encourage collaborative learning, where students share their approaches, discuss their strategies, and learn from each other’s solutions.
In addition to supporting individual learners, CAT's use in classroom settings can enhance formative assessment practices. It allows instructors to monitor progress, identify areas of difficulty, and offer timely feedback to students, ensuring that each learner is appropriately challenged and supported. This ongoing interaction between assessment and learning can help bridge the gap between what students know and what they still need to learn, enhancing the overall effectiveness of the learning experience.

\section{Integrating tutoring capabilities}

A significant limitation of the current study is the absence of a tutoring mechanism, which was suggested during the expert evaluation by pedagogical experts but has not yet been implemented. Integrating tutoring capabilities into the BN-based IAS could transform it into a fully-fledged ITAS,
providing real-time, adaptive support for students and greatly enhancing their learning experiences \citep{millan2000adaptive,soofi2019systematic,Vomlel2004,desmarais2012review,mousavinasab2021intelligent,Hooshyar2016}.
The integration of tutoring functionality is essential for personalising the learning experience, offering tailored feedback and guidance to students in real time. 
The tutoring mechanism would be designed to assess a student’s progress, identify areas of struggle, and provide contextual support to help students overcome difficulties. 
This adaptive tutoring would ensure that the support offered is neither too basic nor too advanced but is instead aligned with the individual’s current understanding and learning needs.

An important aspect of integrating tutoring capabilities lies in determining when and how much tutoring to provide.
Establishing thresholds for identifying when a student is struggling or needs additional assistance to progress will be essential. This could be based on performance patterns, such as consistently incorrect answers, a lack of progress in completing tasks, or extended periods of inactivity. Recognising these signs would enable the system to provide timely support to help the student overcome challenges and continue their learning journey.
The amount of tutoring offered needs to be calibrated to avoid overwhelming the student with excessive guidance. The system should aim to strike a balance, offering enough support to aid the learner’s understanding without diminishing the opportunity for independent problem-solving.

In the context of the CAT, tutoring capabilities could guide students through the activity in various ways. For instance, the system could suggest adjustments to the type of interaction with the platform, such as recommending a shift from the gesture interface to the programming interface or vice versa. It could also activate visual feedback or suggest restarting a task if the student reaches a dead-end, helping them move forward in the learning process.
Additionally, the system could suggest specific changes to the algorithm, such as recommending which blocks or commands to use, helping students refine their approaches and deepen their understanding of AT.

Alternatively, another promising avenue could be transforming this assessment into an adaptive test, where tasks are adjusted based on the student's performance. By measuring the information gained after each task, the system could identify areas of uncertainty and present the most relevant tasks to assess those areas instead of continuing to test already mastered skills. This approach would ensure that students are always challenged according to their current level, making the assessment more efficient and accurately measuring their algorithmic skills \citep{antonucci2021}.

Currently, our system employs BNs with noisy gates for assessment purposes, where the probabilistic relationships between competencies are used to evaluate a student's performance and progress. 
To extend this framework for tutoring capabilities, the existing BN-based noisy gates would need to be adapted to provide real-time, personalised support. This could involve using the probabilistic model to identify when a student is struggling and dynamically adjust the intervention level, such as offering hints, feedback, or suggesting alternative problem-solving strategies.
An alternative approach can rely on Structural Causal Models (SCMs) to explicitly model causal relationships between student actions, task difficulty, and learning outcomes. Unlike BNs, which model probabilistic dependencies, SCMs allow for a more direct understanding of how specific interventions, such as providing feedback or adjusting task difficulty, impact student performance. By modelling these causal effects, the tutoring system can predict the likely outcomes of different actions and select the most effective intervention based on the student's current state. 
Finally, we developed a method for another project to find optimal solutions to the CAT problem \cite{Corecco2023}. This approach, which combines clustering, random search, and reinforcement learning techniques such as Proximal Policy Optimization (PPO), can also be leveraged to integrate tutoring functionalities into our instrument. The method could be incorporated into our developed app to suggest the best move for the student based on their progress so far. Since the system evaluates all possible moves and selects the one that maximises the number of coloured dots, it can be used to provide real-time guidance, recommending the most effective strategies and actions to help the student advance and solve the problem more efficiently. 



\section[Evaluating predictive power of the IAS]{Evaluating predictive power of the Intelligent Assessment System}
In our study, to gauge the effectiveness of the IAS, we have primarily compared the model's outcomes with expert assessments. 
While this provides some validation of the system's utility in capturing relevant skills, a more robust evaluation of its predictive capabilities remains an important area for future development.
One potential approach is to simulate the model's ability to predict answers to new questions based on prior responses. 
For instance, by randomly selecting 8 out of 12 schemas and conditioning on those responses, we could predict the student's answers to the remaining tasks.
To achieve this, we would need to modify our current BN-based noisy gates model, shifting from skill-based inference to answer-based inference. After predicting the answers, we would reapply skill-based inference to generate the assessment based on these predicted answers.
Finally, to assess the accuracy of this predictive approach, we could compute a loss function that measures the discrepancy between the predicted and actual answers for the remaining schemas. This evaluation would provide valuable insights into the model’s ability to anticipate student performance and improve its overall predictive capabilities.


%
%
%

\section{Instrument extension and validation}

A key limitation of this study is that the instrument has not been fully validated to assess its effectiveness in capturing and measuring AT skills within real-world educational settings. Without proper validation, we cannot make definitive claims about its ability to assess AT skills reliably across diverse contexts.
To address this limitation, future work should focus on validating the instrument by assessing its reliability in terms of face validity, content validity, and construct validity, ensuring it accurately measures the relevant AT skills.
Additionally, comparisons of the assessment results with those obtained from other established tools should be conducted to evaluate the instrument's relative effectiveness.
Expert evaluations from professionals in the field of AT are also necessary to ensure the instrument accurately measures the skills it aims to assess.
Additionally, pre-post intervention analyses can be conducted to examine the tool's effectiveness over time, assessing how well it captures changes in AT skills before and after an intervention. 
A large-scale study would also be crucial for evaluating the instrument's generalisability and reliability across various educational environments.

Overall, the results of this research should be interpreted in light of the limited sample size and the specific region of Switzerland where data was collected. 
Further investigations with larger samples from different school systems are needed to confirm the observed development of algorithmic skills. 
Additionally, the study did not explore socio-economic factors, such as parental income or education, which could influence performance. 
It also did not account for students' prior digital education, which might affect their ability to engage with the assessment and benefit from the instructional strategies tested. 
Future research should consider these factors to understand better how prior digital experience impacts AT development.

Finally, future research should explore how students' self-perception, interest, and motivation affect their performance in AT, as there is a strong link between high student engagement, a positive perception of the learning environment, and increased academic success \citep{
Bellino2023,Tai2022,Rachmatullah2022,Hinckle2020,Sun2022,ElHamamsy2023stem,Wang2022influence,Guran2020,Master2021,Beyer2014,KONG2018,Sevin2016,Olivier2018}. 
Additionally, administering questionnaires to the teachers of these students could provide valuable insights into other contextual factors that may influence performance. 
Teachers' perspectives on students' learning habits, classroom dynamics, and individual challenges could shed light on external influences that might affect students' AT development. Moreover, it is important to consider that teachers may have biases, conscious or unconscious, which could impact how pupils learn and develop skills and, consequently, their assessment.
Understanding these potential biases is crucial for interpreting the results accurately and ensuring that such factors do not influence the assessment tools and methods used.

\section{Instrument integration}




Integrating our assessment instrument into schools requires careful planning to ensure its effectiveness and sustainability. 
First, it is essential to collaborate with educators to ensure that the tool aligns with the curriculum and educational goals. We have already involved teachers in the development and testing phases, gathering valuable feedback on how the tool can be used effectively in real classroom settings. 
However, providing professional development for teachers will be crucial, as they will need to understand how to interpret the results and use the tool to inform their teaching practices. 
Additionally, the integration should be seamless, causing minimal disruption to existing classroom routines. Technical support should also be readily available to ensure that both teachers and students can easily use the tool. To enhance usability, it may be necessary to make the tool more adaptable, allowing flexibility in its use and enabling teachers to customise it according to their students' needs.

Limitations related to access to technology still pose a challenge, especially for students without regular access to electronic devices or stable internet connectivity, as well as those with limited technological skills. 
In the studies conducted so far, we have already provided the necessary devices and infrastructure to ensure all participants can engage with the platform. However, integrating the platform into regular classroom settings could present difficulties, particularly in schools with limited or inconsistent access to technology. 
To address this, future studies could explore strategies for ensuring equitable access, such as collaborating with schools to provide devices or designing the platform to be more compatible with a variety of devices and internet conditions. Additionally, offering training to students with limited technological skills could help reduce disparities and facilitate more equitable participation in digital assessments.





\chapter{Conclusions} \label{sec:conclusions}

This thesis presented a comprehensive effort to advance the field of CT by addressing critical gaps in its assessment and analysis. The main objectives were to (i) develop an age-based competence model for CT, (ii) create a large-scale assessment instrument for AT, (iii) design an IAS, and (iv) examine AT competencies in Swiss educational settings, as measured using the developed tools.
Each of these objectives has been systematically addressed through the work presented in this thesis.

The development of two distinct frameworks has provided structured approaches for defining and analysing CT competencies and designing related assessment activities. These frameworks offer a foundation for age-specific competence modelling and have been applied in the creation of assessment activities tailored to different developmental stages.
Secondly, a large-scale assessment instrument for AT was developed in unplugged and virtual formats to ensure accessibility and adaptability in diverse educational contexts, facilitating widespread implementation in real-world settings.
Thirdly, we designed an IAS powered by BN-based noisy gates to enhance traditional assessment methods based on experts' knowledge, introducing a probabilistic approach to assessing AT and providing detailed insights into students’ reasoning processes and performance.
Finally, the comprehensive analysis of data collected through this research has provided a detailed overview of AT competencies in Swiss educational settings, their development, and the factors that may influence learning outcomes.

By addressing these objectives, this work lays the foundation for further research and applications in CT and AT education. The tools and frameworks developed have the potential to be integrated into classroom practices, supporting educators in assessing and fostering CT skills.
Moreover, it also opens pathways for future studies to validate and expand upon the developed tools, explore their applicability in different educational systems, and investigate additional factors influencing AT development.

	\appendix 
	\addtocontents{toc}{\protect\setcounter{tocdepth}{1}}
	\part{Appendices}

\chapter{Participant information sheets and parental consent forms}\label{app:informative}

This appendix presents the documents related to the study's participant information and parental consent. 
The participant information sheets detail the purpose of the study, the procedures involved, and the participants' rights. First, the information sheet for the unplugged CAT study is presented, followed by the corresponding parental consent form. Then, the information sheet for the virtual CAT study and its parental consent form are included.
The actual signed consent forms are kept on file separately for documentation purposes.


\includepdf[pages=-,frame,pagecommand={\thispagestyle{plain}},width=1\linewidth]{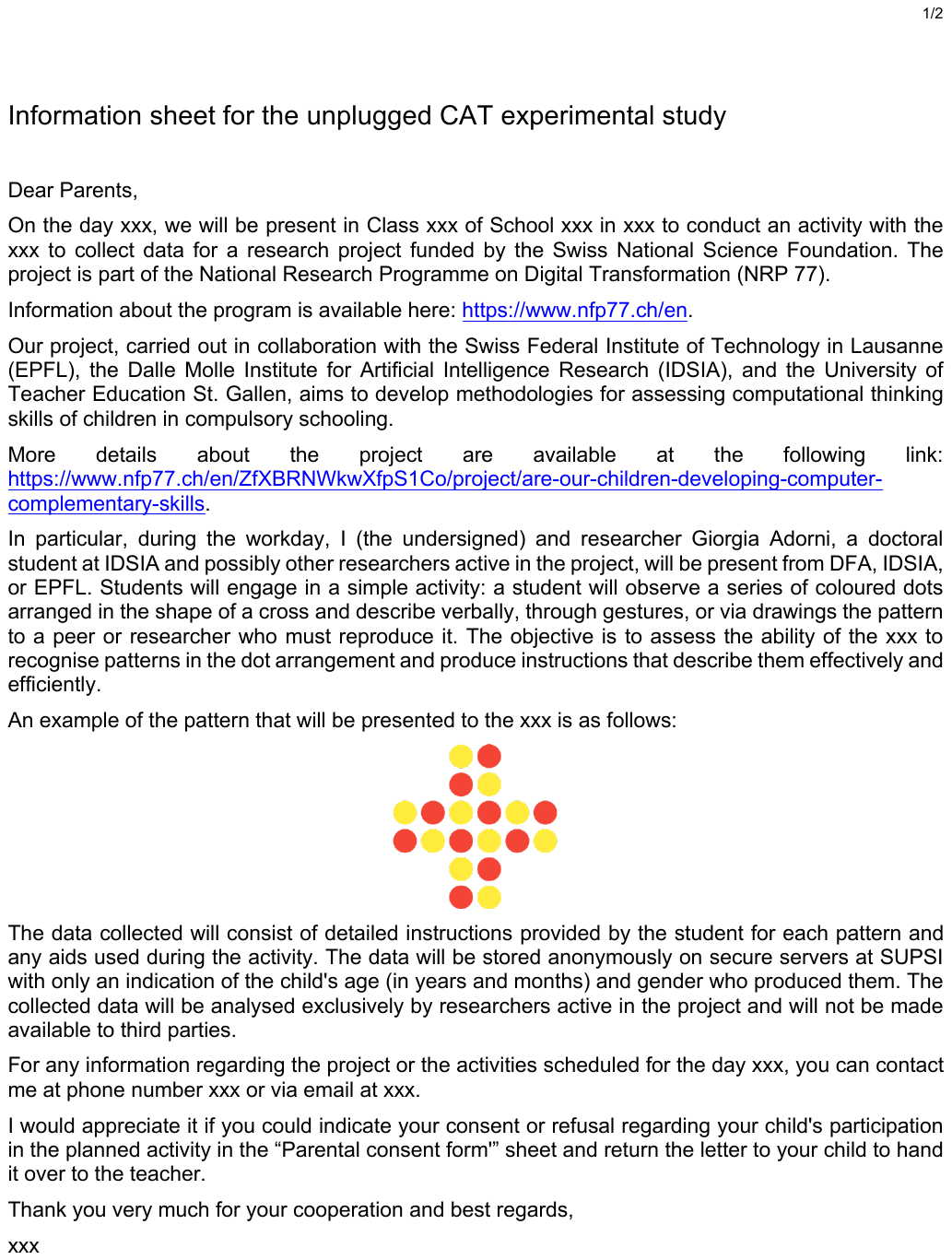}

\includepdf[pages=-,frame,pagecommand={\thispagestyle{plain}},width=1\linewidth]{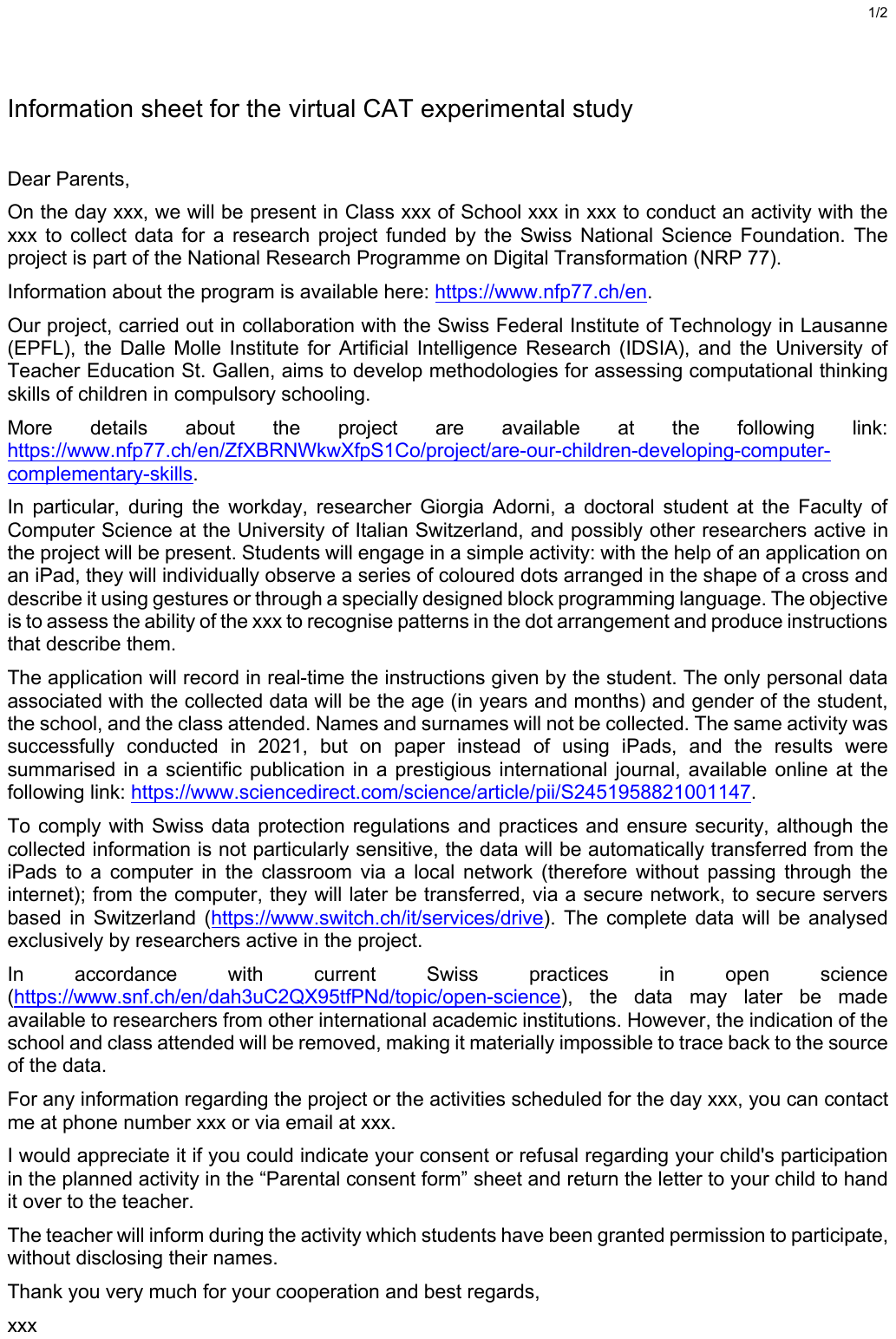}

	\chapter{Linking CTP characteristics to CT competencies} \label{app:tknl_appA_connections_supplementary}

This appendix provides a detailed analysis of our framework, FADE-CTP, represented in \Cref{tab:static-framework-appendix}, illustrating the rationale behind mapping CTP characteristics to CT competencies.

We have organised this analysis according to the main levels of the activity dimension: problem setting, algorithm, and assessment. 
For each dimension, we first describe the link between skills and the required features, then the link between the skills and the characteristics that act as catalysts. 

The CTP features we considered are the functionalities allowed to the problem solver by the tools, the property of the system, such as resettability and observability, and finally, the trait of the task, including the elements required to be found, the type of cardinality of the elements given and to be found, the presence of constraints and the type of representation of the algorithm.

\begin{table*}[ht]
\centering
\mycaption{Comprehensive overview of the relationship between different CTP characteristics and CT competencies.}{
The table shows the relationship between the characteristics of CTPs (columns) and CT competencies (rows). 
The CTP features considered include the tools' functionalities, the system's properties, and the task trait.
}
\label{tab:static-framework-appendix}
\includegraphics[width=\textwidth]{TKNL/static-framework.pdf}
\\\vspace{1em}
\begin{minipage}{\linewidth}
{{\color[HTML]{00B050}\ding{51}} indicates that the characteristics is required for the development of the competence.}

{{\color[HTML]{00B050}\ding{51}$^*$} indicates that at least one of several characteristics in a group is required for the development of the competence.
}

{{\color[HTML]{FF0000}\ding{55}} indicates that the characteristic prevent the development of the competence.}

{{\color[HTML]{0078BF}$+$} indicates that the characteristic can support the development of the competence.}

{Blank cells indicate that the characteristic is irrelevant for the development of the competence.}
\end{minipage}
\end{table*}


\section{Problem setting competencies}

\subsection{Characteristics required for competencies development}
Starting from the problem setting skills, to activate the ``data collection'' competence, the only requirement is that the tools available allow the use and recognition of variables. Without variables, there would be nothing to collect data on.
The ``pattern recognition'' competence requires the presence of repetitions or functions since they allow the identification of repeating patterns in the data.
The ``decomposition'' competence requires the presence of functions or sequences that can be used to break down a complex problem into smaller more manageable components. 
The ``abstraction'' competence demands the presence of variables to represent key concepts
and functions to encapsulate and reuse specific behaviour within a single, self-contained unit, simplifying the original task and allowing the problem solver to reason about the problem at a higher level of abstraction.
Finally, the ``data representation'' competence requires only variables to represent data. 

\subsection{{Characteristics supporting competencies development}}
Generally, the attributes of the problem not required directly to activate the skills can influence them in some way.
In the case of the characteristics of the tools, for example, variables also play a role in ``pattern recognition'' and ``decomposition'', as they can be used to store patterns or parts of a complex problem. Then, operators can be useful for the ``decomposition'' of the problem into smaller parts. At the same time, sequences can contribute to the processes of ``pattern recognition'' and ``abstraction'', helping the problem solver to identify patterns or regularities in the data, as well as the key concepts or essential elements of a problem, but also in ``data representation'' to organise and present data in a clear and meaningful way. Repetitions can influence the activation of the problem setting skills of ``decomposition'', ``abstraction'', and ``data representation'' because they can make the task more complex thus requiring the problem solver to use these practices. Similarly, conditionals can help to structure and simplify a problem, making it more manageable and easier to solve, enabling ``pattern recognition'', ``decomposition'', ``abstraction'' and ``data representation''.
Functions can influence the activation of ``data representation'' by helping the problem solver organise and structure data. Parallelism can influence the activation of the problem-solving skill of ``decomposition'' as it allows for breaking the problem into independent subtasks that can be executed simultaneously. Finally, events can trigger ``data collection'' at a specific point in time.
For the sake of the characteristic of the system, resettability allows the problem solver to start over and try different approaches to solving the problem, thus stimulating can problem setting skills such as ``data collection'', ``pattern recognition'', and ``decomposition'', as they can test different strategies and collect data on their effectiveness. On the other hand, if the system is not resettable, the problem solver may have to rely more on ``abstraction'' and ``data representation'' skills to find a solution, as they cannot try different approaches and must work with the information they have available. In general, a resettable system allows more freedom for the problem solver, giving a chance to explore different solutions. In contrast, a non-resettable system may require more creativity to find a solution.
If the system is observable, the problem solver would likely use skills related to ``data collection'', as he can directly perceive and then gather information about the system's state and properties. 
Additionally, he may use skills related to ``pattern recognition'', such as identifying patterns or trends in the data collected. These skills can help the problem solver understand the system's current state and make informed decisions about how to solve the task.
Conversely, suppose the system is not observable. In that case, the problem solver may need to rely on abstract and hypothetical reasoning to devise a solution, activating ``pattern recognition'', ``decomposition'', and ``abstraction'' to understand the problem and identify possible solutions. Also ``data collection'' may be necessary to gather information about the system and its behaviour, even if that information is not directly observable. Additionally, the ``data representation'' skill can be used to organise and interpret the information they have collected to make sense of the problem and develop a solution.
When there is a many-to-one cardinality in the system, it means that there is a large amount of data that needs to be processed, and multiple inputs or sources of information can be used to achieve a single goal or outcome. In this scenario, the ``data collection'' skill will likely be activated because the problem solver needs to gather a large amount of information to understand the problem and find a solution. Since there are multiple instances of a certain element or pattern, recognising the commonalities and differences among them would be essential to understand the overall system, leading to the use of more complex data collection and analysis strategies, thereby activating the ``pattern recognition'' competence. The ``decomposition'' and the ``abstraction'' skills will also likely be activated as the problem solver needs to break down the problem into smaller manageable parts and find the underlying principles and concepts in the problem to understand the overall system and find a solution. Finally, as there are multiple instances of a specific element, it would be essential to communicate them clearly and concisely, thus activating the ``data representation'' skill.
By contrast, if there is a one-to-one cardinality in the system, the competencies of problem setting that are likely to be activated include ``data collection'', ``pattern recognition'', and ``decomposition''. The skills ``abstraction'' and ``data representation'' are less likely to be activated since the direct correspondence between the system elements means there is less need to abstract or represent the information.
It can be assumed that with implicit elements, the ``data collection'', ``pattern recognition'', ``decomposition'' and ``abstraction'' competencies may be activated as the problem solver needs to infer information from the context or the environment, understand the underlying concepts or patterns in the task, decompose the problem into smaller sub-problems, and create abstract representations of the system.
The same reasoning can be applied to constrained elements. Moreover, it is possible that the competence ``data representation'' may be activated as implicit or constrained elements may require the problem solver to think about how to represent the data in a way that accurately reflects the underlying information or constraints. 
Likewise, with explicit elements, the ``pattern recognition'', ``decomposition'', and ``abstraction'' competencies may also be activated, as the problem solver needs to understand and make sense of the given information, and the presence of unconstrained elements to be found may allow for more flexibility and creativity in problem-solving, potentially activating these skills, as problem solvers may need to find novel ways to organise or make connections among the elements. In this scenario, the ``data collection'' and ``data representation'' competencies may be more straightforward and not as crucial, especially when the elements are explicit and thus the information is already provided in a structured format.
Regarding the representation of the algorithm, overall all problem setting competencies may be activated. Nevertheless, a manifest algorithm makes the problem solver's task easier by providing a clear set of instructions and reducing the need for ``pattern recognition'' and ``decomposition''. However, a not manifest algorithm can promote more ``pattern recognition'', ``decomposition'', and ``abstraction'' as the problem solver needs to infer the algorithm from the problem statement and available information and cannot represent it. 


\section{Algorithmic competencies}
\subsection{{Characteristics required for competencies development}}
For the algorithm dimension, each competence to be activated requires that the corresponding characteristic of the tool is enabled. For example, to activate the ``variable'' skill, the tools used by the problem solver should include variables.
Moreover, in a formal artefactual environment, the task requires that the algorithm is not given but has to be found. Otherwise, it is possible only to assess the problem solver ability to recognise these skills and apply them, but not create an algorithm from scratch.

\subsection{{Characteristics supporting competencies development}}
Again, some characteristics can also influence the activation of algorithmic competencies. 
Regarding the characteristics of the tools, for example, the presence of variables may influence the activation of all the other algorithmic skills, since they provide a fundamental building block for creating algorithms and can be used in conjunction with other algorithmic structures.
Similarly, operators influence the activation of all algorithmic skills.
The presence of sequences may influence the activation of ``variables'', ``operators'', ``repetitions'' and ``functions''; repetitions may influence the activation of ``variables'', ``operators'', ``sequences'' and ``functions''; the presence of conditionals may influence the activation of ``variables'', ``operators'' and ``events''; functions may influence the activation of ``variables'', ``operators'', ``sequences'' and ``repetitions''; the presence of parallelism may influence the activation of ``variables'' and ``operators''; while events may influence the act ``variables'', ``operators'' and ``conditionals''.
The resettability or non-resettability of a system is not relevant for activating or not algorithmic competencies.
The system's observability, or the ability to observe the agent's actions and the system's state, allows tracking of how the algorithm is executing and makes it easier for the problem solver to identify these procedures used by the agent. Instead, a non-observable system may activate the skills of ``variables'', ``operators'', ``sequences'', and ``conditionals'', since the problem solver may need to rely more heavily on their ability to reason about the system and make inferences based on limited information.
Regarding the ratio of elements given and to be found, from one side, a one-to-one cardinality may influence the activation of the algorithmic skill ``variables'' that can be used to define the direct correspondence between the elements in the system and their representations, but also of ``operators'' and ``conditionals'' proper to manipulate them and necessary to ensure the correct mapping. On the other side, a many-to-one cardinality can make it more challenging to understand the relationship between the given elements and those to be found, impacting the ability to understand the algorithm and its parts and enforcing the use of certain types of structures. For example, the problem solver can keep track and map multiple instances to a single object using ``variables''. If the task at hand involves processing multiple pieces of data and producing a single result, a ``repetition'' can be used to iterate over the inputs. Similarly, ``conditionals'' can be used if the task requires selecting one output out of multiple possibilities based on certain conditions. In contrast ``functions'' can be used to modularise the code and make it more organised and maintainable. Finally, ``parallelism'' can be used to speed up the processing of multiple inputs by running multiple iterations simultaneously.
Further, explicit elements provide clear and specific information about the task that must be solved, allowing the problem solver to use all the algorithmic structures to manipulate and work with that information to achieve the desired outcome. Besides, the presence of implicit elements in the task makes it more difficult for the problem solver to understand and determine the necessary steps to solve the task, thus some algorithmic structures may need to be used to compensate for this shortcoming. For example, ``variables'' would be necessary to store and track the values of implicit elements, ``operators'', ``sequences'', ``repetitions'', ``conditionals'', and ``functions'' would help make decisions and perform actions based on the values of these variables. These algorithmic structures would allow the problem solver to explain the implicit elements effectively and develop more sophisticated and efficient solutions.
Similarly, the space for possible solutions is limited when constrained elements are involved in the task and it may be necessary to use some algorithmic structures to ensure those constraints are met. 
For example, while solving a puzzle, the final state and algorithm have to be found, and they have constraints: the problem solver has to fit several pieces together to form a complete image, pieces must fit together to form a specific figure, and certain pieces can only be placed in certain orientations. 
To solve this task, the problem solver might use a combination of algorithmic structures such as ``variables'' to keep track of the current state of the puzzle and the position of the pieces, ``operators'' to manipulate the pieces and move them around, ``sequences'' to try different combinations of pieces, ``repetitions'' to keep trying different combinations until the puzzle is complete, and ``conditionals'' to check if the current combination of pieces meets the constraints. Additionally, ``functions'' could also be used to group sets of repeated actions.
Finally, how the algorithm is represented can affect the activation of various algorithmic structures depending on the type of representation used. Considering different types of tools, each can be more suited to activating one skill rather than another. If the algorithm is represented in a mathematical notation, the use of ``operators'' may be more prominent. On the one hand, if the algorithm is represented in a visual block-based programming language, the use of ``sequences'', ``repetitions'' and ``conditionals'' may be more intuitive and easier to activate. On the other hand, if the algorithm is represented in a text-based programming language, the use of ``variables'' and ``functions'' may be more natural to activate. Finally, robotic programming languages are usually designed for detecting and responding to ``events'', such as sensor readings or other inputs. They often have built-in functionalities for concurrent execution of multiple instructions, allowing ``parallelism''.
Overall, the choice of representation can affect the ease and familiarity of activating different algorithmic structures and may also shape the problem solver's understanding and ability to apply them effectively.


\section{Assessment competencies}
\subsection{{Characteristics required for competencies development}}
Finally, in the assessment category, all skills have in common the need for the system to be resettable for the skill to be activated. For example, in ``algorithm debugging'', if the instruction cannot be reversed, it is impossible to revise and test the previous code versions. Thus, resettability is necessary to debug the algorithm in a controlled and repeatable environment.
The same applies to correcting errors in the state and constraints and improving the solution's performance or generalising it.
In the specific case of ``algorithm debugging'', this skill can be activated in all the artefactual environments if the algorithm has to be found and if it is manifest because it allows the user to understand and check the logic and the flow of the algorithm. This is essential to identify and fix any bugs or errors in the algorithm. 
While it becomes increasingly important to have a written algorithm as the difficulty level of the artifactual environment rises, it may still be possible to solve the problem without one. However, the absence of a written algorithm may make it more challenging to analyse or modify the solution in a formal setting, as the artefactual environment is more abstract and requires a more in-depth understanding. For this reason, we considered the skill required in this context.
The ``system state verification'' competence can be activated in all three artefactual environments if at least one between the initial and final states must be found. 
In embodied environments, direct physical interactions with the system provide a way to observe its state without needing a manifest algorithm.
However, in symbolic and formal environments, a manifest representation of the algorithm, written in the case of formal environments, is crucial to fully understand its logical flow, verify the system state, and perform formal reasoning about its correctness. This may involve analysing the symbolic representation to understand how it impacts the system state. 
To activate the ``constraints validation'' competence, it is blatant that the other necessary characteristic is having constraints on the states to be found.
To enable ``optimisation'', additional features are not required, while for ``generalisation'', variables and functions are necessary to reuse and apply the task solution to different problems.  

\subsection{{Characteristics supporting competencies development}}
Each tool functionality available to the problem solver can be a potential cause of error in the algorithm. For example, if the problem solver is unfamiliar with one of them or does not understand how to use it correctly, he may not use it at all or misuse it. This can lead to errors in the algorithm and potentially result in the problem not being solved correctly. This is why functionalities of the tools if available can activate ``algorithm debugging''.
Also for ``constraints validation'', all the characteristics of the tools are influential. Above all, variables, operators, conditional and functions may allow the problem solver to perform various calculations and comparisons to check if the values assigned to the variables meet the specified constraints. Further, it could be that the constraint imposed is precisely on the algorithm and prohibits using some of these structures.
The functionality of the tools available to the problem solver can greatly impact the ``optimisation'' of the algorithm in several ways. Parallelism allows for multiple tasks or processes to be executed simultaneously, which can greatly reduce the overall time required to complete a task. Sequences and other structures, such as loops, can also help to improve efficiency by allowing for the automation of repetitive tasks and the ability to perform actions in a specific order. Additionally, using functions and subroutines can improve the readability and maintainability of the algorithm, making it easier to identify and fix any errors that may occur.
However, having access to a wide range of functionalities can make it challenging for the problem solver to choose the appropriate one for a specific task, leading to a revision of the solution to increase efficiency and performance.
The competence ``generalisation'' can also be influenced by other characteristics of the tools. The presence of sequences and repetitions in the toolset enables the problem solver to apply the same algorithm to different parts of a problem or task. Similarly, the inclusion of conditionals allows for the application of different algorithms depending on the specific conditions of the task. Furthermore, the presence of events in the toolset allows for creating algorithms that can respond to different triggers within the problem, leading to a greater generalisation of the solution and the ability to adapt to changes within the problem.
In terms of observability, an observable system allows the problem solver to have a clear understanding of the system's state and the output of the algorithm, which can aid in identifying and addressing errors and inefficiencies and performance issues, as well as recognising patterns or regularities that can be generalised to new or different situations. However, it is essential to note that while observability can aid in all assessment skills, it is not strictly necessary for their activation. For example, one could still perform ``algorithm debugging'' and ``system state verification'' on a non-observable system, though it may be more difficult. Similarly, ``generalisation'' can still occur without perfect observability, but it may be harder to identify patterns and regularities without direct access to the system state.
If the system has a many-to-one cardinality, the competence of ``generalisation'' may be activated as it would be necessary to apply the same algorithm to different inputs or outputs.
If the system contains implicit elements, the competencies ``algorithm debugging'' and ``system state verification'' may be activated as the problem solver may need to identify and troubleshoot any issues with the algorithm that are not immediately apparent or infer the current state of the system based on the implicit information provided. Also ``generalisation'' may be activated as the problem solver may need to apply the algorithm to different situations based on the implicit information provided. 
Finally, suppose in the system there are elements to be found with constraints. In that case, the ``generalisation'' skill may be activated because it requires the problem solver to adapt the task to the specific constraints and can be intended as solving a new problem using the knowledge acquired in a previous situation and adapting it to a new one.
	
\chapter{Main study with the unplugged CAT}

This appendix provides detailed documentation related to the main study with the unplugged CAT assessment, in particular we included: (i) the protocol template that outlines the experimental setup and recording procedures for the unplugged CAT assessment and is intended to facilitate replication of the study and ensure transparency in the methodology employed \cite{templateprotocolCAT}; (ii) the illustrations of all the algorithms conceived by students for each schema used in the assessment.

\section{Protocol template for administering the activity}\label{app:chbr_appE}
This section outlines the protocol template used to guide the administration of the unplugged CAT assessment. It provides a detailed framework for the experimental setup, ensuring consistency and reliability in data collection.

For each participant, a separate protocol should be filled in, capturing general information such as Session ID, School ID/Name, Pupil ID, Date, Administrator, Class/Grade, Pupil Age, and Pupil Sex.

For each schema presented to the participants, the type of interaction used and the algorithm dimensions produced should be recorded. Finally, additional notes or observations related to each schema can be included to provide further context and insights into the participant's approach.

\includepdf[pages=-,frame,pagecommand={\thispagestyle{plain}},width=1\linewidth]{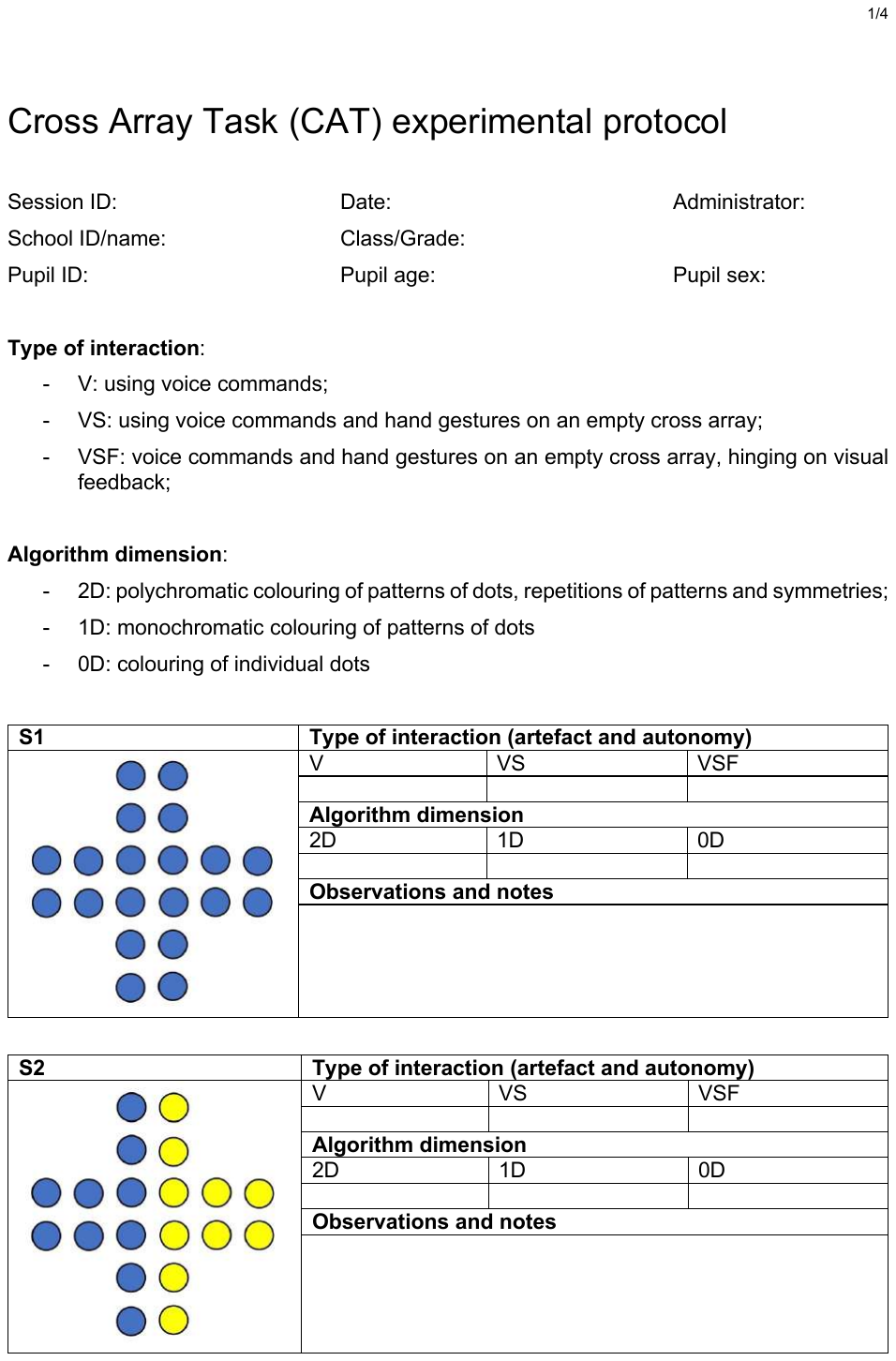}


\section{Algorithmic solutions developed by students}
\label{app:chbr_appF}

This section focuses on the exploration of the algorithms conceived by the students for each of the cross array schemas used in the unplugged CAT assessment. For each schema, we present illustrations of the different algorithms designed by students. The first algorithm for each schema is always the zero-dimensional algorithm, which describes the array point-by-point. All other algorithms are assigned randomly and do not necessarily share any specific commonality. By examining the diversity of algorithms and their features, this section helps illustrate how students engage with algorithmic thinking in an unplugged context.

\def\firstschema{{Schema1Algorithm1/Schema 1 Algorithm 1: Point by point}, 
	{Schema1Algorithm2/Schema 1 Algorithm 2: All blue}, 
	{Schema1Algorithm3/Schema 1 Algorithm 3: Five squares all blue}, 
	{Schema1Algorithm4/Schema 1 Algorithm 4: Two columns and two 
		rows}, 
	{Schema1Algorithm5/Schema 1 Algorithm 5: Four L split in two}}

\def\secondschema{{Schema2Algorithm1/Schema 2 Algorithm 1: Point by 
		point}, 
	{Schema2Algorithm2/Schema 2 Algorithm 2: Split vertically, blue on 
		the left yellow on the right}, 
	{Schema2Algorithm3/Schema 2 Algorithm 3: Two columns and two 
		rows}, 
	{Schema2Algorithm4/Schema 2 Algorithm 4: Two columns and two 
		squares on the sides}, 
	{Schema2Algorithm5/Schema 2 Algorithm 5: Half of one colour, one 
		square and one column with the other}, 
	{Schema2Algorithm6/Schema 2 Algorithm 6: Two columns and the 
		sides row by row}, 
	{Schema2Algorithm7/Schema 2 Algorithm 7: Half point-by-point, the 
		remaining of one colour}}

\def\thirdschema{{Schema3Algorithm1/Schema 3 Algorithm 1: Point by point}, 
	{Schema3Algorithm2/Schema 3 Algorithm 2: Alternate columns}, 
	{Schema3Algorithm3/Schema 3 Algorithm 3: Column by column}, 
	{Schema3Algorithm4/Schema 3 Algorithm 4: Column by column (with 
		redundancy)}, 
	{Schema3Algorithm5/Schema 3 Algorithm 5: Column by column 
		symmetrical sides},
	{Schema3Algorithm6/Schema 3 Algorithm 6: Two rows column by 
		column, point-by-point up and down}, 
	{Schema3Algorithm7/Schema 3 Algorithm 7: Two columns, squares 
		point-by-point},
	{Schema3Algorithm8/Schema 3 Algorithm 8: One colour column by 
		column, the remaining of the other colour},
	{Schema3Algorithm9/Schema 3 Algorithm 9: Two rows column by 
		column, the remaining column by column},
	{Schema3Algorithm10/Schema 3 Algorithm 10: One column with pairs of 
		two, the remaining column by column}}

\def\fourthschema{{Schema4Algorithm1/Schema 4 Algorithm 1: Point by point}, 
	{Schema4Algorithm2/Schema 4 Algorithm 2: Repeat three columns}, 
	{Schema4Algorithm3/Schema 4 Algorithm 3: Column by column}, 
	{Schema4Algorithm4/Schema 4 Algorithm 4: Two columns, two rows 
		(with redundancy)}, 
	{Schema4Algorithm5/Schema 4 Algorithm 5: Column by column, red all 
		together},
	{Schema4Algorithm6/Schema 4 Algorithm 6: Two rows column by 
		column, point-by-point up and down}, 
	{Schema4Algorithm7/Schema 4 Algorithm 7: Two columns, squares 
		point-by-point},
	{Schema4Algorithm8/Schema 4 Algorithm 8: Half point-by-point, the 
		remaining column by column}}

\def\fifthschema{{Schema5Algorithm1/Schema 5 Algorithm 1: Point by point}, 
	{Schema5Algorithm2/Schema 5 Algorithm 2: Two squares in both sides, 
		two columns}, 
	{Schema5Algorithm3/Schema 5 Algorithm 3: Column by column}, 
	{Schema5Algorithm4/Schema 5 Algorithm 4: Column by column, one 
		square}, 
	{Schema5Algorithm5/Schema 5 Algorithm 5: Two squares in both sides, 
		columns point-by-point}}

\def\sixthschema{{Schema6Algorithm1/Schema 6 Algorithm 1: Point by point}, 
	{Schema6Algorithm2/Schema 6 Algorithm 2: Four L},
	{Schema6Algorithm3/Schema 6 Algorithm 3: Four L split in two}, 
	{Schema6Algorithm4/Schema 6 Algorithm 4: Two columns row by row, 
		on the sides row by row},
	{Schema6Algorithm5/Schema 6 Algorithm 5: Column by 
		column},
	{Schema6Algorithm6/Schema 6 Algorithm 6: Two L, two couples, six 
		point-by-point}, 
	{Schema6Algorithm7/Schema 6 Algorithm 7: Two L, two L split in two}}

\def\seventhschema{{Schema7Algorithm1/Schema 7 Algorithm 1: Point by 
		point}, 
	{Schema7Algorithm2/Schema 7 Algorithm 2: Chessboard}, 
	{Schema7Algorithm3/Schema 7 Algorithm 3: Five squares composed 
		of two diagonals}, 
	{Schema7Algorithm4/Schema 7 Algorithm 4: Five squares point by 
		point},
	{Schema7Algorithm5/Schema 7 Algorithm 5: Alternated columns},
	{Schema7Algorithm6/Schema 7 Algorithm 6: Alternated columns and 
		rows (with redundancy)},
	{Schema7Algorithm7/Schema 7 Algorithm 7: Alternated starting 
		from the edges},	
	{Schema7Algorithm8/Schema 7 Algorithm 8: Repeated and alternated 
		columns and row},
	{Schema7Algorithm9/Schema 7 Algorithm 9: Two columns alternated 
		with zig zag, point-by-point on the sides},
	{Schema7Algorithm10/Schema 7 Algorithm 10: Two rows alternated, 
		diagonals in the squares up and down},
	{Schema7Algorithm11/Schema 7 Algorithm 11: Diagonal by diagonal},
	{Schema7Algorithm12/Schema 7 Algorithm 12: Two columns 
		alternated, two squares point-by-point}}

\def\seventhschemat{
	{Schema7Algorithm13/Schema 7 Algorithm 13: Two diagonals, the 
		remaining point-by-point},
	{Schema7Algorithm14/Schema 7 Algorithm 14: Four diagonals, central 
		square point-by-point},
	{Schema7Algorithm15/Schema 7 Algorithm 15: Two rows point by 
		point, two squares equals},
	{Schema7Algorithm16/Schema 7 Algorithm 16: Red diagonals, the 
		remaining yellow},
	{Schema7Algorithm17/Schema 7 Algorithm 17: Point by 
		point, one square composed of diagonals},
	{Schema7Algorithm18/Schema 7 Algorithm 18: Vertical and 
		horizontal zig zag},
	{Schema7Algorithm19/Schema 7 Algorithm 19: Diagonals of two 
		points with intersection (with redundancy)}}

\def\eightthschema{{Schema8Algorithm1/Schema 8 Algorithm 1: Point by 
		point}, 
	{Schema8Algorithm2/Schema 8 Algorithm 2: Five squares with red 
		diagonal and two points}, 
	{Schema8Algorithm3/Schema 8 Algorithm 3: Five squares point by 
		point}, 
	{Schema8Algorithm4/Schema 8 Algorithm 4: Column by column 
		alternated},
	{Schema8Algorithm5/Schema 8 Algorithm 5: Alternated 
		columns and rows (with redundancy)},
	{Schema8Algorithm6/Schema 8 Algorithm 6: Repeated and alternated 
		column by column and row by row},
	{Schema8Algorithm7/Schema 8 Algorithm 7: Two rows alternated, 
		two identical squares point-by-point},
	{Schema8Algorithm8/Schema 8 Algorithm 8: Red alternated one yes 
		and one no vertically and horizontally, yellow among reds in the first 
		columns and in the first row, the others blue},
	{Schema8Algorithm9/Schema 8 Algorithm 9: Red alternated one yes 
		and one no vertically and horizontally, the remaining point-by-point},
	{Schema8Algorithm10/Schema 8 Algorithm 10: Diagonal by                  diagonal},
	{Schema8Algorithm11/Schema 8 Algorithm 11: Two rows alternated, 
		two squares point-by-point},
	{Schema8Algorithm12/Schema 8 Algorithm 12: Repeated and 
		alternated columns and row}}
\def\eightthschemat{
	{Schema8Algorithm13/Schema 8 Algorithm 13: Two rows point by 
		point, two identical squares point-by-point},
	{Schema8Algorithm14/Schema 8 Algorithm 14: Two rows point by 
		point, two identical squares with diagonal},
	{Schema8Algorithm15/Schema 8 Algorithm 15: Red diagonals, blue 
		point-by-point, remaining yellow},
	{Schema8Algorithm16/Schema 8 Algorithm 16: Red diagonals, the 
		remaining point-by-point},
	{Schema8Algorithm17/Schema 8 Algorithm 17: Two rows alternated, 
		two squares with diagonals},
	{Schema8Algorithm18/Schema 8 Algorithm 18: Vertical and horizontal 
		zig zag},
	{Schema8Algorithm19/Schema 8 Algorithm 19: One red diagonal of 
		four points, the remaining point-by-point},
	{Schema8Algorithm20/Schema 8 Algorithm 20: Four pairs of red, the 
		remaining point-by-point},
	{Schema8Algorithm21/Schema 8 Algorithm 21: Two squares with 
		diagonal, the remaining point-by-point}}

\def\ninethschema{{Schema9Algorithm1/Schema 9 Algorithm 1: Point by 
		point}, 
	{Schema9Algorithm2/Schema 9 Algorithm 2: Five squares point by 
		point}, 
	{Schema9Algorithm3/Schema 9 Algorithm 3: Repeated and alternated 
		column by column}, 
	{Schema9Algorithm4/Schema 9 Algorithm 4: Column by column 
		alternated}, 
	{Schema9Algorithm5/Schema 9 Algorithm 5: Alternated columns and 
		rows (with redundancy)},
	{Schema9Algorithm6/Schema 9 Algorithm 6: Two rows alternated, 
		two identical squares point-by-point},
	{Schema9Algorithm7/Schema 9 Algorithm 7: Two rows alternated, two
		squares point-by-point},
	{Schema9Algorithm8/Schema 9 Algorithm 8: Repeated and alternated 
		columns and row},
	{Schema9Algorithm9/Schema 9 Algorithm 9: Diagonal by diagonal 
		with pairs}}

\def\tenthschema{{Schema10Algorithm1/Schema 10 Algorithm 1: Point by 
		point}, 
	{Schema10Algorithm2/Schema 10 Algorithm 2: Inverted symmetry for 
		the two columns, two symmetric squares on the sides composed by 
		diagonal}, 
	{Schema10Algorithm3/Schema 10 Algorithm 3: Two columns with two 
		rows and two squares, two symmetric squares on the sides composed 
		by diagonal}, 
	{Schema10Algorithm4/Schema 10 Algorithm 4: Two identical columns 
		with two pairs, two symmetric squares on the sides composed by 
		diagonal}, 
	{Schema10Algorithm5/Schema 10 Algorithm 5: Two columns with two 
		rows and two squares, two symmetric squares on the sides point by 
		point},
	{Schema10Algorithm6/Schema 10 Algorithm 6: Two identical columns 
		point-by-point, two sides point-by-point},
	{Schema10Algorithm7/Schema 10 Algorithm 7: Two columns row by 
		row, two sides point-by-point},
	{Schema10Algorithm8/Schema 10 Algorithm 8: Column by column},
	{Schema10Algorithm9/Schema 10 Algorithm 9: Two identical columns 
		with two pairs, two rows with pairs (with redundancy)},
	{Schema10Algorithm10/Schema 10 Algorithm 10: Two identical squares 
		up and down with pairs, two symmetric rows with pair of points},
	{Schema10Algorithm11/Schema 10 Algorithm 11: Two columns row by 
		row, double diagonal on both sides},
	{Schema10Algorithm12/Schema 10 Algorithm 12: Two identical columns 
		point-by-point, double diagonal on both sides (one side redundant point 
		by point)}}
\def\tenthschemat{
	{Schema10Algorithm13/Schema 10 Algorithm 13: Two columns with two 
		rows and two squares, one side diagonals, other side point-by-point},
	{Schema10Algorithm14/Schema 10 Algorithm 14: In the central column a 
		square point-by-point, then a row then square and row, double 
		diagonal on both sides},
	{Schema10Algorithm15/Schema 10 Algorithm 15: Two identical columns 
		point-by-point, two symmetric squares point-by-point},
	{Schema10Algorithm16/Schema 10 Algorithm 16: Two columns row by 
		row, two symmetric squares point-by-point},
	{Schema10Algorithm17/Schema 10 Algorithm 17: One 
		colour point by point, the remaining with the other colour},
	{Schema10Algorithm18/Schema 10 Algorithm 18: Two columns with 
		two rows and two squares, two squares on the sides point-by-point}, 
	{Schema10Algorithm19/Schema 10 Algorithm 19: Two identical 
		columns with two pairs, two squares on the sides point-by-point},
	{Schema10Algorithm20/Schema 10 Algorithm 20: Two columns with 
		two rows and two squares, double diagonal on both sides},
	{Schema10Algorithm21/Schema 10 Algorithm 21: Two identical 
		columns with two pairs, double diagonal on both sides},
	{Schema10Algorithm22/Schema 10 Algorithm 22: Blue in pairs and/or 
		diagonals, one diagonal and pair yellow, the remaining point-by-point},
	{Schema10Algorithm23/Schema 10 Algorithm 23: One square blue, 
		other blue in pairs, the remaining yellow},
	{Schema10Algorithm24/Schema 10 Algorithm 24: In the central 
		column a square point-by-point, two rows and a square, two sides point-by-point}}
\def\tenthschematt{
	{Schema10Algorithm25/Schema 10 Algorithm 25: In the 
		central column a square and remaining by rows, two sides point-by-point}}

\def\eleventhschema{{Schema11Algorithm1/Schema 11 Algorithm 1: Point by 
		point}, 
	{Schema11Algorithm2/Schema 11 Algorithm 2: Two identical 
		columns point-by-point, on one side a red line and two points, 
		mirrored on the other side}, 
	{Schema11Algorithm3/Schema 11 Algorithm 3: Two columns row by 
		row, on one side a red line and two points, mirrored on the other 
		side}, 
	{Schema11Algorithm4/Schema 11 Algorithm 4: Two identical 
		columns point-by-point, on the sided point-by-point}, 
	{Schema11Algorithm5/Schema 11 Algorithm 5: Two columns row by 
		row, on the sided point-by-point},
	{Schema11Algorithm6/Schema 11 Algorithm 6: Two identical 
		columns point-by-point, two rows point-by-point (with 
		redundancy)},
	{Schema11Algorithm7/Schema 11 Algorithm 7: Two columns row 
		by row, two rows point-by-point (with redundancy)},
	{Schema11Algorithm8/Schema 11 Algorithm 8: Blue and yellow in 
		pairs, green point-by-point, remaining in red},
	{Schema11Algorithm9/Schema 11 Algorithm 9: Blue, yellow and 
		green in pairs, the remaining red},
	{Schema11Algorithm10/Schema 11 Algorithm 10: Two identical 
		columns point-by-point, two pairs of red, the remaining point by 
		point},
	{Schema11Algorithm11/Schema 11 Algorithm 11: Two columns row 	
		by row, two pairs of red, the remaining point-by-point},
	{Schema11Algorithm12/Schema 11 Algorithm 12: Two identical 
		columns point-by-point, on one side point-by-point, mirrored on 
		the other side}}
\def\eleventhschemat{
	{Schema11Algorithm13/Schema 11 Algorithm 13: Two columns with 
		4 pairs, the remaining point-by-point},
	{Schema11Algorithm14/Schema 11 Algorithm 14: Two identical 
		columns point-by-point, one pair of red, the remaining point by 
		point},
	{Schema11Algorithm15/Schema 11 Algorithm 15: Two columns row 
		by row, one pair of red, the remaining point-by-point}}

\def\twelvethschema{{Schema12Algorithm1/Schema 12 Algorithm 1: Point by 
		point}, 
	{Schema12Algorithm2/Schema 12 Algorithm 2: Point by point (with 
		redundancy)}, 
	{Schema12Algorithm3/Schema 12 Algorithm 3: Point by point 
		except one square with diagonals}, 
	{Schema12Algorithm4/Schema 12 Algorithm 4: L of three green, L 
		of three blue, a square with diagonals, the remaining point by 
		point}, 
	{Schema12Algorithm5/Schema 12 Algorithm 5: Point by point 
		except two squares with diagonals},
	{Schema12Algorithm6/Schema 12 Algorithm 6: Blue and yellow 
		point-by-point, red in pairs, the remaining green}}

\begin{figure*}[htb]
	\foreach \firstname/\firstsubcap in \firstschema {
		\centering
		\hspace*{\fill}\begin{subfigure}[t]{0.20\linewidth}             \centering             \includegraphics[width=\linewidth]{CHBR/algorithms/Schema1/\firstname.pdf}        \mycaptiontitle{\firstsubcap}         \end{subfigure}\hfill\null
	}
	\mycaptiontitle{Algorithms observed for schema S1.}
	\label{fig:schema1}
\end{figure*}

\clearpage
\begin{figure*}[p]	
	\foreach \secondname/\secondsubcap in \secondschema {
		\centering
		\hspace*{\fill}\begin{subfigure}[t]{0.20\linewidth}             \centering             \includegraphics[width=\linewidth]{CHBR/algorithms/Schema2/\secondname.pdf}        \mycaptiontitle{\secondsubcap}         \end{subfigure}\hfill\null
	}
	\mycaptiontitle{Algorithms observed for schema S2.}
	\label{fig:schema2}
\end{figure*}

\clearpage
\begin{figure*}[p]						
	\foreach \thirdname/\thirdsubcap in \thirdschema {
		\hspace*{\fill}\begin{subfigure}[t]{0.20\linewidth}             \centering             \includegraphics[width=\linewidth]{CHBR/algorithms/Schema3/\thirdname.pdf}        \mycaptiontitle{\thirdsubcap}         \end{subfigure}\hfill\null
	}
	\mycaptiontitle{Algorithms observed for schema S3.}
	\label{fig:schema3}
\end{figure*}

\clearpage
\begin{figure*}[p]						
	\foreach \fourthname/\fourthsubcap in \fourthschema {
		\centering
		\hspace*{\fill}\begin{subfigure}[t]{0.20\linewidth}             \centering             \includegraphics[width=\linewidth]{CHBR/algorithms/Schema4/\fourthname.pdf}        \mycaptiontitle{\fourthsubcap}         \end{subfigure}\hfill\null
	}
	\mycaptiontitle{Algorithms observed for schema S4.}
	\label{fig:schema4}
\end{figure*}

\clearpage
\begin{figure*}[p]						
	\foreach \fifthname/\fifthsubcap in \fifthschema {
		\centering
		\hspace*{\fill}\begin{subfigure}[t]{0.20\linewidth}             \centering             \includegraphics[width=\linewidth]{CHBR/algorithms/Schema5/\fifthname.pdf}        \mycaptiontitle{\fifthsubcap}         \end{subfigure}\hfill\null
	}
	\mycaptiontitle{Algorithms observed for schema S5.}
	\label{fig:schema5}
\end{figure*}

\clearpage
\begin{figure*}[p]						
	\foreach \sixthname/\sixthsubcap in \sixthschema {
		\centering
		\hspace*{\fill}\begin{subfigure}[t]{0.20\linewidth}             \centering             \includegraphics[width=\linewidth]{CHBR/algorithms/Schema6/\sixthname.pdf}        \mycaptiontitle{\sixthsubcap}         \end{subfigure}\hfill\null
	}
	\mycaptiontitle{Algorithms observed for schema S6.}
	\label{fig:schema6}
\end{figure*}

\clearpage
\begin{figure*}[p]						
	\foreach \seventhname/\seventhsubcap in \seventhschema {
		\centering
		\hspace*{\fill}\begin{subfigure}[t]{0.20\linewidth}             \centering             \includegraphics[width=\linewidth]{CHBR/algorithms/Schema7/\seventhname.pdf}        \mycaptiontitle{\seventhsubcap}         \end{subfigure}\hfill\null
	}
	\mycaptiontitle{Algorithms observed for schema S7. \small(Continued on the next page).}
	\label{fig:schema7}
\end{figure*}
\begin{figure*}[p]
	\ContinuedFloat						
	\foreach \seventhname/\seventhsubcap in \seventhschemat {
		\centering
		\hspace*{\fill}\begin{subfigure}[t]{0.20\linewidth}             \centering             \includegraphics[width=\linewidth]{CHBR/algorithms/Schema7/\seventhname.pdf}        \mycaptiontitle{\seventhsubcap}         \end{subfigure}\hfill\null
	}
	\mycaptiontitle{Algorithms observed for schema S7. \small(Continued from the previous page).}
\end{figure*}

\clearpage
\begin{figure*}[p]						
	\foreach \eightthname/\eightthsubcap in \eightthschema {
		\centering
		\hspace*{\fill}\begin{subfigure}[t]{0.20\linewidth}             \centering             \includegraphics[width=\linewidth]{CHBR/algorithms/Schema8/\eightthname.pdf}        \mycaptiontitle{\eightthsubcap}         \end{subfigure}\hfill\null
	}
	\mycaptiontitle{Algorithms observed for schema S8. \small(Continued on the next page).}
	\label{fig:schema8}
\end{figure*}
\begin{figure*}[p]					
	\ContinuedFloat	
	\foreach \eightthname/\eightthsubcap in \eightthschemat {
		\centering
		\hspace*{\fill}\begin{subfigure}[t]{0.20\linewidth}             \centering             \includegraphics[width=\linewidth]{CHBR/algorithms/Schema8/\eightthname.pdf}        \mycaptiontitle{\eightthsubcap}         \end{subfigure}\hfill\null
	}
	\mycaptiontitle{Algorithms observed for schema S8. \small(Continued from the previous page).}
\end{figure*}

\clearpage
\begin{figure*}[p]						
	\foreach \ninethname/\ninethsubcap in \ninethschema {
		\centering
		\hspace*{\fill}
        \begin{subfigure}[t]{0.20\linewidth}             
        \centering             \includegraphics[width=\linewidth]{CHBR/algorithms/Schema9/\ninethname.pdf}        
        \mycaptiontitle{\ninethsubcap}         
        \end{subfigure}\hfill\null
	}
	\mycaptiontitle{Algorithms observed for schema S9.}
	\label{fig:schema9}
\end{figure*}

\clearpage
\begin{figure*}[p]						
	\foreach \tenthname/\tenthsubcap in \tenthschema {
		\centering
		\hspace*{\fill}
        \begin{subfigure}[t]{0.20\linewidth}             
        \centering             
        \includegraphics[width=\linewidth]{CHBR/algorithms/Schema10/\tenthname.pdf}        
        \mycaptiontitle{\tenthsubcap}         
        \end{subfigure}\hfill\null
	}
	\mycaptiontitle{Algorithms observed for schema S10. \small(Continued on the next page).}
	\label{fig:schema10}
\end{figure*}

\begin{figure*}[p]						
	\ContinuedFloat
	\foreach \tenthname/\tenthsubcap in \tenthschemat {
		\centering
		\hspace*{\fill}
        \begin{subfigure}[t]{0.20\linewidth}             
        \centering             
        \includegraphics[width=\linewidth]{CHBR/algorithms/Schema10/\tenthname.pdf}        
        \mycaptiontitle{\tenthsubcap}         
        \end{subfigure}\hfill\null
	}
	\mycaptiontitle{Algorithms observed for schema S10. \small(Continued from the previous page).}
\end{figure*}

\begin{figure*}[p]			
	\ContinuedFloat			
	\foreach \tenthname/\tenthsubcap in \tenthschematt {
		\centering
		\hspace*{\fill}
        \begin{subfigure}[t]{0.20\linewidth}             
        \centering             
        \includegraphics[width=\linewidth]{CHBR/algorithms/Schema10/\tenthname.pdf}        
        \mycaptiontitle{\tenthsubcap}         
        \end{subfigure}\hfill\null
	}
	\mycaptiontitle{Algorithms observed for schema S10. \small(Continued from the previous page).}
\end{figure*}

\clearpage
\begin{figure*}[p]						
	\foreach \eleventhname/\eleventhsubcap in \eleventhschema {
		\centering
		\hspace*{\fill}
        \begin{subfigure}[t]{0.20\linewidth}             
        \centering             
        \includegraphics[width=\linewidth]{CHBR/algorithms/Schema11/\eleventhname.pdf}        
        \mycaptiontitle{\eleventhsubcap}         
        \end{subfigure}\hfill\null
	}
	\mycaptiontitle{Algorithms observed for schema S11. \small(Continued on the next page).}
	\label{fig:schema11}
\end{figure*}
\begin{figure*}[p]						
	\ContinuedFloat
	\foreach \eleventhname/\eleventhsubcap in \eleventhschemat {
		\centering
		\hspace*{\fill}
        \begin{subfigure}[t]{0.20\linewidth}             
        \centering             
        \includegraphics[width=\linewidth]{CHBR/algorithms/Schema11/\eleventhname.pdf}
        \mycaptiontitle{\eleventhsubcap}
        \end{subfigure}
        \hfill\null
	}
	\mycaptiontitle{Algorithms observed for schema S11. \small(Continued from the previous page).}
\end{figure*}

\clearpage
\begin{figure*}[p]
    \centering
    \foreach \twelvethname/\twelvethsubcap in \twelvethschema {
     \hspace*{\fill}
        \begin{subfigure}[t]{0.20\linewidth}
            \centering
            \includegraphics[width=\linewidth]{CHBR/algorithms/Schema12/\twelvethname.pdf}
            \mycaptiontitle{\twelvethsubcap} 
        \end{subfigure} \hfill\null
    }
    \mycaptiontitle{Algorithms observed for schema S12.}
    \label{fig:schema12}
\end{figure*}


\section{Algorithmic and interaction strategies by schema}
\label{app:unplugged_strategies}

This section presents an overview of the algorithmic and interaction strategies developed by students for each schema of the unplugged CAT. 
By analysing the approaches, we aim to illustrate how students of different age groups engaged with the task and adapted their problem-solving methods. 
This detailed examination reveals the diversity of strategies employed and highlights patterns specific to each schema, offering insights into the cognitive processes underlying AT.

\def\performance{{S1}, {S2}, {S3}, {S4}, {S5}, {S6}, {S7}, {S8}, {S9}, {S10}, {S11}, {S12}}

\foreach \name in \performance {
	\begin{figure*}[!ht]				
		\centering
		\includegraphics[width=\linewidth]{CHBR/performance_by_age_category_and_schema_\name.pdf}
		\mycaptiontitle{Algorithmic and interaction strategies across age for \name.}
		\label{fig:performance\name}
	\end{figure*}
}
	
	\chapter{Pilot study with the virtual CAT} \label{appendix:pilot}

This appendix presents the details of the pilot study conducted with the virtual CAT, in particular we included: (i) the final application's user interface; (ii) illustrations of the performance for each schema, specifically showing the development of algorithmic and interaction strategies.

\section{{Screens of the final application}}\label{appendix:appA_pilot}

This section provides screenshots of the final application from the pilot study, including all key stages of the virtual CAT assessment. The images cover the initial language selection for the test, the choice between training or validation modes, data entry screens, and the main testing interface with its three different interaction modes. Additionally, it includes visuals of the final results dashboard and the survey screen, where participants provided feedback on the application and its activities. These screenshots offer a detailed overview of the user experience, highlighting the flow and layout of the virtual environment.

\begin{figure}[htb]
    \centering
    \includegraphics[width=.85\textwidth]{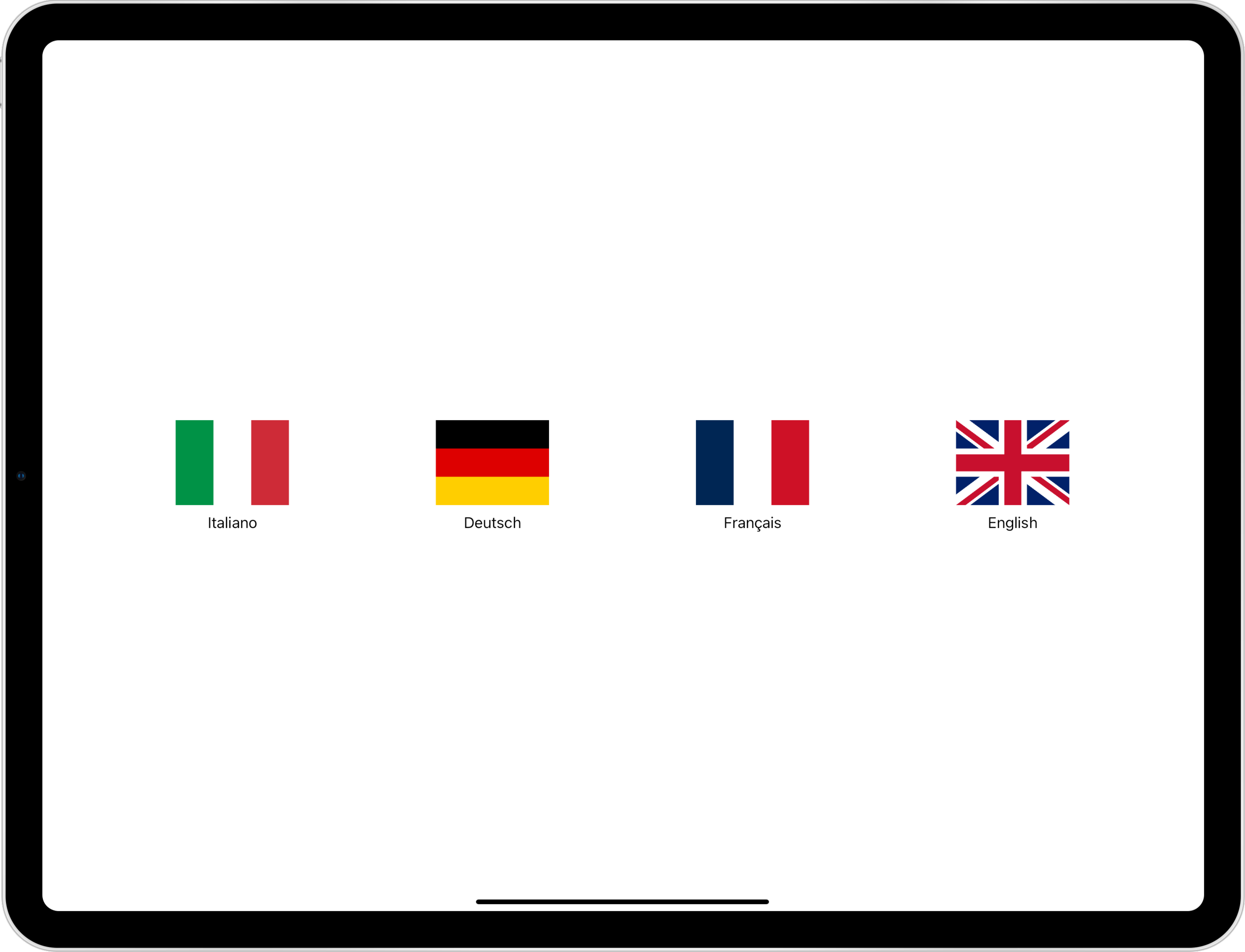}
        \mycaptiontitle{{Language selection.}}
        \label{fig:language}
\end{figure}

\begin{figure}[htb]
    \centering
    \includegraphics[width=.85\textwidth]{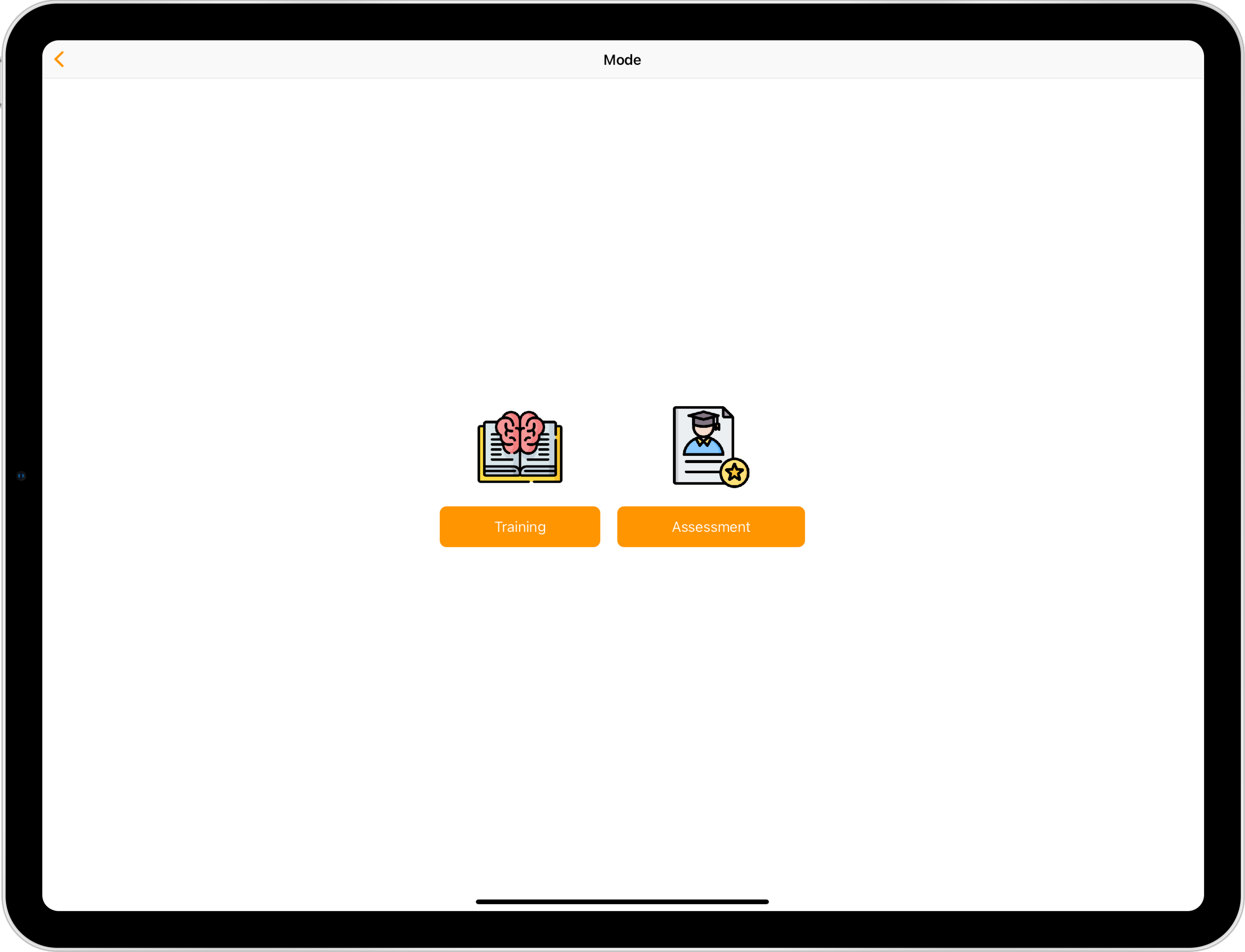}
        \mycaptiontitle{{Module selection.}}
        \label{fig:mode}
\end{figure}

\begin{figure}[htb]
    \centering
    \includegraphics[width=.85\textwidth]{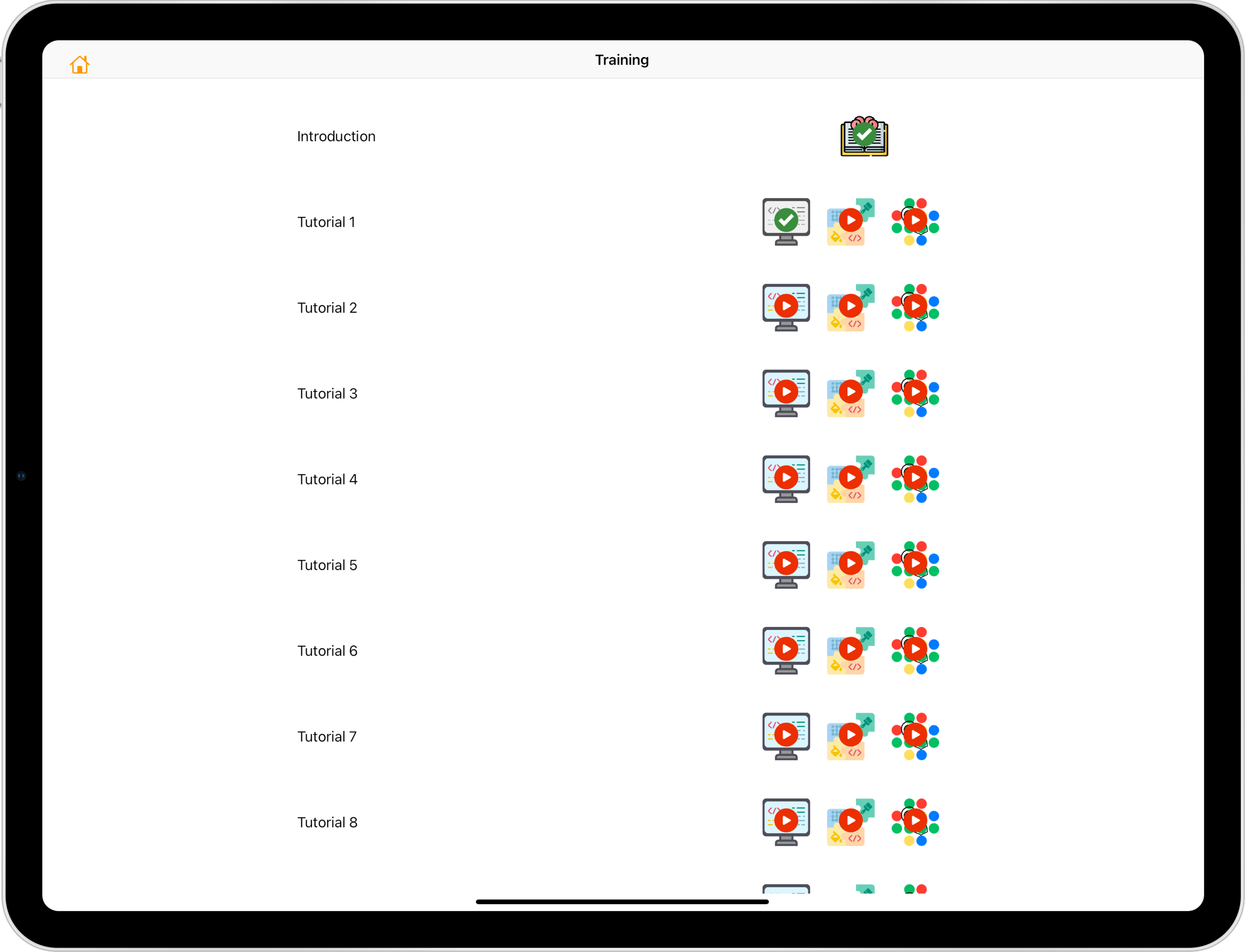}
        \mycaption{Training module.}{An introductory video about the application is provided on the training screen, followed by a series of explanatory videos for all practice tasks in each interface. 
        After watching the video, users can attempt to solve the schema using the provided instructions. 
        When a schema is successfully solved, the video icon is marked with a green checkmark.}
        \label{fig:tutorial}
\end{figure}

\begin{figure}[htb]
    \centering
    \includegraphics[width=.85\textwidth]{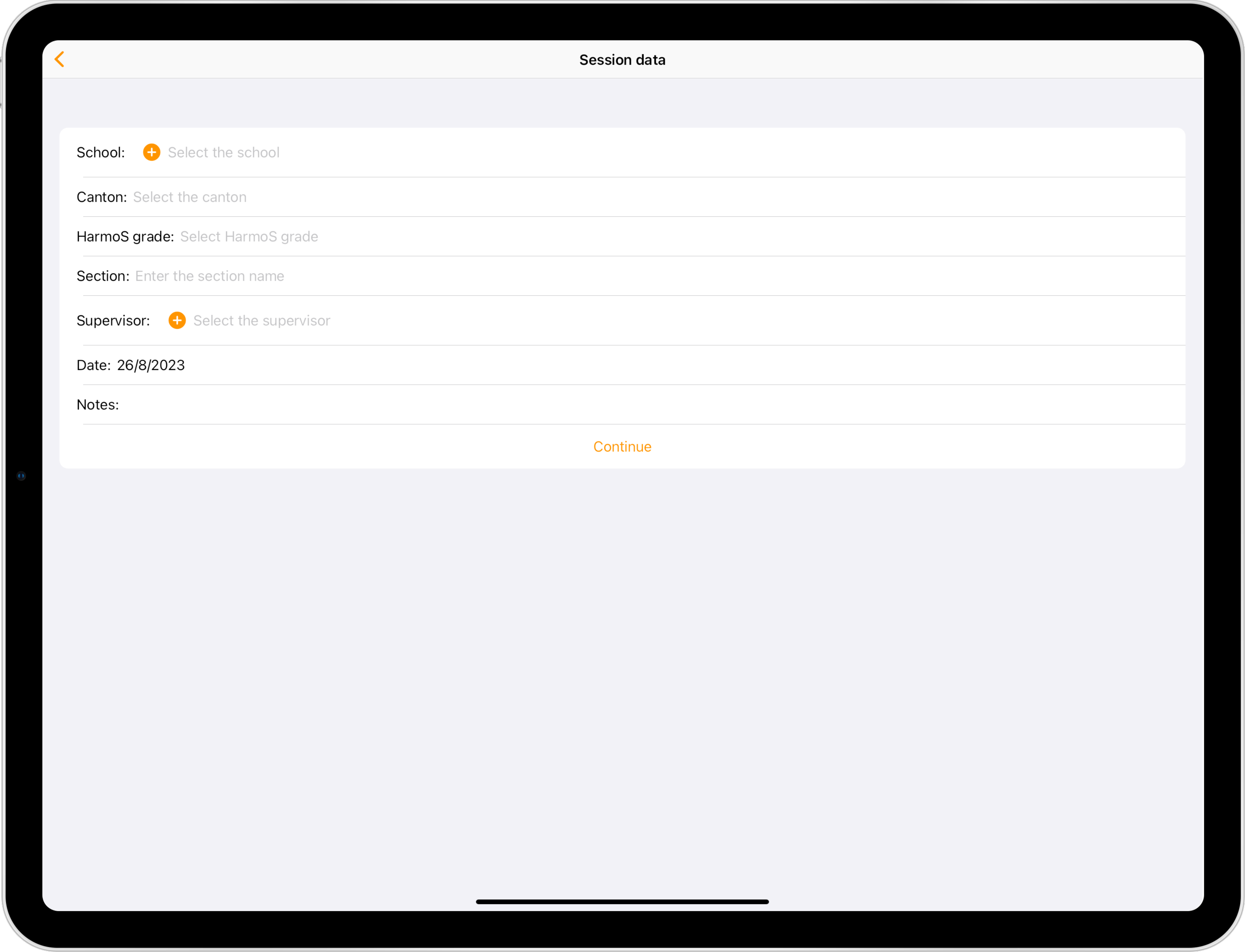}
        \mycaptiontitle{{Session form in the validation module.}}
        \label{fig:session}
\end{figure}

\begin{figure}[htb]
    \centering
    \includegraphics[width=.85\textwidth]{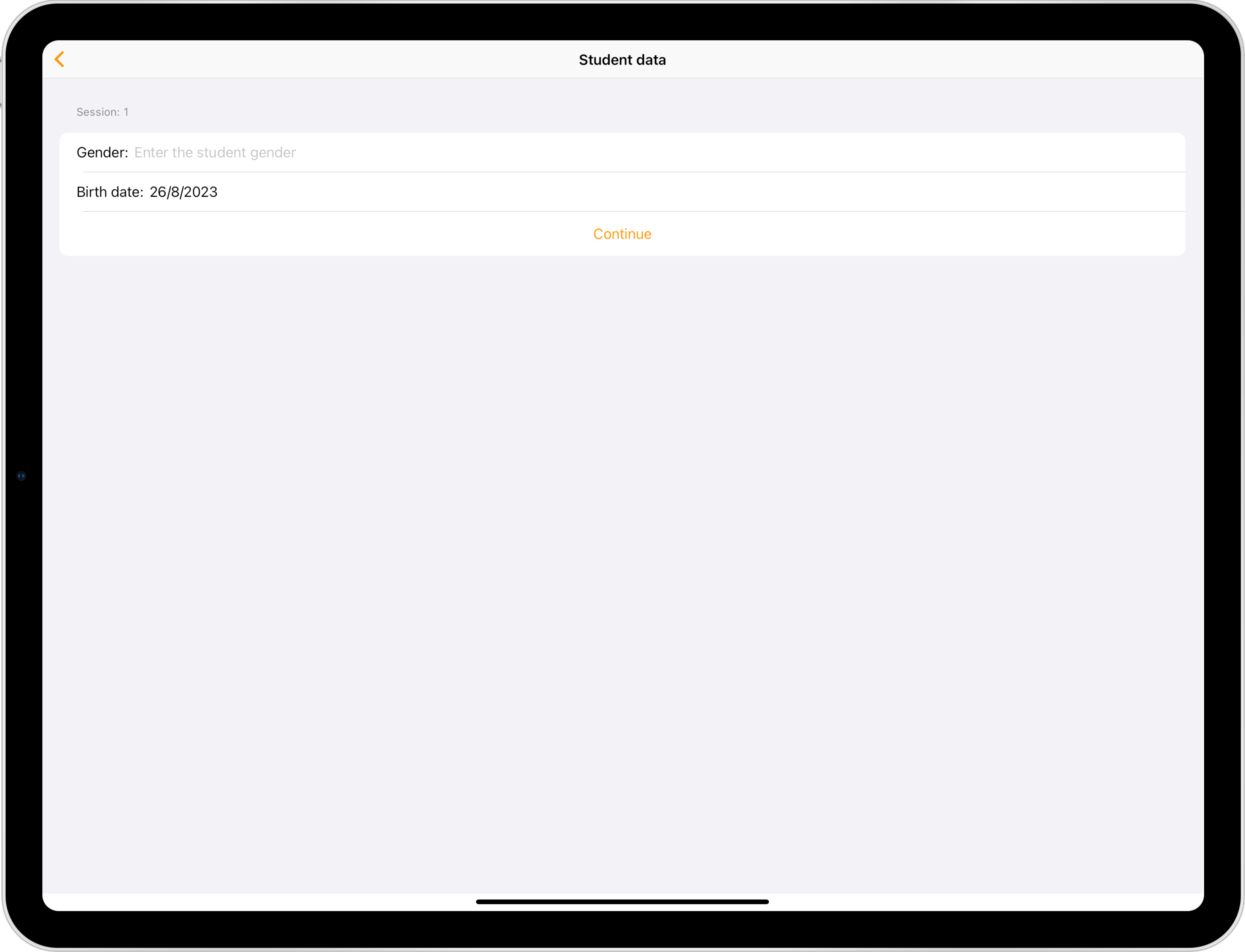}
        \mycaptiontitle{{Student form in the validation module.}}
        \label{fig:student}
\end{figure}

\begin{figure}[htb]
    \centering
    \includegraphics[width=.85\textwidth]{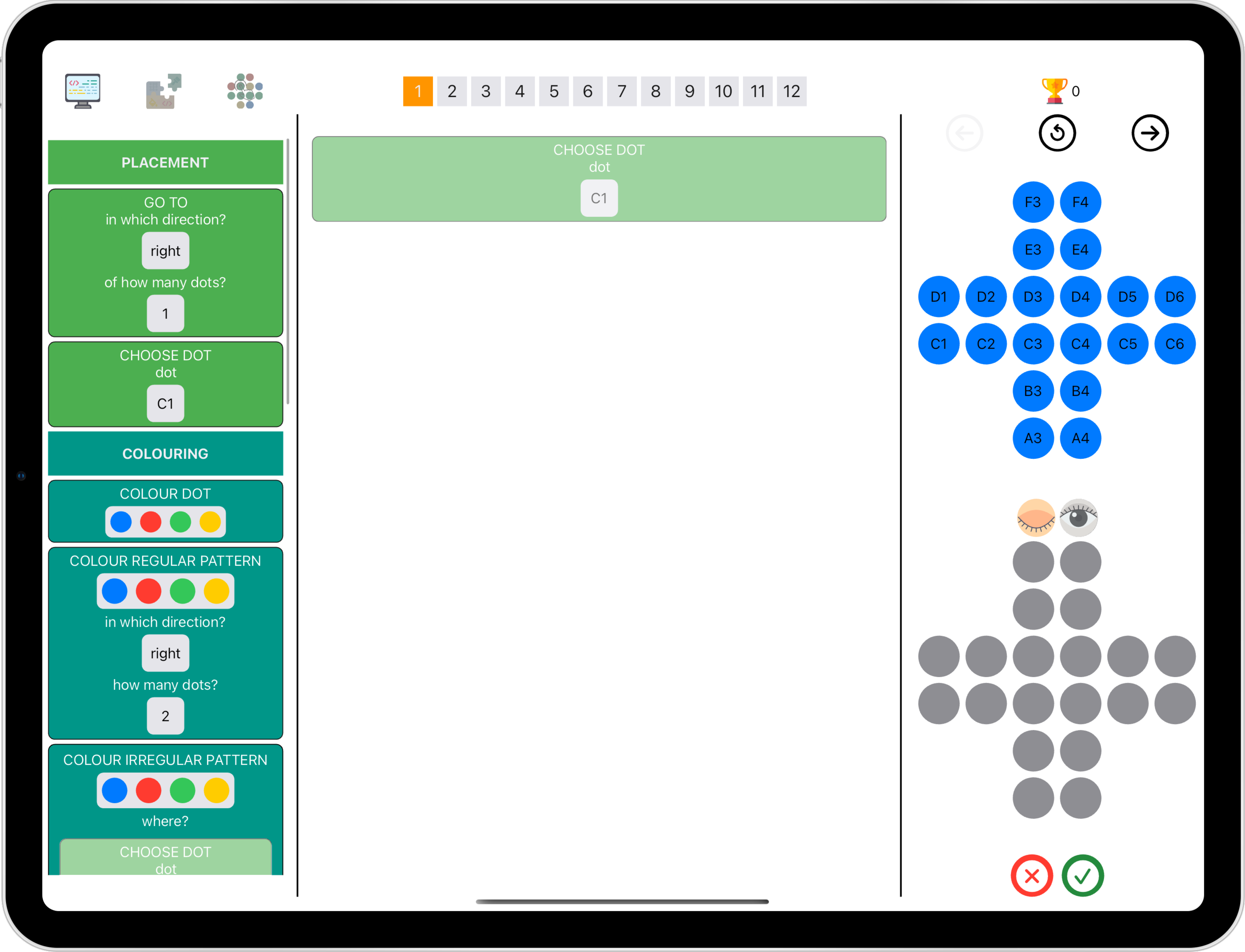}
    \mycaptiontitle{{CAT visual programming interface (CAT-VPI) with textual commands.} }
    \label{fig:programming_final}
\end{figure}

\begin{figure}[htb]
    \centering
    \includegraphics[width=.85\textwidth]{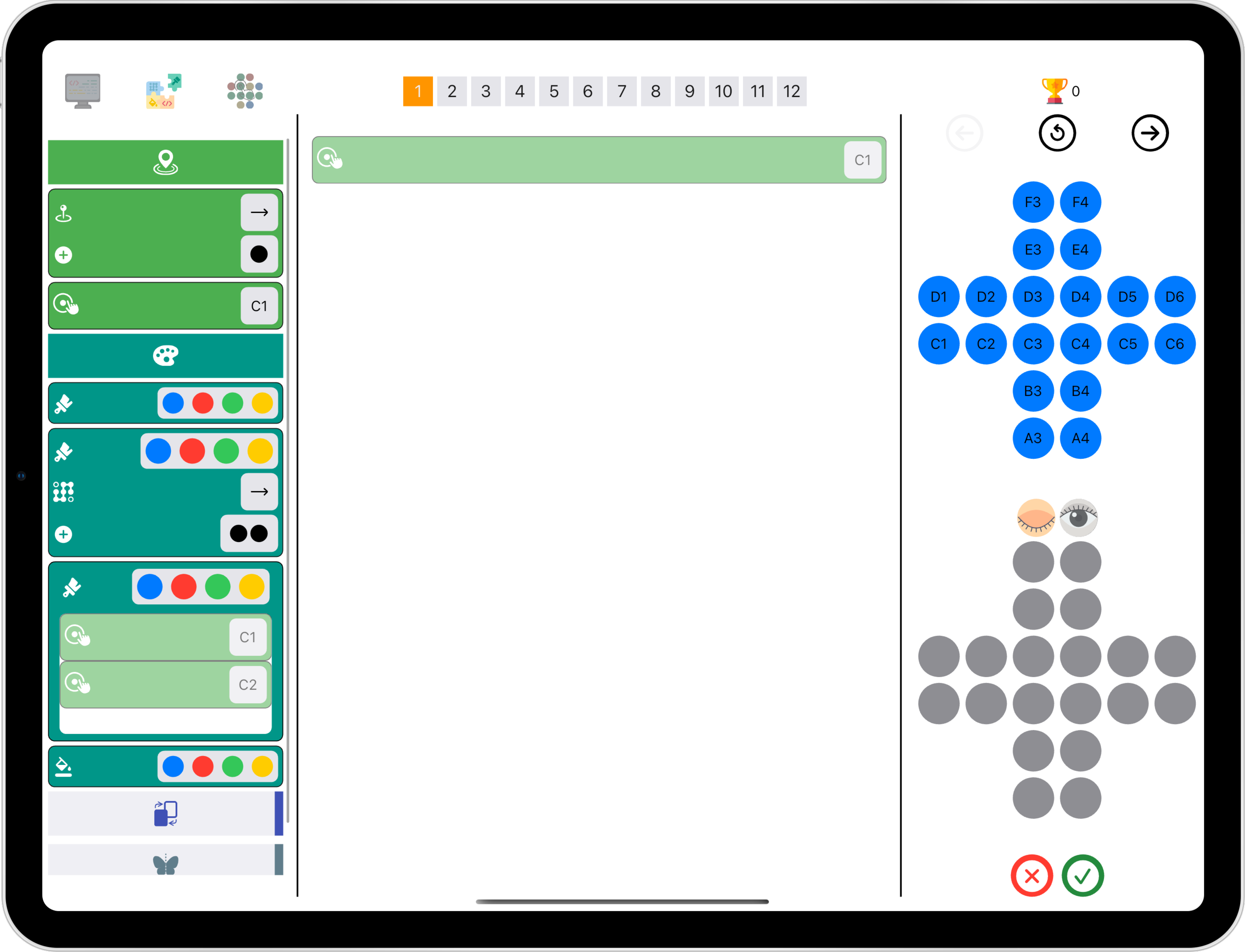}
    \mycaptiontitle{{CAT visual programming interface (CAT-VPI) with symbolic commands.} }
    \label{fig:programming_symbols_final}
\end{figure}

\begin{figure}[htb]
    \centering
    \includegraphics[width=.85\textwidth]{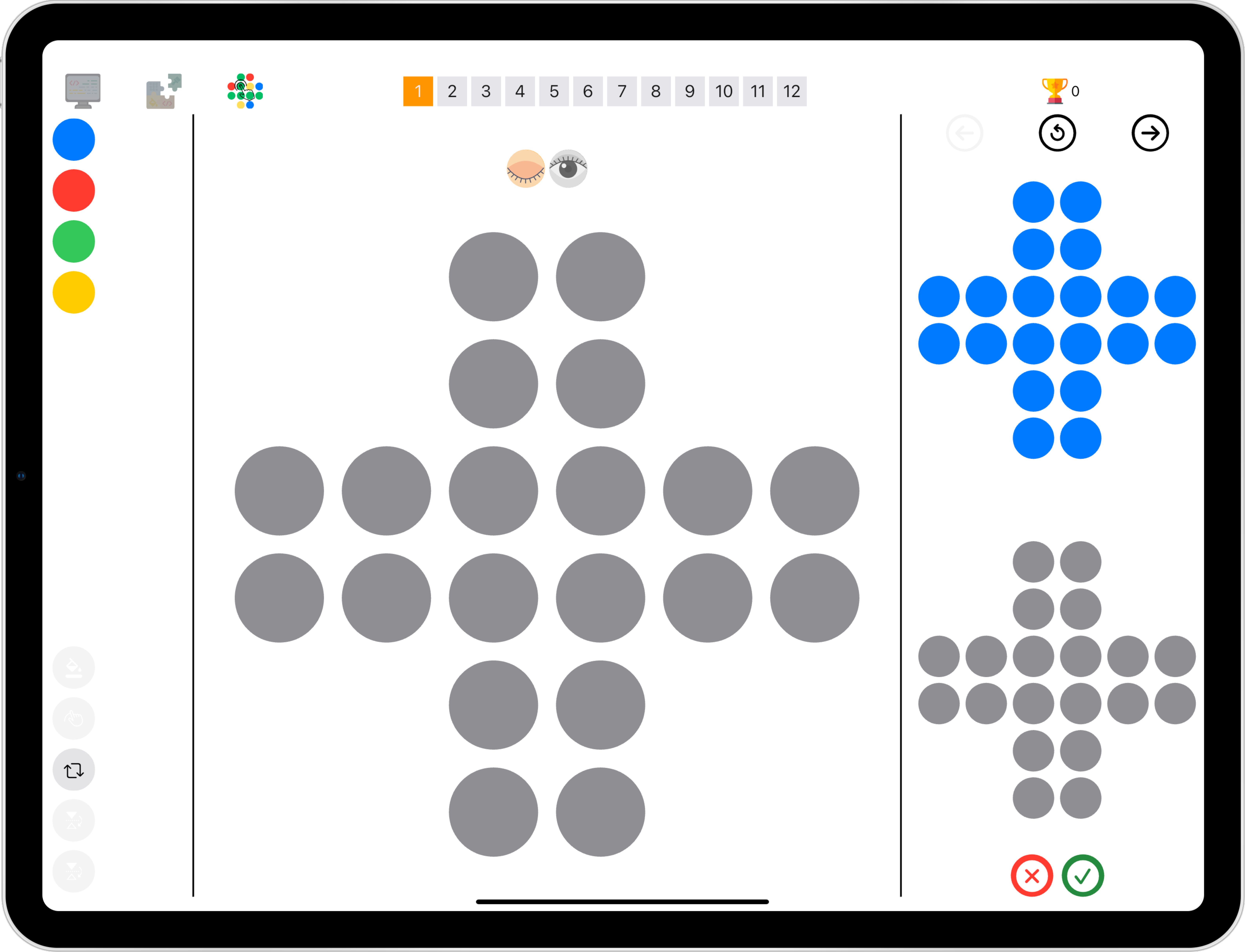}
        \mycaptiontitle{{CAT gesture interface (CAT-GI).}}
        \label{fig:gestures_final}
\end{figure}

\begin{figure}[htb]
    \centering
    \includegraphics[width=.85\textwidth]{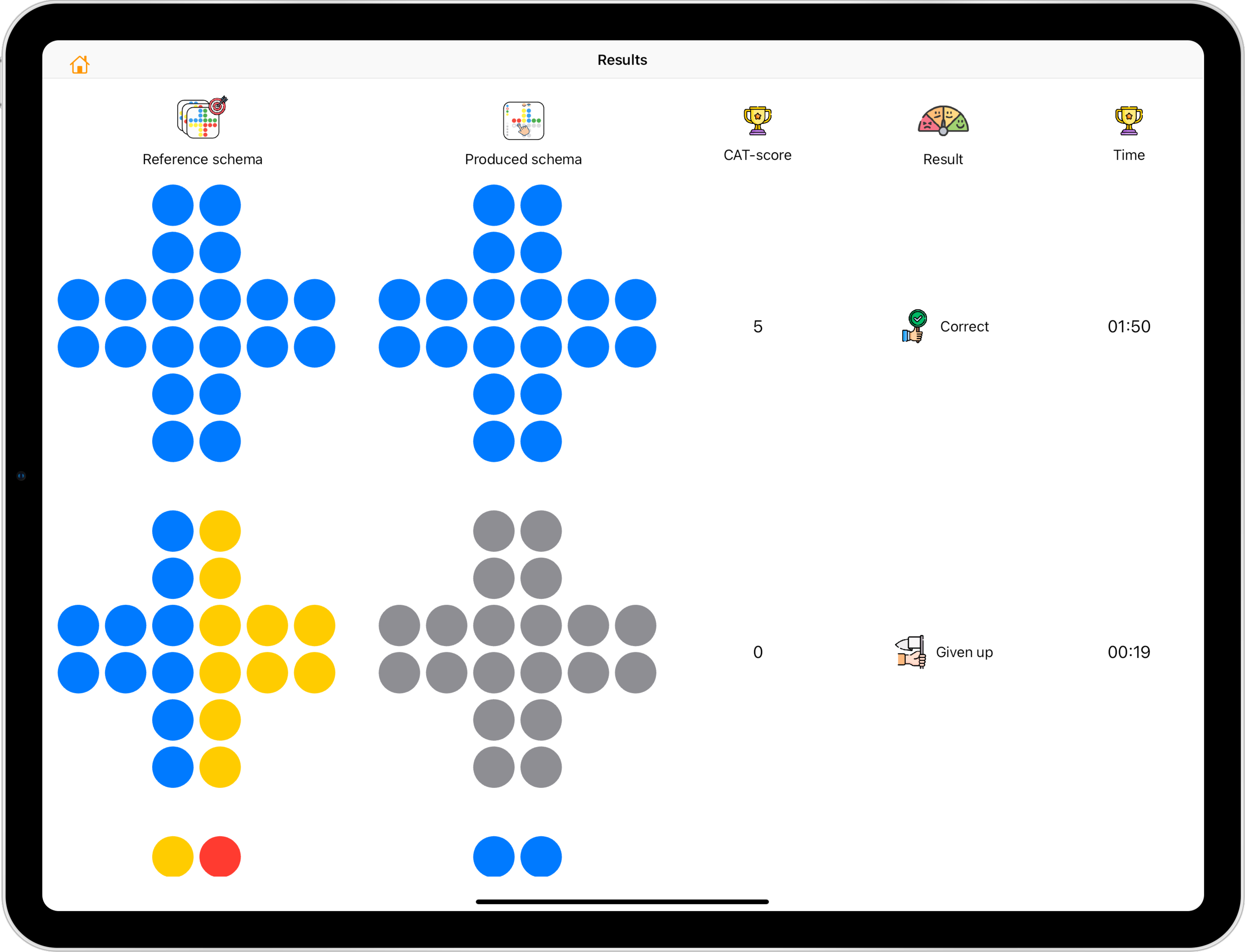}
        \mycaption{Results dashboard.}{It comprehensively summarises pupils' performance across all schemas. 
        This dashboard includes a visual representation of reference schemas alongside those resulting from student instructions, the pupil's score, an indication of whether each schema was completed correctly, incorrectly, or skipped, and the time taken to complete the schema.}
        \label{fig:dashboard}
\end{figure}

\begin{figure}[htb]
    \centering
    \includegraphics[width=.85\textwidth]{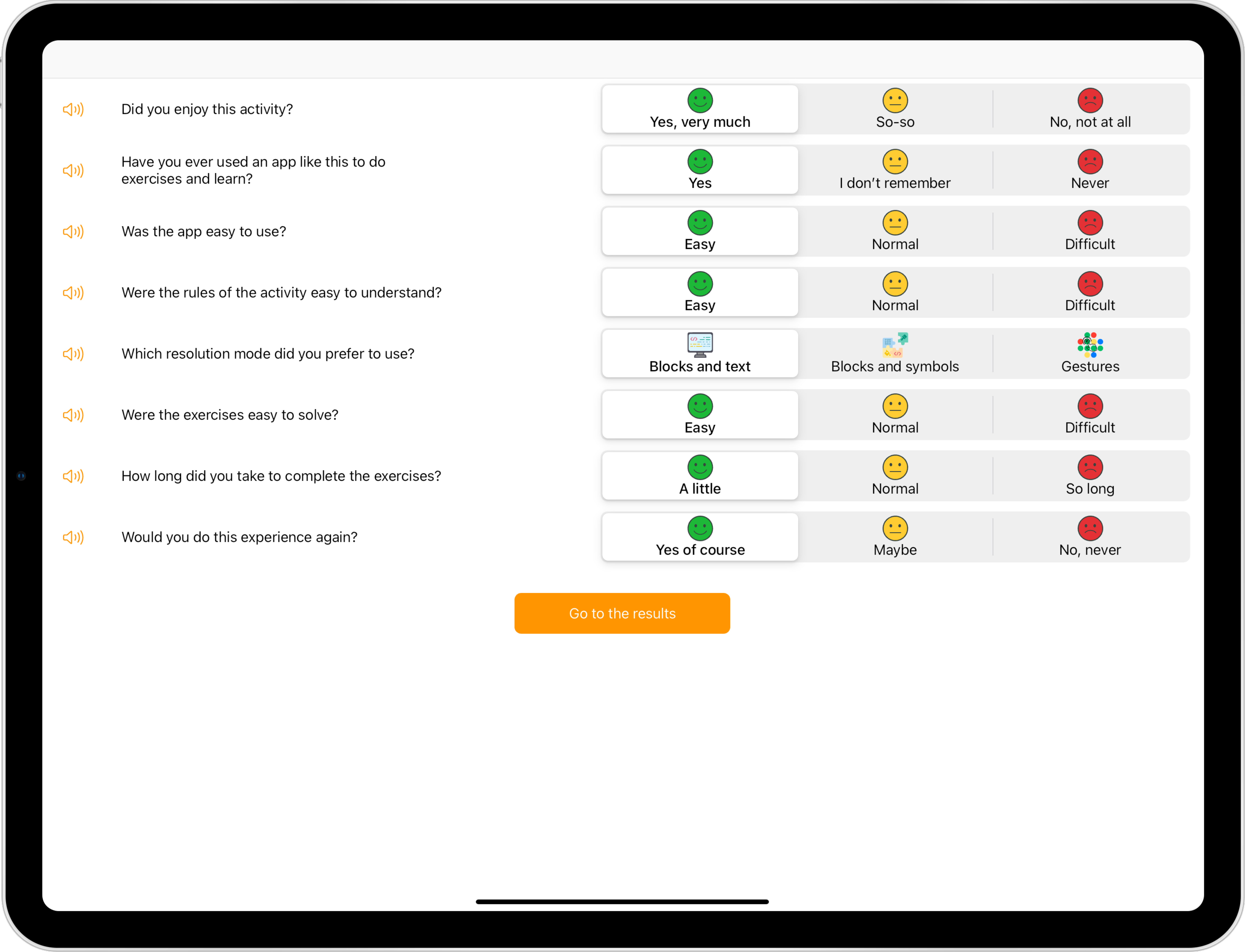}
        \mycaption{Pupil feedback survey.}{The voice-assisted questions 
        evaluate user interactions with the app. Each question is accompanied by three distinct emoticon-style response options: a contented smiling face, a neutral face, and a discontented frowning face. 
        A concluding button invites users to view aggregated results.}
        \label{fig:survey}
\end{figure}

\section{{Algorithmic and interaction strategies by schema}}\label{appendix:appB_pilot}

This section presents and overview of the algorithmic and interaction strategies developed bu students for each schema of the virtual CAT. 
By analysing the approaches, we aim to illustrate how students of different age groups engaged with the task and adapted their problem-solving methods. This detailed examination reveals the diversity of strategies employed and highlights patterns specific to each schema, offering insights into the cognitive processes underlying AT.

\begin{figure}[H]
\centering
    \includegraphics[width=\linewidth]{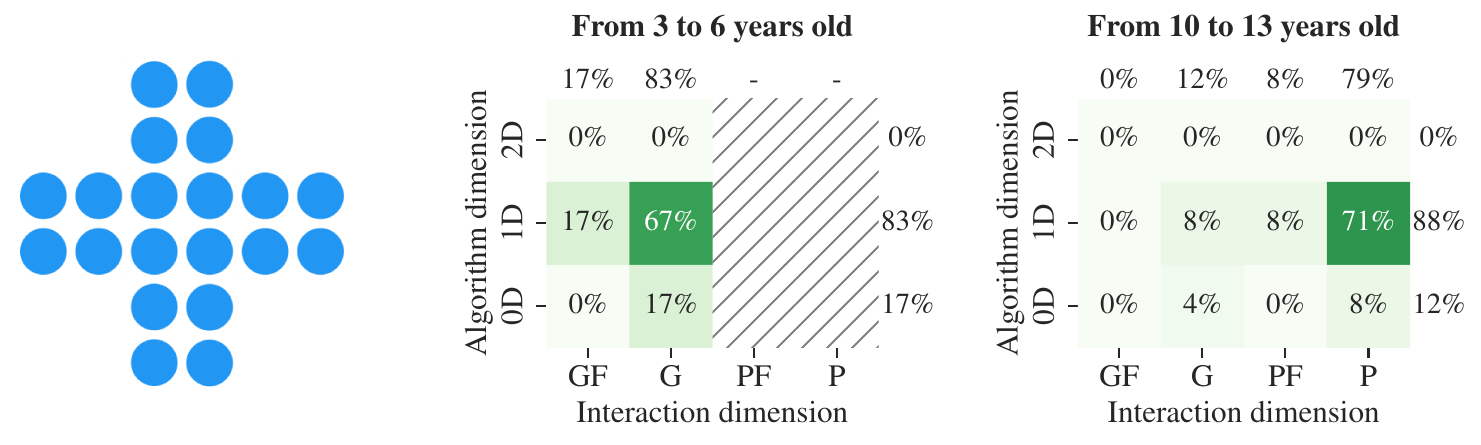}
    \mycaptiontitle{Algorithmic and interaction strategies across age for S1.}
    \label{fig:s1}
\end{figure}

\begin{figure}[H]
    \centering
    \includegraphics[width=\linewidth]{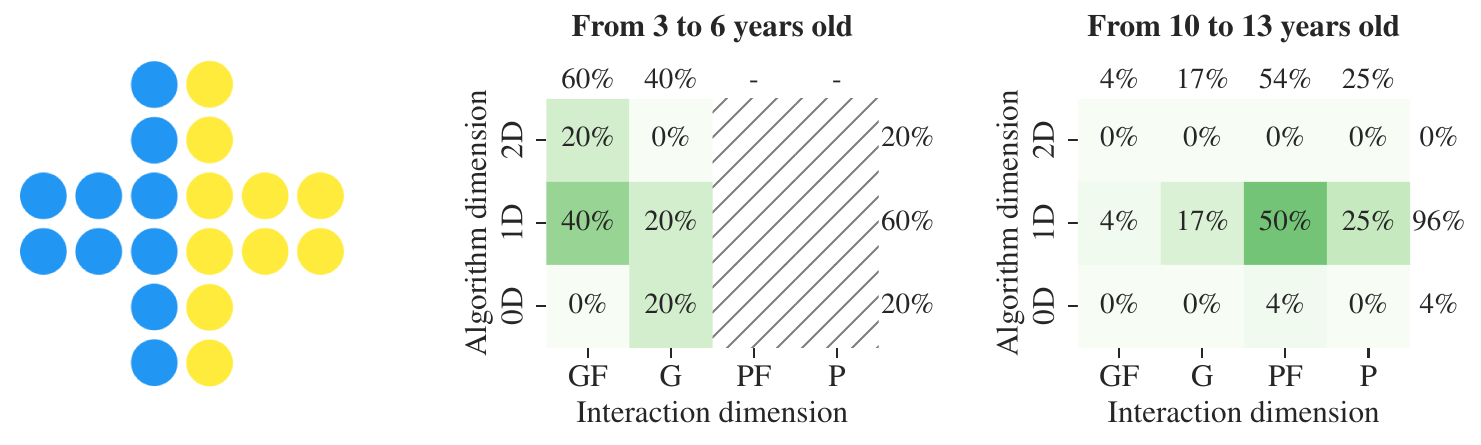}
    \mycaptiontitle{Algorithmic and interaction strategies across age for S2.}
    \label{fig:s2}
\end{figure}

\begin{figure}[H]
    \centering
    \includegraphics[width=\linewidth]{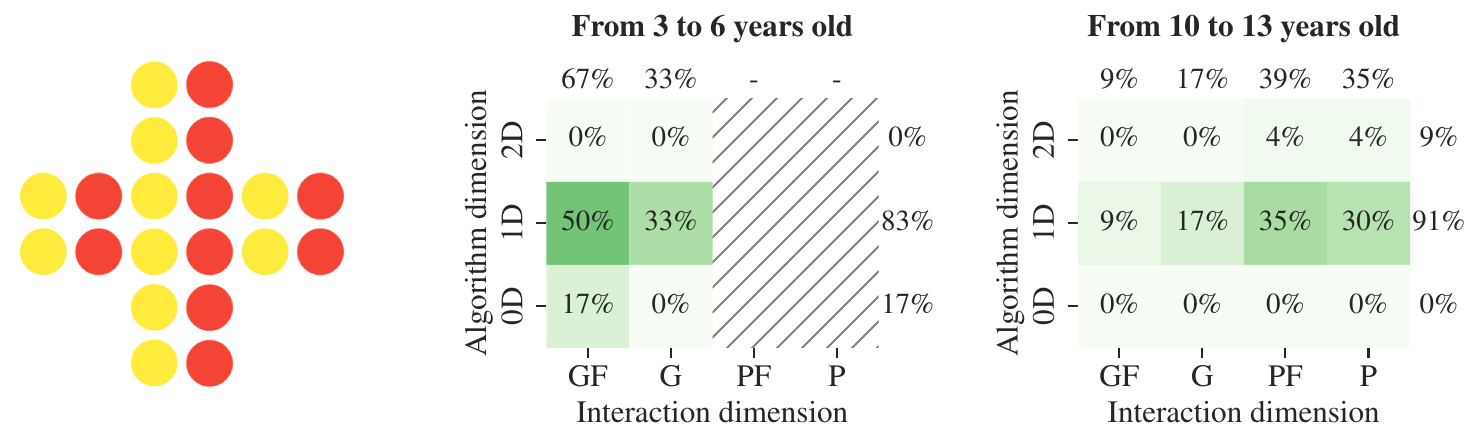}
    \mycaptiontitle{Algorithmic and interaction strategies across age for S3.}
    \label{fig:s3}
\end{figure}

\begin{figure}[H]
    \centering
    \includegraphics[width=\linewidth]{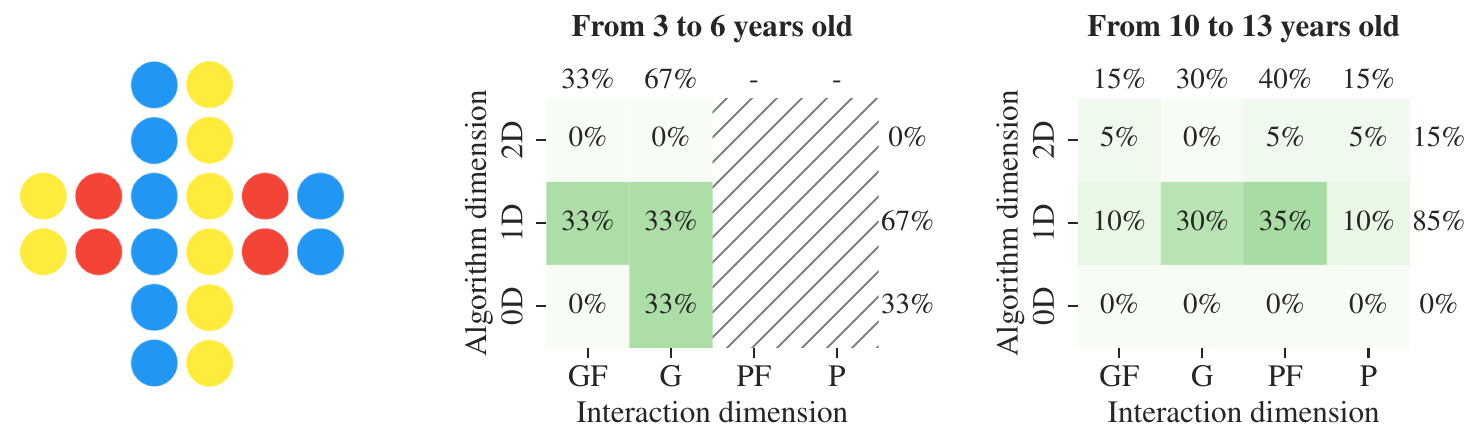}
    \mycaptiontitle{Algorithmic and interaction strategies across age for S4.}
    \label{fig:s4}
\end{figure}

\begin{figure}[H]
    \centering
    \includegraphics[width=\linewidth]{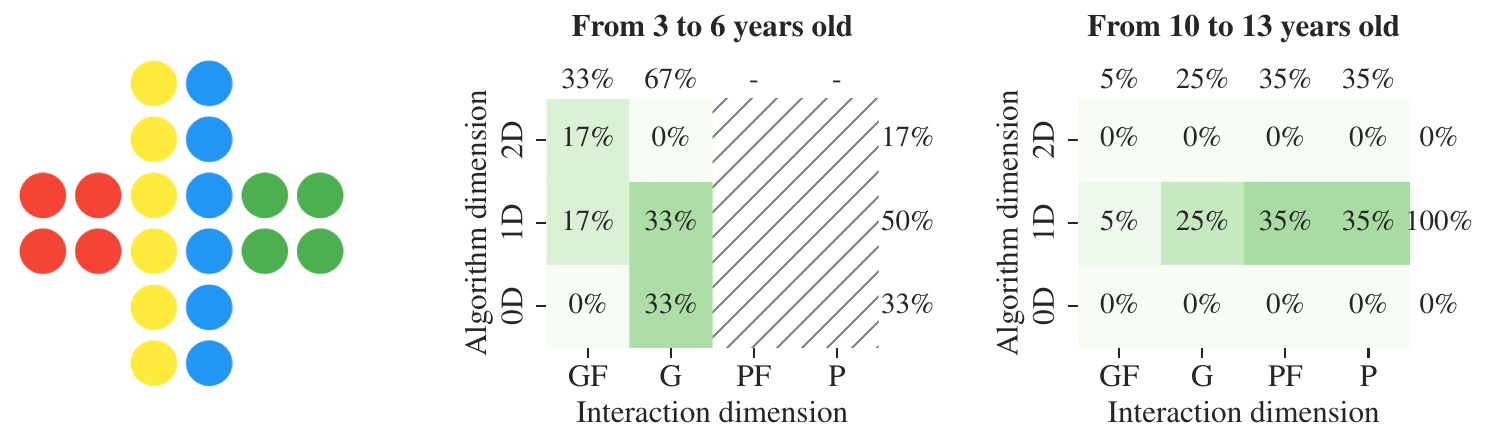}
    \mycaptiontitle{Algorithmic and interaction strategies across age for S5.}
    \label{fig:s5}
\end{figure}

\begin{figure}[H]
    \centering
    \includegraphics[width=\linewidth]{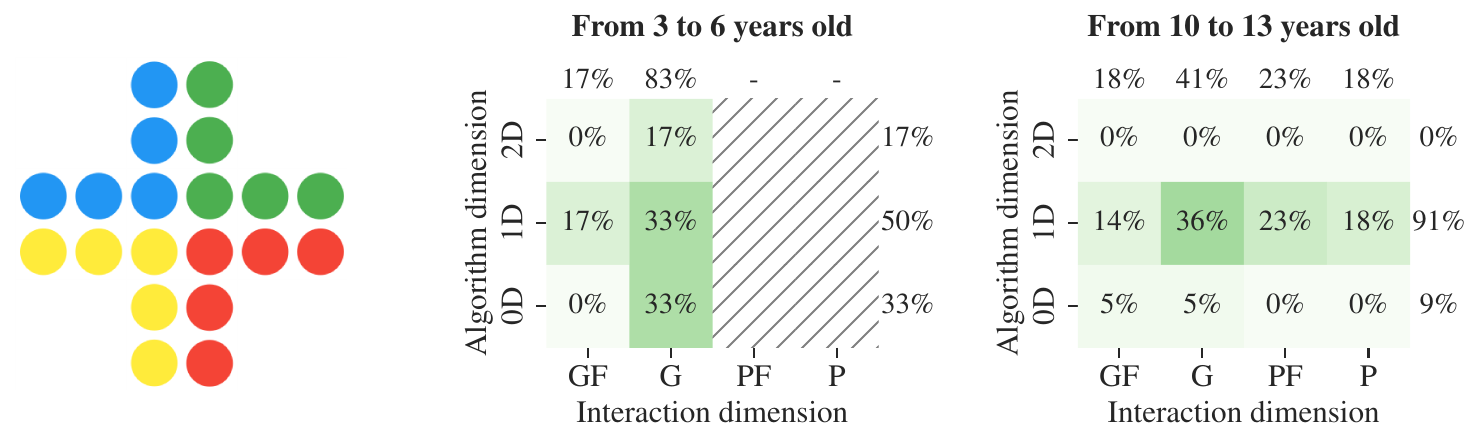}
    \mycaptiontitle{Algorithmic and interaction strategies across age for S6.}
    \label{fig:s6}
\end{figure}

\begin{figure}[H]
    \centering
    \includegraphics[width=\linewidth]{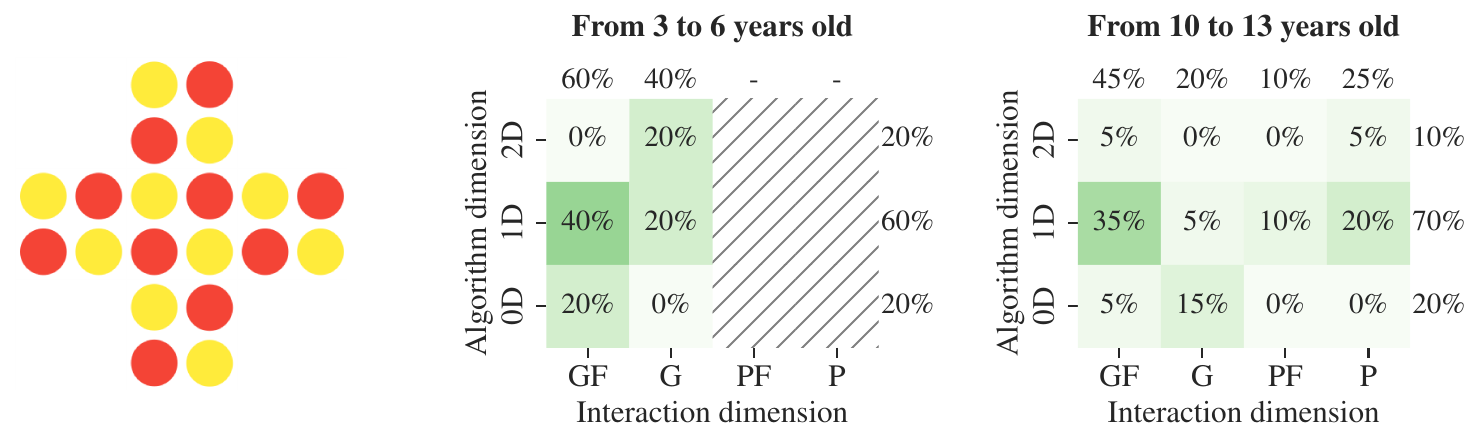}
    \mycaptiontitle{Algorithmic and interaction strategies across age for S7.}
    \label{fig:s7}
\end{figure}

\begin{figure}[H]
    \centering
    \includegraphics[width=\linewidth]{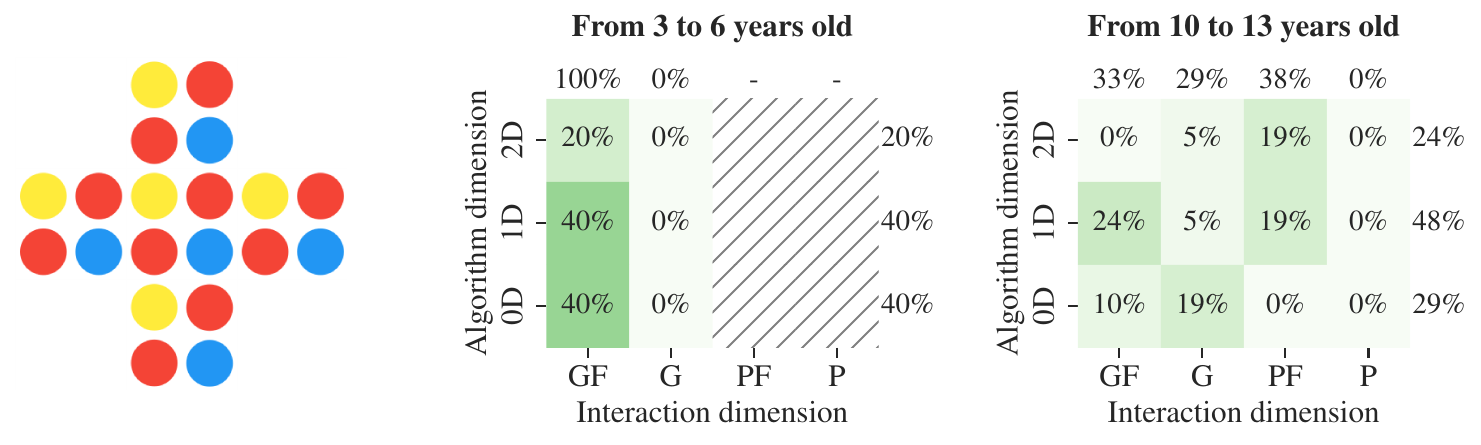}
    \mycaptiontitle{Algorithmic and interaction strategies across age for S8.}
    \label{fig:s8}
\end{figure}

\begin{figure}[H]
    \centering
    \includegraphics[width=\linewidth]{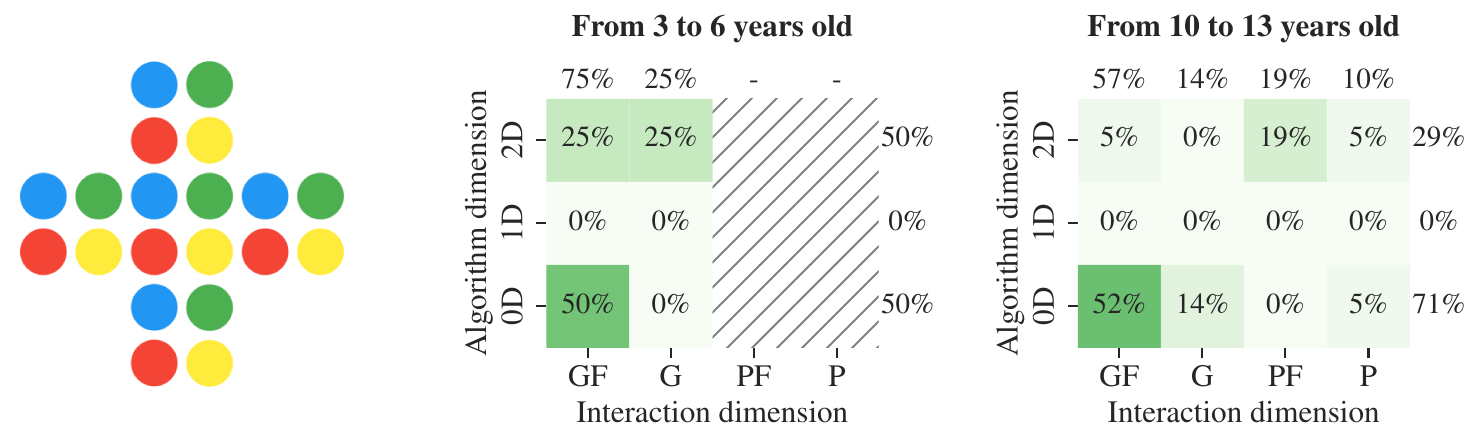}
    \mycaptiontitle{Algorithmic and interaction strategies across age for S9.}
    \label{fig:s9}
\end{figure}

\begin{figure}[H]
    \centering
    \includegraphics[width=\linewidth]{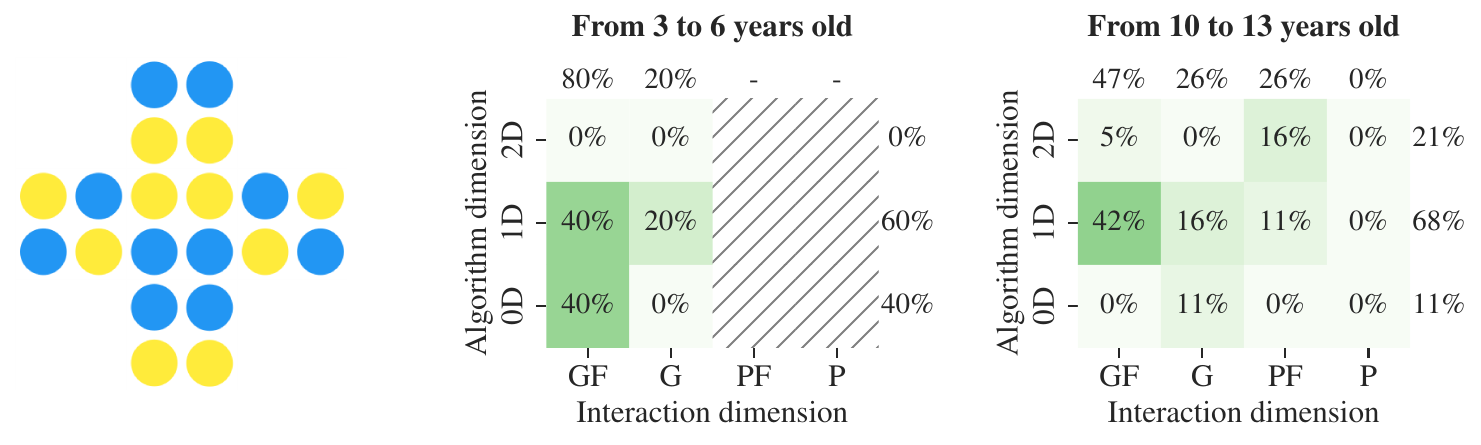}
    \mycaptiontitle{Algorithmic and interaction strategies across age for S10.}
    \label{fig:s10}
\end{figure}

\begin{figure}[H]
    \centering
    \includegraphics[width=\linewidth]{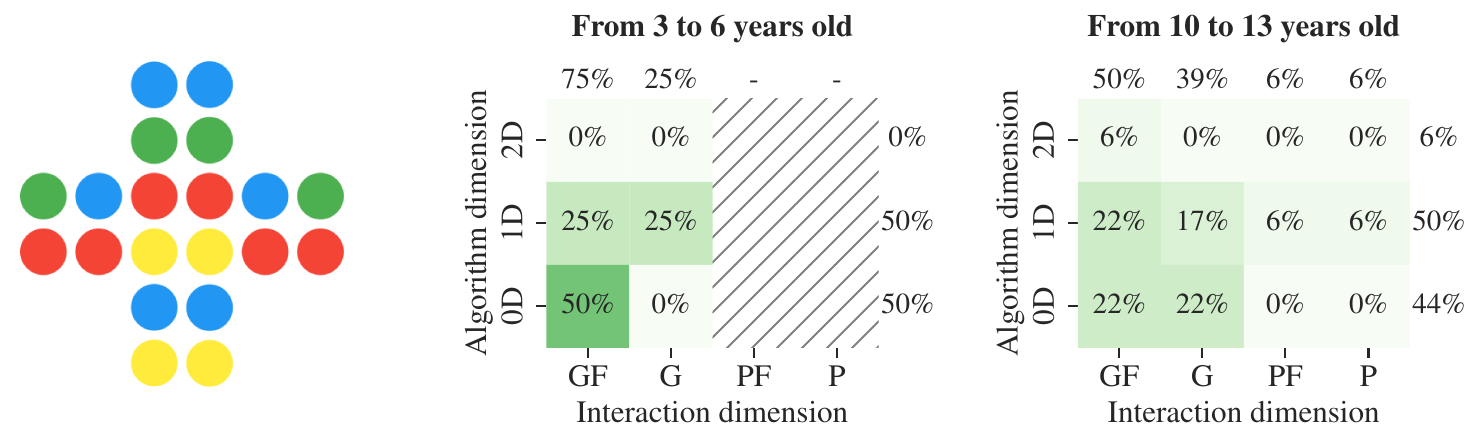}
    \mycaptiontitle{Algorithmic and interaction strategies across age for S11.}
    \label{fig:s11}
\end{figure}

\begin{figure}[H]
    \centering
    \includegraphics[width=\linewidth]{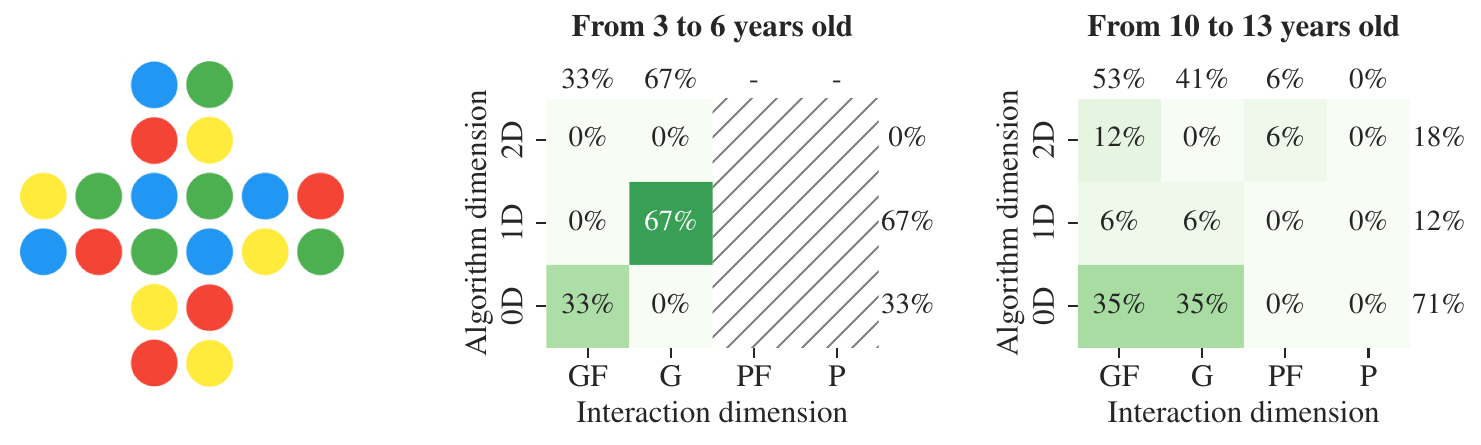}
    \mycaptiontitle{Algorithmic and interaction strategies across age for S12.}
    \label{fig:s12}
\end{figure}


		\backmatter
	\clearpage
	\bibliographystyle{apalike} 
	\bibliography{fullbibliography.bib}

	\cleardoublepage
	
	\phantomsection
	\addcontentsline{toc}{chapter}{Research Contributions}
	\includepdf[pages=-]{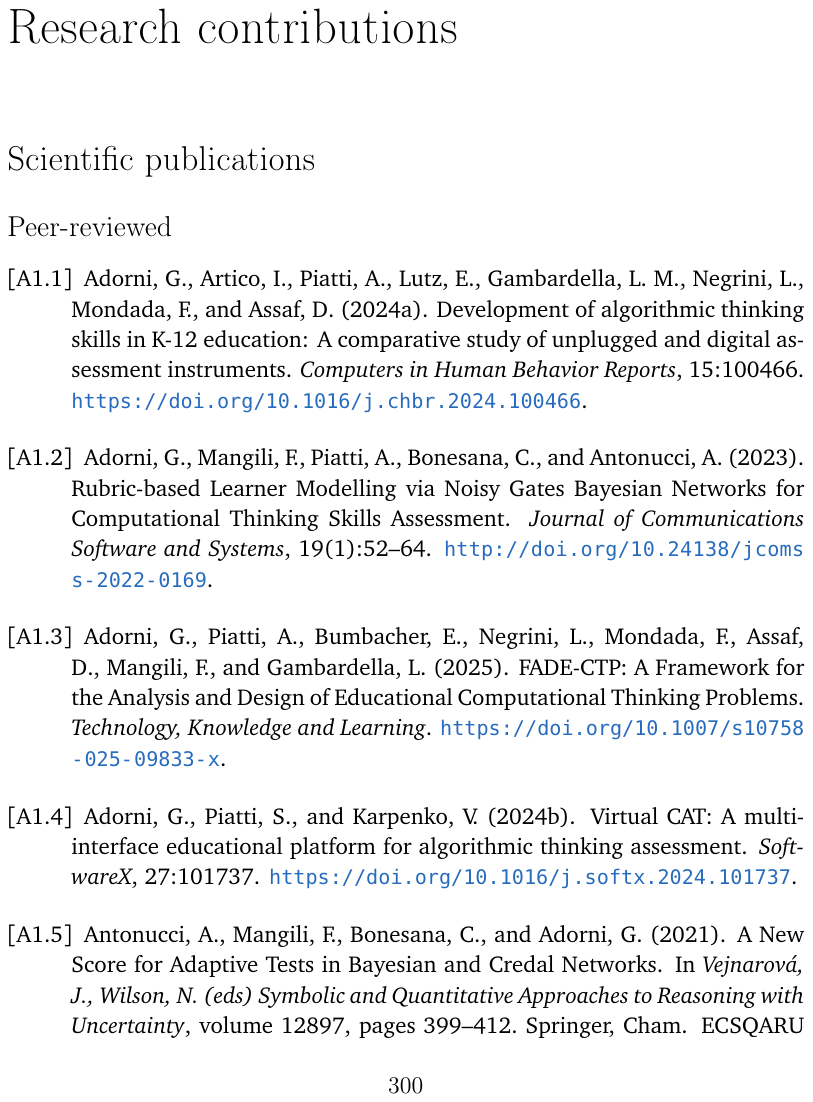}
	


\end{document}